%% file: Thesis.tex
\documentclass[11pt,twoside,a4paper]{book}
\usepackage{amsfonts}
\usepackage{amsmath}
\usepackage{amssymb}
\usepackage{amstext}
\usepackage{appendix}
\usepackage{axodraw4j}
\usepackage{bm}
\usepackage{color}
\usepackage{dsfont}
\usepackage{epsfig}
\usepackage{hyperref}
\usepackage{lineno}
\usepackage{longtable}
\usepackage{multirow}
\usepackage[numbers,sort&compress]{natbib}
\usepackage{pstricks}
\usepackage{sfmath}
\usepackage{slashed}
\usepackage{subfigure}
\usepackage{todonotes}
\usepackage{url}

\usepackage[bindingoffset=1cm,textheight=22cm,textwidth=15cm,hcentering]{geometry}

\usepackage{fancyhdr}
\fancypagestyle{plain}{\fancyhf{} \fancyfoot[LE,RO]{\thepage}  }
\usepackage{titlesec}
\titleformat{\chapter}[display]
{\normalfont\Large\filcenter\sffamily}
{\vspace*{-50pt} \titlerule \vspace{1pc} \LARGE\MakeUppercase{\chaptertitlename} \thechapter}
{1pc}
{\titlerule \vspace{2pc} \huge}

\usepackage{ccaption}			
\captionnamefont{\bfseries}		
\captiontitlefont{\small\sffamily}	

\title{Extra doublets. Global parameter fits of Standard Model extensions in the fermionic or scalar sector}
\author{Otto Eberhardt}

\newcommand*\oline[1]{
  \hspace*{0.2em}
  \vbox{
    \kern-0.35ex
    \hrule height 0.4pt
    \kern0.35ex
    \hbox{
      \kern-0.5em
      \ifmmode#1\else\ensuremath{#1}\fi
      \kern-0.0em
    }
  }
}
\newcommand{\otto}{\Black}
\newcommand{\ottoo}{\Black}
\newcommand{\ottooo}{\Black}
\newcommand{\oto}{\Black}
\begin{document}

\allowdisplaybreaks

\include{titlepage}

\tableofcontents

\include{introduction}
\include{statistics}
\include{sm}
\include{sm4}
\include{2hdm}
\include{conclusions}
\include{inputs}
\include{fitresults}
\include{2hdmrelations}
\include{parallelizer}

\listoffigures

\bibliographystyle{kp}
\bibliography{bibliography}
\clearpage \thispagestyle{plain}
\include{acknowledgements}

\end{document}

%% file: titlepage.tex
\thispagestyle{empty}
\begin{picture}(190,5)(0,0)
 \put(285,-25){\includegraphics[width=130pt]{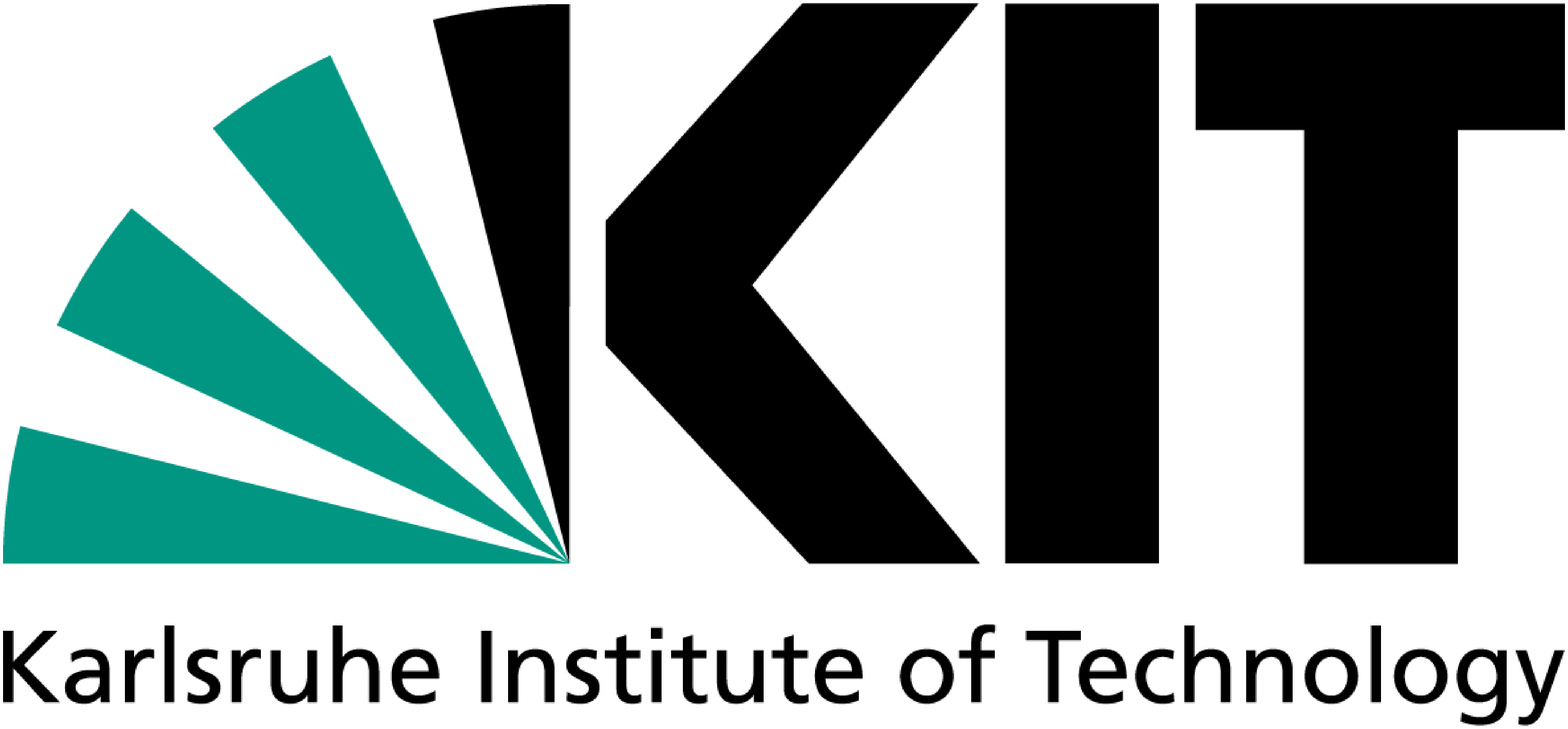}}
\end{picture}
\begin{flushleft}
\vspace*{-40pt}
\hspace*{-15pt}Institut f{\"u}r Theoretische Teilchenphysik\\[5pt]
\hspace*{-15pt}Fakult{\"a}t f{\"u}r Physik\\[5pt]
\hspace*{-15pt}Karlsruher Institut f{\"u}r Technologie\\
\end{flushleft}
\begin{center}
\vspace*{80pt}
\resizebox{250pt}{!}{\textbf{Extra doublets}}\\[20pt]
{\LARGE Global analyses of Standard Model extensions\\[8pt]
in the fermionic or scalar sector\\[140pt]}
Zur Erlangung des akademischen Grades eines\\[5pt]
Doktors der Naturwissenschaften\\[5pt]
von der Fakult{\"a}t f{\"u}r Physik des Karlsruher Instituts f{\"u}r Technologie genehmigte\\[20pt]
\textsc{Dissertation}\\[17pt]
von\\[10pt]

\begin{Large}Dipl.-Phys. \textbf{Otto Eberhardt}\end{Large}\\[10pt]
aus M{\"u}nchen\\[50pt]
\end{center}
Tag der m{\"u}ndlichen Pr{\"u}fung: 12. Juli 2013\\[15pt]
Referent: Prof. Dr. Ulrich Nierste\\[5pt]
Korreferent: Prof. Dr. Dieter Zeppenfeld
\newpage
\clearpage \thispagestyle{empty} \mbox{}
\newpage

\clearpage \thispagestyle{empty} \vspace*{250pt} \hspace*{140pt}\parbox{300pt}{{\Large Des is wia bei jeda Wissenschaft,}\\[5pt] {\Large am Schluss stellt sich dann heraus,}\\[5pt] {\Large dass alles ganz anders war.}\\[15pt] \hspace*{127pt}{\Large (Karl Valentin)}}

\newpage
\clearpage \thispagestyle{empty} \mbox{}

%% file: introduction.tex
\chapter{Introduction}

\begin{figure}
 \centering
 \begin{picture}(190,160)(0,0)
  \put(-36,0){\includegraphics[width=260pt]{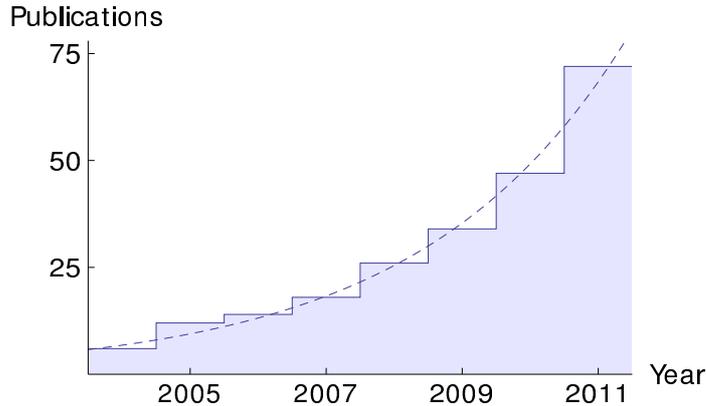}}
 \end{picture}
\caption[SM4 publications from 2004 to 2011.]{The yearly amount of publications about the SM4 between 2004 and 2011 can be described by an exponential function (dashed line).}
 \label{fig:pre2011}
\end{figure}

In the first half of the past century, progress in particle physics was driven by experimental results which physicists struggled to explain with one fundamental theory. In the second half, local gauge symmetries emerged as guiding principle to construct renormalisable relativistic quantum field theories, which finally led to the Standard Model of particle physics (SM). The SM describes three of the four fundamental interactions: the strong and weak nuclear force as well as electromagnetism. Only gravity, the weakest of the interactions, could not be embedded into the framework of a renormalisable quantum field theory up to now. The SM as formulated in the 1970s relies on a special mechanism for generating the masses of the elementary particles \cite{Higgs:1964pj,Englert:1964et,Guralnik:1964eu}.
However, this mechanism require\otto{s} at least one particle that had not been discovered \otto{until recently} \ottooo{\cite{Aad:2012tfa,Chatrchyan:2012ufa}}: the Higgs boson\otto{.}
Even though most experiments in the following decades featured an impressive consistency with \otto{the SM, it} remained incomplete without the detection of this postulated particle.\\
In 2011, when I started working on this PhD thesis, the detectors at the Large Hadron Collider (LHC) at CERN had started to take data, but there was no evidence of the Higgs boson so far. However, several results obtained at the Tevatron collider and other experiments seemed to hint at deviations from the SM, e.g.\ the measurement of the top quark forward-backward asymmetry, the like-sign dimuon charge asymmetry or the magnetic dipole moment of the muon, to mention some of them. Also \otto{cosmological observations} like dark matter, \otto{the} baryon asymmetry in the universe \otto{or} dark energy cannot be explained within the framework of the SM. Furthermore, the SM does not offer solutions to theoretical issues like the hierarchy problem, the unification of the gauge couplings or the origin of the fermion mass hierarchy.
On the search for alternatives, many new models have been discussed. They need to have the SM as \otto{an} effective ``low energy'' limit in order to describe experimental findings. On the other hand a new model has to solve one or more of the shortcomings of the SM. Most of these SM extensions involve additional particles, which have large masses or small couplings to the SM particles to avoid effects on the measured observables that agree with the SM. These properties, however, make it difficult to verify or falsify the theories beyond the SM, especially if many free parameters are introduced.\\
In the first part of this thesis I examine a model that stands out not due to its elegance from theoretical perspective but rather due to its simplicity: the Standard Model extended by a \otto{perturbative} fourth fermion generation (SM4). Contrary to more complex models like supersymmetry, the SM4 is comparably easy to assess because it only has few additional free parameters, which could potentially be determined at the LHC.
Compared to the SM, the SM4 features additional ${\cal CP}$ violating phases, \otto{which were discussed as possible contributions to the} asymmetry between the existing matter and antimatter in our universe \otto{\cite{Hou:2008xd}}.
There have been attempts to explain certain flavour measurements with fourth generation particles \cite{Hou:2005yb,Soni:2008bc}, and the SM4 neutrinos could even contribute to dark matter \cite{Raby:1987ww,Lee:2011jk}. The increased interest in the SM4 since 2004 is illustrated in Fig.\ \ref{fig:pre2011}: For each year, I show the number of particle physics publications containing either ``fourth generation'', ``4th generation'', ``fourth family'', or ``4th family'' in their title \cite{inspirehep:2013}. The popularity growth can be approximated by an exponential function.

But then the first LHC results in 2011 and 2012 revealed that the missing piece of the SM had been found: the discovery of a bosonic resonance in the $\gamma \gamma$ and $ZZ$ invariant mass distribution was probably the most exciting event in particle physics in many years. The properties of this new particle are in good agreement with the predictions for the SM Higgs boson, which imposes strong constraints on many SM extensions. The SM4 happens to be the first popular model that can even be ruled out on the basis of these measurements. To quantify whether a model is excluded or not, one has to carry out a global analysis of its possible parameter constellations, taking into account all relevant experimental constraints, and compare it with the SM using a likelihood ratio test.
Due to the interplay of different observables this comparison is usually non-trivial; for the SM4 it is even more complicated, because it does not decouple from the SM. Therefore, \otto{my collaborators and I have} performed a likelihood ratio test for non-nested models combining Higgs observables and electroweak precision data, and \otto{eventually excluding} the SM4 \cite{Eberhardt:2012gv}.\\
Similar to the extension of the fermionic SM content by a quark and a lepton doublet, one can add a second scalar doublet to its Higgs sector. This so-called Two-Higgs-Doublets model (2HDM) can arise as effective theory from several models, like for instance supersymmetry. I will explain in the second part of this thesis that the 2HDM parameters are also severely constrained by electroweak precision observables as well as LHC and flavour measurements. However, the 2HDM has a decoupling limit and cannot be ruled out completely like the SM4.\\
The SM4 and the 2HDM are comparably simple modifications of the SM; nevertheless, the correct statistical treatment of a model comparison is non-trivial. Therefore, I first want to explain the statistical tools that I need in Chapter \ref{Statistics}, and \otto{discuss the} Standard Model of particle physics in Chapter \ref{SM}. Only then I am prepared to introduce new models and to compare them with the SM. This is done in Chapter \ref{SM4} for the SM4 and in Chapter \ref{2HDM} for the 2HDM, before I conclude in Chapter \ref{Conclusions}.

%% file: statistics.tex
\chapter{Statistics} \label{Statistics}

The challenge of theoretical particle physics is to develop a model that describes all observed phenomena as well as possible and serves to predict the outcome of future experiments. With the Standard Model we already have a powerful theory, but there are a few aspects that indicate that it is only an effective limiting case of a more fundamental theory. Many SM extensions with different virtues have been formulated, of which one might be realized, whereas the others sooner or later will be ruled out by experiments. When we now consider one of these new theoretical models, we want to make a statement on how compatible experimental measurements are with it, and furthermore we want to test the new model against the conventional model, which in our case is the SM. In the frequentist approach -- the statistical method that I \otto{have chosen} for this work -- this model comparison is done by means of a likelihood ratio test.
It provides us with a $p$-value (also called statistical significance) which tells us \otto{to which level a model is excluded}. After \otto{reviewing} the foundations of frequentist statistics following \cite{Beringer:1900zz,Hocker:2001xe,Wiebusch:2012en,Lyons:2013aa}, I will introduce the likelihood ratio test (referring to the discussion in \cite{Wiebusch:2012en}) and its statistical interpretation in Sect.\ \ref{pvalue}, and then address two important aspects in our analyses, the treatment of systematic errors and non-nested model comparisons, in Sect.\ \ref{systematicerrors} and \ref{nonnestedmodels}.

\section{The likelihood} \label{likelihood}

A theoretical model is characterized by its model parameters $\xi _j$ ($j=1,...,N_\xi $), which can only be determined by experimental observations. Let us assume that \otto{there are} $N_X$ observables $X_i$ ($i=1,...,N_X$) with a certain probability density function (p.d.f.)
\otto{$f_i(x_{i}^{\rm exp},\bm{\xi})$. Let us further assume that we have the measurements $x_i^{\rm exp}$ of the observables $X_i$. The probability that one measurement is between $x$ and $x+{\rm d}x$ is given by $f_i(x,\bm{\xi}){\rm d}x$.} Adopting vector notation, I can write the joint p.d.f.\ of all measurements as $f(\bm{x}^{\rm exp},\bm{\xi})$. For a certain set of measurements $\bm{x}_{0}^{\rm exp}$, we can define the likelihood as

\begin{align}
 L(\bm{\xi}) &= f(\bm{x}_0^{\rm exp},\bm{\xi}). \label{eq:likelihood}
\end{align}

If all measurements are statistically independent, we \otto{can write the likelihood as the product of the individual p.d.f.s:}

\begin{align*}
 L(\bm{\xi}) &= \prod\limits_{i=1}^{N_X}f_i(x_{0,i}^{\rm exp},\bm{\xi})
\end{align*}

In order to test the compatibility of the model with the experimental results, we must try to bring the theoretical predictions into agreement with the measurements by adapting the parameters. The most popular method to do this is the maximization \otto{of} the likelihood with respect to the model parameters. I want to refer to this procedure as \textit{fit} in the following and call the parameters which maximize the likelihood \textit{best-fit parameters} $\bm{\xi}_{\rm bf}$.

\section{Gaussian distributions} \label{gaussian}

One essential property of experiments in particle physics is that we cannot obtain reliable information from a single measurement, but rather have to measure many times to extract the p.d.f.\ of an observable. If it is an average of a large number of measurements, the observable $X_i$ will commonly have a Gaussian p.d.f.\ with the statistical error $\sigma _i$ as standard deviation around the central value \otto{$\mu _i$}.
If there are $n$ different measurements $x_{i,k}^{\rm exp}$ for one observable $X_i$, where $k$ is between $1$ and $n$, this corresponds to $n$ draws \ottooo{of} the same random variable. They can be combined to the average $x^{\rm exp}_{i,\text{\tiny comb}}$. This average itself can be interpreted as Gaussian distributed random variable with error $\sigma _{i,\text{\tiny comb}}$. The average and its error can be computed in the following way \cite{Beringer:1900zz}: To each $x_{i,k}^{\rm exp}$, we assign a \textit{weight} $w_{i,k}$, which is defined as the inverse squared error:
\vspace*{-10pt}

\begin{align*}
 w_{i,k} &= \frac{1}{\sigma _{i,k}^2}
\end{align*}

The combined statistical error shrinks according to
\vspace*{-10pt}

\begin{align}
 \sigma _{i,\text{\tiny comb}} &= \left( \sum\limits_{k=1}^n w_{i,k}\right) ^{-\frac{1}{2}}, \label{eq:sigmacombination} 
\end{align}

which means that the combined standard deviation decreases by $1/\sqrt{n}$, if the number of equivalent measurements with equal weights is increased by the factor $n$. For an estimate of the central value of this combination, all individual $x^{\rm exp}_{i,k}$ are weighted:
\vspace*{-10pt}

\begin{align}
 x^{\rm exp}_{i,\text{\tiny comb}} &= \sigma _{i,\text{\tiny comb}}^2 \; \sum\limits_{k=1}^n w_{i,k} x^{\rm exp}_{i,k} \label{eq:obscombination}
\end{align}

Inverting these relations, one can also reconstruct information about parts of the combination.\\
On the theory side, the true value of the observable $X_i$ can be described by the function $x_i^{\rm theo}\left( \bm{\xi}\right)$, assuming the realization of a particular theory. \otto{If this assumption is correct, $x_i^{\rm theo}\left( \bm{\xi}\right)$ will coincide with the central value $\mu _i$ of the p.d.f.} In the $N_X$-dimensional space of observables the images of the expressions $\bm{x}^{\rm theo}\left( \bm{\xi}\right)$ form the so-called theory manifold ${\cal M}$.\\
For every Gaussian distributed observable, we can calculate the \textit{deviation}, which is the difference between the measured value and the theoretical expression around the best-fit point, normalized to the experimental error:

\begin{align}
 \chi_i \left( \bm{\xi}\right) &= \frac{x_i^{\rm exp}-x_i^{\rm theo}\left( \bm{\xi}\right) }{\sigma _i} \label{eq:chi}
\end{align}

With this definition a one-dimensional Gaussian distribution (normal distribution) is given by the function

\begin{align}
 f_{\rm G}(x_i^{\rm exp},x_i^{\rm theo}(\bm{\xi}), \sigma_i) &= \frac{1}{\sqrt{2\pi }\sigma_i}\exp \left[ -\frac{1}{2}\chi_i^2\right]. \label{eq:Gaussianpdf}
\end{align}

\begin{figure}[htbp]
 \centering
 \subfigure[]{\begin{picture}(190,130)(3,0)
              \put(0,0){\includegraphics[width=220pt]{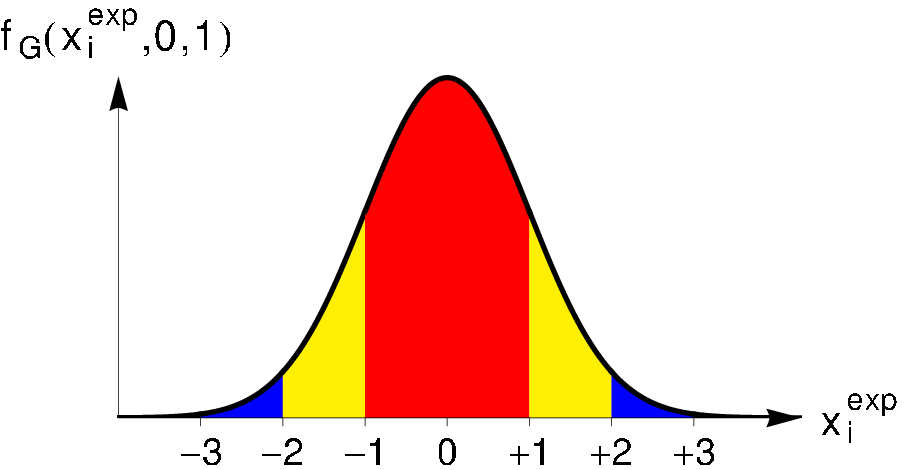}}
	      \end{picture}
	     }
 \qquad
 \subfigure[]{\begin{picture}(190,130)(0,0)
              \put(-20,0){\includegraphics[width=220pt]{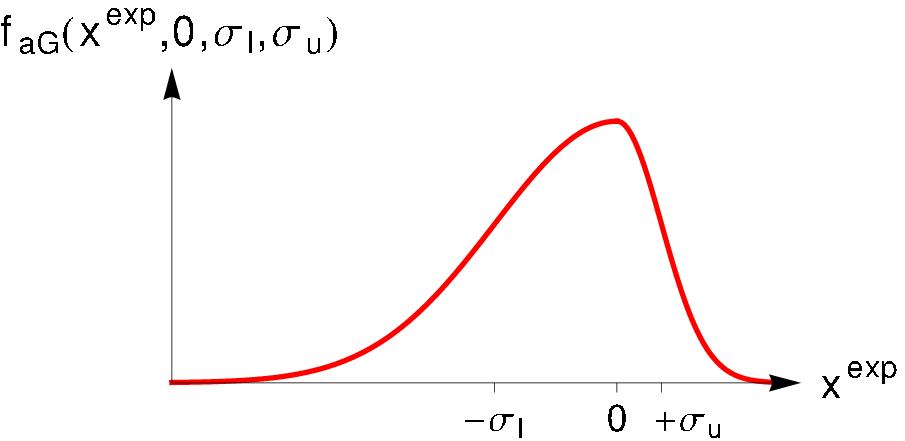}}
	      \end{picture}
	     }
 \caption[Symmetric and asymmetric Gaussian distributions.]{The symmetric Gaussian distribution is shown with a standard deviation of $1$ (a). The lower standard deviation $\sigma_l$ and the upper standard deviation $\sigma_u$ of an asymmetric Gaussian distribution have different values (b). The central value was chosen to be $0$ for both.}
 \label{fig:Gauss}
\end{figure}

This p.d.f.\ is depicted in Fig.\ \ref{fig:Gauss}(a) for $x_i^{\rm theo}=0$ and $\sigma _i=1$.
Some experiments also have an asymmetric Gaussian distribution, which is characterized by different lower and upper Gaussian errors \otto{$\sigma_l$} and \otto{$\sigma_u$}:

\begin{align*}
 f_{\rm aG}(x^{\rm exp},x^{\rm theo}(\bm{\xi}), \sigma_l, \sigma_u) &= \begin{cases} \dfrac{2\sigma_l}{\sigma_l+\sigma_u}\:f_{\rm G}(x^{\rm exp},x^{\rm theo}(\bm{\xi}), \sigma_l), \quad \text{if } x^{\rm exp}<x^{\rm theo}(\bm{\xi})\\[10pt] \dfrac{2\sigma_u}{\sigma_l+\sigma_u}\:f_{\rm G}(x^{\rm exp},x^{\rm theo}(\bm{\xi}), \sigma_u), \quad \text{if } x^{\rm exp}\geq x^{\rm theo}(\bm{\xi})\end{cases}
\end{align*}

An example is shown in Fig.\ \ref{fig:Gauss}(b). If we want to determine the deviation of an observable with an asymmetric Gaussian, we have to choose the appropriate error \otto{in the denominator} \ottooo{of Eq.\ \eqref{eq:chi}} according to the sign of the difference $x^{\rm exp}-x^{\rm theo}(\bm{\xi})$.

For multi-dimensional Gaussian distributions, we need to take into account the correlations between different observables: In general, all observables $x_i^{\rm theo}$ depend on the same set of parameters. Changing the value of a parameter to diminish the squared deviation of one observable and to increase the total likelihood also affects the other observables and their deviations. Around the best-fit point, this correlation is quantified by the covariance matrix $V_c$, which is the inverse of the Hessian matrix of the negative logarithm of the likelihood:

\begin{align*}
 (V_c^{-1})_{nm} &= -\left. \frac{\partial ^2\ln L(\bm{\xi})}{\partial \xi_n\partial \xi_m}\right|_{\bm{\xi}_{\rm bf}}.
\end{align*}

The quadratic expansion of the likelihood around the maximum is proportional to

\begin{align*}
 \chi^2 (\bm{\xi},\bm{x}^{\rm exp}) =\bm{\chi}^T\otto{(\bm{\xi})} V_c^{-1}\: \bm{\chi}\otto{(\bm{\xi})},
\end{align*}

which can be minimized instead of maximizing the likelihood. For independent observables, $V_c = \mathds{1}$, the $\chi^2$ is equal to the sum of the squared deviations, and the likelihood is the product of the individual Gaussian p.d.f.s.\\
The minimal $\chi^2$ of a fit\otto{, which I will call $\chi^2_{\rm min}$,} is a common measure of how well $\bm{\xi}$ describes the data $\bm{x}^{\rm exp}$. 
Often, a first estimate of how well a theory performs is given by the minimal $\chi^2$ per degree of freedom, $\chi^2_{\rm min}/N_{\rm dof}$. The number of degrees of freedom $N_{\rm dof}$ is given by the difference $N_X-N_\xi$ if we do not artificially constrain $\bm{\xi}$ in the fit. By definition, $N_{\rm dof}$ is only positive if we have more observables than fit parameters; a system like that is called overconstrained and all parameters can be fixed by the experimental inputs in a fit. Since the expectation value of the deviation is $1$, the expected outcome of the fit will be $\chi^2_{\rm min}/N_{\rm dof}\approx 1$ if we have many observables which are described well by the theory. The best-fit parameter values $\bm{\xi}_{\rm bf}$ for which the $\chi^2$ is minimal \otto{will become} important in the next section.
Moreover, it is of interest how large the ``allowed'' ranges for the parameters and observables around their best-fit values are. I will also come back to these so-called confidence intervals in the next section.\\
As defined in Eq.\ \eqref{eq:chi}, the deviation is the minimal difference between $x_i^{\rm theo}\left( \bm{\xi}\right)$ and $x_i^{\rm exp}$, i.e.\ the deviation of the best-fit point of a \textit{complete} fit using all $N_X$ measurements $x_i^{\rm exp}$. One can also use the first $N_X-1$ observables to predict the last one; I will refer to this method as \textit{prediction}. The $\chi^2_{\rm min}$ difference $\Delta \chi^2$ between the complete fit and the prediction fit gives us a measure of how large the impact of the measurement of $X_{N_X}$ on the model fit is. Note that $\Delta \chi^2$ and the squared deviation $\chi^2_{N_X}$ are not the same because other observables will in general have a different best-fit deviation in the prediction fit and in the complete fit.\\
An example \otto{in particle physics} for a parameter determination \otto{with the help of a prediction fit} is the Higgs boson mass: Before this last missing parameter of the SM was measured directly in 2012 as mentioned in the introduction, there had been fits that could exclude the Higgs mass outside a certain confidence interval. So under the assumption that the Standard Model was true, experimentalists knew in which mass region to search.
\vspace*{10pt}

\section{The likelihood ratio test} \label{pvalue}

Instead of leaving out or adding information on the experimental side, we can also add new parameters or fix existing ones, hence change the theory. This is especially useful, if we have a model that we do not doubt in general, and we want to check whether a particular realization could exist.
Let us define a \textit{constrained} theory $B$ which differs from the \textit{full} theory $A$ by fixing $\nu$ of the parameters of $A$. To quantify the viability of $B$ or to exclude it we need to perform a hypothesis test. Let us assume that $B$ is realized for a given set of measurements $\bm{x}^{\rm exp}_0$ (``null hypothesis''). To compare the compatibility of the two models with these measurements we could use the ratio of their maximized likelihoods $L^{\rm max}_B$ and $L^{\rm max}_A$.

Instead we can also use

\begin{align}
S(\bm{x}^{\rm exp}) &= -2 \ln \frac{L^{\rm max}_B}{L^{\rm max}_A}. \label{eq:teststatistic}
\end{align}

If $S(\bm{x}^{\rm exp})$ is \otto{greater} than a predefined value $S_0$, we will reject the null hypothesis and accept $A$ as realized. Therefore, this hypothesis test is called \textit{likelihood ratio test}.
$S(\bm{x}^{\rm exp})$ is a random variable called \textit{test statistic} and is itself characterized by a p.d.f.
Since $A$ has more free parameters, $L^{\rm max}_A$ cannot be smaller than $L^{\rm max}_B$, and $S(\bm{x}^{\rm exp})$ is positive \otto{semi-}definite.
Usually, one chooses the rejection condition $S_0$ to be $S(\bm{x}^{\rm exp}_0)$, i.e.\ the value the test statistic would take if $B$ was realized with \otto{its best-fit} parameters $\bm{\xi}_{\rm bf, B}$ for $\bm{x}^{\rm exp}_0$ \otto{and if $A$ was realized with the best-fit parameters of $B$ and the last $\nu$ parameters fitted to $\bm{x}^{\rm exp}_0$.}
If we \otto{denote} the p.d.f.\ of the test statistic \otto{by} $f_{\rm ts}(\bm{x}^{\rm exp}, \bm{\xi}_{\rm bf, B})$, we can calculate the probability for wrongly rejecting $B$, which is called the $p$-value, by the integration of $f_{\rm ts}(\bm{x}^{\rm exp}, \bm{\xi}_{\rm bf, B})$ over the observable regions where $B$ is rejected:

\begin{align}
 p = \int f_{\rm ts}(\bm{x}^{\rm exp}, \bm{\xi}_{\rm bf, B}) \: \Theta \left( S(\bm{x}^{\rm exp})-S_0\right) {\rm d}^{N_X} x^{\rm exp} \label{eq:realpvalue}
\end{align}

\begin{figure}[htbp]
 \centering
 \begin{picture}(180,190)(0,0)
 \put(-40,0){\includegraphics[width=1.7\linewidth]{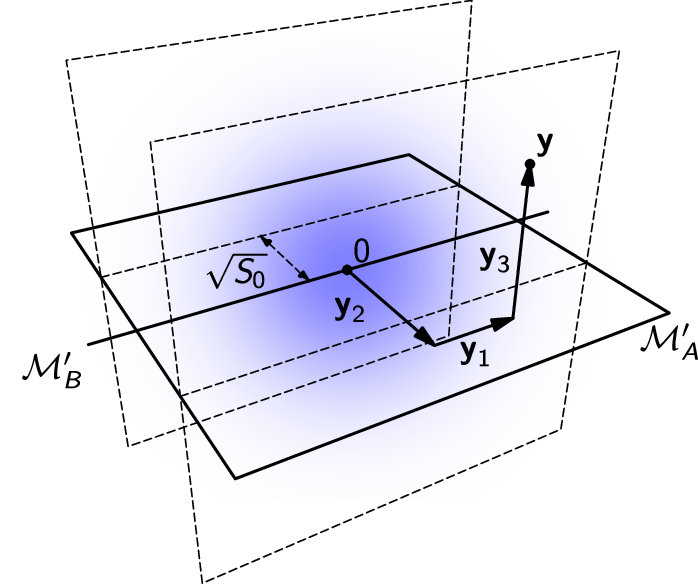}}
 \end{picture}
 \caption[Illustration of nested models in the observable space.]{Nested models in three dimensions (graphic from \cite{Wiebusch:2012en}). The blue shaded region indicates the p.d.f.\ of the toy measurements. The dashed planes are the boundaries for the model rejection of $B$.}
 \label{fig:nested}
\end{figure}

$\Theta $ denotes the Heaviside step function. In our calculations, the measurements $\bm{x}^{\rm exp}$ will be simulated toy measurements, and the integration is done numerically.\\
If we want to depict the calculation of the $p$-value, it is convenient to modify the observable space: In the $N_X$ dimensional observable space we can always find a transformation such that in the new coordinate system $\bm{x}^{\rm exp}_0-\bm{x}^{\rm{theo},B}(\bm{\xi}_{\rm bf, B})$ is mapped to the origin with $\sigma ^{\rm exp}_i=1$ and $V_c=\mathds{1}$. ${\cal M}'_A$ and ${\cal M}'_B$ are the theory manifolds of $A$ and $B$ in the transformed space of observables.
By definition, ${\cal M}'_A\supset {\cal M}'_B$ (because $B$ is $A$ with $\nu$ parameters fixed), and both contain the origin. If $\bm{y}$ is the transformed vector of observables $\bm{x}^{\rm exp}$, we can write the transformed test statistic as $S(\bm{y})$. Under the assumption that ${\cal M}'_A$ and ${\cal M}'_B$ are hyperplanes, $|\bm{y}|^2$ is simply the squared distance between the (toy) measurement\otto{s} and the best-fit value of model $B$, corresponding to the squared deviation $\chi_i^2$ in one dimension. Defining $\bm{y}_2$ as the orthogonal part to ${\cal M}'_B$ of the projection of $\bm{y}$ on ${\cal M}'_A$, the test statistic reads $S(\bm{y})=|\bm{y}_2|^2$. The $p$-value is now the integral over the $\bm{y}$ regions, where $S(\bm{y})>S_0$, i.e.\ outside an infinitely long ``hyper-cylinder'' around ${\cal M}'_B$.
As illustration I show a three-dimensional projection of the observable space with a two-dimensional ${\cal M}'_A$ and a one-dimensional ${\cal M}'_B$ in Fig.\ \ref{fig:nested}.

I will resort to this picture of the transformed observable space in Sect.\ \ref{nonnestedmodels} when I want to discuss the case where $B$ does not emerge from $A$ by fixing some parameters.\\
If all measurements have a Gaussian p.d.f., the test statistic directly translates into the difference $\chi^2_{\rm{min},B}-\chi^2_{\rm{min},A}$ of the respective minimal $\chi^2$ values of the models:

\begin{align}
 S(\bm{x}^{\rm exp}) &=\chi^2_{\rm{min},B}-\chi^2_{\rm{min},A} \label{eq:teststatisticGauss}
\end{align}

The test statistic is distributed according to a $\chi^2$ distribution, and

\begin{align}
 p &= \frac{1}{\Gamma (\nu /2)}\int\limits_{S(\bm{x}_0^{\rm exp})}^\infty \varsigma^{\frac{\nu }{2}-1}{\rm e}^{-\varsigma}{\rm d}\varsigma. \label{eq:naivepvalue}
\end{align}

Commonly, the $p$-value is calculated analytically with this formula in theoretical particle physics. This implies three assumptions \cite{Wiebusch:2012en}: The theory expressions should be linear in the vicinity of the measurements, i.e.\ the theory manifold should be approximately a hyperplane, such that the closest point to some $\bm{x}^{\rm exp}$ is unique. Next, the p.d.f.s should be of Gaussian shape. And the last important condition, which I want to discuss in Sect.\ \ref{nonnestedmodels}, is that the constrained theory $B$ should be embedded in the full theory $A$. Only if these requirements are met, $S(\bm{x}^{\rm exp}_0)$ is given by Eq.\ \otto{\eqref{eq:teststatisticGauss}} and can be converted into a $p$-value by Eq.\ \otto{\eqref{eq:naivepvalue}}.
This is known under the name Wilks' theorem \cite{Wilks:1938aa}. In the following, I will refer to this simplification as \textit{naive} $p$-value. The use of Wilks' theorem is common practice in theoretical particle physics even though the applicability in some cases is questionable. Visualizing Eq.\ \eqref{eq:naivepvalue} in one dimension in the observable space, the $p$-value can be identified with the integral over the ``tails'' of a (Gaussian) distribution. On the contrary, the regions that are not part of the tails, cannot be excluded; they are the above-mentioned confidence intervals.
This is illustrated by the different shaded regions of Fig.\ \ref{fig:Gauss}: while the red region corresponding to deviations smaller than $1$ constitutes more than $68\%$ of the integral over $f_G(x_i^{\rm exp},x_i^{\rm theo},\sigma_i)$, scenarios with large deviations only contribute little to the total integrated surface. According to their maximal deviation the red, yellow and blue shaded regions are referred to as $1\sigma$, $2\sigma$ and $3\sigma$ regions.
I will use the same colour coding in the figures of the next chapters. Stating that a parameter value is excluded at a certain confidence level CL -- where CL is between $0\%$ and $100\%$ -- connotes that the corresponding $p$-value is less than $1-$CL. But this does not only hold for single parameters. By analogy with the one-dimensional Gaussian case, the $p$-value for multi-dimensional problems is also commonly translated into a ``$\sigma$'' statement:
One can claim to exclude model $B$ at $n\sigma$ if its $p$-value corresponds to the one-dimensional picture of the integrated area under the Gaussian tails where \otto{$|\chi_i|>n$}.\\
Observations in particle physics are also based on the hypothesis test idea: Defining a ``background'' model and a ``signal plus background'' model, one can also exclude the former, hence claiming an observation. Over the last decades, certain exclusion and discovery levels evolved as a rule of thumb for particle physicists \cite{Lyons:2013aa}: If a particular point is outside the $2\sigma$ boundary (which roughly corresponds to a confidence level of 95{\%}), it is considered to be disfavoured; everything outside the $3\sigma$ region appears to be excluded.
For the decision between ``signal \otto{plus background}'' and ``background only'', the criteria are a bit different: If a signal has a deviation of $3$ from the background expectation, one speaks of an \textit{evidence}; \ottooo{only} if its deviation is $5$ or more, one can claim a \textit{discovery} (or \textit{observation}).\\
The naive definition of the $p$-value usually gives a good approximation and I will use it to illustrate the confidence intervals of single observables and parameters throughout this thesis; however, for the comparison of two models differing in multiple aspects we rely on the correct formulation from Eq.\ \eqref{eq:realpvalue}, which holds for non-Gaussian p.d.f.s as well as non-linear theory manifolds.
The only requirement that we maintain at this point is that $B$ is \textit{nested} in $A$, meaning that it is a constrained version of $A$. In Chapter \ref{2HDM}, we will find exactly these circumstances when comparing the Two-Higgs-Doublets model to the SM. However, in Chapter \ref{SM4}, we will also need to calculate the $p$-value for non-nested theories. Before we discuss non-nested model comparisons, I want to go into detail about the treatment of systematic uncertainties.

\section{Systematic uncertainties} \label{systematicerrors}

\begin{figure}[htbp]
 \centering
 \subfigure[]{
    \begin{picture}(115,140)(8,0)
    \put(0,0){\includegraphics[width=0.30\linewidth]{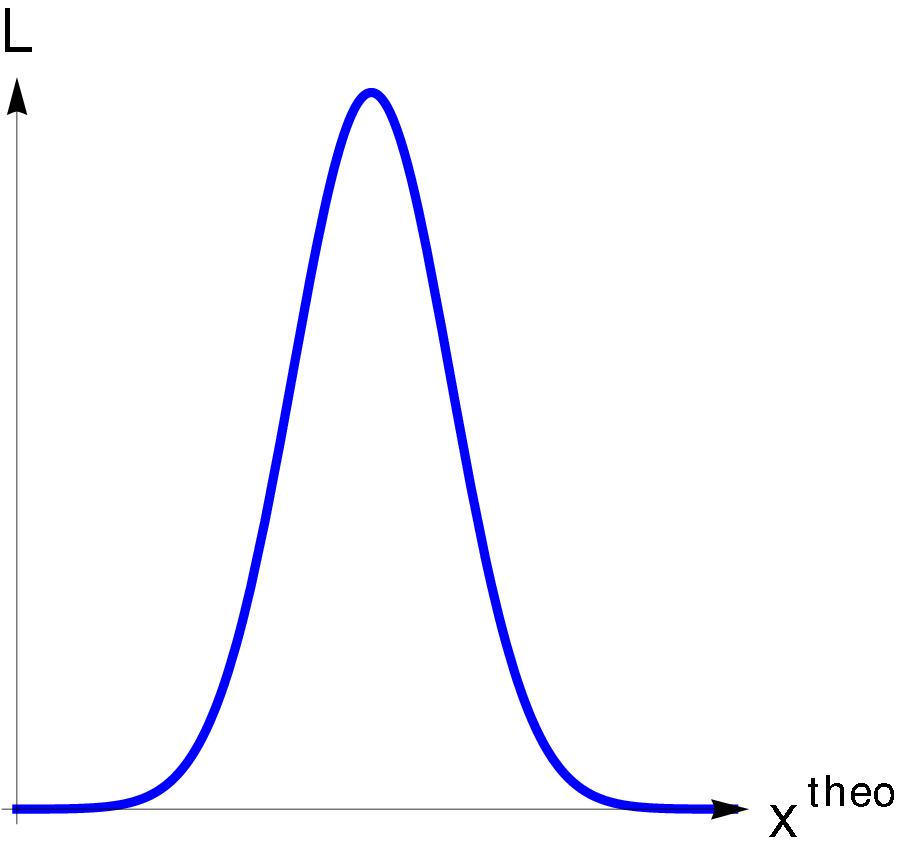}}
    \end{picture}
 }
  \qquad
 \subfigure[]{
    \begin{picture}(115,140)(8,0)
    \put(0,0){\includegraphics[width=0.30\linewidth]{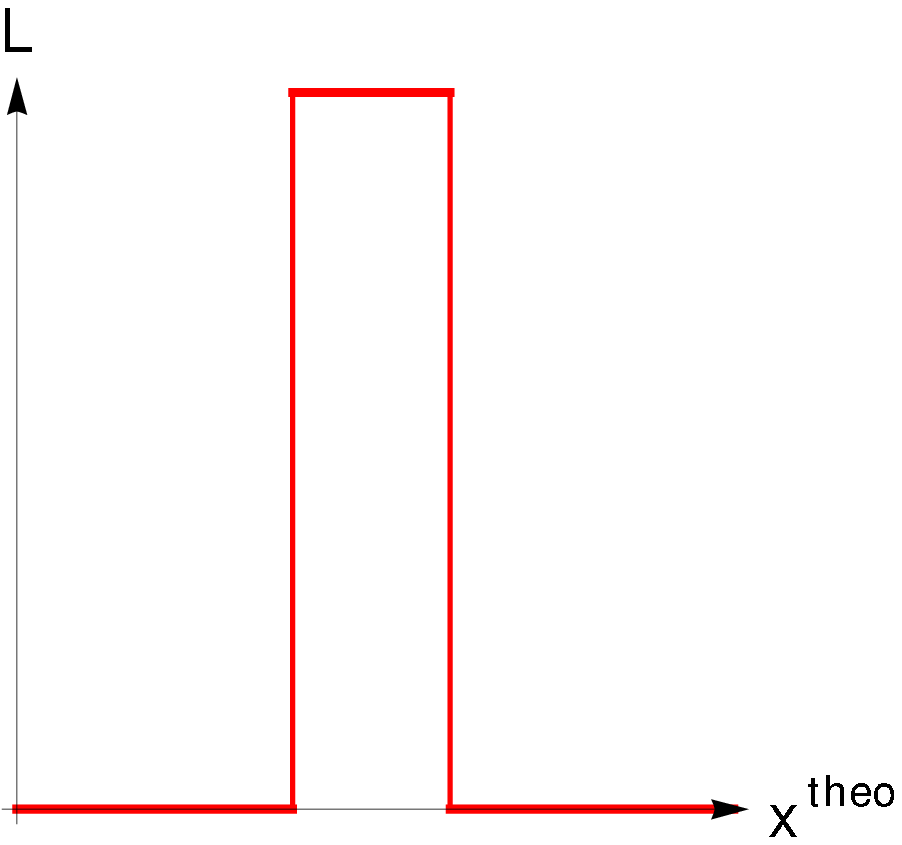}}
    \end{picture}
 }
  \qquad
 \subfigure[]{
    \begin{picture}(115,140)(8,0)
    \put(0,0){\includegraphics[width=0.30\linewidth]{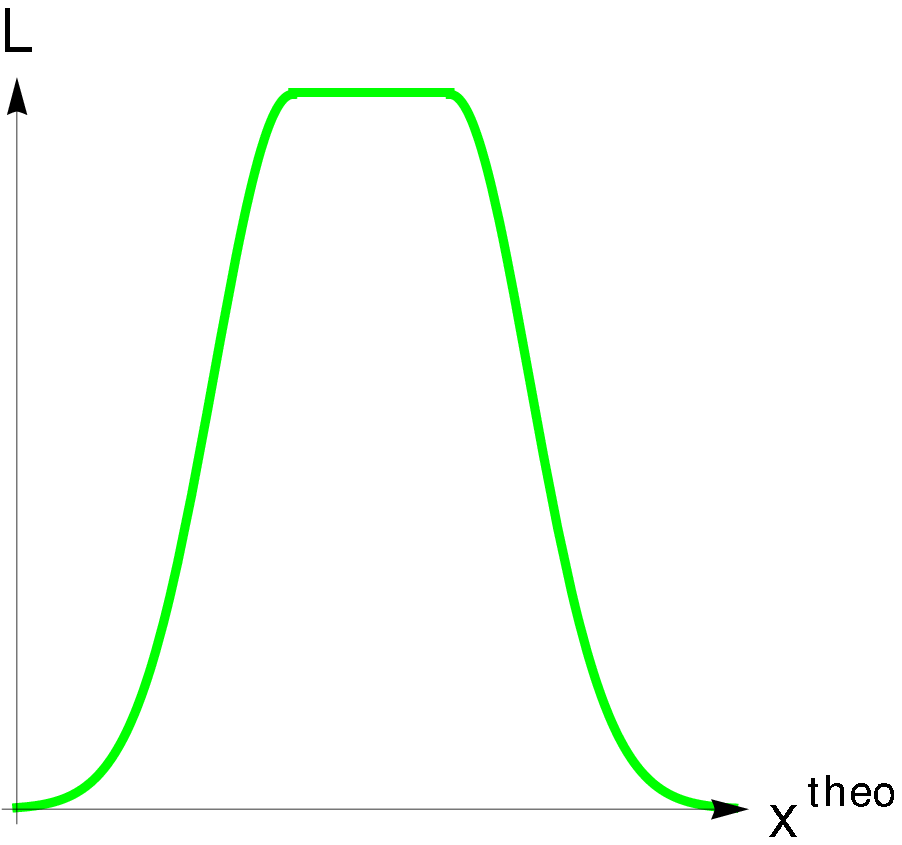}}
    \end{picture}
 }
 \caption[Three different likelihood profiles.]{Three different likelihood profiles: the blue curve (a) is a Gaussian distribution, like in Fig.\ \ref{fig:Gauss}(a), the red line (b) displays a range of likelihood (\textit{R}fit with systematic error), and the green function (c) is the profile of a quantity with both, statistical and systematic errors.}
 \label{fig:Rfitscheme}
\end{figure}

Up to this point, we \otto{have only} \ottooo{treated} p.d.f.s of a Gaussian shape with statistical errors as standard deviation in detail. However, there are as well uncertainties which cannot be diminished by increasing the number of measurements according to Eq.\ \eqref{eq:sigmacombination}. They can originate from the detectors, as for example detector calibration errors, and thus be part of the experimental error, but they also result from theory\otto{, as for instance} uncertainties \otto{on} lattice calculations.
All these errors are offsets between the true value $x_i^{\rm theo}$ and the measured value $x_i^{\rm exp}$ of an observable which do not average out when the experiment is repeated. Irrespective of their origin, they can be subsumed under the name \textit{systematic uncertainties}. In the \textit{R}fit scheme \cite{Hocker:2001xe}, one introduces additional fit parameters, the so-called \textit{nuisance parameters}, which account for systematic uncertainties in the following way: to each observable $X_i$ with a systematic error $\sigma _i^{\rm syst}$, we assign the nuisance parameter $\xi_i^\nu$, which is allowed to float between $-1$ and $1$, and add $\xi^\nu_i\cdot \sigma _i^{\rm syst}$ to $x_i^{\rm theo}$.
In Fig.\ \ref{fig:Rfitscheme}, three different profile likelihood function are shown: the left one is an ordinary Gaussian distribution, the red line displays an observable without statistical error but with a systematic error, and the right curve illustrates the resulting profile likelihood of an observable which has both, statistical and systematic errors, when maximizing with respect to all $\xi_i^\nu$.

The notation implies that all systematic errors (including experimental uncertainties) are converted into bounded theory parameters, where several systematic uncertainties adherent to one observable \otto{can be} combined by adding them linearly.\\
The \textit{R}fit method can also be used to \ottooo{limit} parameters by relating them to an artificial observable with only a systematic error like in Fig.\ \ref{fig:Rfitscheme}(b). On the one hand this can be useful to prevent masses from becoming negative or to limit angles to take values between $0$ and $2\pi$, but on the other hand this smears the definition of $N_{\rm dof}$ because the number of parameters can be arbitrarily increased with this method \cite{Lyons:2013aa}.

\section{CKMfitter} \label{CKMfitter}

For the fits in this thesis I used the CKMfitter package \cite{Hocker:2001xe}, which is based on the \textit{R}fit method. The package is written in WOLFRAM Mathematica files; the minimization is outsourced to Fortran subroutines. A global parameter fit with CKMfitter usually involves two steps: the first is minimizing the $\chi^2$ with respect to all parameters, and the optional second part is the scan over one or two parameters or observables. For the initial minimization, all parameters are treated as free within their allowed ranges, a certain number of starting points is randomly generated, and the $\chi^2$ minimum is searched for using numerical gradients. (Analytical gradients are in principle supported but not applicable if the external routines come into play that I want to use.) For the one-dimensional scans, the specified range for the scan quantity is divided into $N$ equally spaced points for each of which a ``constrained'' minimization is performed, in which the scan quantity is fixed.
($N$ is the granularity defined by the user.) The same applies for two-dimensional scans, respectively. One important feature is that the scans do in general not start at one end of the scan range(s) but at the scan point that is closest to the best-fit point. \otto{All following scan steps use the $\chi^2$ information of the previous step.}
Hence, one usually gains better convergence of the ``constrained'' fits in the sense that the minimum is found more reliably. Especially in two-dimensional fits this behaviour can be an advantage compared to plain ``left-to-right'' scanning. In spite of this virtue I wrote a program that parallelizes 2D scans by partitioning them into 1D scans, which reduces the scan time by a factor $1/N$, provided that one has the possibility to execute the parallelized scans on $N$ processors at the same time. For details, see App.\ \ref{parallelize}. The duration of one fit plus scan procedure is from a second to several days, depending on the complexity of the theory expressions, on the external routines attached to the fitter\otto{,} and on whether parallelization was used or not. I will comment on the performance of the single CKMfitter model implementations when discussing the combined fit results of 
the following chapters.

\section{Non-nested models} \label{nonnestedmodels}

\begin{figure}[htbp]
 \centering
 \begin{picture}(180,180)(0,0)
 \put(-40,0){\includegraphics[width=1.7\linewidth]{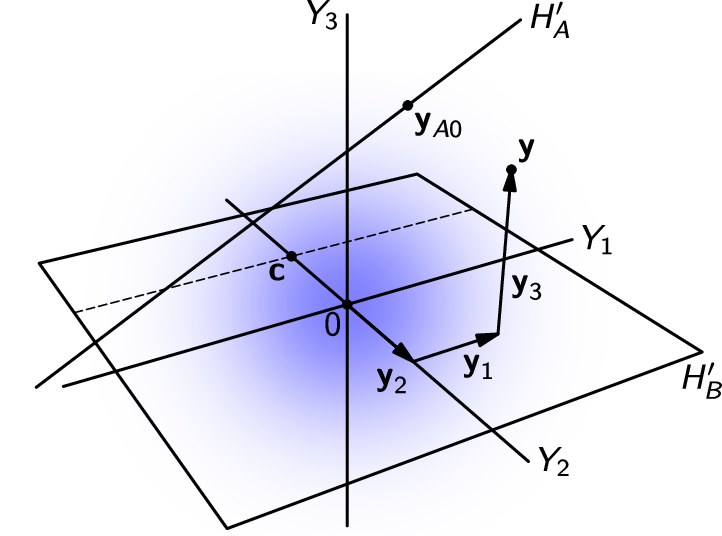}}
 \end{picture}
 \caption[Illustration of non-nested models in the observable space.]{Non-nested models in three dimensions (graphic from \cite{Wiebusch:2012en}). The blue shaded region indicates the p.d.f.\ of the toy measurements.}
 \label{fig:nonnested}
\end{figure}

If two models $A$ and $B$ are not nested, the determination of the $p$-value becomes more complicated. Like in Sect.\ \ref{pvalue}, I will follow \cite{Wiebusch:2012en} to construct a definition based on a geometrical picture: Non-nestedness directly translates into the fact that one theory manifold is not \otto{a} subset of the other. Let us again take $B$ as \otto{the} null hypothesis. The test statistic as defined in \eqref{eq:teststatistic} is not positive \otto{semi-}definite any longer, so we \ottooo{set} $S=0$ if $L^{\rm max}_B>L^{\rm max}_A$.
Assuming Gaussian p.d.f.s, one can find a mapping of the observable space like in Sect.\ \ref{pvalue} such that in the new coordinate system $\bm{x}^{\rm exp}_0-\bm{x}^{\rm{theo},B}(\bm{\xi}_{\rm bf, B})$ corresponds to the origin with $\sigma ^{\rm exp}_i=1$ and $V_c=\mathds{1}$.
If we approximate ${\cal M}'_B$, the transformed \otto{theory} manifold of model $B$, by its tangent hyperplane $H_B'$ in the new observable space, the tangent hyperplane $H_A'$ of ${\cal M}'_A$ will now not be a subset of it, nor vice versa.
We can decompose the transformed vector of a (toy) measurement $\bm{y}$ into a parallel part $\bm{y}_\parallel$ and an orthogonal part $\bm{y}_3$ with respect to $H_B'$ and further divide $\bm{y}_\parallel$ into $\bm{y}_1$ and $\bm{y}_2$, of which the latter is defined to be orthogonal to $H_A'$. Let $Y_1$, $Y_2$ and $Y_3$ be the corresponding subspaces and let $\bm{c}$ be the projection of $\bm{y}_A$ on $Y_2$, where $\bm{y}_A$ is any point on $H_A'$; then the distance of $H_A'$ from the subspace $Y_1$ is given by $|\bm{c}|$. In Fig.\ \ref{fig:nonnested}, a three-dimensional example is given with $H_A'$ being one-dimensional and a two-dimensional $H_B'$.

The \otto{$\chi^2_{\rm{min},B}$} of theory $B$ is now $|\bm{y}_3|^2$, but where in the nested case we would have $\chi^2_{\rm{min},A}=|\bm{y}_2|^2+|\bm{y}_3|^2$, we now only get the lower bound of $|\bm{y}_2-\bm{c}|^2$ for the minimal $\chi^2$ of $A$, which is realized if $\bm{y}-(\bm{y}_2-\bm{c})$ is on $H_A'$. So again requiring $S(\bm{y})$ to be smaller than some reference value $S_0$ we need to integrate over the regions where $|\bm{y}_3|^2 < S_0+|\bm{y}_2-\bm{c}|^2$, but this time there are also contributions of toy measurements with $|\bm{y}_3|^2 > S_0+|\bm{y}_2-\bm{c}|^2$ to the $p$-value integral in \eqref{eq:realpvalue} because we only have a lower bound for $\chi^2_{\rm{min},A}$.
(For instance, take the point $\bm{y}_{A0}$ on $H_A'$ in Fig.\ \ref{fig:nonnested}: if the $Y_3$ components of $\bm{y}_{A0}$ have a different sign than the ones of a toy measurement $\bm{y}$, and at the same time their $Y_1$ and $Y_2$ components agree, the point will always be closer to $H_B'$ than to $H_A'$, regardless of whether $|\bm{y}_3|^2$ is smaller or greater than $S_0$.)
The difficulty now is that these ``special'' regions are not as easy to identify since their boundaries (where $S(\bm{y})=S_0$) are curved, even though we made use of the linear approximation with hyperplanes. Therefore we need to specify a sampling density $\rho(\bm{x}_k^{\rm exp})$ that is tuned to generate Monte Carlo integration points inside and outside the boundaries; the $p$-value expression for the original observable space then reads

\begin{align*}
 p &= \frac{1}{N}\sum\limits_{k=1}^N \frac{f(\bm{x}_k^{\rm exp}, \bm{\xi}_{\rm bf, B})}{\rho (\bm{x}_k^{\rm exp})} \: \Theta \left( S(\bm{x}_k^{\rm exp})-S_0\right) . 
\end{align*}

As our final formulation only depends on \otto{$S(\bm{x}_k^{\rm exp})$} and not on $\chi^2$ values, it is also valid if the p.d.f.s are not Gaussian. And dismissing the hyperplane approximation makes the problem harder to picture, but will not pose a problem to the numerical integration. To obtain a trustworthy result with small uncertainties within reasonable time, it is crucial how $\rho(\bm{x}_k^{\rm exp})$ is chosen. In our fits we use a customized sampling density which \otto{is} adjusted during the calculation.\\
The SM4 model, which I will discuss in Chapter \ref{SM4}, and the SM are not nested; furthermore, linearity is not satisfied and there are systematic errors \otto{involved}.
For our hypothesis test, we used Martin Wiebusch's program \textit{my}Fitter \cite{Wiebusch:2012en} to calculate the $p$-value. To test the performance of the SM4, we need to define our standard first. In particle physics this standard is the Standard Model which will be analyzed thoroughly in the following chapter. Only then we are able to discuss its extensions, where we will need to apply our knowledge about nestedness and linearity.

%% file: sm.tex
\chapter{The Standard Model} \label{SM}

The Standard Model of particle physics (SM) describes three of the four fundamental interactions of all elementary particles we know. (A summary can e.g.\ be found in \cite{Bohm:2001yx} and in \cite{Novaes:1999yn}.) It is based on an $SU(3)\otimes SU(2)\otimes U(1)$ gauge symmetry. The corresponding \otto{quantum numbers} are the colour, the weak isospin $I$ and the hypercharge $Y$, respectively.
The gauge couplings are the strong coupling $g_3$, the weak coupling $g_2$, and the hypercharge coupling $g_1$.
The particle content of the SM consists of elementary fermions and bosons, which transform according to \otto{the} representations of the gauge groups shown in 
Table \ref{tab:particles}. The fermions can be divided into quarks and leptons, which transform in the fundamental and singlet representation of $SU(3)$, respectively. Left-handed quarks and leptons transform as $SU(2)$ doublets $Q_j$ and $L_j$. For the right-handed fields, which are singlets under $SU(2)$ transformations, we differentiate between up-type quarks $u_j$ on the one hand and down-type quarks $d_j$ and the charged leptons $\ell _j$ on the other hand, which correspond to the $SU(2)$ isospin up and down components of the left-handed doublets.
Originally, \otto{the neutrinos $\nu _j$, which are the right-handed equivalent of the isospin up component of $L_j$,} do not belong to the SM fields. However, I will treat them just like the other fermions as massive Dirac particles.
The index $j$ denotes the generation the fields are attributed to; the SM has three generations, which each consist of an $SU(2)$ doublet of an up-type and a down-type quark, and one containing a charged lepton and a neutrino\otto{, as well as the right-handed partners.}\\[10pt]
\begin{tabular}{llll}
&\hspace*{30pt} $\;\;j=1$ & $\;\;j=2$ & $\;\;j=3$ \\[4pt]
Quark doublets: &\hspace*{30pt} $\left( \begin{array}{c} u \\ d \end{array}\right) $, \quad & $\left( \begin{array}{c} c \\ s \end{array}\right) $, \quad & $\left( \begin{array}{c} t \\ b \end{array}\right) $\\[3pt]
Lepton doublets: &\hspace*{30pt} $\left( \begin{array}{c} \nu _e \\ e \end{array}\right) $, \quad & $\left( \begin{array}{c} \nu _\mu \\ \mu \end{array}\right) $, \quad & $\left( \begin{array}{c} \nu _\tau \\ \tau \end{array}\right) $
\end{tabular}\\[12pt]
To every fermion one can attribute a specific quantum number called \textit{flavour}. Whereas the first generation would be sufficient to describe the constituents of atoms and thus explain most aspects of nature as we know it, for this work about high-energy physics primarily the heaviest fermions are of interest: the top quark $t$, the bottom quark $b$ and the $\tau $ lepton.\\
In addition to the elementary fermions, the SM also contains elementary bosons: The gauge vector bosons $G_3^a$, $G_2^a$ and $G_1$ transform as adjoint representations under $SU(3)$, $SU(2)$ and $U(1)$, respectively; they can be combined with the generators $T^a_n$ of the $n$th gauge group to form linear operators $\bm{G}_n=G_n^aT_n^a$ on some arbitrary representations. Finally, \otto{the} Higgs field $\Phi$ is a complex scalar $SU(2)$ doublet.

\begin{table}[htbp]
\centering
 \begin{tabular}{ll|ll}
  Fermion & Representation & Boson & Representation\\
  \hline
  $Q_j$ & $(\bm{3},\bm{2},1/3)$ & $G^a_3$ & $(\bm{8},\bm{1},0)$\\
  $u_j$ & $(\bm{3},\bm{1},4/3)$ & $G^a_2$ & $(\bm{1},\bm{3},0)$\\
  $d_j$ & $(\bm{3},\bm{1},-2/3)$ & $G_1$ & $(\bm{1},\bm{1},0)$\\
  $L_j$ & $(\bm{1},\bm{2},-1)$ & $\Phi$ & $(\bm{1},\bm{2},1)$\\
  $\nu_j$ & $(\bm{1},\bm{1},0)$ &  & \\
  $\ell_j$ & $(\bm{1},\bm{1},-2)$ &  & \\
 \end{tabular}
\caption[The elementary particles of the SM.]{The elementary particles of the SM and their $SU(3)$ and $SU(2)$ representations and hypercharge.} \label{tab:particles}
\end{table}

The SM Lagrangian can be split into four parts:\footnote{I will not list gauge fixing terms.}

\vspace*{-5pt}
\begin{align*}
 {\cal L}_{\text{SM}} &= {\cal L}_G+{\cal L}_F+{\cal L}_H+{\cal L}_Y
\end{align*}

The first term \otto{contains} the Yang-Mills Lagrangians of the gluons $G_3^a$, the weak bosons $G_2^a$, and $G_1$:

\vspace*{-8pt}
\begin{align*}
 {\cal L}_G &= -\frac{1}{4}\sum\limits_{n=1}^3\: \oto{\rm{tr}} \left[(\bm{F}_n)_{\mu \nu }(\bm{F}_n)^{\mu \nu}\right],
\end{align*}

where the field strength tensors \oto{$(\bm{F}_n)^{\mu \nu }$} are defined as

\begin{align*}
 (\bm{F}_n)^{\mu \nu} &= \partial ^\mu (\bm{G}_n)^\nu -\partial ^\nu (\bm{G}_n)^\mu -{\rm i}g_n[\bm{G}_n^\mu,\bm{G}_n^\nu ],
\end{align*}

a sum of gauge boson field derivatives and the commutator of the bosons. (The $U(1)$ commutator is zero.)\\
The next part of the Lagrangian consists of the kinetic terms of the fermions:

\begin{align}
 {\cal L}_F &= {\rm i}\gamma _\mu \sum\limits_{j=1}^3 \left( \oline{Q}_j \bm{D}^\mu Q_j +\oline{L}_j \bm{D}^\mu L_j + \oline{u}_j \bm{D}^\mu u_j + \oline{d}_j \bm{D}^\mu d_j + \oline{\ell}_j \bm{D}^\mu \ell _j + \oline{\nu}_j \bm{D}^\mu \nu _j\right) \label{eq:FermionLagrangian}
\end{align}

Here, the covariant derivatives

\vspace*{-8pt}
\begin{align*}
 \bm{D}^\mu &= \partial ^\mu -{\rm i}\sum\limits_{n=1}^3 g_n \bm{G}_n^\mu
\end{align*}

include the couplings to the gauge bosons.\\[10pt]
After the vector bosons and the fermions and their interactions, we now address the scalar sector of the SM. It contains one $SU(2)$ doublet $\Phi$, the so-called Higgs field:

\begin{align}
 {\cal L}_H &= (\bm{D}\mu \Phi )^\dagger (\bm{D}^\mu \Phi ) +\mu ^2 \Phi ^\dagger \Phi -\frac{\lambda }{4}(\Phi ^\dagger \Phi )^2 \label{eq:HiggsLagrangian}
\end{align}

All massive elementary particles receive their masses by coupling to the Higgs field. However, $\Phi$ needs to have a non-zero, but finite vacuum expectation value $v$ in order to fulfil this feature, which in turn means that the quadratic coupling $\mu^2$ and the quartic coupling $\lambda$ have to be positive. \otto{The fermion mass terms arise from the Yukawa Lagrangian}

\vspace*{-5pt}
\begin{align}
 {\cal L}_Y &= - \sum\limits_{j,k=1}^3 \left[ Y^d_{jk}\left( \oline{Q}_j \Phi \right) d_k +Y^u_{jk} \left( \oline{Q} _j{\rm i}\sigma _2 \Phi ^*\right) u_k \right. \nonumber \\
&\hspace*{60pt}\left. +Y^\ell _{jk} \left( \oline{L}_j \Phi \right) \ell _k +Y^\nu _{jk} \left( \oline{L}_j {\rm i}\sigma _2 \Phi ^*\right) \nu _k +{\rm h.c.}\right] .\label{eq:YukawaLagrangian}
\end{align}

\otto{It} describes the interaction between fermions and the Higgs \otto{field}. Here, the $Y^f_{jk}$ are the Yukawa matrices coupling the Higgs \otto{field} to the right-handed field $f$ of the generation $k$ and to a left-handed partner from generation $j$. In order to introduce mass terms in the SM, the vacuum expectation value of the $SU(2)$ doublet $\Phi$ has to be \otto{non-zero. Exploiting gauge symmetry, it can be brought to the form}

\vspace*{-5pt}
\begin{align*}
 \langle \Phi \rangle _0 &= \frac{1}{\sqrt{2}}\binom{0}{v}
\end{align*}

\otto{with $v>0$}. This \otto{vacuum expectation value} breaks the electroweak symmetry $SU(2)\otimes U(1)$ spontaneously to the electromagnetic symmetry group $U(1)_{\rm em}$, with the conserved electric charge given by

\vspace*{-8pt}
\begin{align*}
 Q &= \oto{I_3} + \frac{Y}{2},
\end{align*}

\oto{where $I_3$ is the third component of the weak isospin.}
The three arising Goldstone bosons \otto{become the} longitudinal components of \otto{three of the four electroweak gauge bosons}. The corresponding charge eigenstates are the electrically charged $W^+$ and $W^-$ with mass $m_W$ as well as two neutral states which can be rotated to the massive $Z$ \ottooo{boson} and to the massless photon $\gamma$ by the so-called weak mixing angle $\theta _w$. This angle also gives us the electrical charge coupling $e=g_2\sin \theta _w$.
The $Z$ and $\gamma$ boson couplings to left-handed and right-handed fermions can be rearranged into a vector part proportional to $\gamma _\mu$ and an axial part with Dirac structure $\gamma _\mu \gamma _5$. Let $g_{Vf}$ and $g_{Af}$ be the corresponding couplings of the $Z$ to fermion $f$, which will be used in the following.\\
Furthermore, we \otto{find one real massive Higgs boson} $H$. \ottooo{Such a \otto{particle} was postulated in 1964 by Peter Higgs \cite{Higgs:1964pj}, but \otto{it has been} found only in 2012 at the LHC (see Sect.\ \ref{SMhiggssearches}), thus completing the particle content of the SM from the experimental point of view.}\\
As for the fermions, ${\cal L}_Y$ contains the mass terms after electroweak symmetry breaking.
Neutrino masses are very small; there are only upper limits on their masses available, \otto{however,} two neutrino mass eigenvalue differences are known to be non-zero \otto{ (see e.g.\ \cite{Beringer:1900zz} for a review)}.
Diagonalizing the $3\times 3$ Yukawa couplings matrices yields the mass eigenvalues; the flavour eigenstates are accordingly transformed to mass eigenstates. The price one has to pay are flavour-changing couplings in ${\cal L}_F$, generated by non-zero off-diagonal elements of the product of the down-type and the up-type transformation matrices.
For the quarks, this product is called quark mixing matrix $V$ or simply CKM matrix, according to its developers Cabibbo, Kobayashi and Maskawa \cite{Cabibbo:1963yz,Kobayashi:1973fv}. It is a $3\times 3$ matrix and mixes the quark flavour eigenstates. As a unitary matrix, it has nine degrees of freedom, of which five are unphysical. The remaining four parameters can be expressed as three rotation angles $\theta _{12}$, $\theta _{13}$ and $\theta _{23}$, which describe the size of the mixing between two generations each, and one \otto{complex} phase $\delta_{13}$. With the abbreviations $c_{ij}=\cos \theta_{ij}$ and $s_{ij}=\sin \theta_{ij}$, the standard parametrization of the CKM matrix reads \cite{Beringer:1900zz}

\begin{align}
 V&\equiv \left(
 \begin{array}{ccc}
  V_{ud} & V_{us} & V_{ub}\\
  V_{cd} & V_{cs} & V_{cb}\\
  V_{td} & V_{ts} & V_{tb}\\
 \end{array}
 \right) 
 = \left( \begin{array}{lll}
  c_{12}c_{13} & s_{12}c_{13} & s_{13}{\rm e}^{-{\rm i}\delta _{13}}\\[13pt]
  -s_{12}c_{23} & c_{12}c_{23} & s_{23}c_{13}\\
  -c_{12}s_{23}s_{13}{\rm e}^{{\rm i}\delta _{13}} & -s_{12}s_{23}s_{13}{\rm e}^{{\rm i}\delta _{13}} & \\[13pt]
  s_{12}s_{23} & -c_{12}s_{23} & c_{23}c_{13}\\
  -c_{12}c_{23}s_{13}{\rm e}^{{\rm i}\delta _{13}} & -s_{12}c_{23}s_{13}{\rm e}^{{\rm i}\delta _{13}} & \\
  \end{array}\right) . \label{eq:CKMmatrixSM}
\end{align}

Also in the lepton sector, such a matrix exists, yet throughout this thesis, I will assume it to be diagonal \otto{and neglect its parameters}.\\
Apart from its gauge symmetry structure, one can formulate three discrete symmetry transformations for ${\cal L}_{\text{SM}}$: replacing particles by their antiparticles is called ${\cal C}$ symmetry, the parity operation ${\cal P}$ is equivalent to mirroring the space coordinates just as ${\cal T}$ inverses the time. ${\cal L}_{\text{SM}}$ is not symmetric under \otto{these} transformations, but as \otto{a} Lorentz transformation invariant \otto{local} quantity, the SM Lagrangian must be invariant under the combination ${\cal CPT}$. While ${\cal P}$ is violated by the weak interaction \cite{Wu:1957my}, the combination ${\cal CP}$ seemed to be conserved until in 1964 an asymmetry in neutral meson decays showed that ${\cal L}_{\text{SM}}$ is not totally ${\cal CP}$ symmetric \cite{Christenson:1964fg}. This result was the original motivation for the postulation of a third generation \cite{Kobayashi:1973fv} \otto{because ${\cal CP}$ violation requires physical complex phases in the Lagrangian. Only if there are at least three fermion generations, the complex phases of the CKM matrix cannot be compensated by phase rotations of the quark fields, and ${\cal L}_F$ contains complex quark couplings}.

\section{Parameters}

After electroweak symmetry breaking, ${\cal L}_{\text{SM}}$ contains 21 real parameters:\footnote{In principle, the $SU(3)$ part of the SM could contain one additional ${\cal CP}$ \otto{violating parameter $\theta$}, but \otto{$|\theta|$ is very small and consistent with zero}.}

\begin{itemize}
 \item 3 gauge couplings $g_1=e/\cos \theta _w$, $g_2=e/\sin \theta _w$, $g_3$,
 \item $\mu $ and the quartic Higgs coupling $\lambda $ from ${\cal L}_H$, connected via $v=2\sqrt{\frac{\mu ^2}{\lambda}}$,
 \item 6 quark masses $m_u$, $m_d$, $m_s$, $m_c$, $m_b$, $m_t$, and 6 lepton masses $m_e$, $m_\mu$, $m_\tau$,$m_{\nu_e}$, $m_{\nu_\mu}$, $m_{\nu_\tau}$, which are identified by $\frac{1}{\sqrt{2}}\tilde Y^f_{jj} v$, where $\tilde Y^f_{jj}$ are the eigenvalues of the diagonalized Yukawa matrices $Y^f_{jk}$ from \eqref{eq:YukawaLagrangian}, and $j$ is the generation index,
 \item 3 quark mixing angles $\theta _{12}$, $\theta _{13}$, $\theta _{23}$, and
 \item 1 quark mixing phase, denoted as $\delta _{13}$.
\end{itemize}

These parameters are a priori free, but \otto{I assume perturbative couplings for the high energy observables that I want to discuss in the following. I}t is useful to switch to a handier set of parameters; all masses \otto{well} below the $Z$ scale (i.e.\ all fermion masses except for the top quark mass) can be treated as fixed since their \otto{uncertainties} are too small to affect our observables.
As next step, I relate $v$ to the well-measured Fermi constant $G_F=1/(\sqrt{2}v^2)$ and also treat it as fixed. Instead of the electromagnetic coupling $e$ and $\theta _w$, I want to use the $Z$ boson mass and the hadronic contribution to the fine structure constant $\alpha _{\rm em}\equiv \frac{e^2}{4\pi }$ at the $Z$ scale. They are related to the original \otto{parameters} by

\begin{align*}
 m_Z &= \frac{ev}{\sin (2\theta _w)} \sqrt{\frac{1}{1-\Delta r}}\\
 \Delta \alpha_\text{had}^{(5)}(m_Z) &= 1-\frac{\alpha _{\rm em}(0)}{\alpha _{\rm em}(m_Z)}-\Delta \alpha_\text{lep}-\Delta \alpha_\text{top},
\end{align*}

where \otto{$\Delta r$ includes higher order corrections, and} $\Delta \alpha_\text{lep}$ and $\Delta \alpha_\text{top}$ denote the leptonic and the $t$ loop contribution to the photon propagator at the $Z$ scale,
of which the error is negligible. Also the fine structure constant is very well known at the $Z$ scale as well as for low energy processes; the main error of $\alpha _{\rm em}(m_Z)$ stems from $\Delta \alpha_\text{had}^{(5)}$ \cite{ALEPH:2005ab}. The $SU(3)$ coupling constant is traded for the strong coupling $\alpha _s=g_3^2/(4\pi)$. Next, the Higgs mass can be derived from the quadratic coupling of \otto{$\Phi$}, $m_H=\sqrt{2}\mu$. Finally, taking the quark mixing parameters as the\oto{y} are, we end up with nine fit parameters:

\begin{align}
 m_t, \quad m_Z, \quad \Delta \alpha_\text{had}^{(5)},\quad \alpha _s,\quad m_H,\quad \theta _{12},\quad \theta _{13},\quad \theta _{23},\quad \delta _{13} \label{eq:SMparams}
\end{align}

Now that we have defined the model, we can compare it with experimental results.
For all fits in this thesis I used the CKMfitter package which treats systematic errors in the \textit{R}fit scheme \cite{Hocker:2001xe} as introduced in Sect.\ \ref{systematicerrors}. The CKM angles and the phase were constrained by taking mainly the PDG values for the CKM matrix elements and a CKMfitter look-up table for the unitarity triangle angle $\gamma $, compare App.\ \ref{inputs}. I will not explicitly mention them in the following. Before discussing the measurements by the detector collaborations at the Large Hadron Collider (LHC), I want to address the results obtained at \otto{three} preceding colliders: the LEP collider\otto{, SLC} and the Tevatron.
The \otto{first two} started in the years 1989 \otto{and 1992, respectively,} to collide electrons and positrons at a centre-of-mass energy of \otto{$\sqrt{s}\approx m_Z$ and later} at LEP up to $\sqrt{s}=209$\:GeV\otto{. The SLC was shut down in 1998 and the LEP collider in 2000 \cite{ALEPH:2005ab}. The Tevatron}, a proton-antiproton collider, ran from 1983 to 2011 at maximally $\sqrt{s}=1.96$\:TeV \otto{\cite{Tevatron:2013tl}}.

\section{Electroweak precision observables} \label{SMEWPO}

Amongst the main results of the LEP collider \otto{and SLC} experiments are precision measurements: After the discovery of the $W$ and $Z$ bosons, their fundamental features and decay properties could be determined with relatively small statistical uncertainties due to the large amount of collected data. \ottooo{Measuring electron-positron collisions}, the LEP \otto{and SLC} experiments benefited from the low background, due to which the systematic errors are small. Over the years, a certain set of observables evolved that was used to test the Standard Model parameters at the $Z$ scale; it is commonly referred to as electroweak precision observables (EWPO). \otto{As their experimental values are mainly based on LEP measurements, I will only mention the SLC data explicitly, where its contribution is important.}
(Most of the EWPO description is taken from \cite{ALEPH:2005ab} and \cite{Novaes:1999yn}). They consist of the following quantities:\\
The $Z$ mass $m_Z$ is assigned to the peak of the bosonic resonance at approximately $91$\:GeV. Its total decay width $\Gamma_Z$ is extracted from a beam energy scan around the centre-of-mass energy $\sqrt{s}=m_Z$.
On the theory side, it is the sum of the hadronic decay width \otto{$\Gamma_Z^{\rm had} = \Gamma_Z^u+\Gamma_Z^d+\Gamma_Z^s+\Gamma_Z^c+\Gamma_Z^b$}, the width of decays into charged leptons \otto{$\Gamma_Z^{\rm lept} = \Gamma_Z^e+\Gamma_Z^\mu+\Gamma_Z^\tau$}, and the invisible decay width $\Gamma_Z^{\rm inv} = N_\nu^{\rm light} \Gamma_Z^\nu$ into light neutrinos. $\Gamma_Z^f$ is the partial decay width of a $Z$ decaying into the final state $f\oline{f}$. It can be related to the real effective vector and axial couplings of fermions to the $Z$ boson, $g_{Vf}$ and $g_{Af}$:

\begin{align*}
 \Gamma_Z^f &= N^f_C\frac{G_F m_Z^3}{6\sqrt{2}\pi}\left( g_{Vf}^2+g_{Af}^2\right)
\end{align*}

The colour factor $N^f_C$ of fermion $f$ is $3$ for quarks and $1$ for leptons. \oto{(I do not explicitly show radiator factors or non-factorisable contributions here.)} Independent measurements of $\Gamma_Z$ and all visible partial decay widths were used to fit the number of neutrinos which have a mass smaller than $m_Z/2$ to

\begin{align}
 N_\nu^{\rm light} &= 2.9840 \pm 0.0082. \label{Nnulight}
\end{align}

The measured total hadronic $Z$ cross section can be compared to the theoretical expectation

\begin{align*}
 \sigma _{\rm had}^0 &= \frac{12\pi}{m_Z^2}\frac{\Gamma_Z^e\Gamma_Z^{\rm had}}{\Gamma_Z^2}.
\end{align*}

Here and in the following, the index $0$ indicates the theoretically corrected pole observables that were extracted from the measurements. (Strictly speaking, these quantities are not observables but merely pseudo-observables, i.e.\ parameters determined by a fit to the observed cross sections and asymmetries.) Another ratio of decay widths is

\begin{align*}
 R_q^0 & = \frac{\Gamma_Z^q}{\Gamma_Z^{\rm had}},
\end{align*}

the partial width of a $Z$ decaying into a specific quark pair, normalized to the total hadronic decay width. The corresponding leptonic quantity is defined reversely:

\begin{align*}
 R_\ell^0 & = \frac{\Gamma_Z^{\rm had}}{\Gamma_Z^\ell}
\end{align*}

It is the ratio of the total hadronic $Z$ width and the width of a $Z$ decaying into a charged lepton pair. Here, lepton universality is assumed, which means that $g_{V\ell}$ and $g_{A\ell}$ are equal for the charged leptons and thus $R_e^0 =R_\mu^0 =R_\tau^0 $.\\
Further important observables are forward-backward asymmetries. They are defined as the difference between \ottooo{the number of} \otto{events with the} final state particles scattered into forward and backward direction as compared to the incoming electron beam, normalized to \otto{the sum of all events}:

\begin{align*}
 A_\text{FB} &= \frac{N_F - N_B}{N_F + N_B}
\end{align*}

At LEP, forward-backward asymmetries have been measured for decays into $c$ and $b$ quarks as well as charged leptons (again assuming lepton universality). On the theory side, they can be related to the asymmetry parameters ${\cal A}_e$ and ${\cal A}_f$ stemming from the tree-level differential cross-section expression for the process $e^+e^-\to f\oline{f}$:

\begin{align*}
 \frac{{\rm d}\sigma _{f\oline{f}}}{{\rm d}\cos \theta } &= \frac{3}{8}\sigma ^{\rm tot}_{f\oline{f}}\left[ \left( 1-{\cal P}_e{\cal A}_e\right) \left( 1+\cos ^2\!\theta \right) + 2 \left( {\cal A}_e - {\cal P}_e\right) {\cal A}_f \cos \theta \right] ,
\end{align*}

where $\theta $ is the scattering angle between the incoming electron and the decay product $f$, and ${\cal P}_e$ is the electron beam polarization. The fermionic asymmetry ${\cal A}_f$ can be expressed in terms of the ratio of the above-mentioned effective couplings of fermions to the $Z$:

\begin{align*}
 {\cal A}_f &= \frac{2g_{Vf}/g_{Af}}{1+(g_{Vf}/g_{Af})^2}
\end{align*}

${\cal A}_e$ can \otto{also} be identified with the left-right asymmetry of electrons, which is defined as \otto{the} difference of the cross sections of left-handed and right-handed electrons in the initial state, normalized to their sum. Once again assuming lepton universality, I take the effective leptonic asymmetry ${\cal A}_\ell$ from \cite{Flacher:2008zq}, which is the combination of all three leptonic asymmetries from the LEP detectors as well as the left-right asymmetry measured by the SLD \otto{detector at the SLC}. In terms of these asymmetries, the expression for the forward-backward asymmetries reads

\begin{align}
 A^{0,f}_\text{FB} &= \frac{3}{4}{\cal A}_e{\cal A}_f.
\end{align}

Another asymmetry is the hadronic forward-backward charge flow: \otto{The difference between charged jet events in forward and backward direction} with respect to the incoming electron beam allows us to extract the squared sine of the effective weak mixing angle, which can be written as

\vspace*{-5pt}
\begin{align}
 \sin^2\theta_\ell^\text{eff} &= \frac{1}{4}\left( 1-\frac{g_{V\ell}}{g_{A\ell}}\right)
\end{align}

Apart from the $Z$ pole observables, I also take into account the $W$ boson mass, which at tree level can be expressed by $m_W=g_2v/2$. It is the first of the mentioned EWPO for which the Tevatron combination is competitive with the LEP value. I also use the total $W$ boson decay width $\Gamma_W$ in my fit, even if its measurement is not precise enough to yield a strong constraint. Furthermore, I take the Tevatron combination of the top \otto{quark} pole mass $m_t^{\text{\tiny pole}}$, the $\Delta \alpha_\text{had}^{(5)} (m_Z)$ determination from \cite{Davier:2010nc} and the $\alpha _s (m_Z)$ extraction from tau decays \cite{Baikov:2008jh} \otto{as inputs}.\\
In short, that amounts to a total of 18 observables, of which the first four coincide with parameters from our above parametrization \eqref{eq:SMparams}:

\begin{align*}
 &m_t^{\text{\tiny pole}}, \quad m_Z, \quad \Delta \alpha_\text{had}^{(5)},\quad \alpha _s,\quad \Gamma _Z,\quad \sigma _{\rm had}^0,\quad R_b^0,\quad R_c^0,\quad R_\ell ^0,\\[5pt] &A_\text{FB}^{0,b},\quad A_\text{FB}^{0,c},\quad A_\text{FB}^{0,\ell },\quad {\cal A}_b,\quad {\cal A}_c,\quad {\cal A}_\ell ,\quad m_W,\quad \Gamma_W, \quad \sin^2\theta_\ell^\text{eff}
\end{align*}

\begin{figure}[htbp]
\centering
\begin{picture}(230,360)(80,50)
    \SetWidth{0.2}
    \SetColor{Gray}
    \Line(170,104)(170,404)
    \Line(170,75)(170,77)
    \Line(200,104)(200,404)
    \Line(200,75)(200,77)
    \Line(230,75)(230,404)
    \Line(290,75)(290,404)
    \Line(320,75)(320,404)
    \SetWidth{5.5}
    \SetColor{Orange}
    \Line(260,394)(260.54,394) 
    \Line(260,376)(260.01,376) 
    \Line(260,358)(258.50,358) 
    \Line(260,340)(260.01,340) 
    \Line(260,322)(253.35,322) 
    \Line(260,304)(311.89,304) 
    \Line(260,286)(328.91,286) 
    \Line(260,268)(258.67,268) 
    \Line(260,250)(288.68,250) 
    \Line(260,232)(179.75,232) 
    \Line(260,214)(231.97,214) 
    \Line(260,196)(282.80,196) 
    \Line(260,178)(242.42,178) 
    \Line(260,160)(262.08,160) 
    \Line(260,142)(298.17,142) 
    \Line(260,124)(268.00,124) 
    \Line(260,106)(255.14,106) 
    \Line(260,88)(283.75,88) 
    \SetWidth{1.0}
    \SetColor{Black}
    \Text(-20,415)[cl]{\Black{\textbf{Observable}}}
    \Text(65,415)[cl]{\Black{\textbf{Best-fit value}}}
    \Text(172,415)[cl]{\Black{\textbf{Deviation}}}
    \Text(360,415)[cl]{\Black{$\bm{\Delta \chi ^2}$}}
    \Text(-10,394)[cl]{\Black{$m_t^{\text{\tiny pole}} $ [GeV]}}
    \Text(-10,376)[cl]{\Black{$m_Z $ [GeV]}}
    \Text(-10,358)[cl]{\Black{$\Delta\alpha_\text{had}^{(5)} $}}
    \Text(-10,340)[cl]{\Black{$\alpha_s $}}
    \Text(-10,322)[cl]{\Black{$\Gamma_Z $ [GeV]}}
    \Text(-10,304)[cl]{\Black{$\sigma^0_\text{had} $ [nb]}}
    \Text(-10,286)[cl]{\Black{$R_b^0 $}}
    \Text(-10,268)[cl]{\Black{$R_c^0 $}}
    \Text(-10,250)[cl]{\Black{$R_\ell^0 $}}
    \Text(-10,232)[cl]{\Black{$A_\text{FB}^{0,b} $}}
    \Text(-10,214)[cl]{\Black{$A_\text{FB}^{0,c} $}}
    \Text(-10,196)[cl]{\Black{$A_\text{FB}^{0,\ell} $}}
    \Text(-10,178)[cl]{\Black{${\cal A}_b $}}
    \Text(-10,160)[cl]{\Black{${\cal A}_c $}}
    \Text(-10,142)[cl]{\Black{${\cal A}_\ell $}}
    \Text(-10,124)[cl]{\Black{$m_W $ [GeV]}}
    \Text(-10,106)[cl]{\Black{$\Gamma_W $ [GeV]}}
    \Text(-10,88)[cl]{\Black{$\sin^2\theta_\ell^\text{eff} $}}
    \Text(80,394)[cl]{\Black{$172.42$}} 
    \Text(80,376)[cl]{\Black{$91.1876$}} 
    \Text(80,358)[cl]{\Black{$0.02758$}} 
    \Text(80,340)[cl]{\Black{$0.1186$}} 
    \Text(80,322)[cl]{\Black{$2.4957$}} 
    \Text(80,304)[cl]{\Black{$41.477$}} 
    \Text(80,286)[cl]{\Black{$0.21477$}} 
    \Text(80,268)[cl]{\Black{$0.1722$}} 
    \Text(80,250)[cl]{\Black{$20.743$}} 
    \Text(80,232)[cl]{\Black{$0.1035$}} 
    \Text(80,214)[cl]{\Black{$0.0740$}} 
    \Text(80,196)[cl]{\Black{$0.0163$}} 
    \Text(80,178)[cl]{\Black{$0.935$}} 
    \Text(80,160)[cl]{\Black{$0.668$}} 
    \Text(80,142)[cl]{\Black{$0.1476$}} 
    \Text(80,124)[cl]{\Black{$80.377$}} 
    \Text(80,106)[cl]{\Black{$2.092$}} 
    \Text(80,88)[cl]{\Black{$0.2315$}} 
    \Text(364,394)[cl]{\Black{$0.09$}} 
    \Text(364,376)[cl]{\Black{$0.42$}} 
    \Text(364,358)[cl]{\Black{$0.26$}} 
    \Text(364,340)[cl]{\Black{$0.00$}} 
    \Text(364,322)[cl]{\Black{$0.10$}} 
    \Text(364,304)[cl]{\Black{$2.82$}} 
    \Text(364,286)[cl]{\Black{$4.61$}} 
    \Text(364,268)[cl]{\Black{$0.16$}} 
    \Text(364,250)[cl]{\Black{$0.66$}} 
    \Text(364,232)[cl]{\Black{$6.40$}} 
    \Text(364,214)[cl]{\Black{$0.50$}} 
    \Text(364,196)[cl]{\Black{$0.63$}} 
    \Text(364,178)[cl]{\Black{$0.10$}} 
    \Text(364,160)[cl]{\Black{$0.01$}} 
    \Text(364,142)[cl]{\Black{$2.09$}} 
    \Text(364,124)[cl]{\Black{$0.21$}} 
    \Text(364,106)[cl]{\Black{$0.03$}} 
    \Text(364,88)[cl]{\Black{$0.64$}} 
    \Text(168,62)[cb]{\Black{\large $-3$}}
    \Text(198,62)[cb]{\Black{\large $-2$}}
    \Text(228,62)[cb]{\Black{\large $-1$}}
    \Text(288,62)[cb]{\Black{\large $+1$}}
    \Text(318,62)[cb]{\Black{\large $+2$}}
    \SetWidth{0.7}
    \Line(165,77)(165,404)
    \Line(165,77)(340,77)
    \Line(165,404)(340,404)
    \Line(340,77)(340,404)
    \Line(260,77)(260,404)
    \put(169,81){\includegraphics[width=0.08\linewidth]{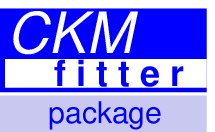}}
\end{picture}
\caption[SM deviations before the Higgs discovery.]{Deviations of the EWPO from the best-fit point in the SM fit before the Higgs discovery.
\otto{(As defined in Eq.\ \eqref{eq:chi}, the deviation of the observable $X_i$ is $\left( x_i^{\rm exp}-x_i^{\rm theo}\right) / \sigma _i$.)}
The experimental inputs can be found in App.\ \ref{inputs}. I also present the individual $\chi^2$ contributions in the last column.}
\label{fig:deviationsSM}
\end{figure}

\oto{For the fits, I linked the subroutine \texttt{DIZET} from the \texttt{Zfitter} code \cite{Bardin:1989tq,Bardin:1999yd,Arbuzov:2005ma} to CKMfitter. \texttt{DIZET} calculates the EWPO including higher order corrections. Moreover,} I created a general multi-purpose interface which can be used to connect any external programs to the CKMfitter program, see App.\ \ref{parallelize}. (\otto{E}ven if \texttt{Zfitter} includes a minimization routine, in my fits the minimization of the parameters was performed by CKMfitter.) The two-loop electroweak corrections for $R_b^0$ from \cite{Freitas:2012sy}, which are not implemented in \texttt{Zfitter}, have been included. They increase the deviation of $R_b^0$ from less than $1$ to more than $2$.
Aside from the four CKM matrix parameters, I used $m_Z$, $\alpha _s(m_Z)$, $\Delta \alpha_\text{had}^{(5)}(m_Z)$, $m_t^{\text{\tiny pole}}$ and $m_H$ as fit parameters. The Higgs mass is not constrained by direct measurements in this section, reproducing the pre-LHC status. All numerical inputs for the electroweak precision observables that I used can be found in App.\ \ref{inputs}. The best-fit values of the single quantities as well as their deviations in the SM fit can be found in Fig.\ \ref{fig:deviationsSM}. I also list the difference $\Delta \chi^2$ between the minimal $\chi^2$ values of the complete fit and the prediction fit as defined in Sect.\ \ref{gaussian}. The $b$ quark forward-backward asymmetry and the $R_b^0 $ ratio exhibit the largest deviations with absolute values greater than $2$. They also yield the largest contributions to the total $\chi^2_{\rm min}$ of the EWPO fit, which is $21.21$.

\begin{figure}[htbp]
 \centering
 \subfigure[]{\includegraphics[width=0.45\linewidth,bb=90 310 500 700,clip]{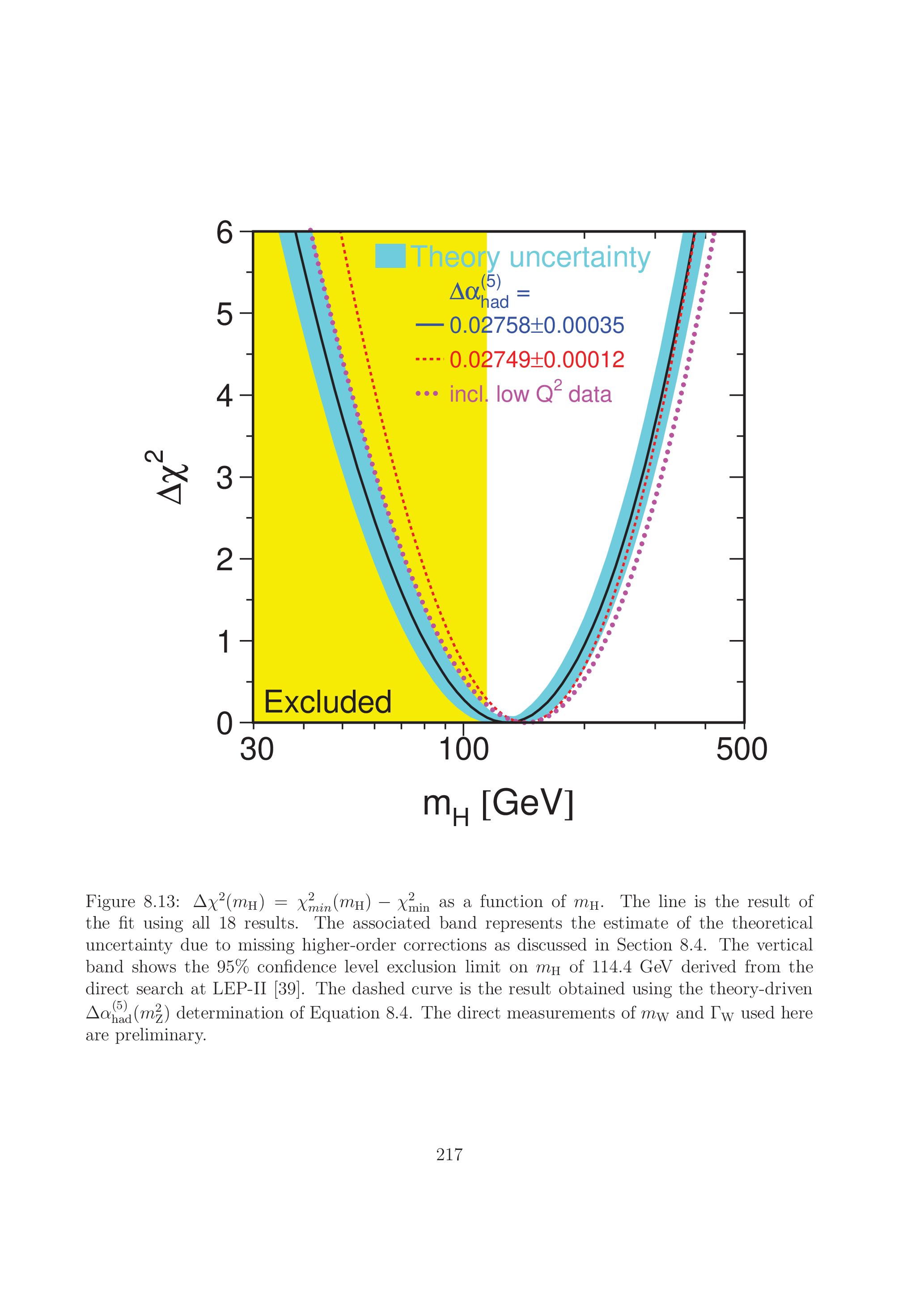}}
 \qquad
 \subfigure[]{\begin{picture}(200,150)(0,0)
	      \put(0,-10.7){\includegraphics[width=0.4762\linewidth]{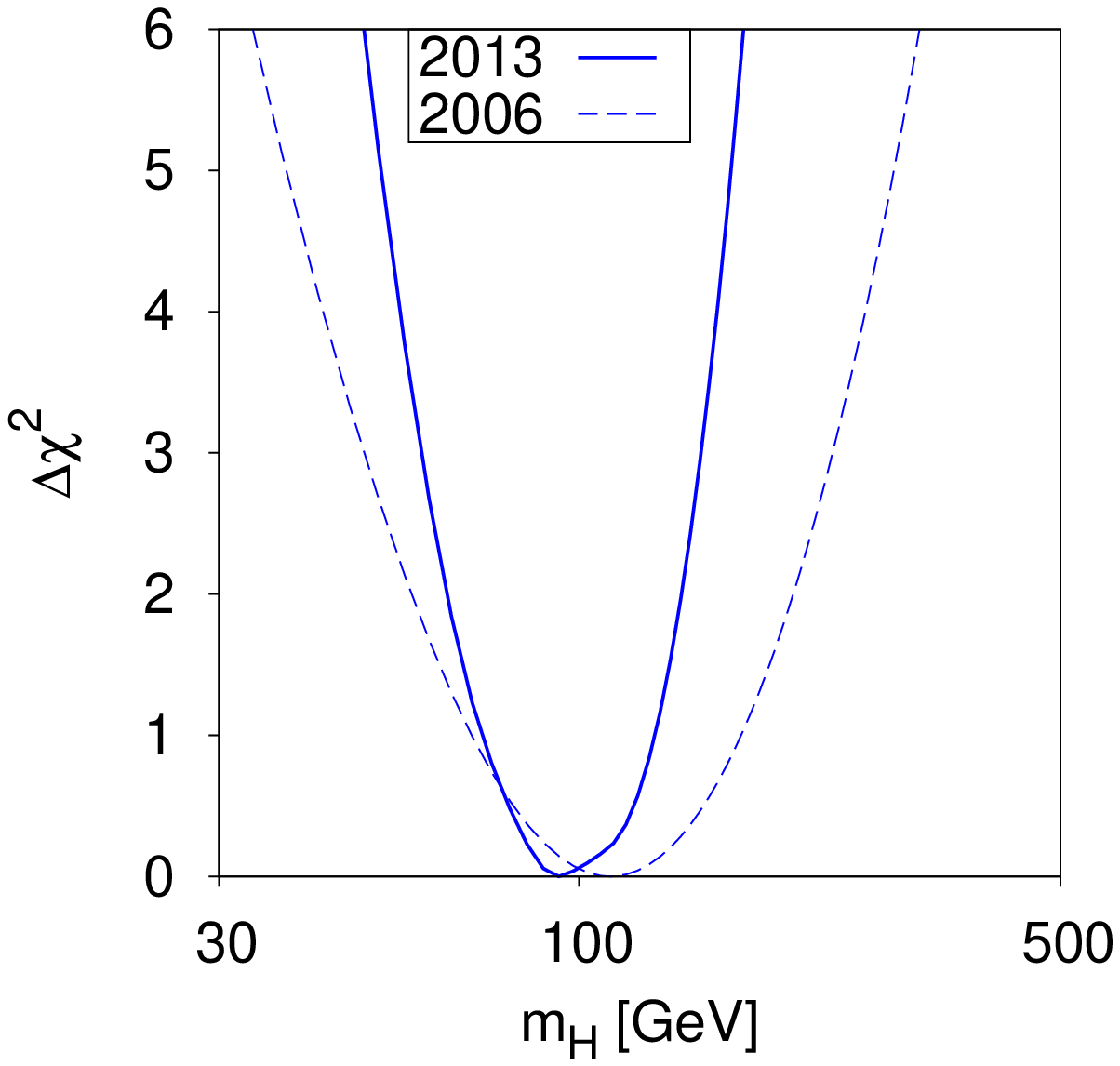}}
	      \put(150.7,32.9){\includegraphics[width=0.08\linewidth]{Images/CKMfitterPackage.eps}}
	      \end{picture}
	     }
 \caption[Historic and new blue-band plot.]{The blue-band plot from \cite{ALEPH:2005ab} on the left shows the $\chi^2$ dependence of the Higgs mass $m_H$ in a global fit to the EWPO. The theoretical errors determine the width of the blue band. The yellow range is excluded by direct Higgs searches at the LEP collider. In my fit on the right, the theoretical errors were taken into account using the Rfit scheme (cf. \otto{Sect.}~\ref{systematicerrors}). The dashed line is a fit using the old inputs from \cite{ALEPH:2005ab}.}
 \label{fig:blueband}
\end{figure}

Here the difference between the squared deviation from the best-fit point and the $\Delta \chi^2$ is obvious: While $\Gamma_Z$ and ${\cal A}_b$ have the very same $\Delta \chi^2$, the best-fit deviation of the $Z$ width is smaller than the one of the $b$ asymmetry. \otto{This means that the larger discrepancy between experiment and theory of ${\cal A}_b$ is accommodated just as well by shifts of the other observables when comparing the prediction fit with the complete fit.}\\
\otto{Due to their small uncertainties, the EWPO yield strong constraints to physics beyond the SM, because effects of heavy particles in loop corrections have not been measured.
Often when new physics models are analyzed,} the electroweak precision fit is taken into account using the so-called oblique parameters $S$, $T$ and $U$, introduced by Peskin and Takeuchi \cite{Peskin:1990zt,Peskin:1991sw}. In the SM, they are zero by definition. They can be used for heavy non-SM particles, if three requirements are met: the electroweak gauge part of the new theory has to be SM-like ($SU(2)\otimes U(1)$ before spontaneous symmetry breaking), the new particles must be heavier than the $Z$ scale, and there must be no vertex corrections from the heavy particles. I will discuss their applicability in the next two chapters in the respective EWPO section.\\
As already stated, until last year the only unknown SM parameter, which could only be roughly extracted from the EWPO, was the Higgs mass. Even if the EWPO depend only logarithmically on \otto{$m_H$} \ottooo{\cite{Veltman:1976rt}}, they constituted the strongest available upper bounds on \otto{it}. This was usually presented in the so-called ``blue-band plot'', the minimal $\chi^2$ value as a function of $m_H$ resulting from a global fit with $m_H$ fixed. The plot before the Higgs discovery from \cite{ALEPH:2005ab} and my own fit are displayed in Fig.\ \ref{fig:blueband}(a) and \ref{fig:blueband}(b), respectively.
Using inputs from \cite{ALEPH:2005ab}, I could reproduce the old blue-band plot. I also show the more restrictive fit including the latest inputs for the EWPO from Table \ref{tab:EWPOinputs} in App.\ \ref{inputs}. Its Higgs mass prediction is $m_H=93^{+35}_{-21}$\:GeV. The historic blue-band plot is also based on \texttt{Zfitter} fits; but since no \textit{R}fit method was used, the theoretical uncertainty is \otto{overlaid in the blue band, while in Fig.\ \ref{fig:blueband}(b) the theoretical uncertainty is contained in the $\Delta \chi^2$ value.}

The LEP detectors could only set a lower bound of $114.4$\:GeV at $95\%$ CL to $m_H$ \cite{Barate:2003sz}, corresponding to the yellow-shaded excluded region in Fig.\ \ref{fig:blueband}(a). In the year 2010, Tevatron Higgs searches could also exclude a Higgs mass range between $162$ and $166$\:GeV at $95\%$ CL \cite{Aaltonen:2010yv}. The actual discovery, however, could be achieved by the LHC detectors ATLAS and CMS.

\section{Higgs searches} \label{SMhiggssearches}

On \otto{4th July} 2012, the ATLAS and CMS collaborations announced the discovery of a new boson ``at a mass near 125 GeV'', for which they \otto{claimed} a local significance of $5.9\:\sigma$ and $5.0\:\sigma$, respectively, in the corresponding publications \cite{Aad:2012tfa,Chatrchyan:2012ufa}. This statement was based on the evaluation of the 2011 data set of around $5$\:fb$^{-1}$, taken at a centre-of-mass energy of $\sqrt{s}=7$\:TeV, as well as \otto{up to} $5$\:fb$^{-1}$ from the 2012 run at $\sqrt{s}=8$\:TeV.
The complete 2011 and 2012 Higgs search data were presented at the Moriond conferences in 2013 and comprise up to $25$\:fb$^{-1}$. Also the Tevatron collaborations of the CDF and D${\emptyset}$ detectors declared a statistical significance of 3.1$\sigma$ at 125 GeV in their latest combined analysis \cite{CDF:2013kxa} which uses up to $10\:fb^{-1}$ \otto{of integrated luminosity}.
In the context of the SM, this new boson is interpreted as the Higgs boson $H$ \otto{introduced} at the beginning of this chapter.
More detailed analyses concentrating on further characteristics like the spin are also in good agreement with SM expectations \cite{ATLAS-CONF-2013-034,ATLAS-CONF-2013-040,CMS-PAS-HIG-13-005}.
Therefore, I will refer to th\otto{is} discovery as the Higgs discovery in this work. Whereas the Tevatron was a proton-antiproton collider with a maximal centre-of-mass energy of $\sqrt{s}=1.96$\:TeV, the LHC used two proton beams with $\sqrt{s}=7$\:TeV in 2011 and with $\sqrt{s}=8$\:TeV in the 2012 run.
At LHC and Tevatron, the main Higgs production processes are gluon-gluon fusion (ggF), vector boson fusion (VBF), $W$ and $Z$ associated production (WH and ZH, sometimes combined to VH), as well as $t\oline{t}$ associated production (ttH). 
I subsume $WW$ and $ZZ$ fusion in one VBF category but mostly consider $WH$ and $ZH$ separately, because in contrast to the two fusion processes the vector boson associated Higgs productions have distinguishable signatures in the final state \otto{and hence can be separated into different observables}.
The five final states of a Higgs decay that \otto{are separable from background} are $b\oline{b}$, $WW^*$, $ZZ^*$, $\tau \tau$ as well as $\gamma \gamma$. Since $m_W$ and $m_Z$ are greater than $m_H/2$, only one of the produced massive bosons can be on the mass shell; the other one is virtual, which is denoted by the asterisk. The overview in Fig.\ \ref{fig:higgsproddecay} shows the leading order processes and also assigns the relative size of the individual production cross sections and the partial decay widths.

\begin{figure}[htbp]
 \begin{picture}(380,75)(-25,0)
\SetWidth{0.5}
\Text(-25,26)[lb]{\Black{Production:}}
\Gluon(66,10)(88,27){2}{4}
\Gluon(66,50)(88,33){2}{4}
\DashLine(93,30)(112,30){3}
\Vertex(90,30){5}
\Photon(139,46)(159,31){1.5}{3.5}
\Photon(159,29)(139,14){1.5}{3.5}
\DashLine(163,30)(182,30){3}
\Vertex(160,30){2}
\Photon(229,30)(205,30){1.5}{4}
\Photon(231,31)(243,50){1.5}{3}
\DashLine(233,30)(252,30){3}
\Vertex(230,30){2}
\Photon(299,30)(275,30){1.5}{4}
\Photon(301,31)(313,50){1.5}{3}
\DashLine(303,30)(322,30){3}
\Vertex(300,30){2}
\Gluon(347,47)(370,47){2}{3}
\Gluon(370,13)(347,13){2}{3}
\Line[arrow,arrowpos=0.5,arrowlength=1,arrowwidth=0.3,arrowinset=0.2](390,47)(370,47)
\Line[arrow,arrowpos=0.5,arrowlength=1,arrowwidth=0.3,arrowinset=0.2](370,47)(370,30)
\Line[arrow,arrowpos=0.5,arrowlength=1,arrowwidth=0.3,arrowinset=0.2](370,30)(370,13)
\Line[arrow,arrowpos=0.5,arrowlength=1,arrowwidth=0.3,arrowinset=0.2](370,13)(390,13)
\DashLine(373,30)(392,30){3}
\Vertex(370,30){2}
\Text(-15,-15)[lb]{\Black{LHC}}
\Text(90,-15)[cb]{\Black{$87.5\:\%$}}
\Text(160,-15)[cb]{\Black{$7.1\:\%$}}
\Text(230,-15)[cb]{\Black{$3.1\:\%$}}
\Text(300,-15)[cb]{\Black{$1.7\:\%$}}
\Text(370,-15)[cb]{\Black{$0.6\:\%$}}
\Text(-15,-30)[lb]{\Black{Tevatron}}
\Text(90,-30)[cb]{\Black{$74.8\:\%$}}
\Text(160,-30)[cb]{\Black{$5.8\:\%$}}
\Text(230,-30)[cb]{\Black{$11.8\:\%$}}
\Text(300,-30)[cb]{\Black{$7.3\:\%$}}
\Text(370,-30)[cb]{\Black{$0.3\:\%$}}
\Text(90,58.3)[cb]{\Black{$gg\to H$}}
\Text(160,59)[cb]{\Black{$WW,ZZ\to H$}}
\Text(230,60.3)[cb]{\Black{$W^*\to WH$}}
\Text(300,60.3)[cb]{\Black{$Z^*\to ZH$}}
\Text(370,58.3)[cb]{\Black{$gg\to t\oline{t}H$}}
\end{picture}\\[40pt]

\begin{picture}(395,100)(-25,-20)
\SetWidth{0.5}
\Text(-25,26)[lb]{\Black{Decay:}}
\Photon(115,10)(93,27){1.5}{3}
\Photon(115,50)(93,33){1.5}{3}
\DashLine(68,30)(87,30){3}
\Vertex(90,30){5}
\Photon(180,50)(159,30){1.5}{3.5}
\Photon(159,30)(180,10){1.5}{3.5}
\DashLine(138,30)(157,30){3}
\Vertex(160,30){2}
\Photon(250,50)(229,30){1.5}{3.5}
\Photon(229,30)(250,10){1.5}{3.5}
\DashLine(208,30)(227,30){3}
\Vertex(230,30){2}
\Line[arrow,arrowpos=0.5,arrowlength=1,arrowwidth=0.3,arrowinset=0.2](300,30)(320,50)
\Line[arrow,arrowpos=0.5,arrowlength=1,arrowwidth=0.3,arrowinset=0.2](320,10)(300,30)
\DashLine(278,30)(297,30){3}
\Vertex(300,30){2}
\Line[arrow,arrowpos=0.5,arrowlength=1,arrowwidth=0.3,arrowinset=0.2](370,30)(390,50)
\Line[arrow,arrowpos=0.5,arrowlength=1,arrowwidth=0.3,arrowinset=0.2](390,10)(370,30)
\DashLine(348,30)(367,30){3}
\Vertex(370,30){2}
\Text(20,-16.5)[cb]{\Black{$m_H=126$\:GeV}}
\Text(90,-15)[cb]{\Black{$0.2\%$}}
\Text(160,-15)[cb]{\Black{$23.1\%$}}
\Text(230,-15)[cb]{\Black{$2.9\%$}}
\Text(300,-15)[cb]{\Black{$6.2\%$}}
\Text(370,-15)[cb]{\Black{$56.1\%$}}
\Text(90,58.3)[cb]{\Black{$H\to \gamma \gamma $}}
\Text(160,60.3)[cb]{\Black{$H\to WW^*$}}
\Text(230,60.3)[cb]{\Black{$H\to ZZ^*$}}
\Text(300,60.3)[cb]{\Black{$H\to \tau \tau $}}
\Text(370,60.3)[cb]{\Black{$H\to b\oline{b}$}}
\end{picture}
 \caption[The most important Higgs production and decay channels.]{The most important Higgs production and decay modes with corresponding relative \otto{contributions} and branching ratios in the SM at leading order. Gluons and photons only couple via loop processes to the Higgs boson, which are illustrated as inclusive vertices. The LHC values are taken from \cite{LHCXSWG:2013aa} for a centre-of-mass energy of $8$\:TeV, the Tevatron values are for $1.96$\:TeV \cite{Baglio:2010um}, both assuming $m_H\approx 126$\:GeV. the branching ratios and \cite{Dittmaier:2012vm}}
 \label{fig:higgsproddecay}
\end{figure}

Due to different initial states and different collision centre-of-mass energies, the LHC production fractions differ from the ones at Tevatron; at both colliders, the gluon-gluon fusion is dominant.
Almost $8.5\%$ of the produced Higgs bosons decay into two gluons; however, those decays cannot be separated from the background. The missing $3\%$ of the decays (e.g.\ $H\to Z\gamma $) are also not (yet) separable from background events. In the SM, gluons and photons couple to the Higgs boson only via loop processes; the main contributions originate from top quark and $W$ boson loops at leading order.
In other models, however, also other particles could be involved already at one-loop level. This fact makes the Higgs particle an excellent probe for physics beyond the SM.\\
As already explained in Chapter \ref{Statistics}, a discovery is defined as a signal that has a statistical significance of at least $5\sigma$. For the Higgs searches, the ``signal'' is quantified by the signal strength $\mu $, the ratio of measured and theoretically expected signal events and can be translated into the ratio of the specific process cross sections $\sigma (X\to H\to Y)$ \otto{of a certain initial state $X$ producing a Higgs $H$ which then decays into the specific final state $Y$}:

\begin{align}
 \mu (X\to H\to Y) &\equiv \frac{N_\text{\tiny observed}}{N_\text{\tiny expected}} = \frac{\sigma _\text{\tiny observed}(X\to H\to Y)}{\sigma _\text{\tiny SM}(X\to H\to Y)} \label{eq:signalstrength}
\end{align}

$\mu (X\to H\to Y)$ is mass dependent and should be $1$ at the Higgs mass and $0$ everywhere else in the SM, if one neglects the Higgs decay width. The signal cross section $\sigma (X\to H\to Y)$ \otto{factorizes in narrow-width approximation:}

\begin{align}
 \sigma (X\to H\to Y) &= \sigma (X\to H) \cdot \epsilon _{XY}\cdot {\cal B}(H\to Y), \label{eq:specificcrosssection}
\end{align}

where $\sigma (X\to H)$ is the Higgs production cross section of the production mode $X$, and ${\cal B}(H\to Y)$ is the branching ratio of a Higgs boson decaying into the final state $Y$. The efficiencies $\epsilon _{XY}$ incorporate all detector, selection and reconstruction efficiencies; they are different for each $X\to H\to Y$ process. Effective Higgs couplings to fermions and vector bosons, like e.g.\ in \cite{Azatov:2012rd,Klute:2012pu,Espinosa:2012im,Carmi:2012in}, can be used to test the details of the SM, but for our SM extension fits they are not applicable; therefore, I will not use them in this thesis. A combination of the four best Higgs mass measurements in the decays $H\to \gamma \gamma $ and $H\to ZZ^*$ at ATLAS and CMS gives us $m_H=125.96 ^{+0.18} _{-0.19}$. The inputs can be found in Table \ref{tab:mHinputs}.
All available relevant Higgs signal strengths that were used in the fits were measured at the LHC, where the initial state is $pp$, and Tevatron with $p\oline{p}$ in the initial state, and can be found in Fig.\ \ref{fig:signalstrengths} in App.\ \ref{inputs} together with a detailed description.
Combining them yields a Higgs signal strength of $1.007 ^{+0.099} _{-0.098}$ at $m_H\approx 126$\:GeV. If one is especially interested in the decay properties, the experimental results can be combined for the five decay channels from Fig.\ \ref{fig:higgsproddecay}\otto{.}

\begin{table}[htbp]
\centering
 \begin{tabular}{lllll}
  Decay channel & $\mu _{\text{\tiny comb}}$ & Number of   & $\Delta \chi^2$ & Deviation\\
		&	      & observables &			       & from $0$\\
  \hline
  $H\to \gamma \gamma $ & $1.16\pm 0.18$ & \quad $39$ & $53.94$ & $6.44$\\
  $H\to WW^*$ & $0.75\pm 0.16$ & \quad $\:5$ & $7.77$ & $4.69$\\
  $H\to ZZ^*$ & $1.13\pm 0.24$ & \quad $\:4$ & $2.03$ & $4.71$\\
  $H\to b\oline{b}$ & $1.30 ^{+0.45} _{-0.44}$ & \quad $\:6$ & $2.26$ & $2.95$\\
  $H\to \tau \tau $ & $1.12\pm 0.29$ & \quad $\:9$ & $8.57$ & $3.86$\\
  \hline
  all channels & $1.007 ^{+0.099} _{-0.098}$ & \quad $63$ & $72.76$ & $10.17$\\
 \end{tabular}
\caption[]{Higgs signal strength combinations at $m_H\approx 126$\:GeV, ordered by decay products. The $N_{\rm dof}$ and the compatibility with signal and background are given in the third, fourth and fifth column.} \label{tab:higgschannels}
\end{table}

\begin{figure}[htbp]
\centering
\begin{picture}(230,450)(80,-30)
    \SetWidth{0.2}
    \SetColor{Gray}
    \Line(170,14)(170,404)
    \Line(170,-15)(170,-13)
    \Line(200,14)(200,404)
    \Line(200,-15)(200,-13)
    \Line(230,-15)(230,404)
    \Line(290,-15)(290,404)
    \Line(320,-15)(320,404)
    \SetWidth{5.5}
    \SetColor{Orange}
    \Line(260,397)(260.54,397) 
    \Line(260,379)(260.01,379) 
    \Line(260,361)(258.50,361) 
    \Line(260,343)(260.01,343) 
    \Line(260,325)(253.35,325) 
    \Line(260,307)(311.89,307) 
    \Line(260,289)(328.91,289) 
    \Line(260,271)(258.67,271) 
    \Line(260,253)(288.68,253) 
    \Line(260,235)(179.75,235) 
    \Line(260,217)(231.97,217) 
    \Line(260,199)(282.80,199) 
    \Line(260,181)(242.42,181) 
    \Line(260,163)(262.08,163) 
    \Line(260,145)(298.17,145) 
    \Line(260,127)(268.00,127) 
    \Line(260,109)(255.14,109) 
    \Line(260,91)(283.75,91) 
    \SetColor{Blue}
    \Line(260,391)(255.71,391) 
    \Line(260,373)(257.14,373) 
    \Line(260,355)(263.90,355) 
    \Line(260,337)(260.01,337) 
    \Line(260,319)(255.17,319) 
    \Line(260,301)(311.97,301) 
    \Line(260,283)(331.32,283) 
    \Line(260,265)(258.53,265) 
    \Line(260,247)(290.48,247) 
    \Line(260,229)(189.31,229) 
    \Line(260,211)(235.40,211) 
    \Line(260,193)(287.51,193) 
    \Line(260,175)(242.53,175) 
    \Line(260,157)(262.43,157) 
    \Line(260,139)(310.17,139) 
    \Line(260,121)(283.20,121) 
    \Line(260,103)(255.41,103) 
    \Line(260,85)(281.53,85) 
    \Line(260,67)(286.67,67) 
    \Line(260,49)(213.13,49) 
    \Line(260,31)(276.25,31) 
    \Line(260,13)(280.00,13) 
    \Line(260,-5)(272.41,-5) 
    \SetWidth{1.0}
    \SetColor{Black}
    \Text(-20,415)[cl]{\Black{\textbf{Observable}}}
    \Text(65,415)[cl]{\Black{\textbf{Best-fit value}}}
    \Text(172,415)[cl]{\Black{\textbf{Deviation}}}
    \Text(360,415)[cl]{\Black{$\bm{\Delta \chi ^2}$}}
    \Text(-10,394)[cl]{\Black{$m_t^{\text{\tiny pole}} $ [GeV]}}
    \Text(-10,376)[cl]{\Black{$m_Z $ [GeV]}}
    \Text(-10,358)[cl]{\Black{$\Delta\alpha_\text{had}^{(5)} $}}
    \Text(-10,340)[cl]{\Black{$\alpha_s $}}
    \Text(-10,322)[cl]{\Black{$\Gamma_Z $ [GeV]}}
    \Text(-10,304)[cl]{\Black{$\sigma^0_\text{had} $ [nb]}}
    \Text(-10,286)[cl]{\Black{$R_b^0 $}}
    \Text(-10,268)[cl]{\Black{$R_c^0 $}}
    \Text(-10,250)[cl]{\Black{$R_\ell^0 $}}
    \Text(-10,232)[cl]{\Black{$A_\text{FB}^{0,b} $}}
    \Text(-10,214)[cl]{\Black{$A_\text{FB}^{0,c} $}}
    \Text(-10,196)[cl]{\Black{$A_\text{FB}^{0,\ell} $}}
    \Text(-10,178)[cl]{\Black{${\cal A}_b $}}
    \Text(-10,160)[cl]{\Black{${\cal A}_c $}}
    \Text(-10,142)[cl]{\Black{${\cal A}_\ell $}}
    \Text(-10,124)[cl]{\Black{$m_W $ [GeV]}}
    \Text(-10,106)[cl]{\Black{$\Gamma_W $ [GeV]}}
    \Text(-10,88)[cl]{\Black{$\sin^2\theta_\ell^\text{eff} $}}
    \Text(-10,70)[cl]{\Black{$\mu _{\text{\tiny comb}}(H\to \gamma \gamma )$}}
    \Text(-10,52)[cl]{\Black{$\mu _{\text{\tiny comb}}(H\to WW^*)$}}
    \Text(-10,34)[cl]{\Black{$\mu _{\text{\tiny comb}}(H\to ZZ^*)$}}
    \Text(-10,16)[cl]{\Black{$\mu _{\text{\tiny comb}}(H\to b\oline{b})$}}
    \Text(-10,-2)[cl]{\Black{$\mu _{\text{\tiny comb}}(H\to \tau \tau )$}}
    \Text(90,394)[cl]{\Black{$174.01$}} 
    \Text(90,376)[cl]{\Black{$91.1878$}} 
    \Text(90,358)[cl]{\Black{$0.02756$}} 
    \Text(90,340)[cl]{\Black{$0.1188$}} 
    \Text(90,322)[cl]{\Black{$2.4956$}} 
    \Text(90,304)[cl]{\Black{$41.477$}} 
    \Text(90,286)[cl]{\Black{$0.21472$}} 
    \Text(90,268)[cl]{\Black{$0.1722$}} 
    \Text(90,250)[cl]{\Black{$20.742$}} 
    \Text(90,232)[cl]{\Black{$0.1030$}} 
    \Text(90,214)[cl]{\Black{$0.0736$}} 
    \Text(90,196)[cl]{\Black{$0.0162$}} 
    \Text(90,178)[cl]{\Black{$0.935$}} 
    \Text(90,160)[cl]{\Black{$0.668$}} 
    \Text(90,142)[cl]{\Black{$0.1469$}} 
    \Text(90,124)[cl]{\Black{$80.369$}} 
    \Text(90,106)[cl]{\Black{$2.091$}} 
    \Text(90,88)[cl]{\Black{$0.2315$}} 
    \Text(90,70)[cl]{\Black{$1$}} 
    \Text(90,52)[cl]{\Black{$1$}} 
    \Text(90,34)[cl]{\Black{$1$}} 
    \Text(90,16)[cl]{\Black{$1$}} 
    \Text(90,-2)[cl]{\Black{$1$}} 
    \Text(364,394)[cl]{\Black{$0.35$}} 
    \Text(364,376)[cl]{\Black{$0.24$}} 
    \Text(364,358)[cl]{\Black{$0.33$}} 
    \Text(364,340)[cl]{\Black{$0.00$}} 
    \Text(364,322)[cl]{\Black{$0.14$}} 
    \Text(364,304)[cl]{\Black{$2.80$}} 
    \Text(364,286)[cl]{\Black{$4.87$}} 
    \Text(364,268)[cl]{\Black{$0.17$}} 
    \Text(364,250)[cl]{\Black{$0.86$}} 
    \Text(364,232)[cl]{\Black{$4.03$}} 
    \Text(364,214)[cl]{\Black{$0.42$}} 
    \Text(364,196)[cl]{\Black{$0.91$}} 
    \Text(364,178)[cl]{\Black{$0.10$}} 
    \Text(364,160)[cl]{\Black{$0.01$}} 
    \Text(364,142)[cl]{\Black{$2.89$}} 
    \Text(364,124)[cl]{\Black{$0.72$}} 
    \Text(364,106)[cl]{\Black{$0.02$}} 
    \Text(364,88)[cl]{\Black{$0.52$}} 
    \Text(364,70)[cl]{\Black{$1.38$}} 
    \Text(364,52)[cl]{\Black{$1.55$}} 
    \Text(364,34)[cl]{\Black{$0.51$}} 
    \Text(364,16)[cl]{\Black{$0.38$}} 
    \Text(364,-2)[cl]{\Black{$0.95$}} 
    \Text(168,-28)[cb]{\Black{\large $-3$}}
    \Text(198,-28)[cb]{\Black{\large $-2$}}
    \Text(228,-28)[cb]{\Black{\large $-1$}}
    \Text(288,-28)[cb]{\Black{\large $+1$}}
    \Text(318,-28)[cb]{\Black{\large $+2$}}
    \SetWidth{0.7}
    \Line(165,-13)(165,404)
    \Line(165,-13)(340,-13)
    \Line(165,404)(340,404)
    \Line(340,-13)(340,404)
    \Line(260,-13)(260,404)
    \put(169,-8){\includegraphics[width=0.08\linewidth]{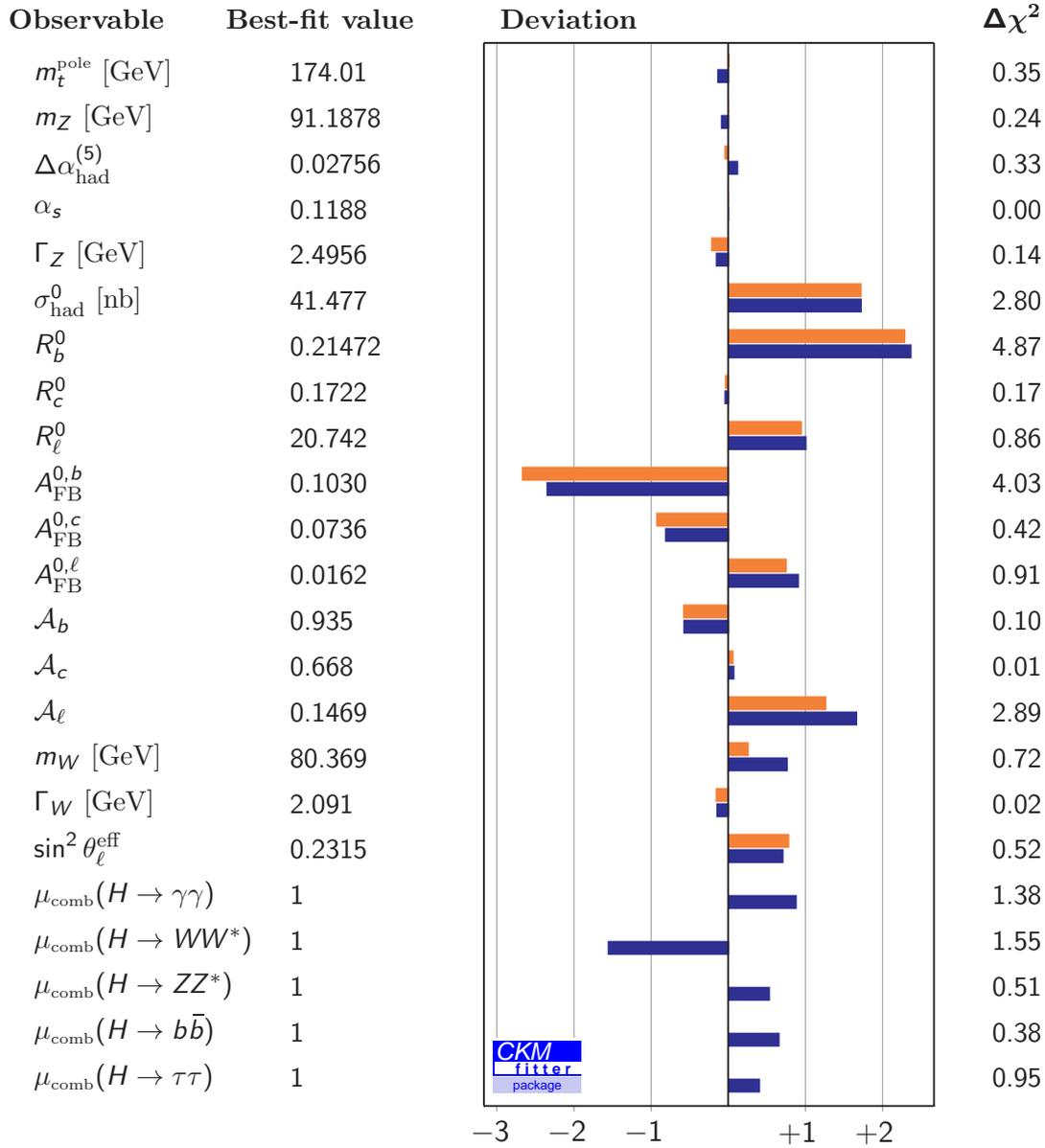}}
\end{picture}
\caption[SM deviations after the Higgs discovery.]{Deviations of the EWPO and the Higgs signal strengths from the best-fit point in the SM fit before (orange) and after (blue) the Higgs discovery. \otto{(The deviations are defined as in Fig.\ \ref{fig:deviationsSM}.)} For the individual $\Delta \chi^2$ of the signal strength combinations in the last column, I give the average contribution per observable (compare Table \ref{tab:higgschannels}).}
\label{fig:deviationsSMwithHiggs}
\end{figure}

\otto{Table \ref{tab:higgschannels}} is only for illustration; in the fits, I used all individual observables rather than these combinations. However, when discussing particular features of one of the decay channels, this picture will be useful.\\
The inclusion of Higgs mass measurements to the precision SM fit means at the same time basically fixing the last unknown parameter of the SM; it raises the $\chi^2$ of the EWPO fit from $21.21$ to $26.93$.
The CKMfitter performance of a fit with $100$ minimizations and a 1D scan with granularity $20$ was $9$ to $14$ minutes.
The impact on the single electroweak observables can be seen in Fig.\ \ref{fig:deviationsSMwithHiggs}, where I show the deviations with and without Higgs mass input.

\begin{figure}[htbp]
 \centering
 \subfigure[]{\begin{picture}(180,150)(-2,0)
	      \put(0,0){\includegraphics[width=0.5\linewidth]{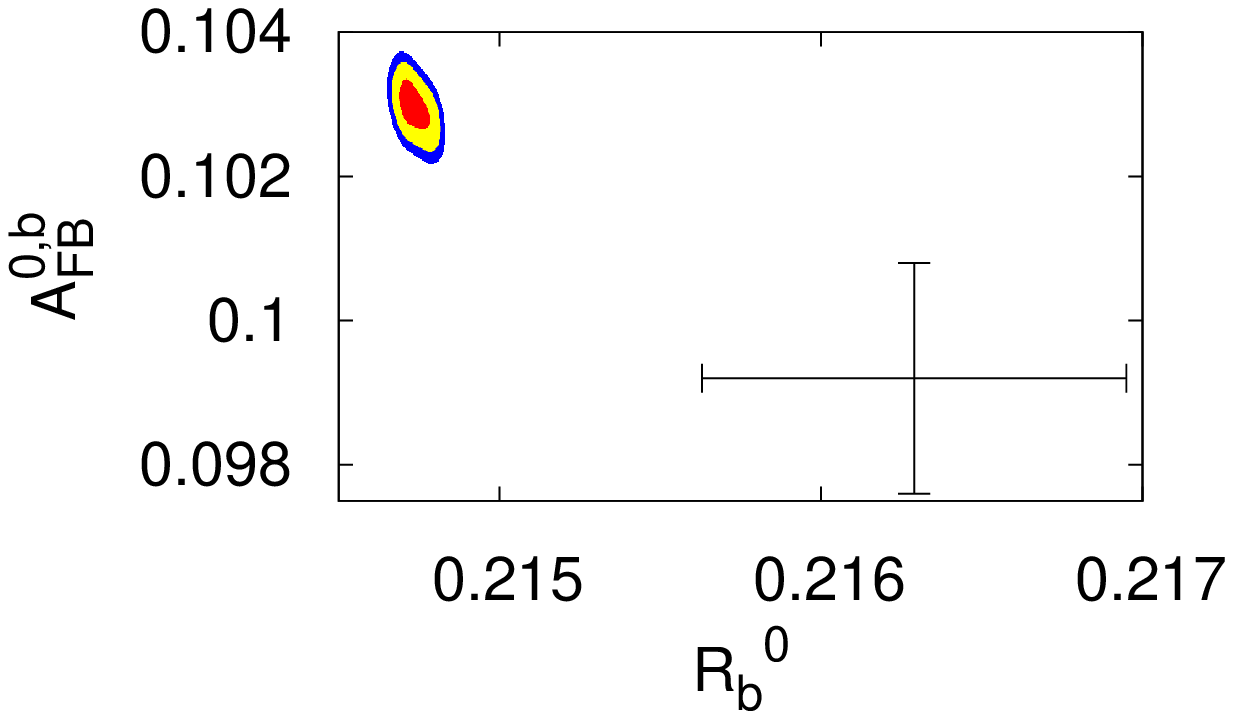}}
	      \put(140.4,98.9){\includegraphics[width=0.08\linewidth]{Images/CKMfitterPackage.eps}}
	      \end{picture}
	     }
 \qquad
 \subfigure[]{\begin{picture}(180,150)(-14,0)
	      \put(0,0){\includegraphics[width=0.5\linewidth]{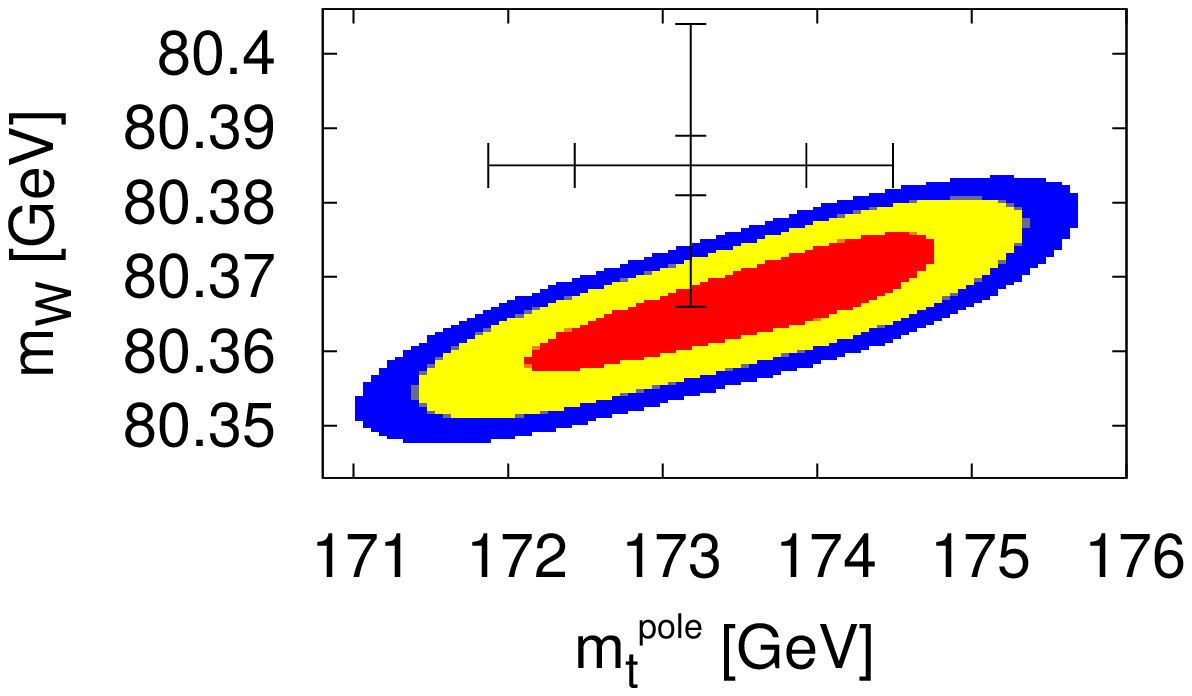}}
	      \put(140.6,40.2){\includegraphics[width=0.08\linewidth]{Images/CKMfitterPackage.eps}}
	      \end{picture}
	     }
 \caption[SM scans over the $R_b^0$-$A_{FB}^{0,b}$ plane and the $m_t^{\text{\tiny pole}}$-$m_W$ plane.]{The $1$, $2$ and $3\sigma$ regions are shown in the $R_b^0$-$A_{FB}^{0,b}$ plane (a) and in the $m_t^{\text{\tiny pole}}$-$m_W$ plane (b), shaded in red, yellow and blue, respectively. The direct measurements are marked by the cross, where the inner error bars are the theoretical errors, and the outer error bars denote the statistical uncertainties. A \otto{plot similar to} (b) can be found in \cite{Eberhardt:2012gv}.}
 \label{fig:mtvsmw}
\end{figure}

Before the Higgs discovery, $A_\text{FB}^{0,b}$ featured the strongest tension between theory and experiment. In the new fit this tension is a bit relaxed, but the deviation is still greater than two and now equal to the slightly increased deviation of $R_b^0$. In Fig.\ \ref{fig:mtvsmw}(a), I show the measured values and the \otto{$1$, $2$ and $3\sigma$ regions surrounding the best-fit point of these two observables. This is the most dramatic illustration of SM incompatibility present in the EWPO.} A much weaker discrepancy is exhibited by the comparison of fit and experiment in the $m_t^{\text{\tiny pole}}$-$m_W$ plane next to it in Fig.\ \ref{fig:mtvsmw}(b). This figure has always been used to illustrate the EWPO fit and its dependence on the Higgs mass, compare \cite{ALEPH:2005ab,Flacher:2008zq,Beringer:1900zz}. And indeed, by fixing $m_H$, the best-fit central values of the top quark mass and 
the $W$ mass receive the largest shifts of all 
EWPO with respect to their experimental errors.
However, this is only visible for the deviation of the bosonic mass (which nevertheless is smaller than one); while in the ``old'' fit the central value of $m_t^{\text{\tiny pole}}$ was set to the lower end of the range assigned by the \textit{R}fit treatment due to the systematic uncertainty, it is now fixed at its upper end, so the difference between both best-fit values of almost $3\sigma$ does not result in a sizeable change of the deviation nor in a noteworthy increase of the $\Delta \chi^2$.\\
Also the discrepanc\ottoo{y} of the ${\cal A}_\ell $ measurement with the SM increased. In total, however, all EWPO data are basically consistent with SM expectations. Also the combined signal strengths are in good agreement with the theory. While the first published data on the $H\to \gamma \gamma$ signal strengths seemed to indicate an excess around $126$\:GeV even too large for the SM and caused a lot of excitement, the deviation of $\mu _{\text{\tiny comb}}(H\to \gamma \gamma )$ has dropped below one if we take the latest data into account. \ottoo{Only $\mu _{\text{\tiny comb}}(H\to WW^*)$ has a deviation greater than one because less events than expected in the SM have been detected and the experimental uncertainty is the smallest of all combined signal strengths.} For all electroweak precision observables and combined signal strengths I also performed prediction fits, where the latter were assumed to have the same efficiencies as the corresponding ATLAS categories at $8$\:TeV.
The \ottooo{predictions} illustrate the compatibility with the other observables as well as the effect of adding the \ottoo{respective measurement} to the fit. The (naive) $p$-value \ottoo{scans} in the SM can be found in App.\ \ref{fitresults} together with the measurement\ottoo{s} and the prediction fit results in the other models which I want to discuss in the next two chapters.

Our analysis \otto{is} the first global fit to all available EWPO and Higgs data \otto{performed} after the discovery of the Higgs boson \cite{Eberhardt:2012gv}. The results largely agree with the electroweak fit by the Gfitter collaboration \cite{Baak:2012kk} \ottoo{and the Bayesian fit by Ciuchini et al. \cite{Ciuchini:2013pca}}.\\
As the Higgs observables are hardly influenced by little modifications of all SM parameters except for the Higgs mass, and the EWPO only have a logarithmic dependence on $m_H$ \cite{Veltman:1976rt},
both analyses can to a good approximation be performed separately\ottoo{. This in turn means that the individual $\Delta \chi^2$ contributions of the signal strength fit can simply be added to the EWPO fit $\chi^2$, amounting} to a total of $\chi^2_{\rm min}=99.69$.
Four of the best-fit parameters can be \ottoo{found in} Fig.\ \ref{fig:deviationsSMwithHiggs}; the remaining values are

\begin{align*}
 m_H=125.96\text{\:GeV}, \quad \theta _{12}=0.227, \quad \theta _{13}=0.00405, \quad \theta _{23}=0.0409, \quad \delta _{13}=1.15.
\end{align*}

\otto{The Higgs mass is simply the combination of the mass extractions presented in Table \ref{tab:mHinputs}, and the mixing angles are mainly constrained by the CKM matrix elements from Table \ref{tab:CKMmatrixinputs} in App.\ \ref{inputs}.}\\
The consistency of the precisely measured experimental data with the SM is astonishing and gives us the opportunity to severely constrain or even rule out physics models beyond the SM. In the next two chapters, I want to shed light on two of such SM extensions. For \ottoo{both}, Higgs search measurements and EWPO depend on the same set of parameters, such that combined fits of the Higgs signal strengths and EWPO are mandatory. The primary goal of this thesis was originally to perform a combined fit of the SM4, which will be discussed in the next chapter.

%% file: sm4.tex
\chapter{The Standard Model with four fermion generations} \label{SM4}

At the beginning of Chapter \ref{SM}, I mentioned that there are three generations of fermions in the Standard Model. The number of generations, however, cannot be deduced from \otto{a fundamental} theoretical \otto{principle}, nor is there a direct indication from \otto{the} experimental side \otto{pointing at} exactly three generations. \otto{The SM4 is the SM amended by a sequential fermion generation with the same quantum numbers as the knwon three generations.}
Adding a new generation of four fermions is equivalent to replacing the $3$ in \eqref{eq:FermionLagrangian} and \eqref{eq:YukawaLagrangian} by a $4$, and the complete fermion content ordered in $SU(2)$ doublets looks like the following:\\[10pt]
\begin{tabular}{llll}
\hspace*{90pt} $\left( \begin{array}{c} u \\ d \end{array}\right) $, \quad & $\left( \begin{array}{c} c \\ s \end{array}\right) $, \quad & $\left( \begin{array}{c} t \\ b \end{array}\right) $, \quad & $\left( \begin{array}{c} t' \\ b' \end{array}\right) $\\[3pt]
\hspace*{90pt} $\left( \begin{array}{c} \nu _e \\ e \end{array}\right) $, \quad & $\left( \begin{array}{c} \nu _\mu \\ \mu \end{array}\right) $, \quad & $\left( \begin{array}{c} \nu _\tau \\ \tau \end{array}\right) $, \quad & $\left( \begin{array}{c} \nu _4 \\ \ell _4 \end{array}\right) $
\end{tabular}\\[12pt]
\otto{An} assumption that I impose on the model I want to analyze in this chapter is the one of \ottoo{\textit{perturbativity}}: increasing the mass of a fermion means enlarging its Yukawa coupling, cf.\ Eq.\ \eqref{eq:YukawaLagrangian}. But like for the other couplings we require a converging power series if we \ottoo{want to} treat Yukawa interactions perturbatively. It has \ottoo{only} been \ottoo{very roughly} quantified how large fermion masses can be before perturbativity breaks down. \otto{Tree-level} partial-wave unitarity arguments limit the quark masses to be lighter than $500$\:GeV and the lepton masses to be lighter than $1$\:TeV if we assume that the corresponding doublets are almost mass degenerate \cite{Chanowitz:1978uj,Chanowitz:1978mv}, meaning that their mass eigenvalues are almost equal.
Analyses of electroweak next-to-leading order contributions to Higgs production with a fourth generation also reveal an incipient breakdown of the perturbation expansion at mass scales around $600$ GeV \cite{Denner:2011vt}.
However, a precise value at which perturbativity fails cannot be deduced easily; one might be able to apply perturbative methods well above these scales. Nevertheless, it is important to state clear\otto{ly} that for the fourth generation model discussed in the following perturbative behaviour is assumed. So my definition of a Standard Model with four generations (SM4) is an SM extension by a complete and perturbative generation of \otto{chiral} fermions.\\
Now why should I add this fourth fermion generation? As already stated, the number of fermion generations cannot be related to any other parameter of the SM. It was shown that there need to be at least three generations in order to have ${\cal CP}$ violation in the Yukawa sector \ottoo{in Eq.\ }\eqref{eq:YukawaLagrangian} \cite{Kobayashi:1973fv}. From LEP data can be inferred that there are only three types of neutrinos lighter than $m_Z/2$, cf.\ \ottoo{Eq.\ }\eqref{Nnulight}. However, heavier neutrinos could still exist and were even discussed as possible dark matter candidates \cite{Raby:1987ww,Lee:2011jk}.
But also the quarks of the fourth generation could possibly contribute to the solution of a cosmological puzzle: the apparent excess of matter in our universe as compared to antimatter discloses a disproportion which cannot be explained by the SM phase $\delta _{13}$ alone; further ${\cal CP}$ violating mechanisms are necessary to account for baryogenesis in order to satisfy the Sakharov criteria \cite{Sakharov:1967dj}.
A sizeable contribution might stem from the two additional phases contained in the $4\times 4$ CKM matrix of the SM4 \cite{Hou:2008xd}\otto{, even if it is not sufficient to account for the entire asymmetry}. Furthermore, a fourth generation of fermions could solve some discrepancies between measured and calculated flavour observables, such as ${\cal CP}$ violation in $B$ meson measurements \cite{Hou:2005yb,Soni:2008bc,Buras:2010pi}.\\
In this chapter, I will introduce the four generations model thoroughly in Sect.\ \ref{SM4Params}, I will discuss the same constraints as in the previous chapter, and finally show that after combining the most important observables in a global fit, the SM4 can be excluded at a statistically significant level.

\section{Parameters} \label{SM4Params}

In addition to the nine SM parameters from \eqref{eq:SMparams} we have nine new parameters in the SM4: the four fermion masses $m_{t'}$, $m_{b'}$, $m_{\ell _4}$ and $m_{\nu _4}$, three more quark mixing angles $\theta _{14}$, $\theta _{24}$, $\theta _{34}$, and two additional quark mixing phases $\delta _{14}$ and $\delta _{24}$.\\
In this work, I will use $800$\:GeV as somewhat arbitrary upper limit for the fourth generation fermion masses assuming that the mentioned breakdown of perturbativity occurs beyond this threshold. For the lower bounds, the direct searches need to be taken into account. Concerning the quarks, the most powerful constraints to date are bounds from the LHC detectors; the latest measurements state $m_{t'}>656$\:GeV \cite{ATLAS:2012qe} and $m_{b'}>675$\:GeV \cite{Chatrchyan:2012af} at $95\%$ CL.
However, both analyses assume specific decay properties: the $t'$ ($b'$) is supposed to decay to $Wb$ ($Wt$) with a branching ratio of $100\%$. Another analysis shows that mass degenerate fourth generation quarks must be heavier than $685$\:GeV \cite{Chatrchyan:2012fp}. The Particle Data Group still lists $m_{t'}>420$\:GeV and $m_{b'}>372$\:GeV at $95\%$ CL as largest exclusion limits in their latest review \cite{Beringer:1900zz}. Since there are no LHC exclusion bounds without any assumption on the branching ratios or mass degeneracy available a conservative lower mass limit of $400$\:GeV for the $t'$ and $b'$ quarks seems to be appropriate in order to not exclude any possible scenario from the beginning. The \otto{most stringent} direct search limits for the fourth generation leptons were obtained at the LEP detectors.
They are $m_{\ell _4}>100.8$ GeV \cite{Achard:2001qw} from $W\nu$ decays and $m_{\nu _4}>45$ GeV from invisible $Z$ decays \cite{Abreu:1991pr}, again at $95\%$ CL.
There are no observables that constrain the \otto{new} CKM matrix parameters directly, so in principle the angles are allowed to take any value between $-\frac{\pi}{2}$ and $\frac{\pi}{2}$ and the phases are varied between $0$ and $2\pi$. However, the $4\times 4$ CKM matrix needs to be unitary, which severely constrains its off-diagonal elements in the fourth row and in the fourth column as already the $3\times 3$ SM part satisfies the unitarity conditions to a good approximation. In this context, it is important to stress that the SM parameters of the CKM matrix have to be reinterpreted in the SM4 \cite{Bobrowski:2009ng}. The CKM matrix parametrization of \otto{Eq.}~\eqref{eq:CKMmatrixSM} is only valid in the SM4 if all \otto{extra} mixing angles are set to zero; in general the SM $3\times 3$ part of the SM4 $4\times 4$ CKM matrix does not have to be unitary.
In principle, one could also parametrize the leptonic Yukawa sector in a similar way; the resulting mixing matrix is called \ottooo{Pontecorvo-Maki-Nakagawa-Sakata (PMNS) matrix \cite{Pontecorvo:1957cp,Maki:1962mu,Pontecorvo:1967fh}}.
However, leptonic mixing effects \otto{on} the observables that I want to discuss here are of minor importance and only occur in higher order corrections. Therefore, I will assume a diagonal lepton mixing matrix in the following, setting all mixing angles and ${\cal CP}$ phases of the PMNS matrix to zero. The neutrinos are supposed to be Dirac particles.\\
Whereas the EWPO and Higgs measurements in the SM have turned out to be effectively orthogonal in the parameter space, we will see that in the SM4 both sets of observables are entangled with respect to the parameters they depend on. A relation between Higgs mass and heavy fermions is also indicated by lattice studies: analyses of the non-perturbative parameter region up to $1$\:TeV have shown that for a Higgs mass of $126$\:GeV fermions with masses of about $300$\:GeV or more can destabilize the Higgs potential \cite{Gerhold:2010wv,Bulava:2013ep}. \otto{However, t}his does not necessarily mean that heavy fermions are excluded\otto{, but may merely indicate that ``our'' vacuum is only metastable}. \otto{Here,} I want to focus on the perturbative part and discuss the interplay of the EWPO and the Higgs signal strengths in the following sections.

\section{Electroweak precision observables} \label{SM4EWPO}

\begin{figure}[htbp]
 \centering
 \subfigure[]{\begin{picture}(180,140)(20,0)
	      \put(0,0){\includegraphics[width=0.5\linewidth]{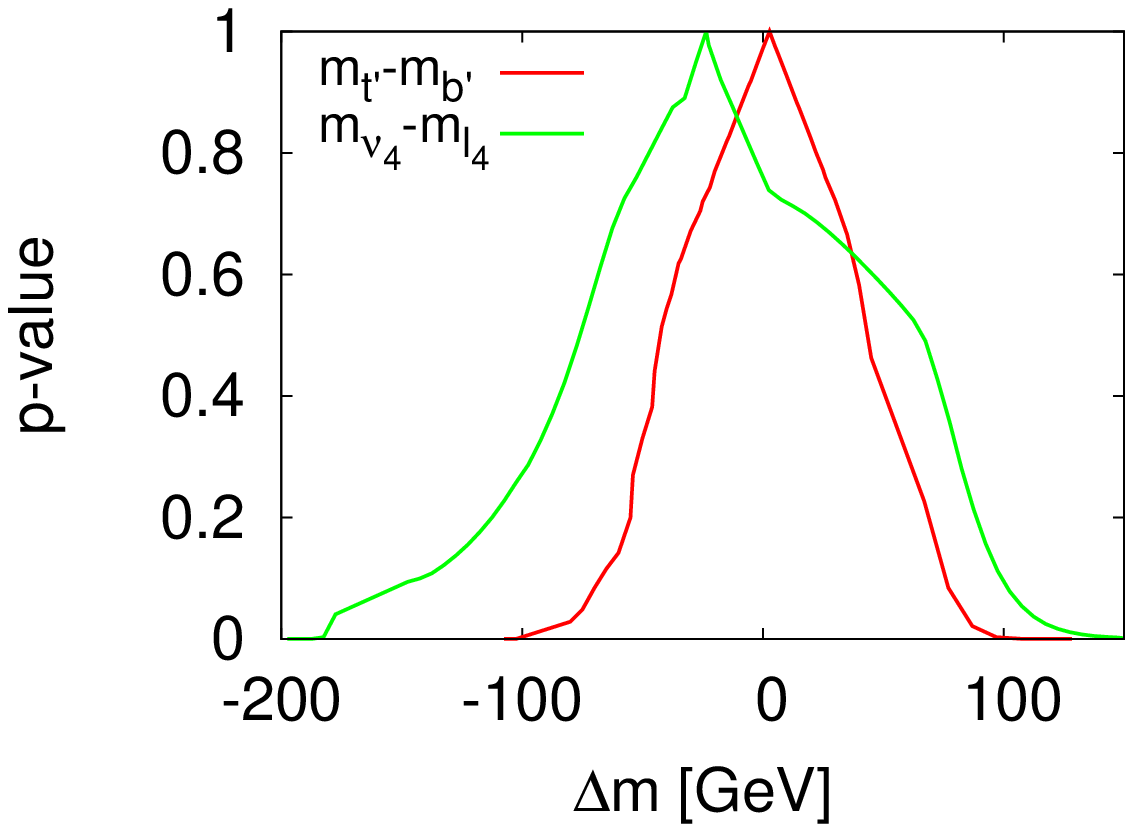}}
	      \put(160.0,116.1){\includegraphics[width=0.08\linewidth]{Images/CKMfitterPackage.eps}}
	      \end{picture}
	     }
 \qquad
 \subfigure[]{\begin{picture}(180,140)(3,0)
	      \put(0,0){\includegraphics[width=0.5\linewidth]{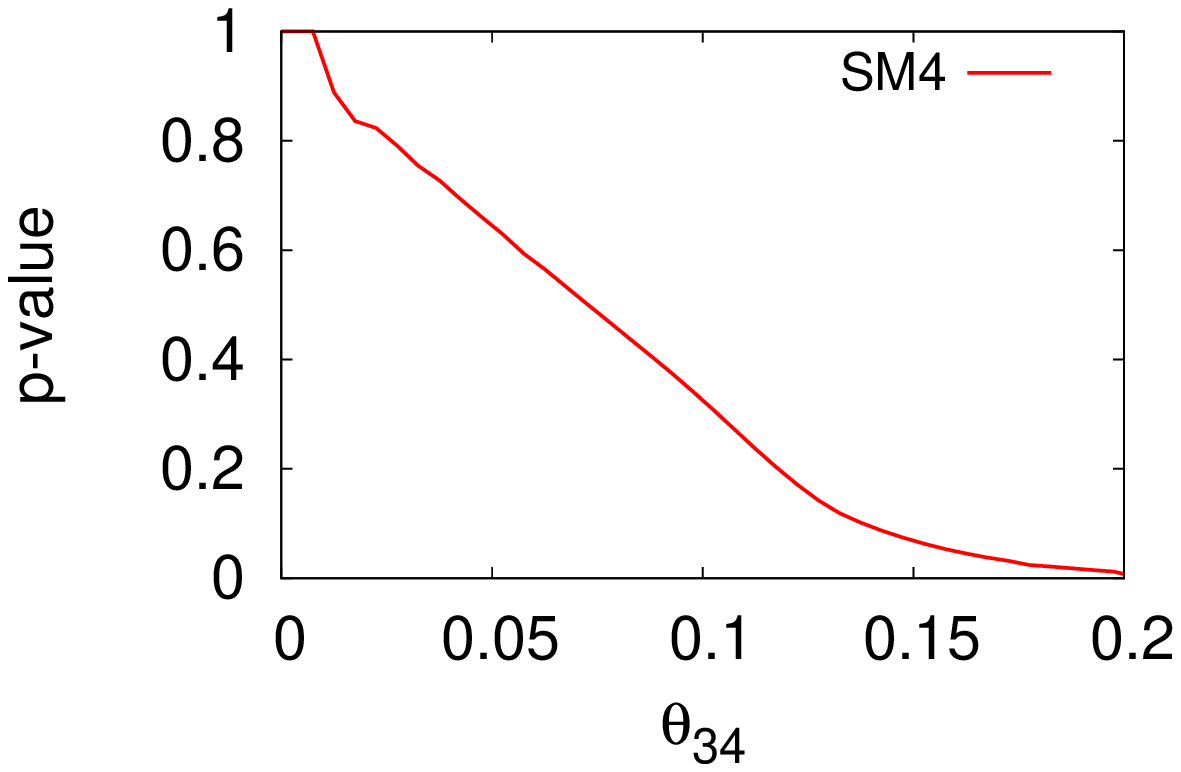}}
	      \put(51.1,43.9){\includegraphics[width=0.08\linewidth]{Images/CKMfitterPackage.eps}}
	      \end{picture}
	     }
 \caption[SM4 scans of the fourth generation mass splittings and of $\theta_{34}$.]{The mass splitting of the fourth generation $SU(2)$ doublet partners is strongly constrained by the EWPO: the absolute quark mass difference cannot exceed \ottoo{$82$}\:GeV at $95\%$ CL, while the lepton mass splitting is somewhat less limited to $167$\:GeV at most at $95\%$ CL (a). The scan over $\theta_{34}$ shows that scenarios are favoured where fourth generation quark decays into SM quarks are suppressed.}
 \label{fig:SM4masssplittings}
\end{figure}

Many studies \otto{have} investigated the viable parameter space of the SM4; I will discuss the most prominent ones in Sect.\ \ref{SM4Combination}. \otto{The majority} of them used the oblique parameters (cf. Sect.\ \ref{SMEWPO}) instead of the full set of electroweak precision observables. It was shown in \cite{He:2001tp,Kribs:2007nz,Eberhardt:2010bm} that in the SM4 the oblique parameter $U$ is negligible if one uses $S$ and $T$; so one ends up with only two pseudo-observables. Only taking into account $S$ and neglecting $T$, the mass degenerate SM4 was even said to be excluded at $6\sigma$ in former PDG reviews \cite{Amsler:2008zzb}. But this approximation has proven to be illegitimate once a non-trivial CKM matrix is assumed \cite{Eberhardt:2010bm}. Any way, the preconditions for the use of the oblique parameters are not satisfied: not only could the new \otto{leptons} be as light as the $Z$ \otto{scale}, the neutrino could even be lighter.
Furthermore, in our general formulation of the model the heavy quarks are allowed to mix with the light species, such that vertex corrections are possible.
This is why we decided to use the complete information from the EWPO instead of the oblique parameters.\\
To this end, we chose \ottooo{the} ``hybrid'' approach introduced in \cite{Gonzalez:2011he}: as in Chapter \ref{SM}, the SM expressions for the EWPO (except for $R_b^0$) are calculated with \texttt{Zfitter} on the level of the effective $Z$--fermion couplings $g_{Vf}$ and $g_{Af}$, while the one-loop SM4 corrections to the effective couplings, $\delta g_{Vf}^{}\equiv g_{Vf}^{\rm SM4}-g_{Vf}^{}$ and $\delta g_{Af}^{}\equiv g_{Af}^{\rm SM4}-g_{Af}^{}$ are obtained using \texttt{FeynArts}, \texttt{FormCalc} and \texttt{LoopTools} \cite{Hahn:1998yk,Hahn:2000kx,Hahn:2006qw}. We can now reduce all SM4 expressions to the SM formulae and correction terms depending on $\delta g_{Vf}$ and $\delta g_{Af}$.\\
In the SM4, the EWPO fit alone is slightly better than in the SM: the total $\chi^2_{\rm min}$ amounts to $20.82$ without fixing $m_H$, and to $26.26$ taking its measurement into account. (The SM fit produced $\chi^2_{\rm min}$ values of $21.21$ and $26.93$ respectively.)
If the Higgs boson had not been found in the mass region below $200$\:GeV, the EWPO fit would have been much better in the SM4 than in the SM, \otto{compare the prediction fit in} Fig.\ \ref{fig:FitEWPOIII}(g). Before the Higgs discovery, this feature was a frequently quoted argument to motivate the SM4. With the Higgs mass around $126$\:GeV, many deviations of the individual observables do not change a lot (cf.\ Fig.\ \ref{fig:deviationsSM4}). The deviations of the $W$ and $Z$ boson mass change the sign, and while the largest deviations in the SM, i.e.\ the ones of $A_\text{FB}^{0,b}$ and $R_b^0$, are diminished in their absolute values, the ones of $\sigma^0_\text{had}$ and ${\cal A}_\ell$ are increased.
There are two major features of the EWPO fit: The first is that large mass splittings in the fourth generation fermion doublets are excluded: Fig.\ \ref{fig:SM4masssplittings}(a) shows that the allowed mass difference \ottooo{$\Delta m$} at $95\%$ CL is between \ottoo{$-75$} and \ottoo{$82$}\:GeV for the quarks and between $-167$ and $109$\:GeV for the leptons. This will be important for the combination with the Higgs signal strength fit that I want to present in the next section. The second important result of the electroweak fit is that mixing between the fourth generation and the SM quarks is disfavoured\otto{. $\theta_{34}$ is smaller than $0.16$ at $95\%$ CL with a best-fit value of $0$, which can be seen in Fig.\ \ref{fig:SM4masssplittings}(b), where a $p$-value scan over $\theta_{34}$ is shown. Quark mixing between the fourth and the third generation would even be stronger suppressed if I used a larger central value for $V_{tb}$ (cf.\ Table \ref{tab:CKMmatrixinputs}).
The constraints on the other two angles are even stronger.}

\section{Higgs searches}

The Higgs content of the SM4 is the same as in the Standard Model: there is one scalar $SU(2)$ doublet that \otto{acquires} a non-zero vacuum expectation value by electroweak symmetry breaking. The only free parameter of the scalar sector of the SM4 is the Higgs mass $m_H$. The measured signal strengths are assumed to be the SM4 ones. I consider the same five production \otto{mechanisms} as in the SM (ggF, VBF, WH, ZH, ttH), but if $m_{\nu _4}$ is small enough, we get an additional, invisible decay channel to the five SM decay modes \cite{Khoze:2001ug}. From Eqns. \eqref{eq:signalstrength} and \eqref{eq:specificcrosssection} one can see that \otto{in narrow-width approximation} the SM4 signal strength splits up into a production and a decay ratio, if one attributes the efficiencies $\epsilon _{XY}$ to the production part:

\begin{align*}
 \mu (X\to H\to Y) &= \frac{\sigma _\text{\tiny SM4}(X\to H) \cdot \epsilon _{XY}}{\sigma _\text{\tiny SM}(X\to H) \cdot \epsilon _{XY}} \cdot \frac{{\cal B}_\text{\tiny SM4}(H\to Y)}{{\cal B}_\text{\tiny SM}(H\to Y)}
\end{align*}

I assume that the efficiencies of the SM4 are the same as in the SM. Instead \otto{of} $\epsilon _{XY}$ I use the \textit{percentage contributions} $\varrho_{XY}$ in the following, which \otto{incorporate the relative admixture of the respective production process $X\to H$ in a particular decay channel to $Y$:}

\begin{align}
 \varrho_{XY} &= \frac{\sigma _{SM}(X\to H) \; \epsilon _{XY} \; {\cal B}_\text{\tiny SM}(H\to Y)}{\sum\limits_i \sigma _{SM}(i\to H) \; \epsilon _{iY} \; {\cal B}_\text{\tiny SM}(H\to Y)}
\label{eq:pc}
\end{align}

\otto{If we sum over the five possible initial states $X$, the} $\varrho_{XY}$ add up to $1$. The percentage contributions are convenient when we want to use signal strengths from the proton colliders Tevatron and LHC: as we have no information about the particular production on parton level for an individual process, we have to sum over all possible production channels. For some signal strengths \ottoo{measurements}, additional signatures in the final state indicate a predominance of a specific production process. In that case one can simply set all other percentage contributions to zero. The $\varrho_{XY}$ inputs used in this thesis can be found in Tables \ottooo{\ref{tab:pcvaluestautau}}, \ref{tab:pcvalues} and \ref{tab:pcvaluesrest}.
If no values were provided by the detector collaborations, I simply take the relative proportions from Fig.\ \ref{fig:higgsproddecay}, thus assuming efficiencies of $100\%$. \ottoo{We can sum over the five production modes, attribute the corresponding $\varrho_{XY}$, and then trade the production cross sections for the total decay widths and the branching ratios, which gives us the convenient expression}

\begin{align}
 \mu (H\to Y) &= \left( \sum\limits_i\frac{{\cal B}_\text{\tiny SM4}(H\to i)}{{\cal B}_\text{\tiny SM}(H\to i)}\varrho_{iY}\right) \cdot \frac{\Gamma _H ^{\text{\tiny SM4}}}{\Gamma _H ^{\text{\tiny SM}}} \cdot \frac{{\cal B}_\text{\tiny SM4}(H\to Y)}{{\cal B}_\text{\tiny SM}(H\to Y)}. \label{eq:sm4signalstrength}
\end{align}

For our fits, we used the publicly available code \texttt{HDecay} v. 4.45 \cite{Djouadi:1997yw} to generate branching ratios and total decay width in the SM as well as in the context of the SM4. The program includes higher order corrections from \cite{Djouadi:1994gf,Djouadi:1994ge,Passarino:2011kv,Denner:2011vt}\ottoo{. The calculated quantities depend on the Higgs mass and -- in case of the SM4 -- on the fourth generation fermion masses}, but it is implicitly assumed that $\theta_{14}=\theta_{24}=\theta_{34}=0$. As for the last angle, we have seen in Fig.\ \ref{fig:SM4masssplittings}(b) that this simplification is supported by the EWPO\otto{.
Since the matrix elements of the first two rows of the SM CKM matrix have been determined quite precisely, and they fulfil the corresponding $3\times 3$ unitarity conditions, we expect $|V_{ub'}|\ll 1$ and $|V_{cb'}|\ll 1$, and thus} also the other two angles to be small, so the suppression of quark mixing \otto{between the fourth generation and the SM particles} seems to be justified. The \texttt{HDecay} output was saved in look-up tables and during the fit procedure an algorithm was used to interpolate their entries.
The \otto{small interpolation} error of this treatment was neglected.\\
In the SM4, the predominant production and decay channels have quite different attributes: the gluon-gluon fusion, which already in the SM is the main production process at proton colliders, is naively enhanced by roughly a factor of $10$ at leading order due to the additional heavy quark contributions in the loop \cite{Gunion:1995tp}. On the other hand, the decay $H\to \gamma \gamma$ is heavily suppressed: Already in the SM, the diphoton production exhibits a destructive interference; at leading order, the $t$ and $W$ loop enter calculations with a different sign and partially cancel each other. In the SM4, this cancellation is \ottooo{even stronger \cite{Gunion:1993bj}. Taking into account next-to-leading order corrections, the suppression is almost perfect; the Higgs decay to two photons is very unlikely \cite{Denner:2011vt}}.
However, as discussed in Chapter \ref{SM}, the LHC detectors at first even measured a larger excess of the signal strengths than they would have expected \ottoo{in the SM}.
Even if the \ottooo{latest} $\mu (H\to \gamma \gamma)$ \ottooo{values} are more SM-like now, their experimental errors have also decreased, and these observables thus strongly contradict the SM4 hypothesis. 
(The deviation of the diphoton signal strength can roughly be estimated by the deviation from $0$ in Table \ref{tab:higgschannels}.)
Also the other signal strengths are affected: the fit favours a neutrino mass which is lighter than \otto{$m_H/2$}, so the total decay width of the Higgs boson $\Gamma _H ^{\text{\tiny SM4}}$ is larger than in the SM. But if ${\cal B}_\text{\tiny SM4}(H\to \nu_4\nu_4)$ and ${\cal B}_\text{\tiny SM4}(H\to gg)$ are increased, all other branching ratios are diminished at the same time, which in turn means that the Higgs decay into vector bosons is suppressed compared to the SM. This also affects the decay to $b$ quarks because they are assumed to be produced in association with vector bosons\ottoo{, and additionally because ${\cal B}_\text{\tiny SM4}(H\to b\oline{b})$ is also reduced}. On the other hand, the $H\to \tau \tau$ signal strength receives such a large boost by the ggF enhancement and the invisible decay width that its predicted value is larger than $4$ at $99\%$ CL.
\oto{This circumstance legitimates the assumption of a diagonal PMNS matrix because lepton mixing effects would additionally increase $\mu (H\to \tau \tau)$.}\\
\otto{Next, I briefly comment on effective Higgs couplings approaches in the literature like e.g.\ \cite{Azatov:2012rd,Klute:2012pu,Espinosa:2012im,Carmi:2012in}, which for our purposes are} not applicable for the SM4 for several reasons: \otto{The mentioned higher order corrections have different effects on the effective couplings for different decay products. Some of the approaches are oversimplified in the sense that they do not leave room for invisible decay channels as present in the SM4 case. Apart from the effective couplings, we would furthermore need their correlations. Moreover, the errors on the Higgs signal strengths are relatively large, such that a quadratic $\chi^2$ distribution is only a vague approximation. Finally, the numerical $p$-value determination with toy-measurements would not be applicable.}

\section{Combined analysis} \label{SM4Combination}

\begin{figure}[htbp]
\centering
\begin{picture}(230,450)(80,-30)
    \SetWidth{0.7}
    \Line(300,372)(300,404)
    \Line(300,372)(358,372)
    \SetWidth{1.0}
    \SetColor{Black}
    \Text(331,394)[cl]{\Black{$\rm SM4$}}
    \Text(331,382)[cl]{\Black{$\rm SM$}}
    \SetWidth{5.5}
    \SetColor{Red}
    \Line(305,393)(325,393)
    \SetColor{Blue}
    \Line(305,381)(325,381)
    \SetWidth{0.2}
    \SetColor{Gray}
    \Line(170,-15)(170,404)
    \Line(190,-15)(190,404)
    \Line(210,-15)(210,404)
    \Line(250,-15)(250,404)
    \Line(270,-15)(270,404)
    \Line(290,-15)(290,404)
    \Line(310,-15)(310,372)
    \Line(330,14)(330,372)
    \Line(330,-15)(330,-13)
    \Line(350,14)(350,372)
    \Line(350,-15)(350,-13)
    \SetWidth{5.5}
    \SetColor{Red}
    \Line(230,397)(230.01,397) 
    \Line(230,379)(234.57,379) 
    \Line(230,361)(237.36,361) 
    \Line(230,343)(230.01,343) 
    \Line(230,325)(249.13,325) 
    \Line(230,307)(267.68,307) 
    \Line(230,289)(273.64,289) 
    \Line(230,271)(229.83,271) 
    \Line(230,253)(250.24,253) 
    \Line(230,235)(189.06,235) 
    \Line(230,217)(217.37,217) 
    \Line(230,199)(253.20,199) 
    \Line(230,181)(216.68,181) 
    \Line(230,163)(232.15,163) 
    \Line(230,145)(276.11,145) 
    \Line(230,127)(217.73,127) 
    \Line(230,109)(227.62,109) 
    \Line(230,91)(240.83,91) 
    \Line(230,73)(350.00,73) 
    \Line(230,55)(213.75,55) 
    \Line(230,37)(249.48,37) 
    \Line(230,19)(284.09,19) 
    \Line(230,1)(214.83,1) 
    \SetColor{Blue}
    \Line(230,391)(227.14,391) 
    \Line(230,373)(228.10,373) 
    \Line(230,355)(231.72,355) 
    \Line(230,337)(230.01,337) 
    \Line(230,319)(226.78,319) 
    \Line(230,301)(264.65,301) 
    \Line(230,283)(277.55,283) 
    \Line(230,265)(229.02,265) 
    \Line(230,247)(250.32,247) 
    \Line(230,229)(182.88,229) 
    \Line(230,211)(213.60,211) 
    \Line(230,193)(248.34,193) 
    \Line(230,175)(218.36,175) 
    \Line(230,157)(231.62,157) 
    \Line(230,139)(263.44,139) 
    \Line(230,121)(245.47,121) 
    \Line(230,103)(226.94,103) 
    \Line(230,85)(244.35,85) 
    \Line(230,67)(247.78,67) 
    \Line(230,49)(198.75,49) 
    \Line(230,31)(240.83,31) 
    \Line(230,13)(243.33,13) 
    \Line(230,-5)(238.28,-5) 
    \SetWidth{1.0}
    \SetColor{Black}
    \Text(-20,415)[cl]{\Black{\textbf{Observable}}}
    \Text(65,415)[cl]{\Black{\textbf{Best-fit value}}}
    \Text(172,415)[cl]{\Black{\textbf{Deviation}}}
    \Text(370,415)[cl]{\Black{$\bm{\Delta \chi ^2}$}}
    \Text(-10,394)[cl]{\Black{$m_t^{\text{\tiny pole}} $\:[GeV]}}
    \Text(-10,376)[cl]{\Black{$m_Z $ [GeV]}}
    \Text(-10,358)[cl]{\Black{$\Delta\alpha_\text{had}^{(5)} $}}
    \Text(-10,340)[cl]{\Black{$\alpha_s $}}
    \Text(-10,322)[cl]{\Black{$\Gamma_Z $\:[GeV]}}
    \Text(-10,304)[cl]{\Black{$\sigma^0_\text{had} $ [nb]}}
    \Text(-10,286)[cl]{\Black{$R_b^0 $}}
    \Text(-10,268)[cl]{\Black{$R_c^0 $}}
    \Text(-10,250)[cl]{\Black{$R_\ell^0 $}}
    \Text(-10,232)[cl]{\Black{$A_\text{FB}^{0,b} $}}
    \Text(-10,214)[cl]{\Black{$A_\text{FB}^{0,c} $}}
    \Text(-10,196)[cl]{\Black{$A_\text{FB}^{0,\ell} $}}
    \Text(-10,178)[cl]{\Black{${\cal A}_b $}}
    \Text(-10,160)[cl]{\Black{${\cal A}_c $}}
    \Text(-10,142)[cl]{\Black{${\cal A}_\ell $}}
    \Text(-10,124)[cl]{\Black{$m_W $ [GeV]}}
    \Text(-10,106)[cl]{\Black{$\Gamma_W $ [GeV]}}
    \Text(-10,88)[cl]{\Black{$\sin^2\theta_\ell^\text{eff} $}}
    \Text(-10,70)[cl]{\Black{$\mu _{\text{\tiny comb}}(H\to \gamma \gamma )$}}
    \Text(-10,52)[cl]{\Black{$\mu _{\text{\tiny comb}}(H\to WW^*)$}}
    \Text(-10,34)[cl]{\Black{$\mu _{\text{\tiny comb}}(H\to ZZ^*)$}}
    \Text(-10,16)[cl]{\Black{$\mu _{\text{\tiny comb}}(H\to b\oline{b})$}}
    \Text(-10,-2)[cl]{\Black{$\mu _{\text{\tiny comb}}(H\to \tau \tau )$}}
    \Text(90,394)[cl]{\Black{$172.65$}} 
    \Text(90,376)[cl]{\Black{$91.1871$}} 
    \Text(90,358)[cl]{\Black{$0.02756$}} 
    \Text(90,340)[cl]{\Black{$0.1196$}} 
    \Text(90,322)[cl]{\Black{$2.4930$}} 
    \Text(90,304)[cl]{\Black{$41.471$}} 
    \Text(90,286)[cl]{\Black{$0.21485$}} 
    \Text(90,268)[cl]{\Black{$0.1721$}} 
    \Text(90,250)[cl]{\Black{$20.742$}} 
    \Text(90,232)[cl]{\Black{$0.1025$}} 
    \Text(90,214)[cl]{\Black{$0.0729$}} 
    \Text(90,196)[cl]{\Black{$0.0159$}} 
    \Text(90,178)[cl]{\Black{$0.936$}} 
    \Text(90,160)[cl]{\Black{$0.667$}} 
    \Text(90,142)[cl]{\Black{$0.1458$}} 
    \Text(90,124)[cl]{\Black{$80.398$}} 
    \Text(90,106)[cl]{\Black{$2.090$}} 
    \Text(90,88)[cl]{\Black{$0.2318$}} 
    \Text(90,70)[cl]{\Black{$0.08$}} 
    \Text(90,52)[cl]{\Black{$0.88$}} 
    \Text(90,34)[cl]{\Black{$0.90$}} 
    \Text(90,16)[cl]{\Black{$0.11$}} 
    \Text(90,-2)[cl]{\Black{$1.34$}} 
    \Text(374,394)[cl]{\Black{$5.26$}} 
    \Text(374,376)[cl]{\Black{$2.94$}} 
    \Text(374,358)[cl]{\Black{$0.86$}} 
    \Text(374,340)[cl]{\Black{$0.08$}} 
    \Text(374,322)[cl]{\Black{$0.00$}} 
    \Text(374,304)[cl]{\Black{$2.90$}} 
    \Text(374,286)[cl]{\Black{$4.24$}} 
    \Text(374,268)[cl]{\Black{$0.11$}} 
    \Text(374,250)[cl]{\Black{$1.20$}} 
    \Text(374,232)[cl]{\Black{$2.55$}} 
    \Text(374,214)[cl]{\Black{$0.27$}} 
    \Text(374,196)[cl]{\Black{$1.25$}} 
    \Text(374,178)[cl]{\Black{$0.00$}} 
    \Text(374,160)[cl]{\Black{$0.00$}} 
    \Text(374,142)[cl]{\Black{$5.89$}} 
    \Text(374,124)[cl]{\Black{$1.08$}} 
    \Text(374,106)[cl]{\Black{$0.03$}} 
    \Text(374,88)[cl]{\Black{$0.30$}} 
    \Text(374,70)[cl]{\Black{$2.31$}} 
    \Text(374,52)[cl]{\Black{$1.14$}} 
    \Text(374,34)[cl]{\Black{$0.74$}} 
    \Text(374,16)[cl]{\Black{$1.57$}} 
    \Text(374,-2)[cl]{\Black{$2.68$}} 
    \Text(168,-28)[cb]{\Black{\large $-3$}}
    \Text(188,-28)[cb]{\Black{\large $-2$}}
    \Text(208,-28)[cb]{\Black{\large $-1$}}
    \Text(248,-28)[cb]{\Black{\large $+1$}}
    \Text(268,-28)[cb]{\Black{\large $+2$}}
    \Text(288,-28)[cb]{\Black{\large $+3$}}
    \Text(308,-28)[cb]{\Black{\large $+4$}}
    \Text(328,-28)[cb]{\Black{\large $+5$}}
    \Text(348,-28)[cb]{\Black{\large $+6$}}
    \SetWidth{0.7}
    \Line(165,-13)(165,404)
    \Line(165,-13)(358,-13)
    \Line(165,404)(358,404)
    \Line(358,-13)(358,404)
    \Line(230,-13)(230,404)
    \put(320,-10){\includegraphics[width=0.08\linewidth]{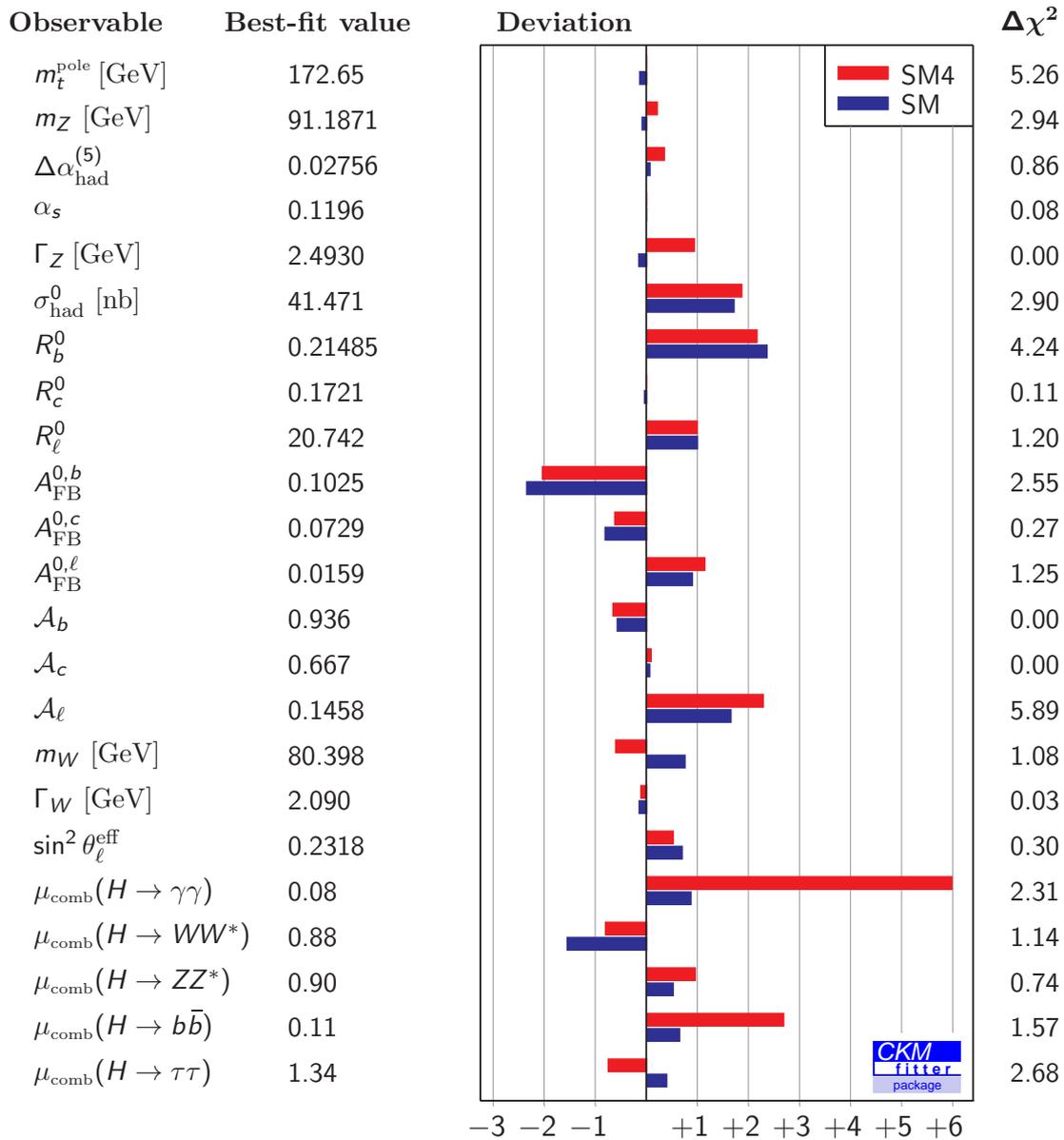}}
\end{picture}
\caption[SM4 deviations after the Higgs discovery.]{Deviations of the EWPO and the Higgs signal strengths from the best-fit point in the SM4 (red) and in the SM (blue).}
\label{fig:deviationsSM4}
\end{figure}

Before combining the discussed constraints, I want to briefly sum up the most important literature on the SM4, which is a story of premature exclusion and resurrection: After the success of the third fermion generation postulated by Kobayashi and Maskawa \cite{Kobayashi:1973fv} a fourth generation was a self-evident next step. However, the SM4 was put under pressure by the oblique parameters in the 1990s as $T$ was fitted to be negative while it is by definition larger than zero in the SM4. This lead to an exclusion statement in the PDG review of $99.2\%$ CL \cite{Caso:1998tx}. Around the year 2000, updated EWPO fits hinted at positive $T$ values, and several \ottoo{authors} inferred that a fourth generation was no longer excluded by electroweak precision data\ottooo{, see e.g.\ }\cite{Frampton:1999xi,He:2001tp,Novikov:2001md,Novikov:2002tk}.
Furthermore, the Higgs boson was still not found, neither at the LEP detectors nor at the Tevatron, and the fact that in the SM4 the Higgs boson could in principle be much heavier than the upper bound deduced from the blue-band plot of the SM made the fourth generation a quite popular model \cite{Kribs:2007nz}, see also Fig.\ \ref{fig:FitEWPOIII}(g) in App.\ \ref{fitresults}. The SM4 analyses became more intricate and also considered quark mixing effects between the SM fermion content and the fourth generation \cite{Bobrowski:2009ng,Chanowitz:2009mz,Buras:2010pi,Erler:2010sk,Eberhardt:2010bm}. But as soon as the first LHC data \ottoo{were} published, the strong interest in this revived model was damped because the first Higgs search results posed serious problems to the SM4 \cite{Djouadi:2012ae,Kuflik:2012ai,Buchkremer:2012yy,Eberhardt:2012ck}.
However, if one wants to exclude the model, all possible \ottoo{realizations} need to be analyzed, such as e.g.\ the possibility of ``light'' fourth generation neutrinos.
Before the ICHEP conference 2012, where the Higgs discovery was proclaimed, our collaboration presented the first quantitative exclusion statement at $3.1\sigma$, taking into account available Higgs search data and the EWPO and combining them with a newly developed $p$-value calculation method for non-nested models \cite{Eberhardt:2012sb}.
The update of this analysis, which belongs to the main achievements presented in this thesis, finally sealed the fate of the SM4, excluding it to more than $5\sigma$ \cite{Eberhardt:2012gv}.\\
In the following, I want to elaborate on the details of this work. In particular, I will repeat parts of the analyses with up-to-date inputs and show that the constraints from the direct Higgs searches \otto{have become} even stronger \ottoo{since the Higgs discovery}.
The SM4 parametrization including the quark mixing matrix was implemented into CKMfitter by Andreas Menzel. On this basis, I added the discussed observables and performed global fits that were cross-checked by Martin Wiebusch using \textit{my}Fitter. One CKMfitter fit with $100$ minimizations and a 1D scan with granularity $20$ takes between one hour and several hours. The bottleneck is the call of the \otto{\texttt{Zfitter} subroutine \texttt{DIZET}} as well as the calculation of the SM4 contributions for the EWPO.\\
The $\chi^2_{\rm min}$ of the SM4 EWPO fit ($20.82$) is comparable with the one of the SM ($21.21$); the one of the signal strengths \ottoo{fit} alone amounts to $124.05$. The combination of both \ottoo{sets of} constraints in the fit yields $\chi^2_{\rm min}=145.33$, corresponding to a $\chi^2_{\rm min}/N_{\rm dof}$ greater than $2.0$. The best-fit parameter values that are not listed in Fig.\ \ref{fig:deviationsSM4} are

\begin{align*}
 &m_H=125.97\text{\:GeV}, \quad \theta _{12}=0.227, \quad \theta _{13}=0.00415, \quad \theta _{23}=0.0410, \quad \delta _{13}=1.17,\\
 &m_{t'}=401\text{\:GeV}, \quad m_{b'}=407\text{\:GeV}, \quad m_{\nu_4}=56.7\text{\:GeV}, \quad m_{\ell_4}=105\text{\:GeV},\\
 &\theta _{14}=0.01, \quad \theta _{24}=0.08, \quad \theta _{34}=0.00.
\end{align*}

\begin{figure}[htbp]
 \centering
 \subfigure[]{\begin{picture}(180,150)(13,0)
	      \put(-5,-10.7){\includegraphics[width=0.5\linewidth]{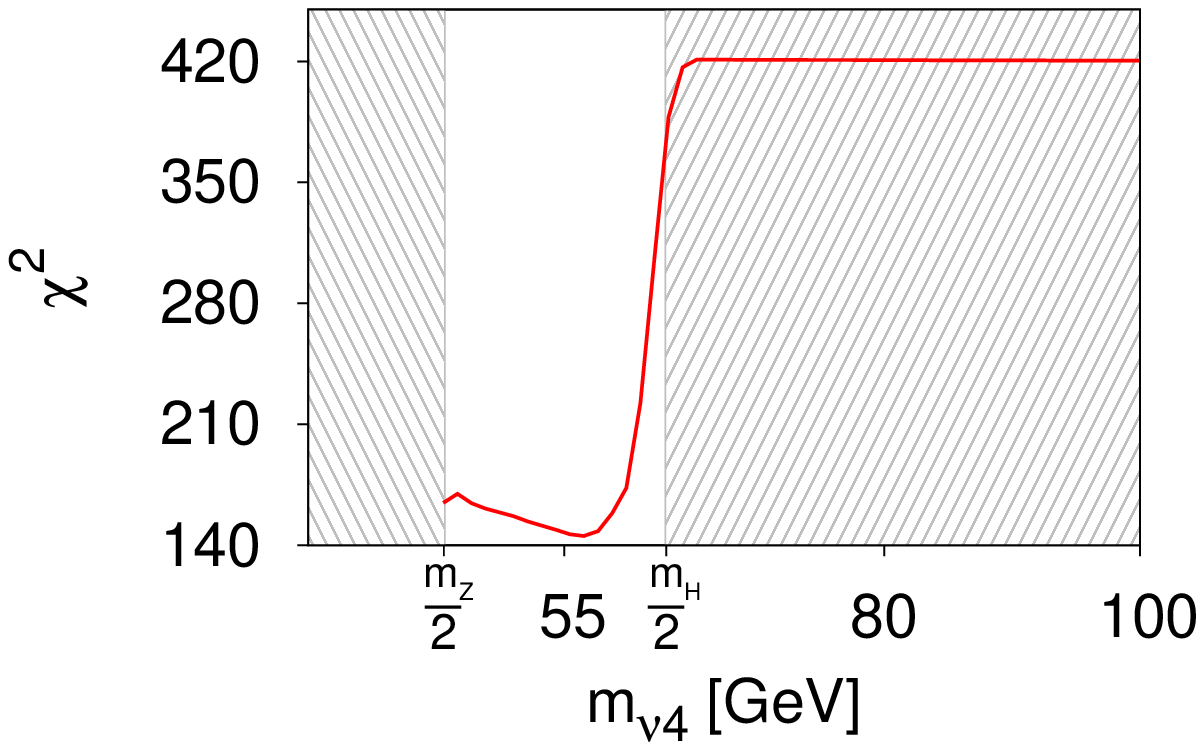}}
	      \put(155.0,35.0){\includegraphics[width=0.08\linewidth]{Images/CKMfitterPackage.eps}}
              \end{picture}
	     }
 \qquad
 \subfigure[]{\begin{picture}(180,150)(-5,0)
	      \put(-2,-10.7){\includegraphics[width=0.5\linewidth]{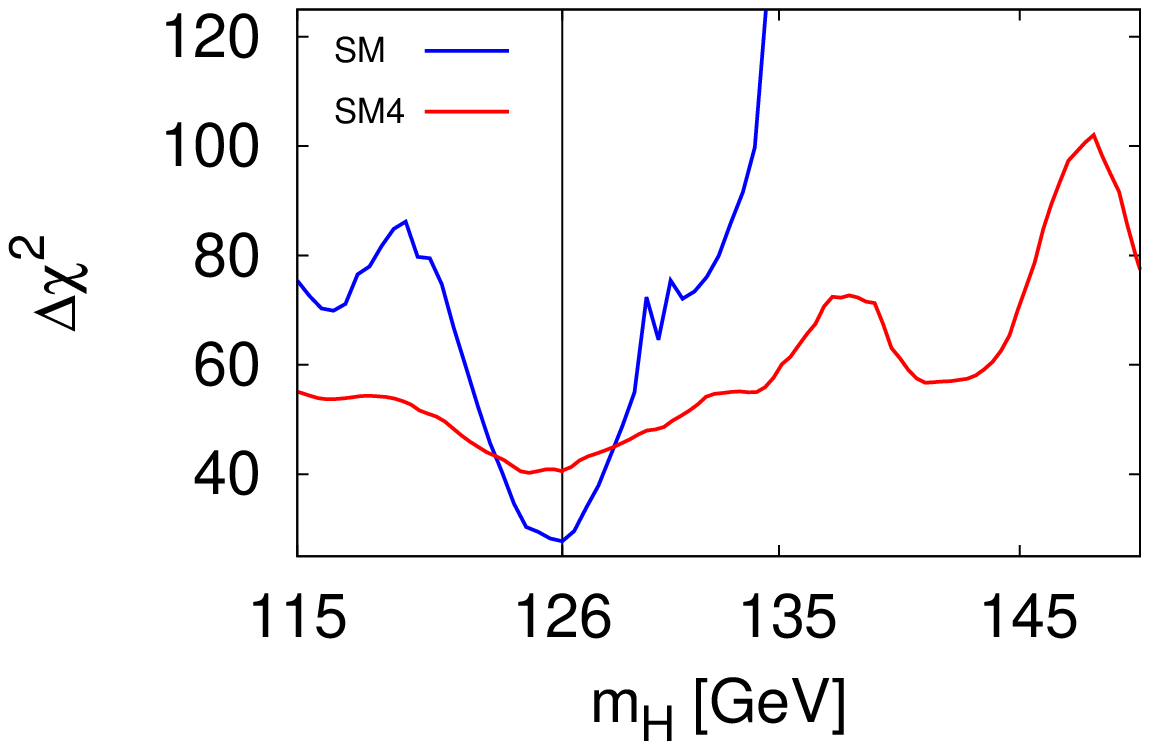}}
	      \put(157.9,33.2){\includegraphics[width=0.08\linewidth]{Images/CKMfitterPackage.eps}}
	      \end{picture}
	     }
 \caption[SM4 scans of $m_{\nu_4}$ and $m_H$.]{On the left side I show a scan over the fourth generation neutrino mass (a); the hatched areas are the regions below the LEP limit on light neutrinos and above $m_H/2$. On the right, the blue-band plot for the SM and the SM4 including the signal strength information from July 2012 is displayed (b). It depicts the best-fit discrepancy between the SM and the SM4: While the SM is compatible with the signal strengths, the $\chi^2_{\rm min}$ of the SM4 is much larger.}
 \label{fig:bluebandSM4}
\end{figure}

The two mixing phases $\delta _{14}$ and $\delta _{24}$ are not constrained by the fit. Exploring best-fit point in detail shows that the \ottoo{SM4} Higgs data fit wants to compensate for the large diphoton and the small $\tau \tau$ signal strength by allowing for large fourth generation mass splittings. And exactly this is forbidden by the EWPO:
In Fig.\ \ref{fig:bluebandSM4}(a), I show the $\chi^2$ depending on $m_{\nu _4}$. One can see that the best-fit value of \ottoo{$56.7$\:GeV} is slightly below $m_H/2$. This is due to the fine\otto{-}tuned Higgs decay width, which in this neutrino mass range is highly sensitive to the invisible decay $H\to \nu_4 \oline{\nu}_4$. Above the Higgs threshold, this invisible decay is kinematically excluded; the enormous $\chi^2$ is basically flat.
If one leaves away the EWPO in the fit, the strong exclusion of an SM4 without invisible decays is relaxed by the permission of large mass splittings. While this effect was clearly visible in our publications \cite{Eberhardt:2012sb,Eberhardt:2012ck}, it is hardly recognizable now that the $\chi^2$ has become that huge: The $\chi^2$ difference between a fit without and with EWPO at the best-fit neutrino mass (which is equal \otto{in both cases}) is about $20$; the difference above the Higgs threshold is $40$.
If one assumes a Majorana character of the fourth neutrino, the result will almost be the same, since the signal strength fit does not change and the effect on the EWPO can be neglected.

The most important parameter of the SM4 exclusion is, of course, the Higgs mass. Whereas in the SM, $m_H$ can be treated as fixed, it is not a priori clear that this also holds for the SM4, so a short comment on this is necessary. For most of the fits in this thesis, I used the four mass extractions from Table \ref{tab:mHinputs}, but in \cite{Eberhardt:2012gv} we compare the explicit $\chi^2$ dependence of the Higgs mass in the SM and SM4. To this end, I digitalized the plots showing the individual signal strengths as functions of $m_H$ that were provided by the detector collaborations at the ICHEP 2012 conference (see Fig.\ \ref{fig:signalstrengths2012}). Furthermore, the signal information was used to disfavour non-signal Higgs mass regions: As already mentioned, the signal strength would be one at the Higgs mass and zero everywhere else, if the experimental errors were negligible \ottoo{and $\Gamma _H^{\text{\tiny SM4}}$ is small}.
To find the signal region, one would have to choose different Higgs mass bins and check for each bin whether the signal strength deviates from zero or not. If we attributed this deviation from zero to each bin, this would translate into a constant shift of the $\chi^2$.
Now we would perform several hypothesis tests, each comparing two non-nested models which only differ in their Higgs mass values. \ottoo{Assuming the realization of a specific Higgs mass (our hypothesis)} we would have to subtract from the $\chi^2$ the squared deviation from zero and add the squared signal strength deviation \ottoo{for the considered mass bin} instead. \ottoo{Finally,} the hypothesis with the smallest $\chi^2$ prevails, which in the Higgs case is the mass bin around $126$\:GeV. (In a \otto{more} thorough analysis of this problem, one would of course rather determine the $p$-value.) Since for the calculation of the $p$-value only the $\chi^2$ difference with respect to the best-fit point is important, I simply subtracted the squared deviation from zero at the respective Higgs mass value. That is why not the total $\chi^2$, but rather a $\Delta \chi^2$ depending on $m_H$ is shown in Fig.\ \ref{fig:bluebandSM4}(b), approximately reproducing the figure from \cite{Eberhardt:2012gv}.
In both models, the SM and the SM4, the best-fit Higgs mass value is about $126$\:GeV. 
Therefore, I will only use the four direct mass measurements from Table \ref{tab:mHinputs} in the following and exclusively discuss signal strengths at $126$\:GeV.\\
Another difference between our publications \cite{Eberhardt:2012sb,Eberhardt:2012ck,Eberhardt:2012gv} and this work is that I do not set $\theta_{34}=0$ \ottoo{here} because the bounds have relaxed a bit compared to \cite{Eberhardt:2012sb}. Nevertheless, a \ottoo{qualitatively} different outcome of the fit is not expected; at the best-fit point \ottoo{essentially} no mixing between the fourth generation and the SM quarks is favoured.\\
The deviations at the best-fit point of the SM4 can be found in Fig.\ \ref{fig:deviationsSM4}, where they are compared to the SM deviations. Again, the signal strengths from Table \ref{tab:higgschannels} are represented by the ATLAS $8$\:TeV quantities. As already mentioned, the EWPO fit is roughly as good as in the SM. An interesting feature is that the top quark mass has the second largest $\Delta \chi^2$, whereas its deviation is zero. \ottoo{This shows the dependence of the other observables on $m_t^{\text{\tiny pole}}$:} Other than in the SM, the SM4 \ottoo{fit} has the freedom to choose \ottoo{relatively} light top quarks in order to compensate for the Higgs mass measurement if direct top quark measurements are not taken into account; the predicted value is even $m_t^{\text{\tiny pole}}=148$\:GeV. (All prediction fits can be found in App.\ \ref{fitresults}, where they are also compared to the SM.)
As soon as the measured $t$ quark mass is included to the fit, $m_t^{\text{\tiny pole}}$ takes a value at the lower end of the allowed \textit{R}fit range. On the contrary, $\Gamma_Z$ has a larger deviation than in the SM, but its $\Delta \chi^2$ is equally small.
While the tensions of $A_\text{FB}^{0,b} $ and $R_b^0$, which \ottoo{feature} the largest deviations of the SM, are a bit ameliorated in the SM4, the discrepancy of ${\cal A}_\ell$ is enhanced.\\
What poses serious problems on the SM4 fit are the signal strengths: As discussed above, tauonic Higgs decays would be seen more often than expected in the SM, and the decays to two photons or $b$ quarks would hardly be visible. However, exactly the contrary seemed to be the case considering the first published Higgs data.
Even if the deficit of $H\to \tau \tau$ events and the apparent excess of the diphoton signal strength seem to converge to their SM expectations with more evaluated and published measurements, more data means at the same time that the experimental errors become smaller, which in turn increases the deviations of the fermionic decay signal strengths\ottoo{. T}he largest impact is the one on the $H\to \gamma \gamma$ signal strength. The SM4 parameters cannot by any means be adapted in such a way that they can account for the accidental cancellation of fermionic and bosonic contributions to the decay amplitude. The largest potential value for $\mu (H\to \gamma \gamma)$ is \ottoo{$0.18$ at $95\%$ CL}, and that is by far not enough to explain the measurement. \ottoo{Its best-fit value of $0.08$ is even smaller; this is due to the tauonic Higgs decays, which determine the behaviour of the SM4 parameters and have the largest $\Delta \chi ^2$ on average.
The dilemma of the SM4 is that a small absolute deviation of $\mu (H\to \tau \tau )$ simultaneously means large deviations of $\mu (H\to \gamma \gamma )$ and $\mu (H\to b\oline{b})$ and vice versa.} This unsolvable discrepancy finally leads to the exclusion of the SM4.

The SM4 and the SM are not nested. For the calculation of the $p$-value, \otto{we use Martin Wiebusch's} program \textit{my}Fitter \cite{Wiebusch:2012en}, which deals with all the subtleties \otto{described} in Sect.\ \ref{nonnestedmodels}. With the data available after the Higgs discovery, we found a $p$-value of $1.1\cdot 10^{-7}$, corresponding to $5.3\sigma$ ($4.8\sigma$, leaving aside Tevatron data) \cite{Eberhardt:2012gv}. Since then, the signal strengths have become even more stringent: On the one hand, the seeming suppression of $\mu (H\to \tau \tau )$ has been lifted, but this fact cannot compensate for the deviations of $\mu (H\to \gamma \gamma )$ and  $\mu (H\to b\oline{b})$ on the other hand, which have grown by $1.7$ and $2.2$, respectively.
For the determination of the $p$-value in \cite{Eberhardt:2012gv}, the simulations already took several days because one needs to generate a sufficient amount of toy measurements that contribute to the numerical integration of the test statistic to guarantee a small uncertainty on the central value.
Considering the fact that the convergence of the numerical integration can only have worsened with the new inputs, I refrained from re-calculating the $p$-value with the available Higgs data. Naively applying Wilks' theorem today, with $\nu$ being the difference in the number of model parameters, we obtain $7.0\cdot 10^{-7}$ for the $p$-value, which corresponds to $5.0\sigma$. The correct $p$-value is expected to be considerably \ottooo{smaller}. Altogether we can regard the possibility of additional fermion doublets \ottoo{in a perturbative sequential generation as SM extension} as excluded and turn to the analogon in the scalar sector: an additional Higgs doublet.

%% file: 2hdm.tex
\chapter{The Two-Higgs-Doublets model} \label{2HDM}

In generic Two-Higgs-Doublets models, one has two scalar $SU(2)$ Higgs doublets $\Phi_1$ and $\Phi_2$ instead of \otto{a single $\Phi$ as in the SM} \cite{Lee:1973iz}. The most general definition of a Two-Higgs-Doublets model as extension of the SM is given by the replacement of ${\cal L}_H$ by the Higgs Lagrangian

\begin{align}
 {\cal L}_H^{\rm 2HDM}=& (\bm{D}_\mu \Phi_1)^\dagger (\bm{D}^\mu \Phi_1) + (\bm{D}_\mu \Phi_2)^\dagger (\bm{D}^\mu \Phi_2) \nonumber \\
 & -m_{11}^2\Phi_1^\dagger\Phi_1^{\phantom{\dagger}}
  - m_{22}^2\Phi_2^\dagger\Phi_2^{\phantom{\dagger}}
  +m_{12}^2 \left( \Phi_1^\dagger\Phi_2^{\phantom{\dagger}}
  + \Phi_2^\dagger\Phi_1^{\phantom{\dagger}} \right) \nonumber \\
 & -\frac{\lambda_1}{2} \left( \Phi_1^\dagger\Phi_1^{\phantom{\dagger}} \right) ^2
  -\frac{\lambda_2}{2} \left( \Phi_2^\dagger\Phi_2^{\phantom{\dagger}} \right) ^2
  -\lambda_3 \left( \Phi_1^\dagger\Phi_1^{\phantom{\dagger}}\right)  \left( \Phi_2^\dagger\Phi_2^{\phantom{\dagger}}\right) \nonumber \\
 & -\lambda_4 \left( \Phi_1^\dagger\Phi_2^{\phantom{\dagger}}\right)  \left( \Phi_2^\dagger\Phi_1^{\phantom{\dagger}}\right) 
  -\left[ \frac{\lambda_5}{2}\left( \Phi_1^\dagger\Phi_2^{\phantom{\dagger}}\right) ^2 +{\rm h.c.}\right] \nonumber \\
 & -\left\{ \left[ \lambda_6 \left( \Phi_1^\dagger\Phi_1^{\phantom{\dagger}} \right) 
  +\lambda_7 \left( \Phi_2^\dagger\Phi_2^{\phantom{\dagger}} \right)  \right] \left( \Phi_1^\dagger\Phi_2^{\phantom{\dagger}} \right) 
  +{\rm h.c.}\right\} .\label{eq:thdmlagrangian}
\end{align}

(I adopt the notation of Gunion and Haber \cite{Gunion:2002zf}. A general discussion can be found e.g.\ in \cite{Branco:2011iw}.) If one sets $\Phi_2$ to zero, one reobtains the SM Higgs part from \eqref{eq:HiggsLagrangian} identifying $-m^2_{11}$ with $\mu^2$ and $\lambda_1$ with $\lambda/2$. However, this general definition, which goes by the name 2HDM of type III, violates ${\cal CP}$ \ottoo{symmetry} and contains flavour-changing neutral currents at tree level.
\otto{A sufficient condition for a ${\cal CP}$-conserving Higgs sector are real parameters $m_{12}^2$, $\lambda_5$, $\lambda_6$ and $\lambda_7$. (The other parameters are real by hermiticity of ${\cal L}_H^{\rm 2HDM}$.)} Flavour-changing neutral currents are strongly constrained by experimental data; therefore we can impose an additional $Z_2$ symmetry to eliminate the relevant terms from the Lagrangian \cite{Glashow:1976nt,Donoghue:1978cj}:
\ottoo{I}f the full model Lagrangian ${\cal L}^{\rm 2HDM}$ is invariant under the transformation $\Phi_1\to -\Phi_1$, $\Phi_1$ does not couple to the SM particles at all.
This model is called 2HDM of type I. If the $Z_2$ symmetry transformation is $\Phi_1\to -\Phi_1$ and $d_j\to -d_j$, one speaks of the 2HDM of type II. In this scenario, $\Phi_2$ only couples to up-type quarks \ottoo{and neutrinos}, whereas $\Phi_1$ exclusively couples to down-type quarks and charged leptons; the Yukawa Lagrangian \ottoo{of the 2HDM of type II} takes the following shape:

\begin{align*}
 {\cal L}_Y^{\rm 2HDM} &= - \sum\limits_{j,k=1}^3 \left[ Y^d_{jk}\left( \oline{Q}_j \Phi_1 \right) d_k +Y^u_{jk} \left( \oline{Q} _j{\rm i}\sigma _2 \Phi_2 ^*\right) u_k  \right. \\
&\hspace*{60pt}\left. +Y^\ell _{jk} \left( \oline{L}_j \Phi_1 \right) \ell _k +Y^\nu _{jk} \left( \oline{L}_j {\rm i}\sigma _2 \Phi_2 ^*\right) \nu _k +{\rm h.c.}\right]
\end{align*}

With this $Z_2$ symmetry, flavour-changing neutral currents are automatically absent \otto{at} tree level. The $m_{12}^2$, $\lambda_6$ and $\lambda_7$ terms in the Higgs Lagrangian violate this symmetry, but we \otto{keep $m_{12}^2$ to permit soft $Z_2$ breaking without spoiling the desired form of ${\cal L}_Y^{\rm 2HDM}$}.\\
In this work, I want to focus on the ${\cal CP}$\otto{-}conserving 2HDM of type II with soft $Z_2$ breaking term, which I will refer to as 2HDM in the following context. Its Higgs Lagrangian is \eqref{eq:thdmlagrangian} without the last line ($\lambda_6=\lambda_7=0$), and \otto{$m_{12}^2$ and $\lambda_5$} are real. The reason why this specific model is very popular is that it could be a limiting case of supersymmetric models, and the discovery of a second Higgs doublet with the characteristics of a 2HDM type II could be an indication for the realization of supersymmetry in nature\ottoo{, see e.g.\ \cite{Djouadi:2005gj} for a review}. \otto{Moreover, the 2HDM of type II easily complies with most of the constraints from flavour physics because it lacks flavour-changing neutral currents at tree level.}
\vspace*{10pt}

\section{Parameters}

In the above formulation, the 2HDM is parametrized by \ottoo{the} couplings of \ottoo{the} quadratic and quartic Higgs field terms in the Lagrangian. These couplings, however, have to fulfil several requirements to describe a consistent theory of the physical nature. It is therefore convenient to change to a more intuitive basis, which sometimes is called \textit{physical} parametrization. If we expand both $SU(2)$ doublets around their vacuum expectation value, we get

\begin{align}
 \Phi_a (x)=\frac{1}{\sqrt{2}}\binom{\phi_a^+}{v_a + h_a + {\rm i}\chi_a} \qquad (a=1,2).\\ \nonumber
\end{align}

Simultaneous diagonalization of the $2\times 2$ squared mass matrices of the charged components $\phi_a^+$ and the imaginary parts $\chi_a$ is characterized by the rotation angle $\beta $. It gives us two zero eigenvalues that are identified by the longitudinal components of the massive $W$ and $Z$ bosons, and two finite mass eigenvalues $m_{H^+}^2$ and $m_{A^0}^2$, corresponding to the eigenstates $H^+$ and $A^0$, which \otto{are the} charged Higgs boson and ${\cal CP}$-odd Higgs boson.
The vacuum expectation values are related to the diagonalization angle via $\tan \beta=\frac{v_2}{v_1}$. Furthermore, they have to account for electroweak symmetry breaking in the well-known way, so $v_1^2+v_2^2=v^2$.
The squared mass matrix of the ${\cal CP}$-even components $h_a$ are not diagonalized by $\beta $, but we have the additional freedom to perform this rotation, and call the corresponding angle $\alpha $. The mass eigenvalues are called $h^0$ and $H^0$, where the former is by definition the lighter one and \ottoo{in this work} will be interpreted as the new boson discovered at the LHC. The SM Higgs field would be a linear combination of both of them: $H= -h^0 \sin (\beta -\alpha ) -H^0 \cos (\beta -\alpha )$. The ratios of all resulting tree-level couplings of the neutral Higgs bosons to fermions and gauge bosons and the respective SM coupling only depend on $\alpha$ and $\beta$, see Table \ref{tab:Higgscouplings}.

\begin{table}[htbp]
\centering
 \begin{tabular}{l|l|l|l}
  Higgs boson & $d_jd_j,\ell_j\ell_j$ & $u_ju_j$ & $WW,ZZ$\\
  \hline
  $h^0$ & $-\dfrac{\sin \alpha }{\cos \beta }$ & $\dfrac{\cos \alpha }{\sin \beta }$ & $\sin (\beta-\alpha)$\\[10pt]
  $H^0$ & $\dfrac{\cos \alpha }{\cos \beta }$ & $\dfrac{\sin \alpha }{\sin \beta }$ & $\cos (\beta-\alpha)$\\[5pt]
  $A^0$ & $\tan \beta$ & $\cot \beta$ & $0$\\
 \end{tabular}
\caption[2HDM neutral Higgs couplings.]{Neutral Higgs couplings to fermions and vector bosons, normalized to the SM Higgs couplings.} \label{tab:Higgscouplings}
\end{table}

Since in the $W$ and $Z$ coupling ratios only the difference between the two diagonalization angles appears, I will choose $\beta -\alpha$ as \otto{the relevant} fit parameter. Moreover, \otto{I will use the quartic coupling $\lambda _5$, which can be related to the soft breaking parameter $m_{12}^2$}. \ottooo{I} fix all SM parameters including $v$ to their best-fit values and set $m_{h^0}=126$\:GeV \ottoo{because varying them is not expected to change the 2HDM fit results qualitatively. W}e \ottoo{then} end up with six free parameters in the 2HDM:

\begin{align*}
 m_{H^+}, \quad m_{A^0}, \quad m_{H^0}, \quad \beta-\alpha, \quad \tan \beta, \quad \lambda_5
\end{align*}

The relations between the couplings and the physical parameters of the 2HDM are listed in App.\ \ref{thdmrelations}.\\
Before completely switching to the physical parametrization, let us explore some theoretical restrictions first: In the SM, the condition that the Higgs potential must be bounded from below can be simply expressed by requiring $\lambda\geq 0$. This feature, called \textit{positivity}, looks a bit more complicated in the 2HDM. Gunion and Haber have showed that the following inequalities must be satisfied if one wants to avoid an unstable Higgs potential \cite{Gunion:2002zf}:

\begin{align}
 \lambda_1&>0, \qquad \lambda_2>0, \qquad \lambda_3>-\sqrt{\lambda_1 \lambda_2}, \qquad \lambda_4>|\lambda_5|-\lambda_3-\sqrt{\lambda_1 \lambda_2}
\end{align}

Furthermore, we want to guarantee that the minimum of the Higgs potential at $246$\:GeV is the global minimum. If it was only a local minimum, it would be metastable and the different vacuum expectation value of the global minimum could be attained one day such that all particle masses would change. However, this scenario is strongly constrained by the lifetime of our universe, and therefore I want to make the assumption that we live in the global minimum of the Higgs potential. This topic was extensively discussed in \cite{Barroso:2013awa}; it was found that requiring that the global minimum is the one around $246$\:GeV is equivalent to the validity of the following inequality:

\begin{align}
 m_{12}^2 & \left( m_{11}^2 - \sqrt{\frac{\lambda_1}{\lambda_2}} m_{22}^2\right) \left( \tan \beta -\sqrt[4]{\frac{\lambda_1}{\lambda_2}} \right) >0
\end{align}

I will refer to this property as \textit{stability} condition.\\
The last aspect on the theoretical side will be a similar one \otto{to that} in the SM4: as with the fourth generation Yukawa couplings, I want to \otto{require perturbativity of} the 2HDM quartic couplings. In principle, higher order calculations are necessary to determine the maximal absolute values of the $\lambda_i$ above which perturbativity fails. Since this is beyond the scope of this thesis, I simply choose a universal upper limit on the quartic couplings and impose

\begin{align}
 |\lambda_i|\leq 2\pi .
\end{align}

The most conservative estimate of the perturbative breakdown is $|\lambda_i|\leq 4\pi$, but it was shown in \cite{Nierste:1995zx} that in the SM \otto{$|\lambda|\leq 2\pi$ is a more appropriate upper bound for} the quartic coupling $\lambda$. Since the same applies in the 2HDM, where we have five $\lambda_i$, I adopt this convention. The consequences of a different choice of the perturbativity bound will be discussed in Sect.\ \ref{2HDMCombination}. The main effect of the limitation of the $|\lambda_i|$ values is that the mass splitting between $m_{H^0}$, $m_{A^0}$ and $m_{H^+}$ has to be of order of the vacuum expectation value $v\approx 246$\:GeV. In all fits in the following sections, the three constraints of positivity, stability, and perturbativity are implicitly employed.\\
Like for the SM4 particles we also have to take direct Higgs particle searches into account in the 2HDM. For the neutral heavy Higgs particles almost the same bounds apply as for the SM Higgs. The charged Higgs, however, is of particular interest for experimental searches, as it would contribute to flavour-changing neutral current processes at loop level. At LEP, $m_{H^+}$ has found to be larger than $79.3$\:GeV at $95\%$ CL \cite{Heister:2002ev}. For the lower bounds on the masses of the neutral Higgs bosons, I refer to the discussion in Sect.\ \ref{2HDMhiggssearches}. In the fits I use $10$ TeV as arbitrary upper mass limit for the heavy Higgs masses. (Heavier Higgs particles would be out of reach for the current LHC even if it ran at its design energy of $\sqrt{s}=14$ TeV.)
In our parametrization, we assume $\beta$ to be between $0$ and $\frac{\pi }{2}$ without loss of generality, which translates into a range from $0$ to $\infty$ for $\tan \beta$. $\alpha$ is supposed to be in the fourth quadrant (between $-\frac{\pi }{2}$ and $0$), so $\beta -\alpha $ is in the interval from $0$ to $\pi$.\\
In contrast to the SM4, the \otto{SM is nested in the 2HDM}. The SM can be reobtained by fixing $\beta -\alpha =\frac{\pi }{2}$ and sending all heavy Higgs masses to infinity. This limit is called the \textit{decoupling limit} \cite{Gunion:2002zf}. As 2HDM realizations in the vicinity of this limiting case are hardly distinguishable from the SM, I want to focus on the parameter regions where the 2HDM phenomenology deviates from the SM, i.e.\ primarily analyze the low mass scenarios with masses below $1$\:TeV. Let us again start the discussion of experimental model constraints with the EWPO.
\vspace*{10pt}

\section{Electroweak precision observables}

\begin{figure}[htbp]
 \centering
 \subfigure[]{
 \begin{picture}(210,150)(0,0)
	      \put(7,0){\includegraphics[width=0.5\linewidth]{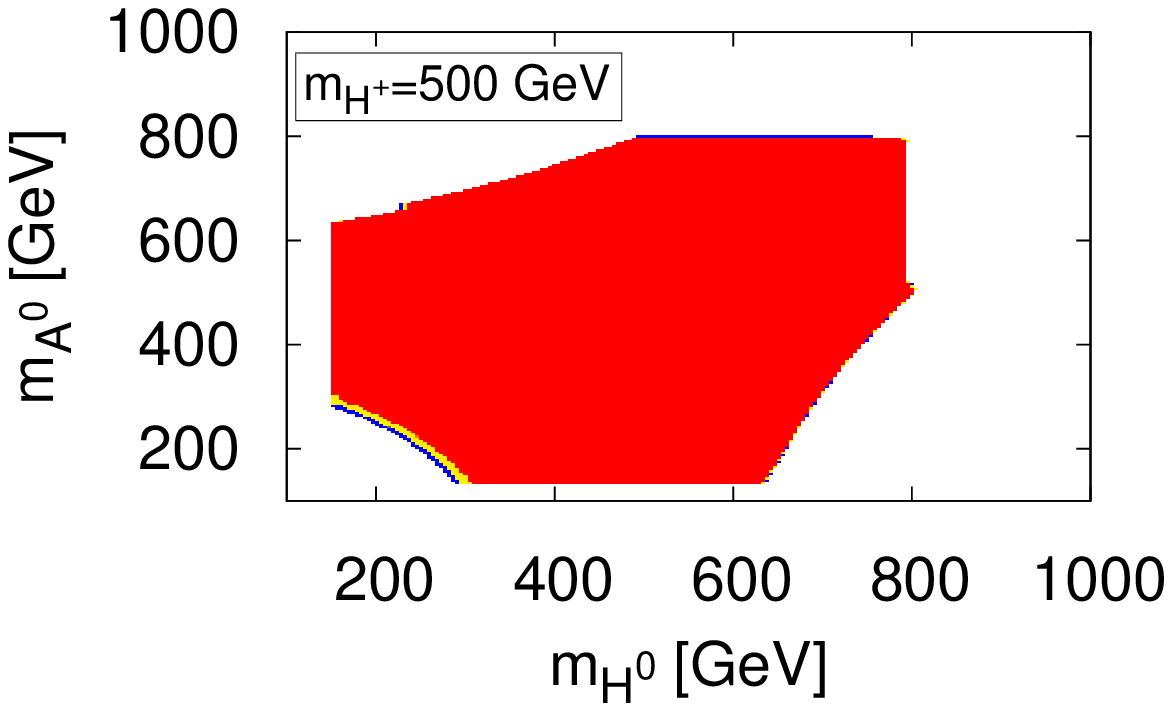}}
	      \put(147.6,40.3){\includegraphics[width=0.08\linewidth]{Images/CKMfitterPackage.eps}}
	      \end{picture}
	     }
 \qquad
 \subfigure[]{\begin{picture}(180,150)(0,0)
	      \put(-5,0){\includegraphics[width=0.5\linewidth]{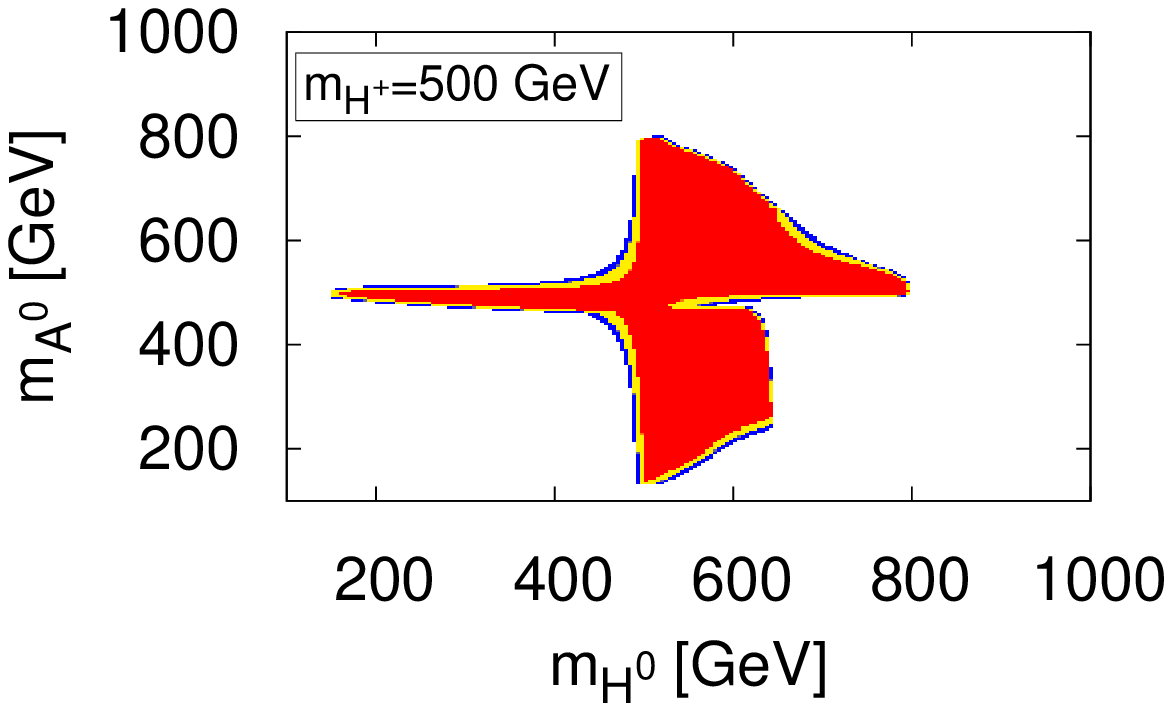}}
	      \put(135.5,40.3){\includegraphics[width=0.08\linewidth]{Images/CKMfitterPackage.eps}}
	      \end{picture}
	     }
 \caption[The difference between perturbativity and EWPO constraints.]{I show scans over the $m_{H^0}$-$m_{A^0}$ plane for $m_{H^+}\!=\!500$\:GeV \otto{without (a) and with (b) EWPO constraints. The requirement of p}erturbativity cuts away large mass splitting regions at the edges (a), whereas one of the neutral boson masses is forced to stay closer than $150$\:GeV to the charged Higgs mass if we add the EWPO to the fit (b).}
 \label{fig:2HDMmasssplittings}
\end{figure}

\otto{As} in the SM4, we cannot use the oblique parameters due to \otto{$Z$} vertex contributions by the 2HDM particles. Especially for the decay $Z\to b\oline{b}$, heavy Higgs contributions play a role at loop level \cite{Hollik:1986gg,Hollik:1987fg,Haber:1999zh}, which affects e.g.\ $R_b^0$ or ${\cal A}_b$. So once more the method from \cite{Gonzalez:2011he} is applied: we obtain the SM values using \texttt{Zfitter} and add the 2HDM contributions calculated at one-loop level with FeynArts, FormCal\otto{c} and LoopTools. For details, see Sect.\ \ref{SM4EWPO}. Compared to the SM4 case we now need to call the \texttt{Zfitter} routine \texttt{DIZET} only once for each fit, since we fixed the SM parameters, and thus avoid the bottleneck that took most of the SM4 fit running time.
\otto{(Varying SM parameters might have an impact on the 2HDM EWPO fit, but the best-fit parameter region will be near the decoupling limit even without EWPO constraints, so an SM-like EWPO fit outcome is expected.)}
The EWPO modify the heavy Higgs mass spectrum of the 2HDM in the following way: While \otto{perturbativity of the} quartic couplings limit\otto{s} the mass splittings to a few hundred GeV, they now additionally have to fulfil the condition that either the $H^0$ or the $A^0$ mass have to be closer to $m_{H^+}$ than about $150$\:GeV.
\otto{Since} both, the perturbativity bound and the EWPO constrain the heavy Higgs masses, I illuminate the differences in Fig.\ \ref{fig:2HDMmasssplittings} for a fixed charged Higgs mass of $500$\:GeV with a small Gaussian error of $3$\:GeV. On the left side I only use the above-mentioned theoretical constraints, which forbid that the neutral Higgs masses exceed $m_{H^+}$ by more than $300$\:GeV and that the mass difference $m_{H^0}-m_{A^0}$ becomes too large in magnitude. The right figure shows the impact on the same fit if the EWPO are included: For $m_{H^0}<m_{H^+}$, the mass of the ${\cal CP}$-odd Higgs is basically also fixed to $m_{H^+}$. If $m_{H^0}$ is larger than the charged Higgs mass, one of the neutral Higgs particles needs to have a mass which differs from $m_{H^+}$ at most by $150$\:GeV.

\vspace*{10pt}

\section{Higgs searches} \label{2HDMhiggssearches}

I interpret the new boson discovered at around $126$\:GeV at the LHC to be the lighter ${\cal CP}$-even eigenvalue and thus discard the possibility that it is the heavier $H^0$ and that the $h^0$ was not seen at LEP, as discussed in \cite{Ferreira:2012my}. I also want to dismiss the idea of a mass degenerate Higgs resonance at the LHC implying that $m_{h^0}\approx m_{H^0}\approx 126$\:GeV \cite{Ferreira:2012nv}. In order to clearly separate the heavy Higgs bosons from $h^0$, I impose a lower limit of $150$\:GeV on their masses.\\
To make use of the Higgs signal strength information, I want to change notation: Let us define the factor $r_i^\text{\tiny 2HDM}$ as the ratio of the squared 2HDM vertex coupling of a neutral Higgs $H$ to the particle $i$ and the respective squared SM coupling. This ratio corresponds to the ratio of the partial widths in the corresponding models: $\Gamma_H^{i,\text{\tiny 2HDM}}=r_i^\text{\tiny 2HDM} \Gamma_H^{i,\text{\tiny SM}}$. By analogy with the SM4 expression \eqref{eq:sm4signalstrength}, and taking the percentage contributions from Eq.\ \eqref{eq:pc}, we can rewrite the signal strength in a compact way as

\begin{align}
 \mu (H\to Y) &= \sum \limits_i r_i^\text{\tiny 2HDM} \cdot \varrho_{iY} \cdot \frac{r_Y^\text{\tiny 2HDM}}{\sum \limits_f r_f^\text{\tiny 2HDM} {\cal B}_\text{\tiny SM}(H\to f)}.
\end{align}

So to calculate the theoretical 2HDM expectation for the signal strengths, the only remaining quantities we need are the $r_i^\text{\tiny 2HDM}$ factors, the percentage contributions and the SM branching ratios. All information about the 2HDM contributions are encoded in the \ottooo{$r_i^\text{\tiny 2HDM}$} if we again assume that the efficiencies and thus the percentage contributions in the 2HDM do not differ from their SM values. We only need to know the couplings of the Higgs boson to the heavy fermions and the massive gauge bosons, for which we take the \ottooo{squared} tree-level expressions from Table \ref{tab:Higgscouplings}, as well as the coupling ratios to $g$ and $\gamma $. The latter two couple to $H$ only via loop processes, which also involve the heavy Higgs particles of the 2HDM. In my CKMfitter implementation, $r_g^\text{\tiny 2HDM}$ and $r_\gamma ^\text{\tiny 2HDM}$ are calculated by an external routine written by Martin Wiebusch that I linked to the rest of the code.\\
The above definition of the signal strength may also be applied to the heavy neutral Higgs particles. In this work, I have only used it for $h^0$ and $H^0$. For the latter, I digitalized the $95\%$ confidence level exclusion limits of $H\to ZZ$ searches at CMS up to $1$ TeV \cite{CMS-PAS-HIG-13-002}, for details see App.\ \ref{inputs}. Similar exclusion limits are available for the $H\to WW$ decay, but since the 2HDM Higgs couplings to $W$ and $Z$ are the same, I confined myself to using only the $H\to ZZ$ exclusion information.\\
In the SM4, the combination of the EWPO and the Higgs searches was sufficient to exclude the model. As the 2HDM cannot be excluded because the SM is nested, it is worthwhile to include further observables to the fit. Since we have decided to discuss a flavour conserving Higgs sector, strong constraints can be derived from flavour-changing neutral currents, which can only occur at loop level like in the SM.
\vspace*{10pt}

\section{Flavour observables}

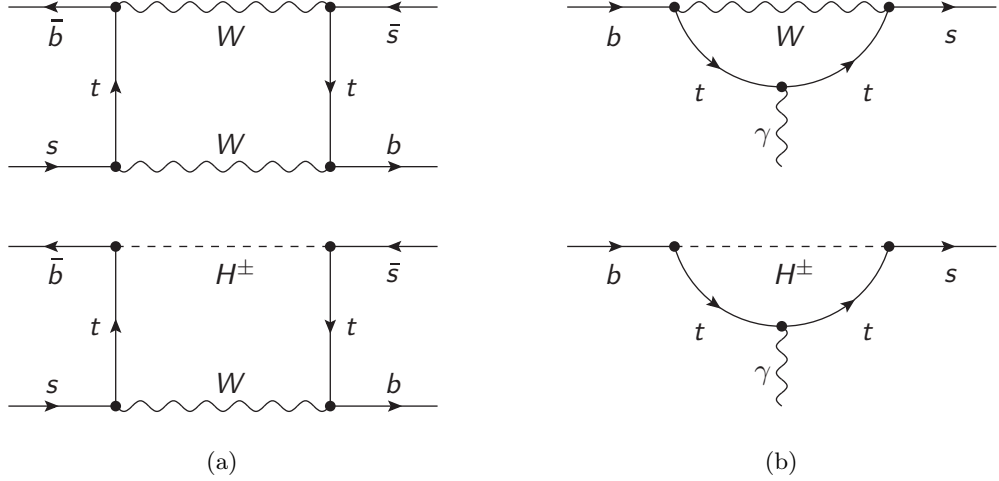
\begin{figure}[htbp]
 \centering
 \subfigure[]{\begin{picture}(180,170)(0,-10)
	       \Line[arrow,arrowpos=0.6,arrowlength=3,arrowwidth=1.2,arrowinset=0.2](50,150)(10,150)
	       \Line[arrow,arrowpos=0.5,arrowlength=3,arrowwidth=1.2,arrowinset=0.2](50,90)(50,150)
	       \Line[arrow,arrowpos=0.4,arrowlength=3,arrowwidth=1.2,arrowinset=0.2](10,90)(50,90)
	       \Vertex(50,150){2}
	       \Vertex(50,90){2}
	       \Photon(50,150)(130,150){2}{6.5}
	       \Photon(130,90)(50,90){2}{6.5}
	       \Vertex(130,150){2}
	       \Vertex(130,90){2}
	       \Line[arrow,arrowpos=0.4,arrowlength=3,arrowwidth=1.2,arrowinset=0.2](170,150)(130,150)
	       \Line[arrow,arrowpos=0.5,arrowlength=3,arrowwidth=1.2,arrowinset=0.2](130,150)(130,90)
	       \Line[arrow,arrowpos=0.6,arrowlength=3,arrowwidth=1.2,arrowinset=0.2](130,90)(170,90)
	       \Text(24,136)[lb]{\Black{$\oline{b}$}}
	       \Text(24,95)[lb]{\Black{$s$}}
	       \Text(40,117)[lb]{\Black{$t$}}
	       \Text(87,136)[lb]{\Black{$W$}}
	       \Text(87,95)[lb]{\Black{$W$}}
	       \Text(151,137)[lb]{\Black{$\oline{s}$}}
	       \Text(151,95)[lb]{\Black{$b$}}
	       \Text(136,117)[lb]{\Black{$t$}}
	       \Line[arrow,arrowpos=0.6,arrowlength=3,arrowwidth=1.2,arrowinset=0.2](50,60)(10,60)
	       \Line[arrow,arrowpos=0.5,arrowlength=3,arrowwidth=1.2,arrowinset=0.2](50,0)(50,60)
	       \Line[arrow,arrowpos=0.4,arrowlength=3,arrowwidth=1.2,arrowinset=0.2](10,0)(50,0)
	       \Vertex(50,60){2}
	       \Vertex(50,0){2}
	       \DashLine(50,60)(130,60){3}
	       \Photon(130,0)(50,0){2}{6.5}
	       \Vertex(130,60){2}
	       \Vertex(130,0){2}
	       \Line[arrow,arrowpos=0.4,arrowlength=3,arrowwidth=1.2,arrowinset=0.2](170,60)(130,60)
	       \Line[arrow,arrowpos=0.5,arrowlength=3,arrowwidth=1.2,arrowinset=0.2](130,60)(130,0)
	       \Line[arrow,arrowpos=0.6,arrowlength=3,arrowwidth=1.2,arrowinset=0.2](130,0)(170,0)
	       \Text(24,46)[lb]{\Black{$\oline{b}$}}
	       \Text(24,5)[lb]{\Black{$s$}}
	       \Text(40,27)[lb]{\Black{$t$}}
	       \Text(87,46)[lb]{\Black{$H^{\pm }$}}
	       \Text(87,5)[lb]{\Black{$W$}}
	       \Text(151,47)[lb]{\Black{$\oline{s}$}}
	       \Text(151,5)[lb]{\Black{$b$}}
	       \Text(136,27)[lb]{\Black{$t$}}
	      \end{picture}
	     }
 \qquad
 \subfigure[]{\begin{picture}(180,170)(0,-10)
	       \Line[arrow,arrowpos=0.45,arrowlength=3,arrowwidth=1.2,arrowinset=0.2](10,150)(50,150)
	       \Vertex(50,150){2}
	       \Arc[arrow,arrowpos=0.5,arrowlength=3,arrowwidth=1.2,arrowinset=0.2](90,161.67)(41.67,270,343.74)
	       \Arc[arrow,arrowpos=0.5,arrowlength=3,arrowwidth=1.2,arrowinset=0.2](90,161.67)(41.67,196.26,270)
	       \Photon(90,90)(90,120){2}{3}
	       \Photon(130,150)(50,150){2}{6.5}
	       \Vertex(90,120){2}
	       \Vertex(130,150){2}
	       \Line[arrow,arrowpos=0.55,arrowlength=3,arrowwidth=1.2,arrowinset=0.2](130,150)(170,150)
	       \Text(24,136)[lb]{\Black{$b$}}
	       \Text(57,115)[lb]{\Black{$t$}}
	       \Text(80,98)[lb]{\Black{$\gamma$}}
	       \Text(87,136)[lb]{\Black{$W$}}
	       \Text(151,137)[lb]{\Black{$s$}}
	       \Text(120,115)[lb]{\Black{$t$}}
	       \Line[arrow,arrowpos=0.45,arrowlength=3,arrowwidth=1.2,arrowinset=0.2](10,60)(50,60)
	       \Vertex(50,60){2}
	       \Arc[arrow,arrowpos=0.5,arrowlength=3,arrowwidth=1.2,arrowinset=0.2](90,71.67)(41.67,270,343.74)
	       \Arc[arrow,arrowpos=0.5,arrowlength=3,arrowwidth=1.2,arrowinset=0.2](90,71.67)(41.67,196.26,270)
	       \Photon(90,0)(90,30){2}{3}
	       \DashLine(50,60)(130,60){3}
	       \Vertex(90,30){2}
	       \Vertex(130,60){2}
	       \Line[arrow,arrowpos=0.55,arrowlength=3,arrowwidth=1.2,arrowinset=0.2](130,60)(170,60)
	       \Text(24,46)[lb]{\Black{$b$}}
	       \Text(57,25)[lb]{\Black{$t$}}
	       \Text(80,8)[lb]{\Black{$\gamma$}}
	       \Text(87,46)[lb]{\Black{$H^\pm$}}
	       \Text(151,47)[lb]{\Black{$s$}}
	       \Text(120,25)[lb]{\Black{$t$}}
	      \end{picture}
	     }
 \caption[Possible $B_s^0\to \oline{B}_s^0$ and $b\to s\gamma$ diagrams.]{\otto{S}ample diagrams for one-loop $B_s^0$--$\oline{B}_s^0$ mixing (a) and $b\to s \gamma$ decays (b). The occurring virtual particles include $W$ and $t$ in the SM (upper diagrams) as well as $H^+$ in the 2HDM (lower diagrams).}
 \label{fig:FCNCdiagrams}
\end{figure}

Many flavour observables are important fit constraints for $\tan \beta <1$ or in the large $\tan \beta$ region. When $\tan \beta \gtrsim 40$, the heavy Higgs couplings to down-type quarks and charged leptons are enhanced, and observables like the branching ratios of the decays $B\to \tau \nu$, $B\to D\tau \nu$ or $B\to D^*\tau \nu$, which receive sizeable contributions from the charged Higgs boson, would have to be included to the fit \cite{Eberhardt:2013uba}. However, I want to concentrate on low $\tan \beta$ scenarios in this thesis for two reasons: I will show that only then sizeable deviations from the decoupling limit condition $\beta -\alpha =\pi /2$ are possible; and furthermore, the measurements of the mentioned (semi-)tauonic $B$ decays are in conflict with each other in the SM, and the disagreement \otto{becomes worse} in the 2HDM of type II \cite{Crivellin:2012ye}.
So the very large $\tan \beta$ regions of the figures in this thesis have to be taken with a grain of salt\otto{, in the sense that the $p$-value of the 2HDM is overestimated}.\\
Yet there are two flavour observables which I want to use in my fits because they are powerful constraints for moderate $\tan \beta$ values: the $B_s^0$--$\oline{B}_s^0$ mixing frequency $\Delta m_{B_s}$ and the branching ratio of $B$ mesons decaying into $X_s\gamma $\ottooo{, where $X_s$ is a hadronic state containing an $s$ or $\oline{s}$ quark}. Both processes can only be described by flavour-changing neutral currents at loop level, to which in the 2HDM also the charged Higgs contributes. In Fig.\ \ref{fig:FCNCdiagrams} I show examples of relevant SM diagrams and possible 2HDM diagrams for $B_s^0$--$\oline{B}_s^0$ mixing and $b\to s\gamma$ transitions.

\otto{The oscillation frequency in the $B_d^0$ system has also been determined experimentally, however the ratio $\Delta m_{B_d}/\Delta m_{B_s}$ is the same in the 2HDM of type II and in the SM, and since the relative error of $\Delta m_{B_s}$ is much smaller, $B_d^0$--$\oline{B}_d^0$ mixing would not additionally constrain our parameters. I now discuss how I include the two flavour observables to my fits and show their impact on the parameter space.}

\subsection{$B_s^0$--$\oline{B}_s^0$ mixing}

\begin{figure}[htbp]
 \centering
 \begin{picture}(180,168)(30,0)
	      \put(0,0){\includegraphics[width=0.63\linewidth]{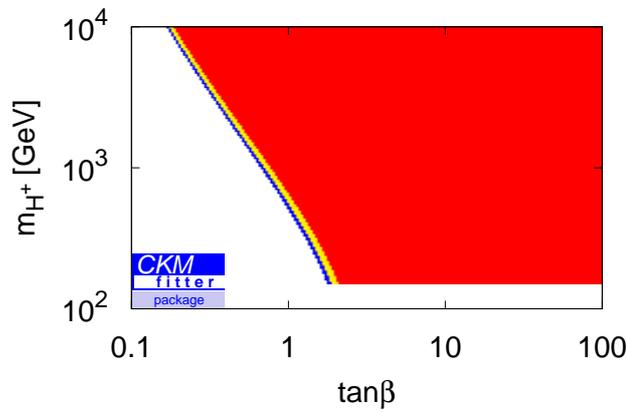}}
	      \put(46.7,46.8){\includegraphics[width=0.08\linewidth]{Images/CKMfitterPackage.eps}}
	      \end{picture}
 \caption[$\Delta m_{B_s}$ effect on the $\tan \beta$--$m_{H^+}$ plane.]{$B_s^0$--$\oline{B}_s^0$ mixing disallows small $\tan \beta $ values. The exact lower bound depends on the mass of the charged Higgs particle. The $1\sigma$, $2\sigma$ and $3\sigma$, regions are the coloured in red, yellow and blue, respectively.}
 \label{fig:DeltaMBsImpact}
\end{figure}

The first measurement of the oscillation frequency of a $B_s^0$ meson changing to its antiparticle $\oline{B}_s^0$ and vice versa dates back to the year 2006 \cite{Abazov:2006dm}. Of all neutral mesons systems, the $B_s^0$--$\oline{B}_s^0$ mixing has the highest oscillation frequency and was determined to unprecedented precision only recently \cite{Aaij:2013mpa}. This frequency is equal to the mass difference between the heavier and the lighter mass eigenstate of the $B_s^0$--$\oline{B}_s^0$ system and in the 2HDM it reads \cite{Abbott:1979dt,Branco:1985pf,Geng:1988bq,Buras:1989ui,Deschamps:2009rh}

\begin{align*}
 \Delta m_{B_s} =&\frac{G_F^2}{24\pi^2}\left| V_{ts}^{*}V_{tb}^{}\right|^2 f_{B_s}^2\eta _{B_s} \hat B_{B_s}m_{B_s}\oline{m}_t^2 \left( S_{WW} +S_{WH} +S_{HH}\right) , \;\; \text{where} \\[10pt]
 S_{WW} &=1 +\frac{9}{1-x_{tW}} -\frac{6}{(1-x_{tW})^2} -6\frac{x_{tW}^2 \ln (x_{tW})}{(1-x_{tW})^3},\\
 S_{WH} &=\frac{x_{tH}}{\tan \beta ^2}\left( \frac{(2x_{HW}-8)\ln (x_{tH})}{(1-x_{HW})(1-x_{tH})^2} +\frac{6x_{HW}\ln (x_{tW})}{(1-x_{HW})(1-x_{tW})^2} +\frac{8-2x_{tW}}{(1-x_{tW})(1-x_{tH})}\right) ,\\
 S_{HH} &=\frac{x_{tH}}{\tan \beta ^4}\left( \frac{1+x_{tH}}{(1-x_{tH})^2} +\frac{2x_{tH} \ln (x_{tH})}{(1-x_{tH})^3}\right) ,
\end{align*}

and $x_{ij}=m_i^2/m_j^2$. In this context, $H$ denotes the charged Higgs boson $H^+$. The SM part is incorporated in the function $S_{WW}$. The additional 2HDM diagrams with one or both $W$ bosons exchanged by $H^+$ lead to the occurrence of $S_{WH}$ and $S_{HH}$, which both are zero in the limit $m_{H^+}\to \infty$ or $\tan \beta \to \infty$. Apart from the Fermi constant and the CKM matrix elements, the appearing prefactors are the decay constant $f_{B_s}$, the bag factor $\hat B_{B_s}$\ottooo{,} the QCD correction $\eta _{B_s}$, the $B_s^0$ mass $m_{B_s}$, and the $\oline{MS}$ top mass $\oline{m}_t$. Since we fixed the top pole mass to its SM best-fit value, also $\oline{m}_t$ is treated as fixed in my fits; the same applies for the CKM matrix elements.
The remaining numerical input values are listed in Table \ref{tab:Flavourinputs} in App.\ \ref{inputs}. $\Delta m_{B_s}$ has a sizeable impact on the allowed $\tan \beta $ values and sets a lower limit of $0.17$ at $95\%$ CL to it, even if $m_{H^+}$ is at the upper end of its \otto{scan} range, see Fig.\ \ref{fig:DeltaMBsImpact}. For $m_{H^+}<1$\:TeV the corresponding requirement is $\tan \beta >0.7$.

Whereas $B_s^0$ meson mixing provides us with a bound on $\tan \beta $, I will now address an observable for which low $m_{H^+}$ contradicts the experimental measurement -- regardless of $\tan \beta $.
\vspace*{10pt}

\subsection{$b\to s\gamma$}

For the flavour-changing neutral current $b\to s\gamma$ as displayed in Fig.\ \ref{fig:FCNCdiagrams}(b) I included to my fit the branching ratio of $\oline{B}$ mesons decaying into hadrons containing an $\oline{s}$ quark and a photon with an energy larger than $1.6$\:GeV. Since for these processes also three-loop effects are important, the analytical formulae are quite complicated.
Therefore, I parametrized the most up-to-date next-to-next-to-leading order result available \cite{Hermann:2012fc}:

\begin{align*}
 {\cal B}(\oline{B}\to X_s\gamma) &= {\cal B}(\oline{B}\to X_c e^- \oline{\nu})_{\text{\tiny exp}} \left| \frac{V_{ts}^*V_{tb}^{}}{V_{cb}}\right|^2 \frac{6 \:\alpha _{\rm em}}{\pi \:C}\\
 & \qquad \cdot \Bigg\{ 0.1271 +2.884\: (\cosh (3.097 - 1.345 L) +15.17)^{-1}\Big. \\
 & \qquad \quad +\frac{1}{\tan ^2\beta }\Big[  \left( 0.05342 + 1.210\:(0.4343 L-1)^{1.9} + 9.125\:(0.4343 L-1)^5\right) \Big. \\
 & \qquad \quad \hspace*{50pt} \Big. (\cosh(6.696 - 2.908 L))^{-1}\Big] \\
 & \qquad \quad +\frac{1}{\tan ^4\beta }\Big[ \left( 186.3 - 810.0L  + 1605 L^2 - 1909 L^3 \right. \Big. \\
 & \qquad \quad \hspace*{50pt} + 1501 L^4 - 811.8 L^5 + 303.9 L^6 \\
 & \qquad \quad \hspace*{50pt} - 77.49 L^7 + 12.86 L ^8 - 1.252 L^9 \\
 & \qquad \quad \hspace*{50pt} \Bigg. \Big. \left. + 0.05439 L ^{10} \right) (\cosh(13.24 - 5.750 L))^{-1} \Big] \Bigg\}
\end{align*}

\begin{figure}[htbp]
 \centering
 \subfigure[]{
 \begin{picture}(210,160)(0,10)
	      \put(-9,-2){\includegraphics[width=260pt]{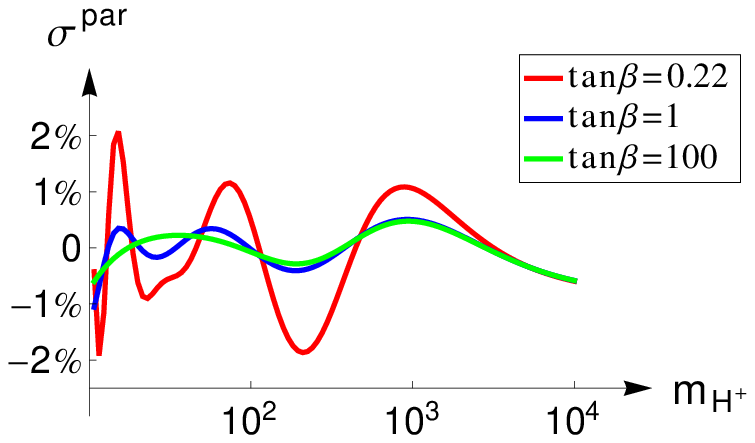}}
	      \end{picture}
	     }
 \qquad
 \subfigure[]{\begin{picture}(180,160)(-15,0)
	      \put(-15,0){\includegraphics[width=0.5\linewidth]{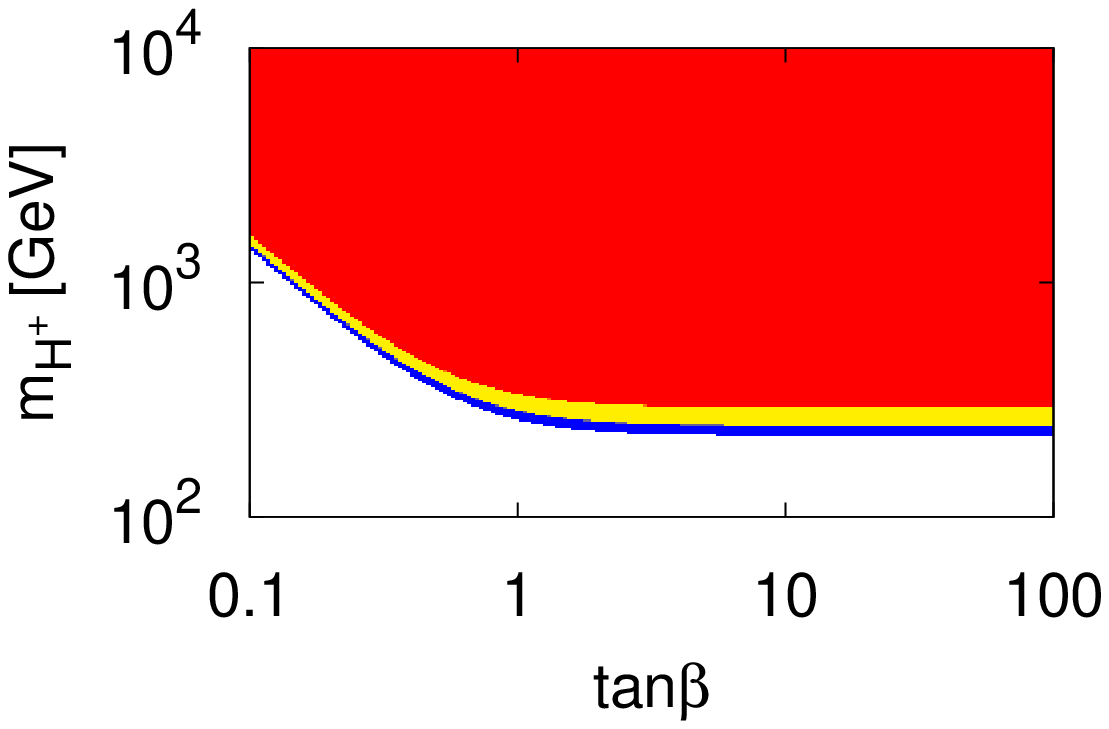}}
	      \put(125.6,99.2){\includegraphics[width=0.08\linewidth]{Images/CKMfitterPackage.eps}}
	      \end{picture}
	     }
 \caption[${\cal B}(\oline{B}\to X_s\gamma)$ error estimation and impact on the $\tan \beta$--$m_{H^+}$ plane.]{The left figure displays the normalized error for my ${\cal B}(\oline{B}\to X_s\gamma)$ approximation depending on $m_{H^+}$ for three different values of $\tan \beta$. On the right, the effect of ${\cal B}(\oline{B}\to X_s\gamma)$ on the charged Higgs mass \ottooo{fit} is shown: $m_{H^+}<230$\:GeV is excluded at $3\sigma $.}
 \label{fig:bsg}
\end{figure}

$L = \log (m_{H^+}/\text{GeV})$ is the logarithm\footnote{Note that ``$\log$'' is the decadic logarithm whereas I denote the natural logarithm by ``$\ln$'' throughout this work.} of the \otto{charged Higgs mass in units of GeV} \ottooo{and} ${\cal B}(\oline{B}\to X_c e^- \oline{\nu})_{\text{\tiny exp}}$ is the measured branching ratio of semileptonic $B$ to $D$ decays, which is corrected with the factor $C$ \otto{to account for its charm quark mass dependence \cite{Misiak:2006ab}}.
The used values can be found in Table \ref{tab:Flavourinputs} in App.\ \ref{inputs}. This parametrized approximation is valid for all $\tan \beta>0.1$ and $m_{H^+}$ up to $10$\:TeV; its error $\sigma^\text{\tiny par}$ is around $2\%$ at most for small $\tan \beta$, compare Fig.\ \ref{fig:bsg}(a). But we are only interested in scenarios which are compatible with the $\Delta m_{B_s}$ bound, for which $\sigma^\text{\tiny par}$ is well below $1\%$.
Together with the uncertainties from the prefactors, which were estimated according to \cite{Hermann:2012fc}, I obtain a total theoretical error of $14\%$. When including ${\cal B}(\oline{B}\to X_s\gamma)$ to the fit, the main effect is that the lower charged Higgs mass bound is increased to $m_{H^+}>250$\:GeV at $95\%$ CL for all $\tan \beta$, compare Fig.\ \ref{fig:bsg}(b). This limit is slightly different from the one obtained in \cite{Eberhardt:2013uba}, because my error estimation is a bit more conservative.

Similar figures like \ref{fig:DeltaMBsImpact} and \ref{fig:bsg}(b) can be found in the CKMfitter analysis of 2HDM effects on flavour observables \cite{Deschamps:2009rh}.

\section{Combined analysis} \label{2HDMCombination}

Like in the previous chapter, I first want to list the relevant literature in order to compare my results with them. The original formulation of the 2HDM dates back to the 1970s, but comprehensive analyses have only been \otto{performed} in the past years, because only now, we are equipped with observables that can sizeably constrain the 2HDM parameter space. Since it is almost impossible to go into detail about all of these studies, I want to restrict myself to a small selection:\\
In 2008, the Gfitter collaboration published a 2HDM fit of $\tan \beta$ and the charged Higgs mass, \otto{using} observables that depend -- apart from SM parameters -- only on those two parameters. With $R_b^0$, ${\cal B}(\oline{B}\to X_s\gamma)$ and $B\to \tau \nu$ as their strongest constraints, they found $m_{H^+}>240$\:GeV at $95\%$ CL \cite{Flacher:2008zq}. Before the LHC announced the first \otto{hints of the Higgs} in 2011, the CKMfitter collaboration performed an extensive  fit which in addition to the Gfitter analysis included more flavour observables like e.g.\ $\Delta m_{B_s}$ \cite{Deschamps:2009rh}, also specializing on $\tan \beta$ and $m_{H^+}$ only.
After the Higgs discovery the general interest in the 2HDM increased: The question arose whether the detected boson was the SM Higgs or a Higgs particle of a different model, possibly followed by the observation of further Higgs bosons. Furthermore, the measured Higgs signal strengths could be used as constraints of the 2HDM parameters. A lot of analyses addressed the compatibility of the discovered Higgs with the light ${\cal CP}$-even Higgs eigenstate of 2HDM of various types \cite{Ferreira:2011aa,Basso:2012st,Cheon:2012rh,Carmi:2012in,Drozd:2012vf,Chen:2013kt,Celis:2013rcs,Giardino:2013bma,Grinstein:2013npa,Barroso:2013zxa,Coleppa:2013dya};
even the possibilities were discussed that the heavy ${\cal CP}$-even \cite{Ferreira:2012my} or the ${\cal CP}$-odd Higgs had been found \cite{Burdman:2011ki}, or that two mass degenerate Higgs particles hide behind the bosonic resonance \cite{Ferreira:2012nv}, as already mentioned. \otto{When} the viability of various 2HDM manifestations was tested, most authors \otto{presented their results of scans over the 2HDM parameter space, which, however}, strongly depends on the chosen parametrization.
For instance in \cite{Coleppa:2013dya}, a parameter scan was performed combining Higgs data, some flavour constraints and the $\rho$ parameter related to the oblique parameter $T$ as well as perturbativity and stability. They did not find allowed parameter sets with $\tan \beta>4$ because they chose a parametrization with disadvantageous scan steps. This shows the crucial virtue of parameter \textit{fits}:
In parameter scans one usually scans with predefined steps over the single model parameter ranges, calculates the $\chi^2$ at each point and finally takes the lowest $\chi^2$ as global minimum. But these scans will only find the correct global $\chi^2_{\rm min}$ if one of the scan points by chance is the best-fit point, which is highly improbable. Parameter fits, however, use the information about the gradient to converge to the $\chi^2$ minimum, so one cannot miss the global $\chi^2$ minimum.\\
In my combined fit, I use the following bounds to constrain the 2HDM parameters: positivity of the Higgs potential, stability of the $246$\:GeV vacuum, perturbativity of the quartic couplings, the EWPO, \ottooo{the} light Higgs \ottooo{signal strengths}, direct $H$ searches up to $1$ TeV, and the measurements of $\Delta m_{B_s}$ and ${\cal B}(\oline{B}\to X_s\gamma)$. The SM parameters were set to their best-fit values from Chapter \ref{SM} because they are known to a good precision compared to the 2HDM parameters and are not expected to change sizeably. As in the SM4 case, there was already an existing implementation of the 2HDM in the CKMfitter package for their above-mentioned publication on flavour constraints \cite{Deschamps:2009rh}.
But apart from the $B_s$ mixing part, I did not use it because in \cite{Deschamps:2009rh} the only free parameters were $\tan \beta$ and the charged Higgs mass, and \otto{in the most general 2HDM of type II further parameters enter the analysis of the Higgs signal strengths and EWPO}. A fit with $100$ minimizations and a 1D scan with granularity $20$ took CKMfitter a few minutes. \ottooo{Having fixed} the SM parameters to their best-fit values, I only have to call the \ottooo{\texttt{Zfitter}} routine \ottooo{\texttt{DIZET}} once for every fit and once for every scan, so the main \ottooo{source of} slow-down of the SM4 fits was avoided here.\\
I do not show the figures from \cite{Eberhardt:2013uba} as they are based on data produced with the \textit{my}Fitter framework and cross-checked by my CKMfitter implementation; the figures in this work stem from my CKMfitter fits and are in some sense the invisible half of our publication. Differences between the publication and this thesis can be traced back to slightly changed inputs in the flavour sector, a different parametrization of the 2HDM (CKMfitter uses the physical basis whereas the original couplings of \eqref{eq:thdmlagrangian} were used in \textit{my}Fitter) and the inclusion of the latest $R_b^0$ expression in the SM, which changed the SM best-fit values. At the best-fit point, $\chi^2_{\rm min}=91.76$; the corresponding parameter values are

\begin{align*}
&m_{H^+}=387\text{\:GeV}, \quad m_{A^0}=394\text{\:GeV}, \quad m_{H^0}=465\text{\:GeV},\\
&\quad \beta-\alpha=1.581=0.5032\pi, \quad \tan \beta=4.42, \quad \lambda_5=0.56
\end{align*}

\begin{figure}[htbp]
\centering
\begin{picture}(230,440)(80,-30)
    \SetWidth{0.7}
    \Line(273,372)(273,404)
    \Line(273,372)(340,372)
    \SetWidth{1.0}
    \SetColor{Black}
    \Text(306,394)[cl]{\Black{$\rm 2HDM$}}
    \Text(306,382)[cl]{\Black{$\rm SM$}}
    \SetWidth{5.5}
    \SetColor{Green}
    \Line(280,393)(300,393)
    \SetColor{Blue}
    \Line(280,381)(300,381)
    \SetWidth{0.2}
    \SetColor{Gray}
    \Line(170,14)(170,404)
    \Line(170,-15)(170,-13)
    \Line(200,14)(200,404)
    \Line(200,-15)(200,-13)
    \Line(230,-15)(230,404)
    \Line(290,-15)(290,372)
    \Line(320,-15)(320,372)
    \SetWidth{5.5}
    \SetColor{Green}
    \Line(260,397)(255.71,397) 
    \Line(260,379)(257.14,379) 
    \Line(260,361)(263.90,361) 
    \Line(260,343)(260.01,343) 
    \Line(260,325)(252.17,325) 
    \Line(260,307)(311.73,307) 
    \Line(260,289)(332.73,289) 
    \Line(260,271)(258.37,271) 
    \Line(260,253)(290.72,253) 
    \Line(260,235)(186.16,235) 
    \Line(260,217)(233.11,217) 
    \Line(260,199)(284.75,199) 
    \Line(260,181)(242.51,181) 
    \Line(260,163)(262.26,163) 
    \Line(260,145)(302.92,145) 
    \Line(260,127)(289.60,127) 
    \Line(260,109)(255.46,109) 
    \Line(260,91)(282.65,91) 
    \Line(260,73)(294.33,73) 
    \Line(260,55)(220.63,55) 
    \Line(260,37)(283.13,37) 
    \Line(260,19)(278.27,19) 
    \Line(260,1)(272.31,1) 
    \SetColor{Blue}
    \Line(260,391)(255.71,391) 
    \Line(260,373)(257.14,373) 
    \Line(260,355)(263.90,355) 
    \Line(260,337)(260.01,337) 
    \Line(260,319)(255.17,319) 
    \Line(260,301)(311.97,301) 
    \Line(260,283)(331.32,283) 
    \Line(260,265)(258.53,265) 
    \Line(260,247)(290.48,247) 
    \Line(260,229)(189.31,229) 
    \Line(260,211)(235.40,211) 
    \Line(260,193)(287.51,193) 
    \Line(260,175)(242.53,175) 
    \Line(260,157)(262.43,157) 
    \Line(260,139)(310.17,139) 
    \Line(260,121)(283.20,121) 
    \Line(260,103)(255.41,103) 
    \Line(260,85)(281.53,85) 
    \Line(260,67)(286.67,67) 
    \Line(260,49)(213.13,49) 
    \Line(260,31)(276.25,31) 
    \Line(260,13)(280.00,13) 
    \Line(260,-5)(272.41,-5) 
    \SetWidth{1.0}
    \SetColor{Black}
    \Text(-20,415)[cl]{\Black{\textbf{Observable}}}
    \Text(65,415)[cl]{\Black{\textbf{Best-fit value}}}
    \Text(172,415)[cl]{\Black{\textbf{Deviation}}}
    \Text(360,415)[cl]{\Black{$\bm{\Delta \chi ^2}$}}
    \Text(-10,394)[cl]{\Black{$m_t^{\text{\tiny pole}} $ [GeV]$^{(*)}$}}
    \Text(-10,376)[cl]{\Black{$m_Z $ [GeV]$^{(*)}$}}
    \Text(-10,358)[cl]{\Black{$\Delta\alpha_\text{had}^{(5)}{}^{(*)}$}}
    \Text(-10,340)[cl]{\Black{$\alpha_s {}^{(*)}$}}
    \Text(-10,322)[cl]{\Black{$\Gamma_Z $ [GeV]}}
    \Text(-10,304)[cl]{\Black{$\sigma^0_\text{had} $ [nb]}}
    \Text(-10,286)[cl]{\Black{$R_b^0 $}}
    \Text(-10,268)[cl]{\Black{$R_c^0 $}}
    \Text(-10,250)[cl]{\Black{$R_\ell^0 $}}
    \Text(-10,232)[cl]{\Black{$A_\text{FB}^{0,b} $}}
    \Text(-10,214)[cl]{\Black{$A_\text{FB}^{0,c} $}}
    \Text(-10,196)[cl]{\Black{$A_\text{FB}^{0,\ell} $}}
    \Text(-10,178)[cl]{\Black{${\cal A}_b $}}
    \Text(-10,160)[cl]{\Black{${\cal A}_c $}}
    \Text(-10,142)[cl]{\Black{${\cal A}_\ell $}}
    \Text(-10,124)[cl]{\Black{$m_W $ [GeV]}}
    \Text(-10,106)[cl]{\Black{$\Gamma_W $ [GeV]}}
    \Text(-10,88)[cl]{\Black{$\sin^2\theta_\ell^\text{eff} $}}
    \Text(-10,70)[cl]{\Black{$\mu _{\text{\tiny comb}}(H\to \gamma \gamma )$}}
    \Text(-10,52)[cl]{\Black{$\mu _{\text{\tiny comb}}(H\to WW^*)$}}
    \Text(-10,34)[cl]{\Black{$\mu _{\text{\tiny comb}}(H\to ZZ^*)$}}
    \Text(-10,16)[cl]{\Black{$\mu _{\text{\tiny comb}}(H\to b\oline{b})$}}
    \Text(-10,-2)[cl]{\Black{$\mu _{\text{\tiny comb}}(H\to \tau \tau )$}}
    \Text(90,394)[cl]{\Black{$174.01$}} 
    \Text(90,376)[cl]{\Black{$91.1878$}} 
    \Text(90,358)[cl]{\Black{$0.02756$}} 
    \Text(90,340)[cl]{\Black{$0.1188$}} 
    \Text(90,322)[cl]{\Black{$2.4958$}} 
    \Text(90,304)[cl]{\Black{$41.477$}} 
    \Text(90,286)[cl]{\Black{$0.21469$}} 
    \Text(90,268)[cl]{\Black{$0.1723$}} 
    \Text(90,250)[cl]{\Black{$20.741$}} 
    \Text(90,232)[cl]{\Black{$0.1031$}} 
    \Text(90,214)[cl]{\Black{$0.0738$}} 
    \Text(90,196)[cl]{\Black{$0.0163$}} 
    \Text(90,178)[cl]{\Black{$0.935$}} 
    \Text(90,160)[cl]{\Black{$0.668$}} 
    \Text(90,142)[cl]{\Black{$0.1473$}} 
    \Text(90,124)[cl]{\Black{$80.366$}} 
    \Text(90,106)[cl]{\Black{$2.091$}} 
    \Text(90,88)[cl]{\Black{$0.2315$}} 
    \Text(90,70)[cl]{\Black{$0.95$}} 
    \Text(90,52)[cl]{\Black{$0.96$}} 
    \Text(90,34)[cl]{\Black{$0.95$}} 
    \Text(90,16)[cl]{\Black{$1.03$}} 
    \Text(90,-2)[cl]{\Black{$1.00$}} 
    \Text(364,394)[cl]{\Black{--}} 
    \Text(364,376)[cl]{\Black{--}} 
    \Text(364,358)[cl]{\Black{--}} 
    \Text(364,340)[cl]{\Black{--}} 
    \Text(364,322)[cl]{\Black{$0.00$}} 
    \Text(364,304)[cl]{\Black{$2.29$}} 
    \Text(364,286)[cl]{\Black{$5.23$}} 
    \Text(364,268)[cl]{\Black{$0.17$}} 
    \Text(364,250)[cl]{\Black{$0.54$}} 
    \Text(364,232)[cl]{\Black{$5.32$}} 
    \Text(364,214)[cl]{\Black{$0.42$}} 
    \Text(364,196)[cl]{\Black{$0.74$}} 
    \Text(364,178)[cl]{\Black{$0.06$}} 
    \Text(364,160)[cl]{\Black{$0.00$}} 
    \Text(364,142)[cl]{\Black{$2.90$}} 
    \Text(364,124)[cl]{\Black{$0.41$}} 
    \Text(364,106)[cl]{\Black{$0.00$}} 
    \Text(364,88)[cl]{\Black{$0.55$}} 
    \Text(364,70)[cl]{\Black{$1.40$}} 
    \Text(364,52)[cl]{\Black{$1.30$}} 
    \Text(364,34)[cl]{\Black{$0.59$}} 
    \Text(364,16)[cl]{\Black{$0.35$}} 
    \Text(364,-2)[cl]{\Black{$0.94$}} 
    \Text(168,-28)[cb]{\Black{\large $-3$}}
    \Text(198,-28)[cb]{\Black{\large $-2$}}
    \Text(228,-28)[cb]{\Black{\large $-1$}}
    \Text(288,-28)[cb]{\Black{\large $+1$}}
    \Text(318,-28)[cb]{\Black{\large $+2$}}
    \SetWidth{0.7}
    \Line(165,-13)(165,404)
    \Line(165,-13)(340,-13)
    \Line(165,404)(340,404)
    \Line(340,-13)(340,404)
    \Line(260,-13)(260,404)
    \put(168,-10){\includegraphics[width=0.08\linewidth]{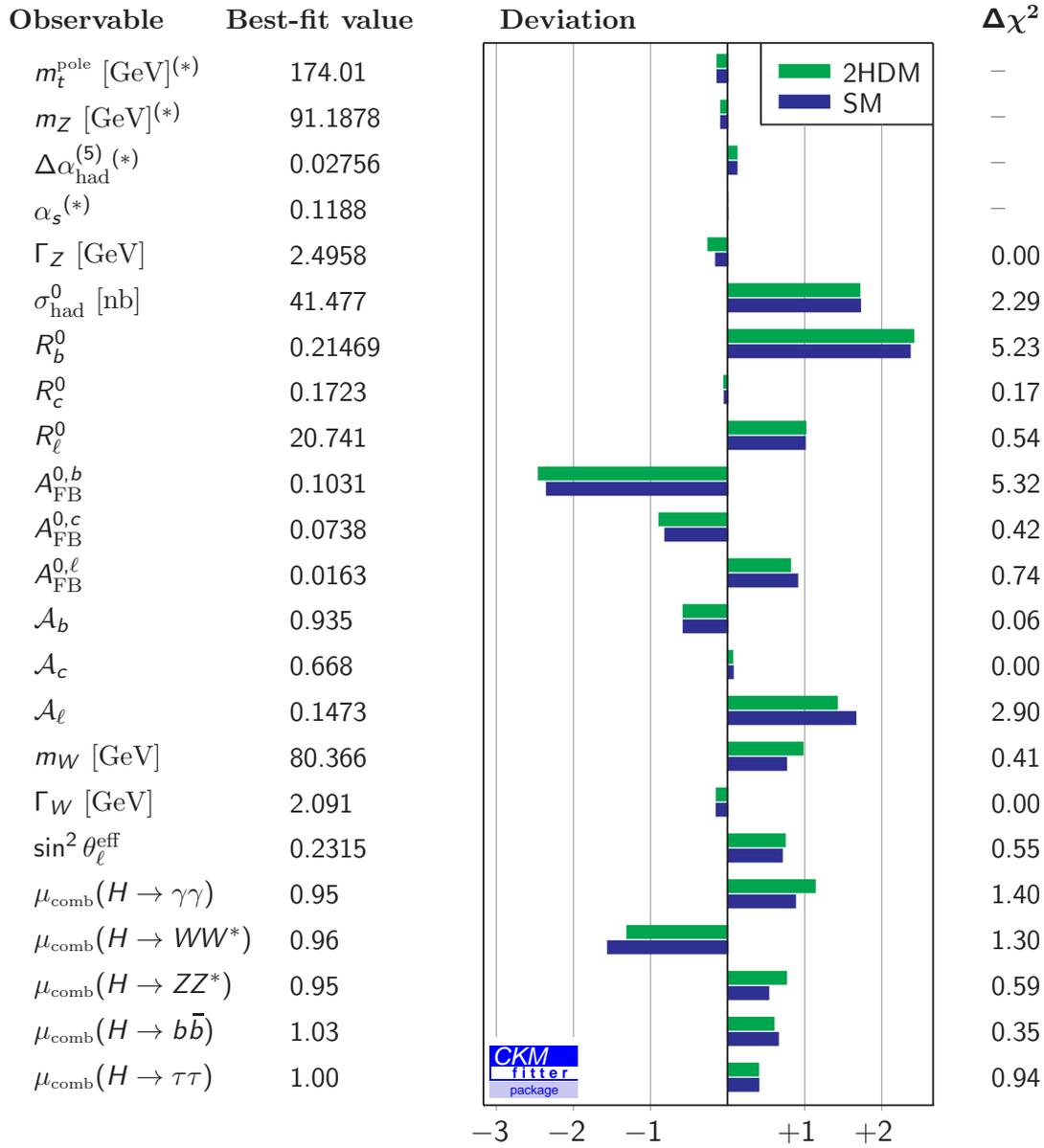}}
\end{picture}
\caption[2HDM deviations after the Higgs discovery.]{Deviations of the EWPO and the Higgs signal strengths from the best-fit point in the 2HDM (green) and in the SM (blue). \otto{$^{(*)}$ The first four parameters have been treated as fixed in the fit, so their 2HDM deviations are the ones from the SM fit.}}
\label{fig:deviations2HDM}
\end{figure}

Most notably is that $\beta-\alpha$ takes approximately the decoupling limit value of $\frac{\pi}{2}$ at the best-fit point, making the 2HDM and the SM hard to distinguish in low energy observables. Hence, the electroweak precision fit only marginally differs from the SM fit as shown in the deviations list in Fig.\ \ref{fig:deviations2HDM}.
The signal strength part is a bit more interesting: all bosonic Higgs decays are slightly suppressed in the 2HDM, which could \ottooo{release the tension of the $\mu (H\to WW^*)$ measurements}, but on the other hand, more Higgs decays to two photons or $Z$ bosons have been \ottooo{observed} than one would expect in the SM, and that is exactly the opposite \ottooo{of} the 2HDM prediction, so the \ottooo{corresponding} deviations are somewhat increased. (The diminished signal strength can also be seen in the one-dimensional observable predictions \ottooo{of the Higgs decays to neutral bosons} in App.\ \ref{fitresults}.)
Although I also use the two mentioned flavour observables, I do not show their deviation because they are zero due to the theoretical errors involved. Furthermore, they do not contribute to the total $\chi^2_{\rm min}$ of the best-fit point.

\begin{figure}[htbp]
 \centering
 \subfigure[]{
 \begin{picture}(210,150)(0,0)
	      \put(7,0){\includegraphics[width=0.5\linewidth]{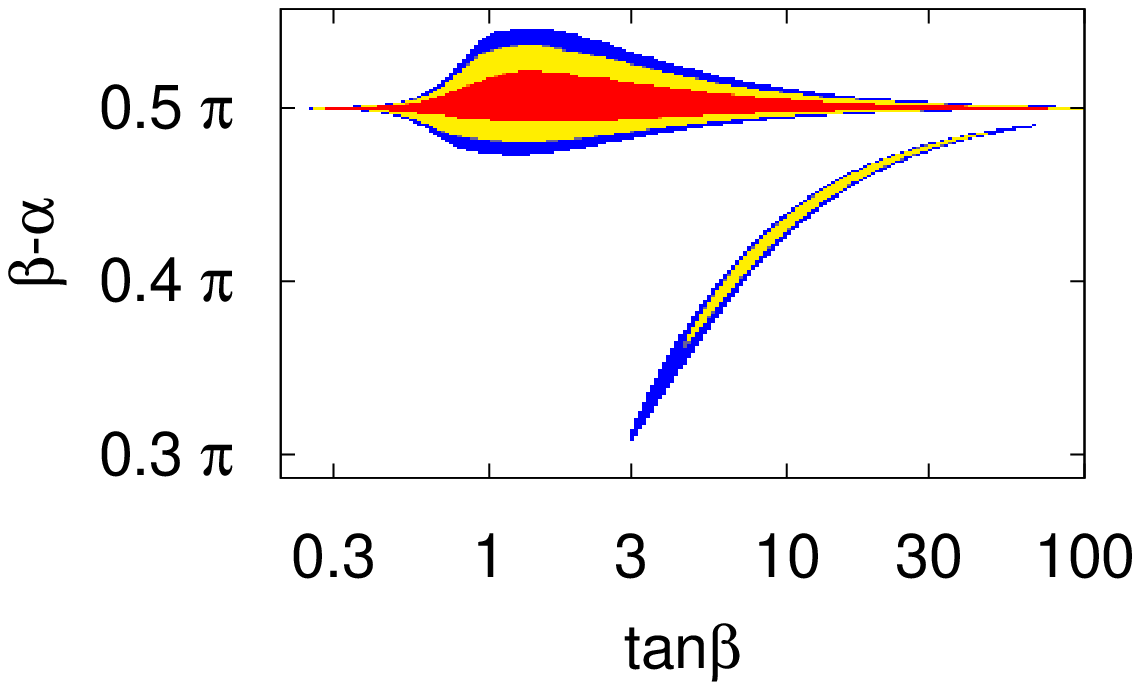}}
	      \put(45.4,40.2){\includegraphics[width=0.08\linewidth]{Images/CKMfitterPackage.eps}}
	      \end{picture}
	     }
 \qquad
 \subfigure[]{\begin{picture}(180,150)(0,0)
	      \put(0,0){\includegraphics[width=0.5\linewidth]{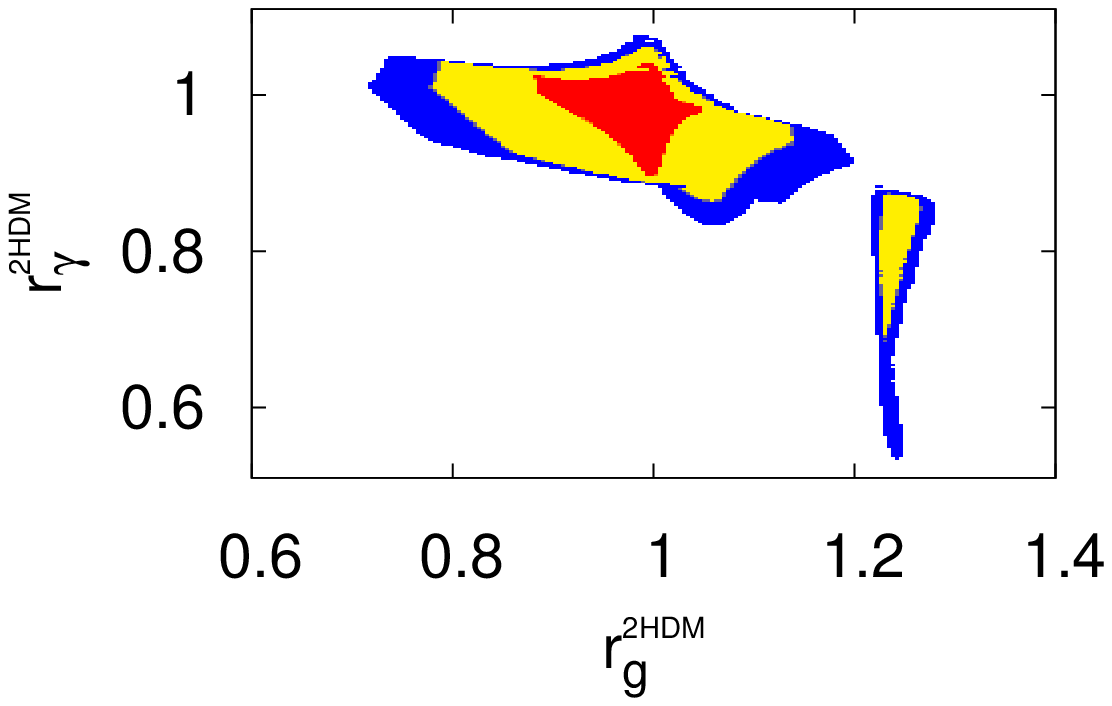}}
	      \put(38.3,40.2){\includegraphics[width=0.08\linewidth]{Images/CKMfitterPackage.eps}}
	      \end{picture}
	     }
 \caption[2HDM scans of the $\tan \beta$-$(\beta-\alpha)$ plane and the $r_g^\text{\tiny 2HDM}$-$r_\gamma^\text{\tiny 2HDM}$ plane.]{Large regions in the $\tan \beta$-$(\beta-\alpha)$ plane are excluded (a). The lower branch is excluded at $1\sigma$. It corresponds to the right area in the $r_g^\text{\tiny 2HDM}$-$r_\gamma^\text{\tiny 2HDM}$ plane (b).}
 \label{fig:ltbbma}
\end{figure}

To further investigate how much $\beta-\alpha$ can deviate from $\frac{\pi}{2}$, I show a scan over the $\tan \beta$-$(\beta-\alpha)$ plane in Fig.\ \ref{fig:ltbbma}(a). One important feature is that $\tan \beta$ is only constrained from below by the $\Delta m_{B_s}$ measurement\ottooo{, and only because I impose $m_{H^+}<10$\:TeV}. At one standard deviation, $\beta-\alpha$ cannot depart from $\frac{\pi}{2}$ by more than $0.02\pi$, independent of $\tan \beta$. For $\tan \beta<0.5$, $\beta-\alpha$ is basically fixed to the decoupling value. This is due to the flavour observables, which can only compensate the large low $\tan \beta$ contributions with a heavy $H^+$. This in turn entails heavy neutral Higgs masses \ottooo{because of the EWPO}, and we obtain the decoupling limit. If $\tan \beta$ is larger than $3$, a second branch appears, and $\beta-\alpha$ can be smaller than $0.4\pi$.
This branch, however, which is characterized by comparably small heavy Higgs masses, is excluded at $1\sigma$ by the same observables that force $\beta-\alpha$ to the decoupling limit for small $\tan \beta$.
For large $\tan \beta$, this strip also approaches the $\beta-\alpha=\frac{\pi}{2}$ limit. In the plane of the coupling ratios $r_g^\text{\tiny 2HDM}$ and $r_\gamma^\text{\tiny 2HDM}$ in Fig.\ \ref{fig:ltbbma}(b), the branch directly corresponds to the disjoint region at the lower right, which features an enhanced ggF Higgs production and at the same time a suppression of diphoton decays. Opposed to this, the $1\sigma$ region prefers SM like Higgs couplings, because $\beta-\alpha=\frac{\pi}{2}$ is equivalent to $r_i^\text{\tiny 2HDM}=1$ for all $i$ (compare Table \ref{tab:Higgscouplings}).

\begin{figure}[ht]
 \centering
 \subfigure[]{
 \begin{picture}(200,150)(0,0)
	      \put(4,0){\includegraphics[width=0.5\linewidth]{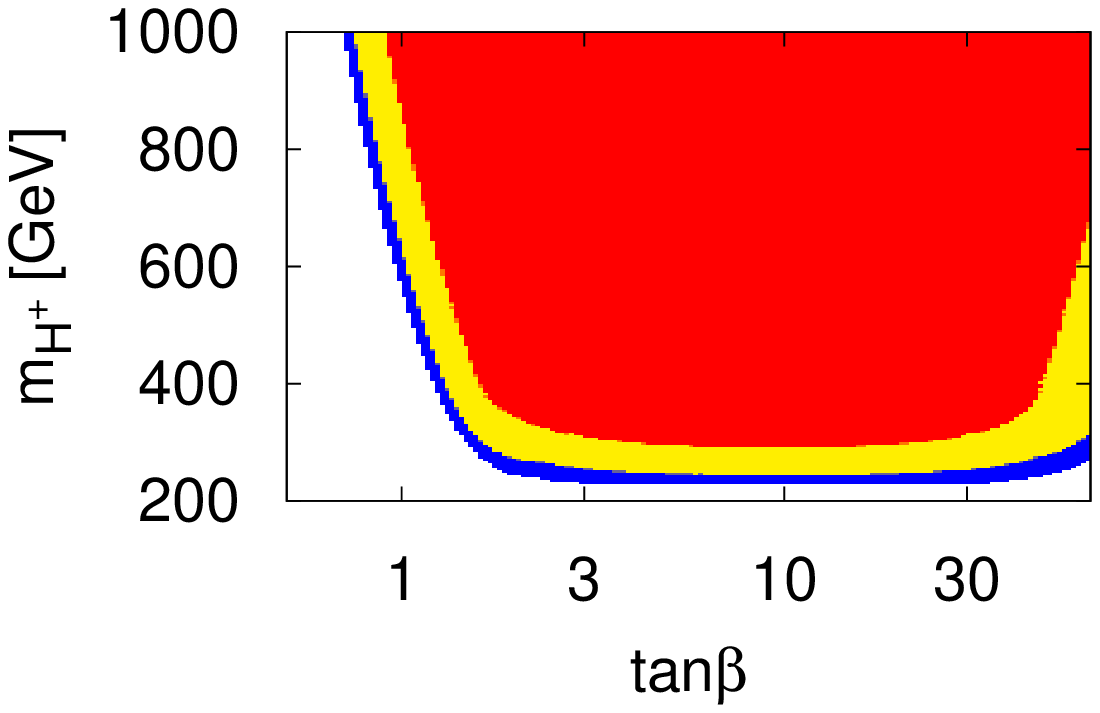}}
	      \put(144.5,98.9){\includegraphics[width=0.08\linewidth]{Images/CKMfitterPackage.eps}}
	      \end{picture}
	     }
 \qquad
 \subfigure[]{
 \begin{picture}(180,150)(0,0)
	      \put(-0.6,0){\includegraphics[width=0.5\linewidth]{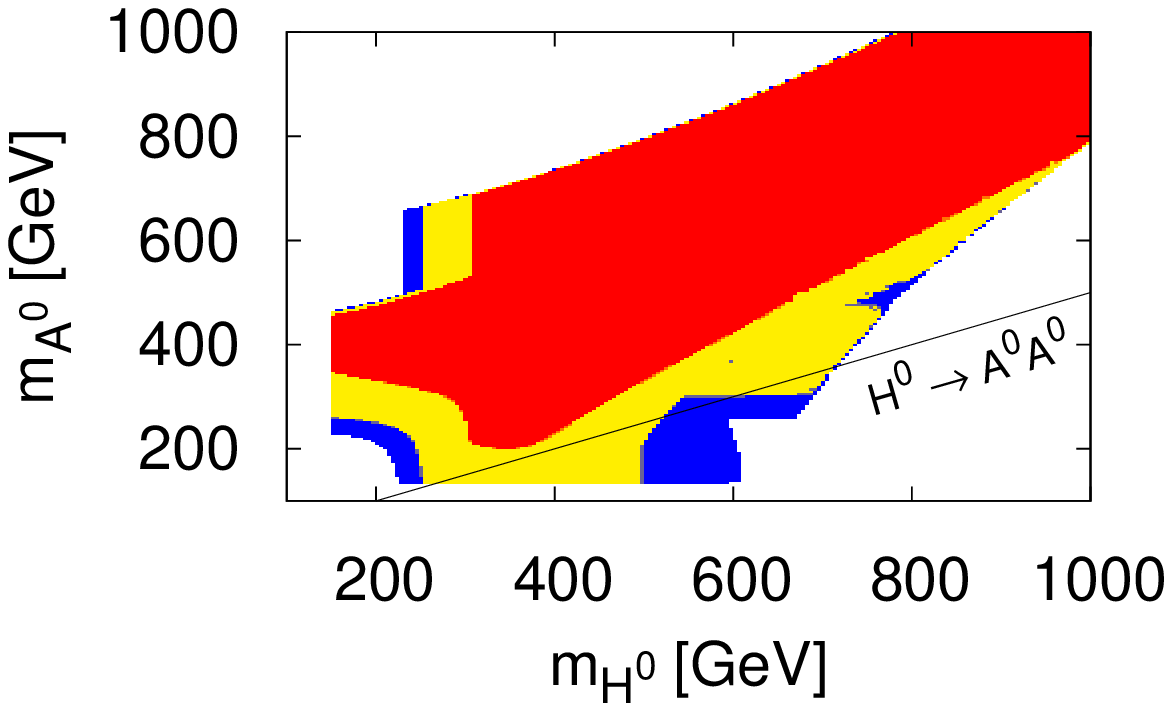}}
	      \put(37.9,98.9){\includegraphics[width=0.08\linewidth]{Images/CKMfitterPackage.eps}}
	      \end{picture}
	     }\\
 \subfigure[]{\begin{picture}(200,150)(0,0)
	      \put(4,0){\includegraphics[width=0.5\linewidth]{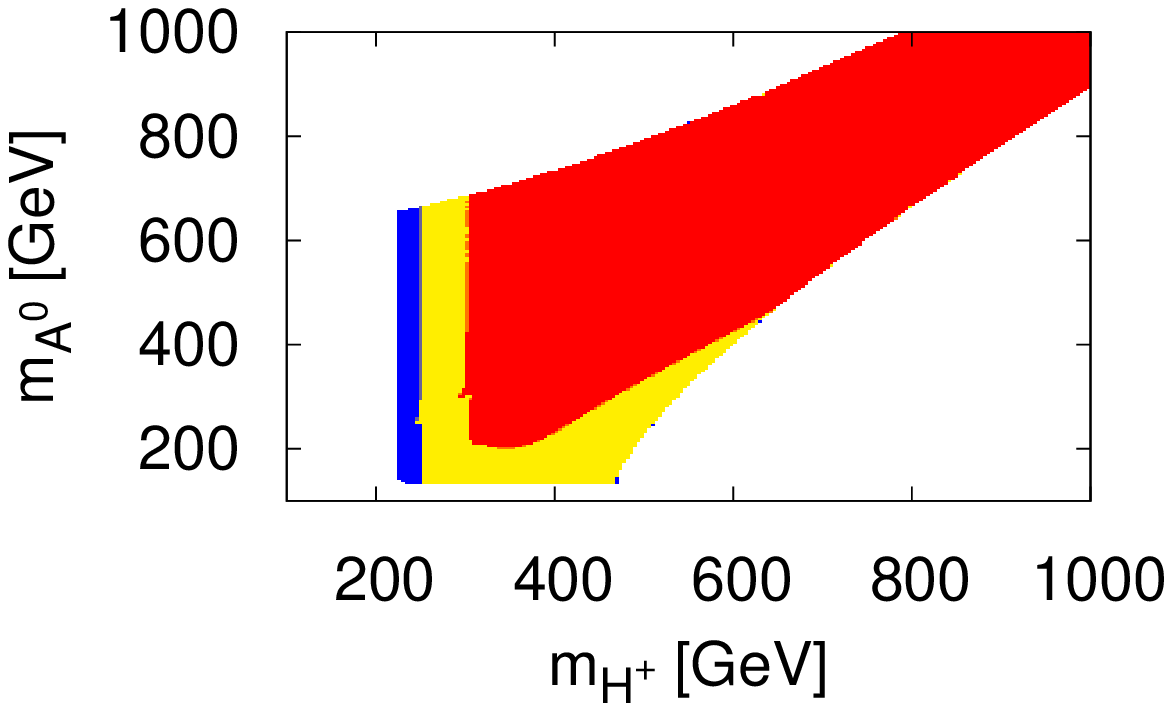}}
	      \put(144.5,40.2){\includegraphics[width=0.08\linewidth]{Images/CKMfitterPackage.eps}}
	      \end{picture}
	     }
 \qquad
 \subfigure[]{\begin{picture}(180,150)(0,0)
	      \put(3,0){\includegraphics[width=0.5\linewidth]{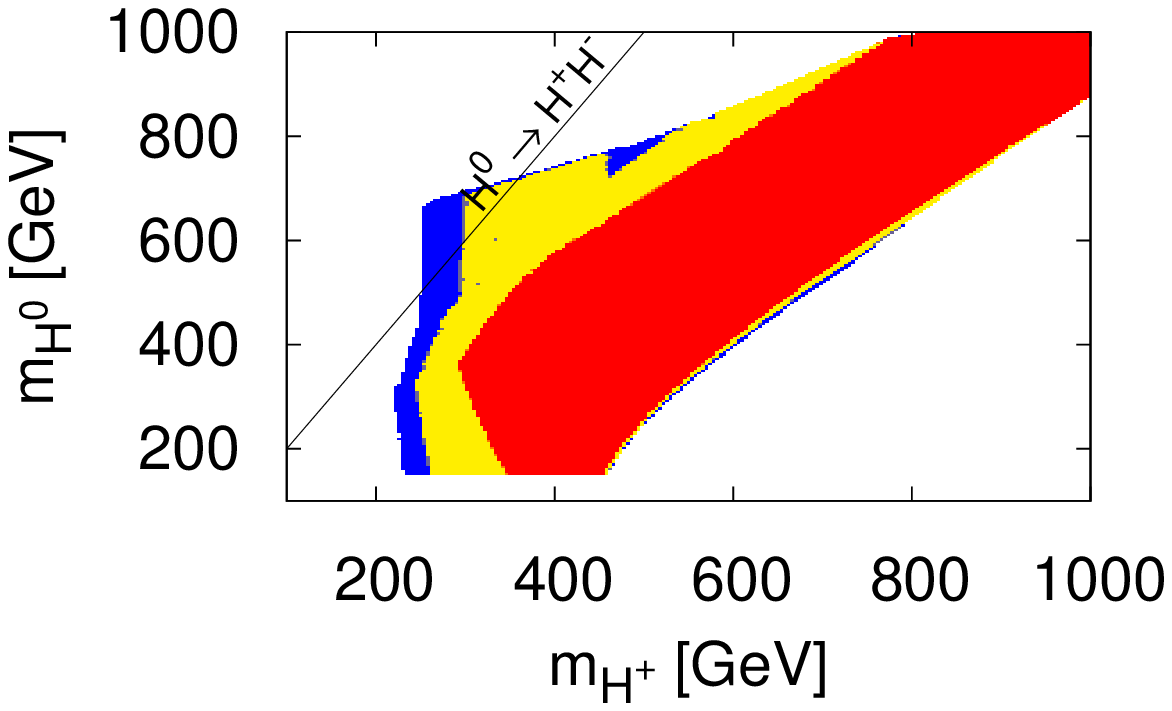}}
	      \put(143.5,40.2){\includegraphics[width=0.08\linewidth]{Images/CKMfitterPackage.eps}}
	      \end{picture}
	     }
 \caption[2HDM scans of the $\tan \beta$-$m_{H^+}$ plane and the three heavy Higgs mass planes.]{Light charged Higgs bosons and low $\tan \beta$ values are excluded by a combination of flavour observables (a). The allowed heavy Higgs mass values are shown in the $m_{H^0}$-$m_{A^0}$ plane (b), the $m_{H^+}$-$m_{A^0}$ plane (c) and the $m_{H^+}$-$m_{H^0}$ plane (d). \ottooo{$m_{A^0}>200$\:GeV and $m_{H^+}>290$\:GeV at $1\sigma $. The magnitude of potential mass splittings decreases with heavier Higgs masses. The decays $H^0\to A^0 A^0$ and $H^0\to H^+ H^-$ can be excluded at $1\sigma $ and for $m_{H^0}>715$\:GeV even at $3\sigma $.}}
 \label{fig:2HDMmasses}
\end{figure}

In Fig.\ \ref{fig:2HDMmasses}(a) I show the $\tan \beta$-$m_{H^+}$ plane\ottooo{,} compared to Fig.\ \ref{fig:DeltaMBsImpact} and Fig.\ \ref{fig:bsg}(b) only for small charged Higgs masses. The flavour observables cut away low $\tan \beta$ and low $m_{H^+}$ regions. If $\tan \beta<1$, the charged Higgs has to be heavier than $600$\:GeV at $95\%$ CL.

Scans over the heavy 2HDM masses can be found in Fig.\ \ref{fig:2HDMmasses}\ottooo{(b)-(d)}. One can see that large mass splittings are disfavoured; especially in the high mass regions, perturbative quartic couplings force the masses of the heavy Higgs particles to be relatively close to each other. In combination with constraints for the low mass scenarios like the flavour observables, this allows us to exclude certain on-shell decays \otto{for $m_{H^0}$ larger than $715$\:GeV: the possibility that an $H^0$ can decay into two charged Higgs bosons is excluded at $3\sigma$, which can be seen in Fig.\ \ref{fig:2HDMmasses}(d), and its decay into two $A^0$ is also excluded at $3\sigma$, compare Fig.\ \ref{fig:2HDMmasses}(b)}. \ottooo{Both decays can be ruled out at $1\sigma$ independent of the 2HDM masses.} Moreover, 2HDM realizations where the $H^0$ and the $A^0$ are simultaneously lighter than $250$\:GeV are excluded at two standard deviation; $m_{A^0}<200$\:GeV is disfavoured at $1\sigma$.
\otto{The heavy ${\cal CP}$-even Higgs mass is not constrained from below.}
If we chose $4\pi$ as a more conservative perturbativity bound on the quartic couplings, the mass splittings can be \otto{larger by $250$\:GeV} \cite{Eberhardt:2013uba}.

As already mentioned, the SM is nested in the 2HDM. However, the assumption of linear behaviour of the Higgs signal strengths is only an approximation since their errors are still sizeable. And especially the flavour observables are not Gaussian distributed, because they have large theoretical uncertainties. As all experimental data pushes the 2HDM towards its decoupling limit, i.e.\ the SM, it is useless to calculate the $p$-value of the SM: it will be one.\\
The next step could be the inclusion of further flavour observables to examine the large $\tan \beta$ regions. As the (semi-)tauonic $B$ decay measurements cannot be explained in the framework of the 2HDM of type II, it would be interesting to consider a more general model of type III, where flavour-changing neutral currents can occur at tree level.\\
As \otto{a} final remark about the 2HDM\ottooo{,} I want to mention that there have been analyses which combine the idea of a fourth fermion generation with an additional Higgs doublet \cite{BarShalom:2011zj,Chen:2012wz}. The signal strengths, which essentially ruled out the SM4, have to be reinterpreted in these scenarios. For example, due to supplementary Higgs loops in this combined model the diphoton decay would not necessarily be suppressed like it was the case in the SM4. If perturbative heavy fermions were to be found at the LHC (to which there are no indications at the moment), it would mean at the same time that some other part of the SM Lagrangian needs to be modified -- the most obvious choice would then probably be the Higgs \ottooo{sector}.

%% file: conclusions.tex
\chapter{Conclusions} \label{Conclusions}

In this thesis I have presented comprehensive analyses of the viable parameter regions of the Standard Model of particle physics and two of its extensions: the Standard Model with four fermion generations and the Two-Higgs-Doublets model of type II. I \otto{have} performed global fits which show that the SM is compatible with electroweak precision data and the Higgs signal strengths, the SM4 is excluded by the experimental results at more than $5\sigma$, and \otto{that in the 2HDM, despite the fit preferring the decoupling limit where it mimics the SM, neutral Higgs boson masses below $200$\:GeV are still allowed}.

\begin{figure}[htbp]
 \centering
 \begin{picture}(190,160)(0,0)
  \put(-36,0){\includegraphics[width=260pt]{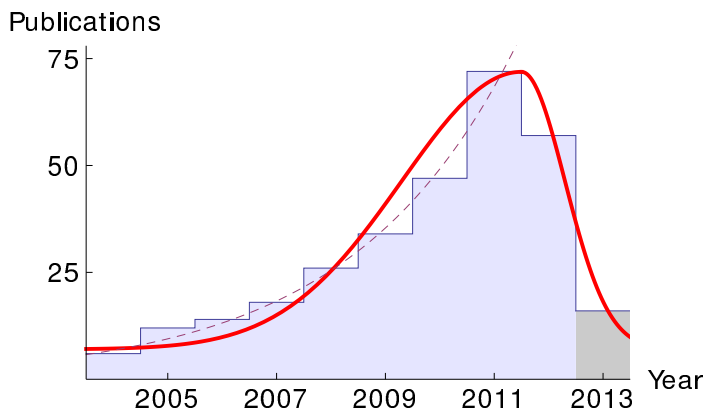}}
 \end{picture}
 \caption[SM4 publications from 2004 to 2013.]{Distribution of SM4 publications over the last decade.}
 \label{fig:SM4Publications}
\end{figure}

The Higgs discovery announced on 4th July 2012 means that for the first time we are equipped with direct measurements \otto{of} \textit{all} SM parameters. In that sense one could say that the SM is complete now. After that date, we accomplished the first global SM fit to EWPO and Higgs data \cite{Eberhardt:2012gv}. Almost all observables of the EWPO and the Higgs signal strengths are in good agreement with the SM; the $b$ quark forward-backward asymmetry and the decay width ratio $R_b^0$ have the largest deviations. If the combined Higgs mass input is included in the fit, the $W$ boson mass and the top quark mass receive \otto{shifts which are sizeable} compared to their experimental uncertainties from their measured central values.\\
The same observables which corroborate the SM are in con\otto{flict} with the SM4 hypothesis. Most notably, the signal strength of Higgs decays to two photons, which should approximately vanish in the SM4, has a deviation of $6\sigma$. Also the signal strength of $H\to b\oline{b}$ decays is expected to be suppressed in the SM4, yet the measurements seem to favour the SM expectation, and the SM4 deviation is almost $3\sigma$. The best-fit neutrino mass is required to be lighter than $m_H/2$, such that the invisible decay $H\to \nu_4 \nu_4$ is possible.
\ottooo{Even then fewer tauonic Higgs decays than expected in the SM4 have been observed.}
The EWPO fit prefers no mixing between the fourth generation and the SM quarks, and it demands small mass splittings in the $SU(2)$ fermion doublets, so the best-fit mass of the fourth generation charged lepton is almost as low as allowed by direct searches. This also holds for the fourth generation quarks: the SM4 particles do not decouple, and the higher their masses the larger is the effect on the discussed observables. This non-nestedness makes it difficult to reliably determine the $p$-value. \otto{However}, exactly this property enables us to rule out a fourth fermion generation. For this purpose, we \otto{have} used the newly developed program \textit{my}Fitter to carry out a likelihood ratio test comparing the SM and the SM4.
With the data that was publicly available shortly after the announcement of the Higgs discovery, we could exclude the SM4 with a $p$-value of $1.1\cdot 10^{-7}$ corresponding to $5.3\sigma$ \cite{Eberhardt:2012gv}.
With the latest Higgs search data, the $p$-value is expected to be even smaller. After the Higgs discovery, the interest in the SM4 dropped almost immediately. The rise and fall of the SM4 in the last decade is illustrated in Fig.\ \ref{fig:SM4Publications}, which can be approximated by the very same asymmetric Gaussian p.d.f.\ that we know already from Fig.\ \ref{fig:Gauss}(b), assuming an offset of $7$. Like in the introduction I display the publication density function of articles containing the expressions ``fourth generation'', ``4th generation'', ``fourth family'', or ``4th family'' in their title \cite{inspirehep:2013}.
Fig.\ \ref{fig:SM4Publications} shows the number of papers released within the last ten years ordered by the year of publication. While the slope seemed to be an exponential at the beginning (dashed line), the publication rate reached a maximum in the year 2011 and has been receding since then. (The 2013 data shaded in grey were extrapolated assuming a constant distribution over the year.)

In contrast to the SM4, the 2HDM and the SM are nested, so that \otto{one can only exclude the parameter regions of the 2HDM} which exhibit a fundamentally different phenomenology than the SM. In order to analyze these regions, I implemented the 2HDM of type II with soft $Z_2$ breaking and with perturbativity, positivity and stability conditions into CKMfitter. Additionally to the EWPO and Higgs signal strengths I also included the flavour-changing neutral current \ottooo{observables} ${\cal B}(\oline{B}\to X_s\gamma)$ and $\Delta m_{B_s}$ in the fit because they severely constrain low mass regions of the parameter space. The discovered Higgs boson is interpreted as the lighter ${\cal CP}$-even mass eigenstate throughout and its signal strengths are treated similar to the SM4. The heavy ${\cal CP}$-even mass eigenstate was constrained by the non-observation of $H\to ZZ$ decays between $150$\:GeV and $1$\:TeV. The result of the fit is that the decoupling limit is 
preferred.
While low $\tan \beta$ and $m_{H^+}$ values are excluded by the flavour observables, perturbativity and the EWPO delimit the mass splittings of the heavy Higgs particles. Only for an intermediate $\tan \beta$ range $\beta-\alpha$ is allowed to deviate substantially from the decoupling value $\frac{\pi}{2}$, but these shifts are excluded at $1\sigma$. \otto{The decays $H^0\to H^+H^-$ and $H^0\to A^0A^0$ are excluded at $3\sigma$, if the $H^0$ is heavier than $715$\:GeV.} The results have been published in \cite{Eberhardt:2013uba}. \otto{Furthermore, I have performed scans predicting the EWPO and the Higgs signal strengths in the three models, which can be found in Appendix \ref{fitresults}.}\\
The main message of the evaluated LHC data from 2011 and 2012 is that the SM is now complete and \otto{con}firmed to an unprecedented extent. Many physicists hope that the next \ottooo{LHC} run, which will start in 2015 with an increased centre-of-mass energy, reveals unexpected signatures, maybe even the discovery of new particles. It will certainly provide us with constraints on many models beyond the SM. However, to determine their viability or even to state an exclusion, it is indispensable to perform statistically correct analyses. \otto{The methodology and the tools developed for this thesis can be used to study a wide class of \ottooo{these} new physics models.}

%% file: inputs.tex
\appendix
\noappendicestocpagenum
\addappheadtotoc

\chapter{Inputs} \label{inputs}

In this appendix I list all numerical inputs that were used for the fits in this thesis. First of all, I present the data that were important for the SM fit in Chapter \ref{SM}. It is followed by a discussion of the Higgs observables, which are crucial for the SM4 fits in Chapter \ref{SM4}, but also for the other two models. Finally, the numerical values for the flavour observables that were used as constraints in Chapter \ref{2HDM} are shown.
For all observables, I give the central input value and the errors as well as the source. The first error is always the statistical uncertainty -- in some cases they are asymmetric --, and the second is the systematic uncertainty, if any.\\
As far as the CKM matrix elements are concerned, I rely on the values from the Particle Data Book.
Yet especially for the SM4 part one needs to be careful not to include any inputs basing on $3\times 3$ unitarity of $V$. For example $|V_{tb}|=1$ is a valid approximation in the SM and in the 2HDM, because we know to good precision that $|V_{ub}|$ and $|V_{cb}|$ are small and the unitarity relation is $|V_{ub}|^2+|V_{cb}|^2+|V_{tb}|^2=1$. But this relation does not necessarily hold in the SM4, so I chose the tree-level values below and let the fitter parts of the respective model account for unitarity. A CKMfitter look-up table for the unitarity triangle angle $\gamma$ was used to constrain the ${\cal CP}$ phase(s).

\begin{center}
\begin{longtable}{lll}
Quantity & Input value & Source\\
\hline
\hline
$|V_{ud}|$ & $0.97425\pm 0.00022$ & \cite{Hardy:2008gy}\\
$|V_{us}|$ & $0.2252\pm 0.0009$ & \cite{Beringer:1900zz}\\
$|V_{ub}|$ & $4.15\cdot 10^{-3}\pm 0.49\cdot 10^{-3}$ & \cite{Beringer:1900zz}\\
$|V_{cd}|$ & $0.230\pm 0.011$ & \cite{Beringer:1900zz}\\
$|V_{cs}|$ & $0.98\pm 0.01\pm 0.10$ & \cite{Beringer:1900zz}\\
$|V_{cb}|$ & $40.9\cdot 10^{-3}\pm 1.1\cdot 10^{-3}$ & \cite{Beringer:1900zz}\\
$|V_{tb}|$ & $0.89\pm 0.07$ & \cite{Beringer:1900zz}\\
$\gamma$ & CKMfitter fit & \cite{CKMfitter:2012mo}\\
\caption{CKM matrix inputs.}
\label{tab:CKMmatrixinputs}
\end{longtable}
\end{center}

\newpage

The EWPO fits presented in Sect.\ \ref{SMEWPO} and \ref{SMhiggssearches}, and also the SM4 and 2HDM fits are based on the following experimental values, mainly stemming from the LEP detectors \ottooo{and SLD}:

\begin{center}
\begin{longtable}{lll}
Quantity & Input value & Source\\
\hline
\hline
\endhead
$m_t^{\text{\tiny pole}}$ & $173.18\pm 0.56\pm 0.75$\:GeV & \cite{Aaltonen:2012ra}\\
$m_Z$ & $91.1876\pm 0.0021$ GeV & \cite{Beringer:1900zz}\\
$\Delta\alpha_\text{had}^{(5)}$ & $0.02757\pm 0.00010$ & \cite{Davier:2010nc}\\
$\alpha_s$ & $0.1202\pm 0.0006\pm 0.0021$ & \cite{Baikov:2008jh}\\
$\Gamma_Z$ & $2.4952\pm 0.0023$ GeV & \cite{ALEPH:2005ab}\\
$\sigma^0_\text{had}$ & $41.541\pm 0.037$ nb & \cite{Beringer:1900zz}\\
$R_b^0$ & $0.21629\pm 0.00066$ & \cite{ALEPH:2005ab}\\
$R_c^0$ & $0.1721\pm 0.0030$ & \cite{ALEPH:2005ab}\\
$R_\ell^0$ & $20.767\pm 0.025$ & \cite{ALEPH:2005ab}\\
$A_\text{FB}^{0,b}$ & $0.0992\pm 0.0016$ & \cite{ALEPH:2005ab}\\
$A_\text{FB}^{0,c}$ & $0.0707\pm 0.0035$ & \cite{ALEPH:2005ab}\\
$A_\text{FB}^{0,\ell}$ & $0.0171\pm 0.0010$ & \cite{ALEPH:2005ab}\\
${\cal A}_b$ & $0.923\pm 0.020$ & \cite{ALEPH:2005ab}\\
${\cal A}_c$ & $0.670\pm 0.027$ & \cite{ALEPH:2005ab}\\
${\cal A}_\ell$ & $0.1499\pm 0.0018$ & \cite{ALEPH:2005ab,Baak:2012kk}\\
$m_W$ & $80.385\pm 0.015\pm 0.004$ GeV & \cite{Group:2012gb,Awramik:2003rn}\\
$\Gamma_W$ & $2.085\pm 0.042$ GeV & \cite{TEW:2010aj}\\
$\sin^2\theta_\ell^\text{eff}$ & $0.2324\pm 0.0012$ & \cite{ALEPH:2005ab}\\
\caption{EWPO inputs.}
\label{tab:EWPOinputs}
\end{longtable}
\end{center}

The systematic uncertainty for $m_W$ was adopted from \ottooo{\cite{Awramik:2003rn} like in the Gfitter publication \cite{Baak:2012kk}, but the} systematic error on $\sin^2\theta_\ell^\text{eff}$ was neglected in this work, since its magnitude is less than 4{\%} of the statistical error. During the last stage of this work, a new combination of Tevatron top mass measurements was released: $m_t^{\text{\tiny pole}}=173.20\pm 0.51\pm 0.71$\:GeV \cite{CDF:2013jga}. Compared to the above value used for my fits, the central value stays approximately the same and the errors decrease slightly; however, I do not expect that the improved uncertainties fundamentally change the results presented in this thesis. Some of the $Z$ pole observables are correlated; their inverse covariance matrix entries are displayed in Table \ref{tab:EWPOcorrelations}.

\newpage

\begin{center}
\begin{longtable}{lllllll}
& $M_Z$ & $\Gamma _Z$ & $\sigma ^0_{\text{had}}$ & $R^0_\ell $ & $A_{FB}^{0,\ell }$ & \\
$M_Z$ & $1$ & $-0.02$ & $-0.05$ & $0.03$ & $0.06$ & \\
$\Gamma _Z$ & & $1$ & $-0.30$ & $0$ & $0$ & \\
$\sigma ^0_{\text{had}}$ & & & $1$ & $0.18$ & $0.01$ & \\
$R^0_\ell $ & & & & $1$ & $-0.06$ & \\
$A_{FB}^{0,\ell }$ & & & & & $1$ & \\[20pt]
& $A_{FB}^{0,c}$ & $A_{FB}^{0,b}$ & $A_c$ & $A_b$ & $R^0_c$ & $R^0_b$\\
$A_{FB}^{0,c}$ & $1$ & $0.15$ & $0.04$ & $-0.02$ & $-0.06$ & $0.07$\\
$A_{FB}^{0,b}$ & & $1$ & $0.01$ & $0.06$ & $0.04$ & $-0.10$\\
$A_c$ & & & $1$ & $0.11$ & $-0.06$ & $0.04$\\
$A_b$ & & & & $1$ & $0.04$ & $-0.08$\\
$R^0_c$ & & & & & $1$ & $-0.18$\\
$R^0_b$ & & & & & & $1$\\[20pt]
\caption{Correlations between $Z$ and leptonic observables and between $b$ and $c$ precision measurements at LEP \cite{ALEPH:2005ab}.}
\label{tab:EWPOcorrelations}
\end{longtable}
\end{center}

The $W$ mass and decay width are also correlated, but this correlation is only small, and since the values for the two observables originate from different experiments, I neglect it.\\
The last missing input on the SM fit was the Higgs mass. I use a combination of the following LHC measurements:\\[-10pt]
\begin{center}
\begin{longtable}{lll}
Quantity & Input value & Source\\
\hline
\hline
$m_H^{\text{\tiny ATLAS}}(H\to \gamma \gamma)$ & $126.8\pm 0.2\pm 0.7$ GeV & \cite{ATLAS-CONF-2013-012}\\
$m_H^{\text{\tiny CMS}}(H\to \gamma \gamma)$ & $125.4\pm 0.5\pm 0.6$ GeV & \cite{CMS-PAS-HIG-13-001}\\
$m_H^{\text{\tiny ATLAS}}(H\to ZZ)$ & $124.3^{+0.6}_{-0.5}\pm 0.4$ GeV & \cite{ATLAS-CONF-2013-013}\\
$m_H^{\text{\tiny CMS}}(H\to ZZ)$ & $125.8\pm 0.5\pm 0.2$ GeV & \cite{CMS-PAS-HIG-13-002}\\
\caption{Mass measurements for the bosonic resonance around $126$\:GeV. The combination yields $m_H=125.96 ^{+0.18} _{-0.19}$.}
\label{tab:mHinputs}
\end{longtable}
\end{center}

\newpage

\begin{figure}[phtb]
\centering
\begin{picture}(270,630)(0,0)
    \SetWidth{0.2}
    \SetColor{Gray}
    \Line(140,4)(140,633)
    \Line(160,4)(160,633)
    \Line(180,4)(180,633)
    \Line(200,4)(200,633)
    \Line(220,4)(220,633)
    \Line(240,4)(240,633)
    \Line(260,4)(260,633)
    \Line(280,4)(280,633)
    \Line(300,4)(300,633)
    \Line(320,4)(320,633)
    \Line(340,4)(340,633)
    \SetWidth{0.1}
    \SetColor{Green}
    \CBox(209.09,7)(211.06,633){Green}{Green}
    \SetWidth{1.1}
    \SetColor{Black}
    \EBox(135,7)(350,633)
    \SetWidth{2.5}
    \SetColor{Blue}
    \Vertex(215.90,630){2}
    \Line(222.80,630)(208.70,630) 
    \Vertex(259.70,620){2}
    \Line(293.60,620)(228.50,620) 
    \Vertex(216.80,610){2}
    \Line(239.60,610)(200.00,610) 
    \Vertex(209.40,600){2}
    \Line(217.90,600)(201.10,600) 
    \Vertex(204.81,590){2}
    \Line(226.66,590)(182.96,590) 
    \Vertex(228.96,580){2}
    \Line(247.83,580)(210.09,580) 
    \Vertex(163.35,570){2}
    \Line(182.51,570)(144.19,570) 
    \Vertex(252.76,560){2}
    \Line(278.30,560)(227.22,560) 
    \Vertex(193.25,550){2}
    \Line(221.70,550)(165.09,550) 
    \Vertex(227.79,540){2}
    \Line(247.24,540)(208.63,540) 
    \Vertex(203.70,530){2}
    \Line(239.98,530)(167.71,530) 
    \Vertex(200.22,520){2}
    \Line(219.09,520)(181.06,520) 
    \Vertex(206.02,510){2}
    \Line(220.24,510)(191.80,510) 
    \Vertex(309.36,500){2}
    \Line(345.93,500)(272.79,500) 
    \Vertex(219.67,490){2}
    \Line(235.64,490)(203.70,490) 
    \Vertex(204.07,480){2}
    \Line(220.37,480)(183.70,480) 
    \Vertex(200.00,470){2}
    \Line(206.00,470)(194.00,470) 
    \Vertex(211.85,460){2}
    \Line(224.07,460)(203.70,460) 
    \Vertex(210.00,450){2}
    \Line(230.00,450)(190.00,450) 
    \Vertex(227.11,440){2}
    \Line(246.91,440)(210.54,440) 
    \Vertex(220.05,430){2}
    \Line(235.24,430)(207.47,430) 
    \Vertex(213.91,420){2}
    \Line(224.34,420)(204.40,420) 
    \Vertex(212.99,410){2}
    \Line(226.19,410)(200.25,410) 
    \Vertex(222.20,400){2}
    \Line(233.86,400)(212.23,400) 
    \Vertex(228.18,390){2}
    \Line(245.06,390)(212.22,390) 
    \Vertex(227.72,380){2}
    \Line(245.52,380)(213.91,380) 
    \Vertex(203.32,370){2}
    \Line(220.66,370)(188.74,370) 
    \Vertex(229.87,360){2}
    \Line(257.03,360)(208.23,360) 
    \Vertex(216.21,350){2}
    \Line(224.50,350)(209.46,350) 
    \Vertex(209.62,340){2}
    \Line(220.52,340)(200.26,340) 
    \Vertex(208.85,330){2}
    \Line(216.06,330)(201.79,330) 
    \Vertex(227.11,320){2}
    \Line(240.61,320)(215.60,320) 
    \Vertex(225.27,310){2}
    \Line(234.48,310)(217.44,310) 
    \Vertex(207.56,300){2}
    \Line(215.31,300)(200.11,300) 
    \Vertex(212.60,290){2}
    \Line(216.10,290)(209.10,290) 
    \Vertex(216.03,280){2}
    \Line(220.26,280)(211.44,280) 
    \Vertex(192.71,270){2}
    \Line(212.89,270)(174.18,270) 
    \Vertex(205.88,260){2}
    \Line(218.23,260)(194.35,260) 
    \Vertex(238.32,250){2}
    \Line(258.74,250)(221.61,250) 
    \Vertex(201.93,240){2}
    \Line(211.95,240)(192.28,240) 
    \Vertex(200.45,230){2}
    \Line(213.07,230)(187.83,230) 
    \Vertex(214.93,220){2}
    \Line(231.27,220)(198.59,220) 
    \Vertex(242.03,210){2}
    \Line(265.42,210)(224.21,210) 
    \Vertex(210.00,200){2}
    \Line(224.41,200)(196.00,200) 
    \Vertex(182.82,190){2}
    \Line(196.00,190)(171.29,190) 
    \Vertex(206.71,180){2}
    \Line(247.47,180)(176.24,180) 
    \Vertex(207.26,170){2}
    \Line(211.43,170)(203.14,170) 
    \Vertex(206.71,160){2}
    \Line(213.71,160)(201.77,160) 
    \Vertex(215.84,150){2}
    \Line(223.55,150)(208.80,150) 
    \Vertex(221.98,140){2}
    \Line(231.26,140)(213.81,140) 
    \Vertex(200.45,130){2}
    \Line(207.50,130)(193.40,130) 
    \Vertex(203.04,120){2}
    \Line(207.87,120)(198.21,120) 
    \Vertex(196.73,110){2}
    \Line(204.90,110)(188.19,110) 
    \Vertex(193.39,100){2}
    \Line(221.24,100)(173.71,100) 
    \Vertex(208.24,90){2}
    \Line(219.01,90)(198.22,90) 
    \Vertex(203.04,80){2}
    \Line(209.72,80)(197.10,80) 
    \Vertex(219.38,70){2}
    \Line(245.37,70)(195.99,70) 
    \Vertex(204.16,60){2}
    \Line(222.35,60)(190.05,60) 
    \Vertex(211.09,50){2}
    \Line(215.25,50)(206.92,50) 
    \Vertex(223.75,40){2}
    \Line(230.49,40)(216.72,40) 
    \Vertex(207.80,30){2}
    \Line(223.93,30)(190.57,30) 
    \Vertex(206.47,20){2}
    \Line(208.72,20)(204.26,20) 
    \Vertex(209.75,10){2}
    \Line(213.07,10)(207.00,10) 
    \SetWidth{1.0}
    \SetColor{Black}
    \Text(-65,630)[cl]{\Black{\scriptsize $\mu ^{\rm T}(VH\to Vb\oline{b})=1.59 ^{+0.69} {}_{-0.72}$}}
    \Text(95,630)[cl]{\Black{\scriptsize \cite{CDF:2013kxa}}}
    \Text(-65,620)[cl]{\Black{\scriptsize $\mu ^{\rm T}(H\to \gamma \gamma )=5.97 ^{+3.39} {}_{-3.12}$}}
    \Text(95,620)[cl]{\Black{\scriptsize \cite{CDF:2013kxa}}}
    \Text(-65,610)[cl]{\Black{\scriptsize $\mu ^{\rm T}(H\to \tau \tau )=1.68 ^{+2.28} {}_{-1.68}$}}
    \Text(95,610)[cl]{\Black{\scriptsize \cite{CDF:2013kxa}}}
    \Text(-65,600)[cl]{\Black{\scriptsize $\mu ^{\rm T}(H\to WW)=0.94 ^{+0.85} {}_{-0.83}$}}
    \Text(95,600)[cl]{\Black{\scriptsize \cite{CDF:2013kxa}}}
    \Text(-65,590)[cl]{\Black{\scriptsize $\mu ^{\rm A7}(VH\to Vb\oline{b})=0.481 \pm 2.185$}}
    \Text(95,590)[cl]{\Black{\scriptsize \cite{Aad:2012an}}}
    \Text(-65,580)[cl]{\Black{\scriptsize $\mu ^{\rm A7}(H\to \gamma \gamma )_{\text{\tiny jj}}=2.896 \pm 1.887$}}
    \Text(95,580)[cl]{\Black{\scriptsize \cite{ATLAS-CONF-2012-091}}}
    \Text(-65,570)[cl]{\Black{\scriptsize $\mu ^{\rm A7}(H\to \gamma \gamma )_{\text{\tiny cch}}=-3.665 \pm 1.916$}}
    \Text(95,570)[cl]{\Black{\scriptsize \cite{ATLAS-CONF-2012-091}}}
    \Text(-65,560)[cl]{\Black{\scriptsize $\mu ^{\rm A7}(H\to \gamma \gamma )_{\text{\tiny ccl}}=5.276 \pm 2.554$}}
    \Text(95,560)[cl]{\Black{\scriptsize \cite{ATLAS-CONF-2012-091}}}
    \Text(-65,550)[cl]{\Black{\scriptsize $\mu ^{\rm A7}(H\to \gamma \gamma )_{\text{\tiny crh}}=-0.675 ^{+2.845} {}_{-2.816}$}}
    \Text(95,550)[cl]{\Black{\scriptsize \cite{ATLAS-CONF-2012-091}}}
    \Text(-65,540)[cl]{\Black{\scriptsize $\mu ^{\rm A7}(H\to \gamma \gamma )_{\text{\tiny crl}}=2.779 ^{+1.945} {}_{-1.916}$}}
    \Text(95,540)[cl]{\Black{\scriptsize \cite{ATLAS-CONF-2012-091}}}
    \Text(-65,530)[cl]{\Black{\scriptsize $\mu ^{\rm A7}(H\to \gamma \gamma )_{\text{\tiny ct}}=0.370 ^{+3.628} {}_{-3.599}$}}
    \Text(95,530)[cl]{\Black{\scriptsize \cite{ATLAS-CONF-2012-091}}}
    \Text(-65,520)[cl]{\Black{\scriptsize $\mu ^{\rm A7}(H\to \gamma \gamma )_{\text{\tiny uch}}=0.022 ^{+1.887} {}_{-1.916}$}}
    \Text(95,520)[cl]{\Black{\scriptsize \cite{ATLAS-CONF-2012-091}}}
    \Text(-65,510)[cl]{\Black{\scriptsize $\mu ^{\rm A7}(H\to \gamma \gamma )_{\text{\tiny ucl}}=0.602 \pm 1.422$}}
    \Text(95,510)[cl]{\Black{\scriptsize \cite{ATLAS-CONF-2012-091}}}
    \Text(-65,500)[cl]{\Black{\scriptsize $\mu ^{\rm A7}(H\to \gamma \gamma )_{\text{\tiny urh}}=10.936 \pm 3.657$}}
    \Text(95,500)[cl]{\Black{\scriptsize \cite{ATLAS-CONF-2012-091}}}
    \Text(-65,490)[cl]{\Black{\scriptsize $\mu ^{\rm A7}(H\to \gamma \gamma )_{\text{\tiny url}}=1.967 \pm 1.597$}}
    \Text(95,490)[cl]{\Black{\scriptsize \cite{ATLAS-CONF-2012-091}}}
    \Text(-65,480)[cl]{\Black{\scriptsize $\mu ^{\rm A7}(H\to \tau \tau )=0.407 ^{+1.630} {}_{-2.037}$}}
    \Text(95,480)[cl]{\Black{\scriptsize \cite{Aad:2012an}}}
    \Text(-65,470)[cl]{\Black{\scriptsize $\mu ^{\rm A7}(H\to WW)=0.0 ^{+0.6} {}_{-0.6}$}}
    \Text(95,470)[cl]{\Black{\scriptsize \cite{ATLAS-CONF-2013-030}}}
    \Text(-65,460)[cl]{\Black{\scriptsize $\mu ^{\rm A7}(H\to ZZ)=1.185 ^{+1.222} {}_{-0.815}$}}
    \Text(95,460)[cl]{\Black{\scriptsize \cite{Aad:2012an}}}
    \Text(-65,450)[cl]{\Black{\scriptsize $\mu ^{\rm A8}(VH\to Vb\oline{b})=1.0 \pm 0.9 \pm 1.1$}}
    \Text(95,450)[cl]{\Black{\scriptsize \cite{ATLAS-CONF-2012-161}}}
    \Text(-65,440)[cl]{\Black{\scriptsize $\mu ^{\rm A8}(H\to \gamma \gamma )_{\text{\tiny lept}}=2.711 ^{+1.980} {}_{-1.657}$}}
    \Text(95,440)[cl]{\Black{\scriptsize \cite{ATLAS-CONF-2013-012}}}
    \Text(-65,430)[cl]{\Black{\scriptsize $\mu ^{\rm A8}(H\to \gamma \gamma )_{\text{\tiny cch}}=2.005 ^{+1.519} {}_{-1.258}$}}
    \Text(95,430)[cl]{\Black{\scriptsize \cite{ATLAS-CONF-2013-012}}}
    \Text(-65,420)[cl]{\Black{\scriptsize $\mu ^{\rm A8}(H\to \gamma \gamma )_{\text{\tiny ccl}}=1.391 ^{+1.043} {}_{-0.951}$}}
    \Text(95,420)[cl]{\Black{\scriptsize \cite{ATLAS-CONF-2013-012}}}
    \Text(-65,410)[cl]{\Black{\scriptsize $\mu ^{\rm A8}(H\to \gamma \gamma )_{\text{\tiny crh}}=1.299 ^{+1.320} {}_{-1.274}$}}
    \Text(95,410)[cl]{\Black{\scriptsize \cite{ATLAS-CONF-2013-012}}}
    \Text(-65,400)[cl]{\Black{\scriptsize $\mu ^{\rm A8}(H\to \gamma \gamma )_{\text{\tiny crl}}=2.220 ^{+1.166} {}_{-0.997}$}}
    \Text(95,400)[cl]{\Black{\scriptsize \cite{ATLAS-CONF-2013-012}}}
    \Text(-65,390)[cl]{\Black{\scriptsize $\mu ^{\rm A8}(H\to \gamma \gamma )_{\text{\tiny ct}}=2.818 ^{+1.688} {}_{-1.596}$}}
    \Text(95,390)[cl]{\Black{\scriptsize \cite{ATLAS-CONF-2013-012}}}
    \Text(-65,380)[cl]{\Black{\scriptsize $\mu ^{\rm A8}(H\to \gamma \gamma )_{\text{\tiny lhmjj}}=2.772 ^{+1.780} {}_{-1.381}$}}
    \Text(95,380)[cl]{\Black{\scriptsize \cite{ATLAS-CONF-2013-012}}}
    \Text(-65,370)[cl]{\Black{\scriptsize $\mu ^{\rm A8}(H\to \gamma \gamma )_{\text{\tiny lmjj}}=0.332 ^{+1.734} {}_{-1.458}$}}
    \Text(95,370)[cl]{\Black{\scriptsize \cite{ATLAS-CONF-2013-012}}}
    \Text(-65,360)[cl]{\Black{\scriptsize $\mu ^{\rm A8}(H\to \gamma \gamma )_{\text{\tiny $\slashed E_{\rm T}$}}=2.987 ^{+2.716} {}_{-2.164}$}}
    \Text(95,360)[cl]{\Black{\scriptsize \cite{ATLAS-CONF-2013-012}}}
    \Text(-65,350)[cl]{\Black{\scriptsize $\mu ^{\rm A8}(H\to \gamma \gamma )_{\text{\tiny thmjj}}=1.621 ^{+0.829} {}_{-0.675}$}}
    \Text(95,350)[cl]{\Black{\scriptsize \cite{ATLAS-CONF-2013-012}}}
    \Text(-65,340)[cl]{\Black{\scriptsize $\mu ^{\rm A8}(H\to \gamma \gamma )_{\text{\tiny uch}}=0.962 ^{+1.090} {}_{-0.936}$}}
    \Text(95,340)[cl]{\Black{\scriptsize \cite{ATLAS-CONF-2013-012}}}
    \Text(-65,330)[cl]{\Black{\scriptsize $\mu ^{\rm A8}(H\to \gamma \gamma )_{\text{\tiny ucl}}=0.885 ^{+0.721} {}_{-0.706}$}}
    \Text(95,330)[cl]{\Black{\scriptsize \cite{ATLAS-CONF-2013-012}}}
    \Text(-65,320)[cl]{\Black{\scriptsize $\mu ^{\rm A8}(H\to \gamma \gamma )_{\text{\tiny urh}}=2.711 ^{+1.350} {}_{-1.151}$}}
    \Text(95,320)[cl]{\Black{\scriptsize \cite{ATLAS-CONF-2013-012}}}
    \Text(-65,310)[cl]{\Black{\scriptsize $\mu ^{\rm A8}(H\to \gamma \gamma )_{\text{\tiny url}}=2.527 ^{+0.921} {}_{-0.783}$}}
    \Text(95,310)[cl]{\Black{\scriptsize \cite{ATLAS-CONF-2013-012}}}
    \Text(-65,300)[cl]{\Black{\scriptsize $\mu ^{\rm A8}(H\to \tau \tau )=0.756 ^{+0.775} {}_{-0.745}$}}
    \Text(95,300)[cl]{\Black{\scriptsize \cite{Aad:2012an,ATLAS-CONF-2012-160}}}
    \Text(-65,290)[cl]{\Black{\scriptsize $\mu ^{\rm A8}(H\to WW)=1.26 \pm 0.35$}}
    \Text(95,290)[cl]{\Black{\scriptsize \cite{ATLAS-CONF-2013-030}}}
    \Text(-65,280)[cl]{\Black{\scriptsize $\mu ^{\rm A8}(H\to ZZ)=1.603 ^{+0.423} {}_{-0.459}$}}
    \Text(95,280)[cl]{\Black{\scriptsize \cite{Aad:2012an,ATLAS-CONF-2013-013}}}
    \Text(-65,270)[cl]{\Black{\scriptsize $\mu ^{\rm C7}(t\oline{t} H\to t\oline{t} b\oline{b})=-0.729 ^{+2.018} {}_{-1.853}$}}
    \Text(95,270)[cl]{\Black{\scriptsize \cite{CMS-PAS-HIG-12-020}}}
    \Text(-65,260)[cl]{\Black{\scriptsize $\mu ^{\rm C7}(VH\to Vb\oline{b})=0.588 ^{+1.235} {}_{-1.153}$}}
    \Text(95,260)[cl]{\Black{\scriptsize \cite{CMS-PAS-HIG-12-020}}}
    \Text(-65,250)[cl]{\Black{\scriptsize $\mu ^{\rm C7}(H\to \gamma \gamma )_{\text{\tiny u0}}=3.832 ^{+2.042} {}_{-1.671}$}}
    \Text(95,250)[cl]{\Black{\scriptsize \cite{CMS-PAS-HIG-13-001}}}
    \Text(-65,240)[cl]{\Black{\scriptsize $\mu ^{\rm C7}(H\to \gamma \gamma )_{\text{\tiny u1}}=0.193 ^{+1.002} {}_{-0.965}$}}
    \Text(95,240)[cl]{\Black{\scriptsize \cite{CMS-PAS-HIG-13-001}}}
    \Text(-65,230)[cl]{\Black{\scriptsize $\mu ^{\rm C7}(H\to \gamma \gamma )_{\text{\tiny u2}}=0.045 \pm 1.262$}}
    \Text(95,230)[cl]{\Black{\scriptsize \cite{CMS-PAS-HIG-13-001}}}
    \Text(-65,220)[cl]{\Black{\scriptsize $\mu ^{\rm C7}(H\to \gamma \gamma )_{\text{\tiny u3}}=1.493 \pm 1.634$}}
    \Text(95,220)[cl]{\Black{\scriptsize \cite{CMS-PAS-HIG-13-001}}}
    \Text(-65,210)[cl]{\Black{\scriptsize $\mu ^{\rm C7}(H\to \gamma \gamma )_{\text{\tiny jj}}=4.203 ^{+2.339} {}_{-1.782}$}}
    \Text(95,210)[cl]{\Black{\scriptsize \cite{CMS-PAS-HIG-13-001}}}
    \Text(-65,200)[cl]{\Black{\scriptsize $\mu ^{\rm C7}(H\to \tau \tau )_{\text{\tiny 0/1 jet}}=1.000 ^{+1.441} {}_{-1.400}$}}
    \Text(95,200)[cl]{\Black{\scriptsize \cite{CMS-PAS-HIG-12-020}}}
    \Text(-65,190)[cl]{\Black{\scriptsize $\mu ^{\rm C7}(H\to \tau \tau )_{\text{\tiny VBF}}=-1.718 ^{+1.318} {}_{-1.153}$}}
    \Text(95,190)[cl]{\Black{\scriptsize \cite{CMS-PAS-HIG-12-020}}}
    \Text(-65,180)[cl]{\Black{\scriptsize $\mu ^{\rm C7}(H\to \tau \tau )_{\text{\tiny VH}}=0.671 ^{+4.076} {}_{-3.047}$}}
    \Text(95,180)[cl]{\Black{\scriptsize \cite{CMS-PAS-HIG-12-020}}}
    \Text(-65,170)[cl]{\Black{\scriptsize $\mu ^{\rm C7}(H\to WW)=0.726 ^{+0.417} {}_{-0.412}$}}
    \Text(95,170)[cl]{\Black{\scriptsize \cite{CMS-PAS-HIG-13-003}}}
    \Text(-65,160)[cl]{\Black{\scriptsize $\mu ^{\rm C7}(H\to ZZ)=0.671 ^{+0.700} {}_{-0.494}$}}
    \Text(95,160)[cl]{\Black{\scriptsize \cite{CMS-PAS-HIG-12-020}}}
    \Text(-65,150)[cl]{\Black{\scriptsize $\mu ^{\rm C8}(VH\to Vb\oline{b})=1.584 ^{+0.771} {}_{-0.704}$}}
    \Text(95,150)[cl]{\Black{\scriptsize \cite{CMS-PAS-HIG-12-020,CMS-PAS-HIG-12-045}}}
    \Text(-65,140)[cl]{\Black{\scriptsize $\mu ^{\rm C8}(H\to \gamma \gamma )_{\text{\tiny u0}}=2.198 ^{+0.928} {}_{-0.817}$}}
    \Text(95,140)[cl]{\Black{\scriptsize \cite{CMS-PAS-HIG-13-001}}}
    \Text(-65,130)[cl]{\Black{\scriptsize $\mu ^{\rm C8}(H\to \gamma \gamma )_{\text{\tiny u1}}=0.045 \pm 0.705$}}
    \Text(95,130)[cl]{\Black{\scriptsize \cite{CMS-PAS-HIG-13-001}}}
    \Text(-65,120)[cl]{\Black{\scriptsize $\mu ^{\rm C8}(H\to \gamma \gamma )_{\text{\tiny u2}}=0.304 \pm 0.483$}}
    \Text(95,120)[cl]{\Black{\scriptsize \cite{CMS-PAS-HIG-13-001}}}
    \Text(-65,110)[cl]{\Black{\scriptsize $\mu ^{\rm C8}(H\to \gamma \gamma )_{\text{\tiny u3}}=-0.327 ^{+0.817} {}_{-0.854}$}}
    \Text(95,110)[cl]{\Black{\scriptsize \cite{CMS-PAS-HIG-13-001}}}
    \Text(-65,100)[cl]{\Black{\scriptsize $\mu ^{\rm C8}(H\to \gamma \gamma )_{\text{\tiny $e$}}=-0.661 ^{+2.785} {}_{-1.968}$}}
    \Text(95,100)[cl]{\Black{\scriptsize \cite{CMS-PAS-HIG-13-001}}}
    \Text(-65,90)[cl]{\Black{\scriptsize $\mu ^{\rm C8}(H\to \gamma \gamma )_{\text{\tiny jj loose}}=0.824 ^{+1.077} {}_{-1.002}$}}
    \Text(95,90)[cl]{\Black{\scriptsize \cite{CMS-PAS-HIG-13-001}}}
    \Text(-65,80)[cl]{\Black{\scriptsize $\mu ^{\rm C8}(H\to \gamma \gamma )_{\text{\tiny jj tight}}=0.304 ^{+0.668} {}_{-0.594}$}}
    \Text(95,80)[cl]{\Black{\scriptsize \cite{CMS-PAS-HIG-13-001}}}
    \Text(-65,70)[cl]{\Black{\scriptsize $\mu ^{\rm C8}(H\to \gamma \gamma )_{\text{\tiny $\slashed E_{\rm T}$}}=1.938 ^{+2.599} {}_{-2.339}$}}
    \Text(95,70)[cl]{\Black{\scriptsize \cite{CMS-PAS-HIG-13-001}}}
    \Text(-65,60)[cl]{\Black{\scriptsize $\mu ^{\rm C8}(H\to \gamma \gamma )_{\text{\tiny $\mu$}}=0.416 ^{+1.819} {}_{-1.411}$}}
    \Text(95,60)[cl]{\Black{\scriptsize \cite{CMS-PAS-HIG-13-001}}}
    \Text(-65,50)[cl]{\Black{\scriptsize $\mu ^{\rm C8}(H\to \tau \tau )_{\text{\tiny 0/1 jet}}=1.109 ^{+0.416} {}_{-0.417}$}}
    \Text(95,50)[cl]{\Black{\scriptsize \cite{CMS-PAS-HIG-12-020,CMS-PAS-HIG-13-004}}}
    \Text(-65,40)[cl]{\Black{\scriptsize $\mu ^{\rm C8}(H\to \tau \tau )_{\text{\tiny VBF}}=2.375 ^{+0.674} {}_{-0.703}$}}
    \Text(95,40)[cl]{\Black{\scriptsize \cite{CMS-PAS-HIG-12-020,CMS-PAS-HIG-13-004}}}
    \Text(-65,30)[cl]{\Black{\scriptsize $\mu ^{\rm C8}(H\to \tau \tau )_{\text{\tiny VH}}=0.780 ^{+1.613} {}_{-1.723}$}}
    \Text(95,30)[cl]{\Black{\scriptsize \cite{CMS-PAS-HIG-12-020,CMS-PAS-HIG-13-004}}}
    \Text(-65,20)[cl]{\Black{\scriptsize $\mu ^{\rm C8}(H\to WW)=0.647 ^{+0.225} {}_{-0.221}$}}
    \Text(95,20)[cl]{\Black{\scriptsize \cite{CMS-PAS-HIG-13-003}}}
    \Text(-65,10)[cl]{\Black{\scriptsize $\mu ^{\rm C8}(H\to ZZ)=0.975 ^{+0.332} {}_{-0.275}$}}
    \Text(95,10)[cl]{\Black{\scriptsize \cite{CMS-PAS-HIG-12-020,CMS-PAS-HIG-13-002}}}
    \Text(139,-3)[cb]{\Black{\scriptsize $-6$}}
    \Text(159,-3)[cb]{\Black{\scriptsize $-4$}}
    \Text(179,-3)[cb]{\Black{\scriptsize $-2$}}
    \Text(200.7,-3)[cb]{\Black{\scriptsize $0$}}
    \Text(221,-3)[cb]{\Black{\scriptsize $2$}}
    \Text(241,-3)[cb]{\Black{\scriptsize $4$}}
    \Text(261,-3)[cb]{\Black{\scriptsize $6$}}
    \Text(281,-3)[cb]{\Black{\scriptsize $8$}}
    \Text(301.7,-3)[cb]{\Black{\scriptsize $10$}}
    \Text(321.7,-3)[cb]{\Black{\scriptsize $12$}}
    \Text(341.7,-3)[cb]{\Black{\scriptsize $14$}}
\end{picture}
\caption[Higgs signal strength inputs.]{Individual signal strengths measured by D${\emptyset}$, CDF, ATLAS and CMS (status April 2013 after the Moriond conferences). The combination yields $\mu _{\text{\tiny comb}}=1.007 ^{+0.099} {}_{-0.098}$ and is depicted by the green band. \otto{A similar figure can be found in \cite{Eberhardt:2013uba}.}}
\label{fig:signalstrengths}
\end{figure}
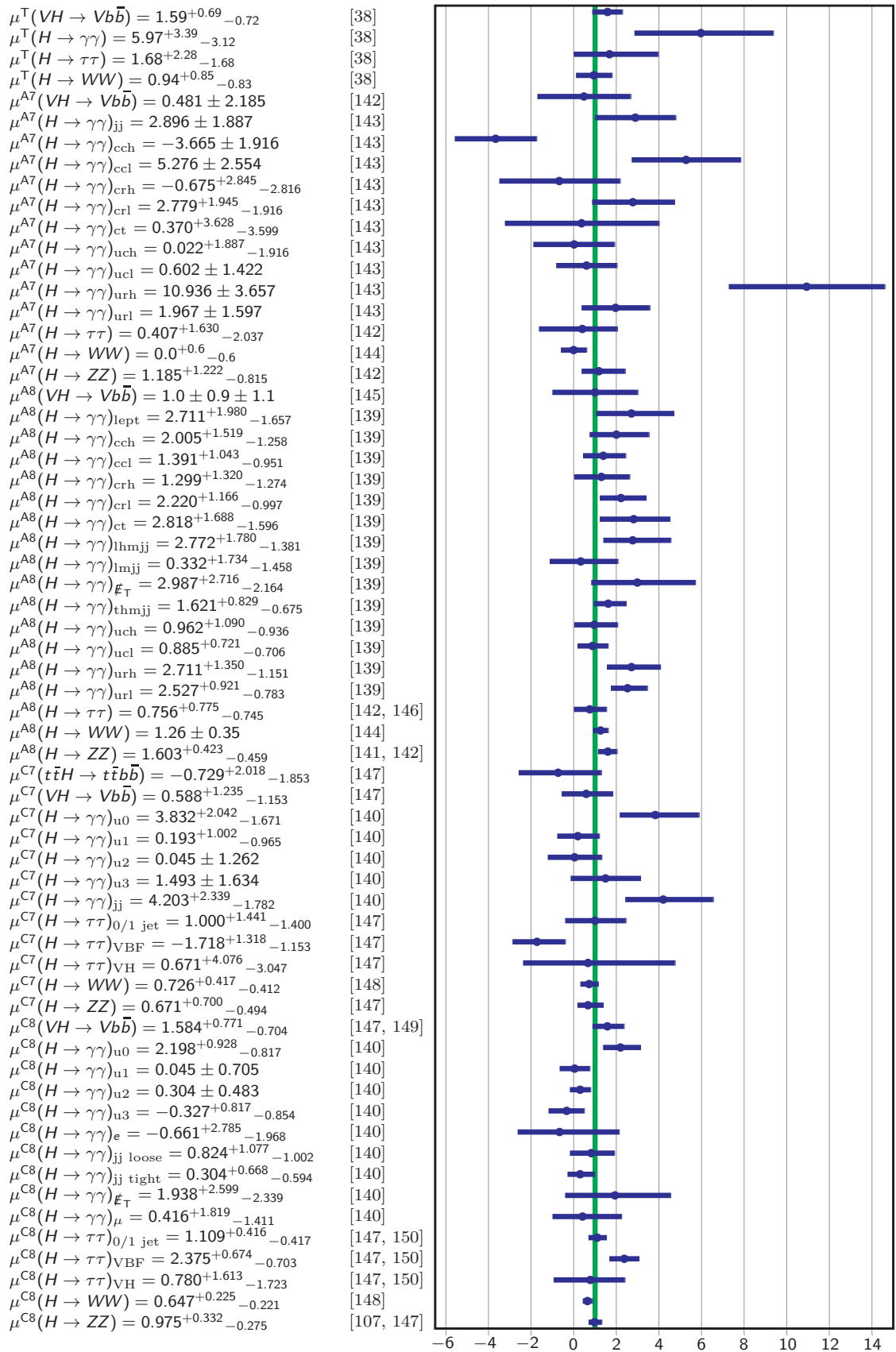

But not only the reconstructed invariant mass of the (light ${\cal CP}$-even) Higgs boson is of importance, also other information like the relative occurrence of decay products compared to the SM expectation are relevant; this is expressed by the collection of signal strengths in Fig.\ \ref{fig:signalstrengths} which represents the status after the Moriond 2013 conferences.

The upper index denotes the data set the measured signal strength relies on. ``$T$'' stands for the complete evaluated Tevatron data of up to 10\:fb$^{-1}$. ``$A7$'' and ``$A8$'' are the publicly available ATLAS signal strengths obtained at $\sqrt{s}=7$\:TeV and at $\sqrt{s}=8$\:TeV, respectively; ``$C7$'' and ``$C8$'' label the corresponding measurements by the CMS detector. The Higgs decay products $XX^{(*)}$ are indicated by the bracket $(H\to XX)$, where $X\in \{ \gamma, \tau, W, Z\}$.
For signal strengths characterized by the decay $H\to b\oline{b}$ I assumed that only vector boson associated production plays a role (except for $\mu ^{\rm C7}(t\oline{t} H\to t\oline{t} b\oline{b})$, where exclusively $t\oline{t}$ associated production is important), therefore I explicitly list the associated particles. $\mu ^{\rm A8}(H\to \tau \tau )$, $\mu ^{\rm A8}(H\to ZZ)$, $\mu ^{\rm C8}(VH\to Vb\oline{b})$, the tauonic signal strengths of CMS at $\sqrt{s}=8$\:TeV and $\mu ^{\rm C8}(H\to ZZ)$ were reconstructed from the combined values in the quoted publications using Equations \eqref{eq:sigmacombination} and \eqref{eq:obscombination} and our knowledge of the $7$\:TeV data. The tags of the tauonic decays measured at CMS denote the reconstructed final states additional to the $\tau$ pair.
$H\to \gamma \gamma$ events were separated into up to fourteen categories, which were provided by the detector collaborations together with the corresponding percentage contributions. I refer to the related publications for further explanation. This figure represents the situation at the beginning of May 2013; the latest CMS updates in the middle of May 2013 have not been included. Even if there are correlations between Higgs signal strengths, I could not take them into account because they were not provided by the detector collaborations.\\
Since we do not know the production mechanism of the individual Higgs candidate events, I chose to use percentage contributions. If they are not given directly, one can derive them from the efficiencies via Eq.\ \eqref{eq:pc}. For the tauonic Higgs decays at CMS the percentage contributions were derived using the summed efficiencies of the five decay sub-channels $\mu \tau_h$, $e\tau_h$, $e\mu$, $\mu \mu $ and $\tau_h \tau_h$ (where $\tau_h$ is a $\tau$ lepton decaying into hadrons); they can be found in Table \ref{tab:pcvaluestautau}. Whereas the ggF and VBF (``$qq\to H$'') dominated production parts are explicitly given in \cite{CMS-PAS-HIG-12-043}, the three less important production channels are subsumed under the tag ``$qq\to Ht\oline{t}$ or $VH$''; they were split to the channels defined in Sect.\ \ref{SMhiggssearches} according to the relative occurrence from Fig.\ \ref{fig:higgsproddecay}. Same efficiencies at $7$ and $8$\:TeV were assumed.

{\small
\begin{longtable}{lllllll}
& & ggF & VBF & WH & ZH & ttH \\
\hline
\hline
CMS & 0/1 jet & $0.808$ & $0.119$ & $0.042$ & $0.024$ & $0.008$\\
& VBF & $0.245$ & $0.742$ & $0.007$ & $0.004$ & $0.001$\\
\caption{Percentage contributions as defined in equation \eqref{eq:pc} for the $H\to \tau \tau$ signal strengths derived from the efficiencies in \cite{CMS-PAS-HIG-12-043}.}
\label{tab:pcvaluestautau}
\end{longtable}
}

\newpage

In the diphoton decay channel, percentage contributions for both, the $7$ and $8$\:TeV data set are stated in the respective ATLAS and CMS publications. They are listed in Table \ref{tab:pcvalues}. The different tags correspond to the sub-channels in Fig.\ \ref{fig:signalstrengths}.

{\small
\begin{longtable}{lllllll}
& & ggF & VBF & WH & ZH & ttH \\
\hline
\hline
\endhead
ATLAS 7 TeV &&&&&&\\
\hline
& jj & $0.225$ & $0.767$ & $0.004$ & $0.002$ & $0.001$\\
& cch & $0.666$ & $0.153$ & $0.100$ & $0.057$ & $0.025$\\
& ccl & $0.928$ & $0.040$ & $0.019$ & $0.010$ & $0.002$\\
& crh & $0.653$ & $0.160$ & $0.110$ & $0.059$ & $0.018$\\
& crl & $0.928$ & $0.038$ & $0.020$ & $0.011$ & $0.002$\\
& ct & $0.894$ & $0.052$ & $0.033$ & $0.017$ & $0.003$\\
& uch & $0.665$ & $0.157$ & $0.099$ & $0.057$ & $0.024$\\
& ucl & $0.929$ & $0.040$ & $0.018$ & $0.010$ & $0.002$\\
& urh & $0.654$ & $0.161$ & $0.108$ & $0.061$ & $0.018$\\
& url & $0.928$ & $0.039$ & $0.020$ & $0.011$ & $0.002$\\
\hline
ATLAS 8 TeV &&&&&&\\
\hline
& lept & $0.022$ & $0.006$ & $0.632$ & $0.154$ & $0.186$\\
& cch & $0.789$ & $0.126$ & $0.043$ & $0.027$ & $0.015$\\
& ccl & $0.936$ & $0.040$ & $0.013$ & $0.009$ & $0.002$\\
& crh & $0.777$ & $0.130$ & $0.052$ & $0.030$ & $0.011$\\
& crl & $0.932$ & $0.041$ & $0.016$ & $0.010$ & $0.001$\\
& ct & $0.907$ & $0.055$ & $0.022$ & $0.013$ & $0.002$\\
& lhmjj & $0.450$ & $0.541$ & $0.005$ & $0.003$ & $0.001$\\
& lmjj & $0.481$ & $0.030$ & $0.297$ & $0.172$ & $0.019$\\
& $\slashed E_{\rm T}$ & $0.041$ & $0.005$ & $0.357$ & $0.476$ & $0.121$\\
& thmjj & $0.238$ & $0.760$ & $0.001$ & $0.001$ & $0.000$\\
& uch & $0.793$ & $0.126$ & $0.041$ & $0.025$ & $0.014$\\
& ucl & $0.937$ & $0.040$ & $0.014$ & $0.008$ & $0.002$\\
& urh & $0.781$ & $0.133$ & $0.047$ & $0.028$ & $0.011$\\
& url & $0.932$ & $0.040$ & $0.016$ & $0.010$ & $0.001$\\[140pt]
CMS 7 TeV &&&&&&\\
\hline
& u0 & $0.614$ & $0.168$ & $0.121$ & $0.066$ & $0.031$\\
& u1 & $0.876$ & $0.062$ & $0.036$ & $0.020$ & $0.005$\\
& u2 & $0.913$ & $0.044$ & $0.025$ & $0.014$ & $0.003$\\
& u3 & $0.913$ & $0.044$ & $0.026$ & $0.015$ & $0.002$\\
& jj & $0.268$ & $0.725$ & $0.004$ & $0.002$ & $0.000$\\
\hline
CMS 8 TeV &&&&&&\\
\hline
& u0 & $0.729$ & $0.116$ & $0.082$ & $0.047$ & $0.026$\\
& u1 & $0.835$ & $0.084$ & $0.045$ & $0.026$ & $0.010$\\
& u2 & $0.916$ & $0.045$ & $0.023$ & $0.013$ & $0.004$\\
& u3 & $0.925$ & $0.039$ & $0.021$ & $0.012$ & $0.003$\\
& $e$ & $0.011$ & $0.004$ & $0.502$ & $0.285$ & $0.198$\\
& jj loose & $0.470$ & $0.509$ & $0.011$ & $0.006$ & $0.005$\\
& jj tight & $0.207$ & $0.789$ & $0.002$ & $0.001$ & $0.001$\\
& $\slashed E_{\rm T}$ & $0.220$ & $0.026$ & $0.407$ & $0.230$ & $0.117$\\
& $\mu$ & $0.000$ & $0.002$ & $0.504$ & $0.286$ & $0.208$\\
\caption{Percentage contributions from \cite{ATLAS-CONF-2012-091a,ATLAS-CONF-2013-012,CMS-PAS-HIG-13-001}.}
\label{tab:pcvalues}
\end{longtable}
}

If neither percentage contributions nor efficiencies were provided by the detector collaborations, I used the relative occurrences as listed in Fig.\ \ref{fig:higgsproddecay}. (In the case of the $H\to b\oline{b}$ signal strengths, I set all other contributions to zero.) They can be found in Table \ref{tab:pcvaluesrest}.

{\small
\begin{longtable}{lllllll}
& & ggF & VBF & WH & ZH & ttH \\
\hline
Tevatron & & $0.748$ & $0.058$ & $0.118$ & $0.073$ & $0.003$\\
Tevatron & VH only & $0.000$ & $0.000$ & $0.619$ & $0.381$ & $0.000$\\
\hline
LHC 7 TeV &  & $0.875$ & $0.070$ & $0.032$ & $0.018$ & $0.005$\\
LHC 7 TeV & VH only & $0.000$ & $0.000$ & $0.644$ & $0.356$ & $0.000$\\
\hline
LHC 8 TeV &  & $0.875$ & $0.071$ & $0.031$ & $0.017$ & $0.006$\\
LHC 8 TeV & VH only & $0.000$ & $0.000$ & $0.638$ & $0.362$ & $0.000$\\
\caption{Percentage contributions as defined in equation \eqref{eq:pc} as extracted from \cite{Baglio:2010um,LHCXSWG:2013aa}.}
\label{tab:pcvaluesrest}
\end{longtable}
}

\newpage

\begin{figure}[hptb]
\centering
\begin{picture}(260,440)(0,7)
\put(-90,322){\includegraphics[scale=0.35]{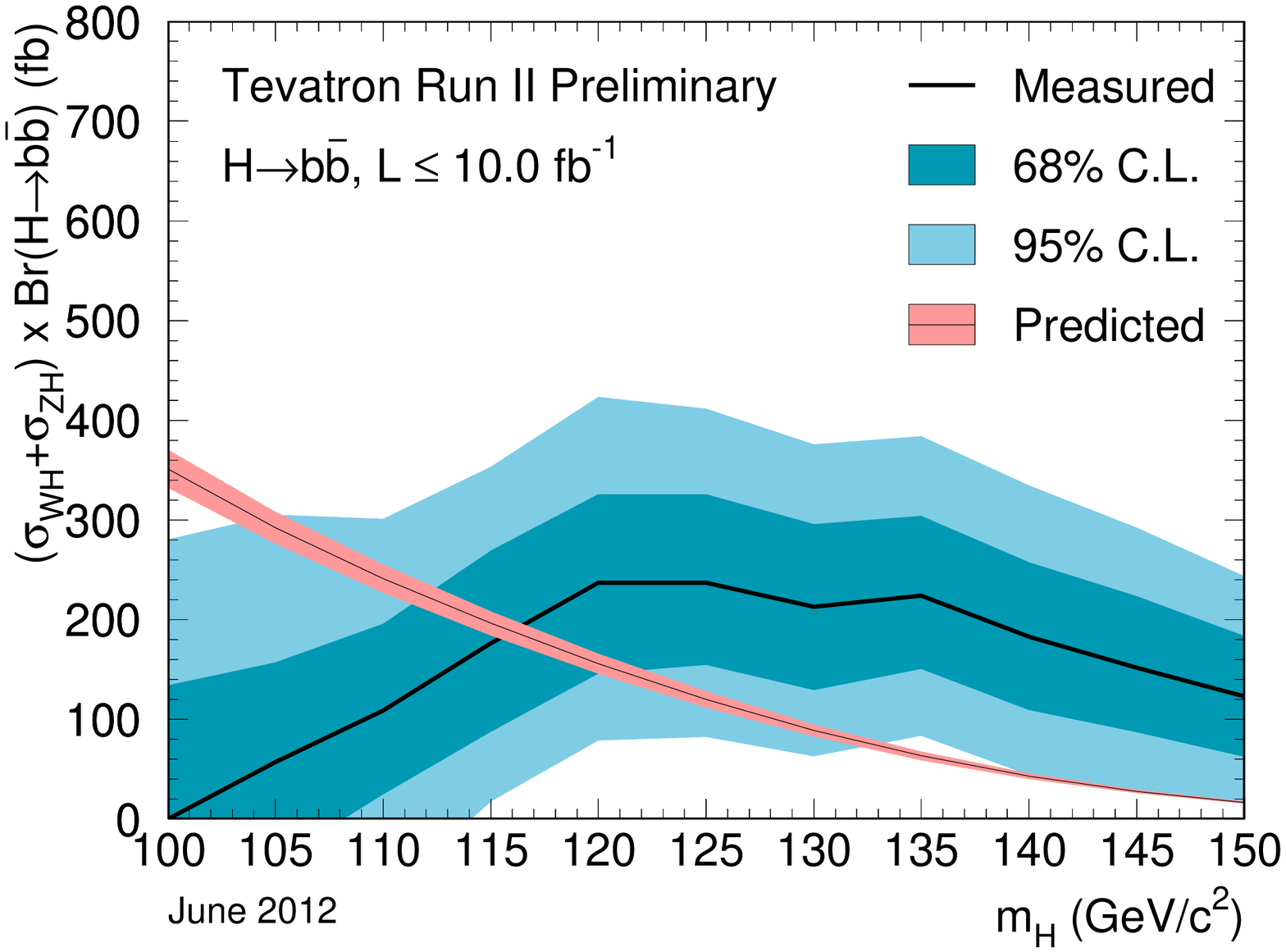}} 
\put(130,317.8){\includegraphics[scale=0.371]{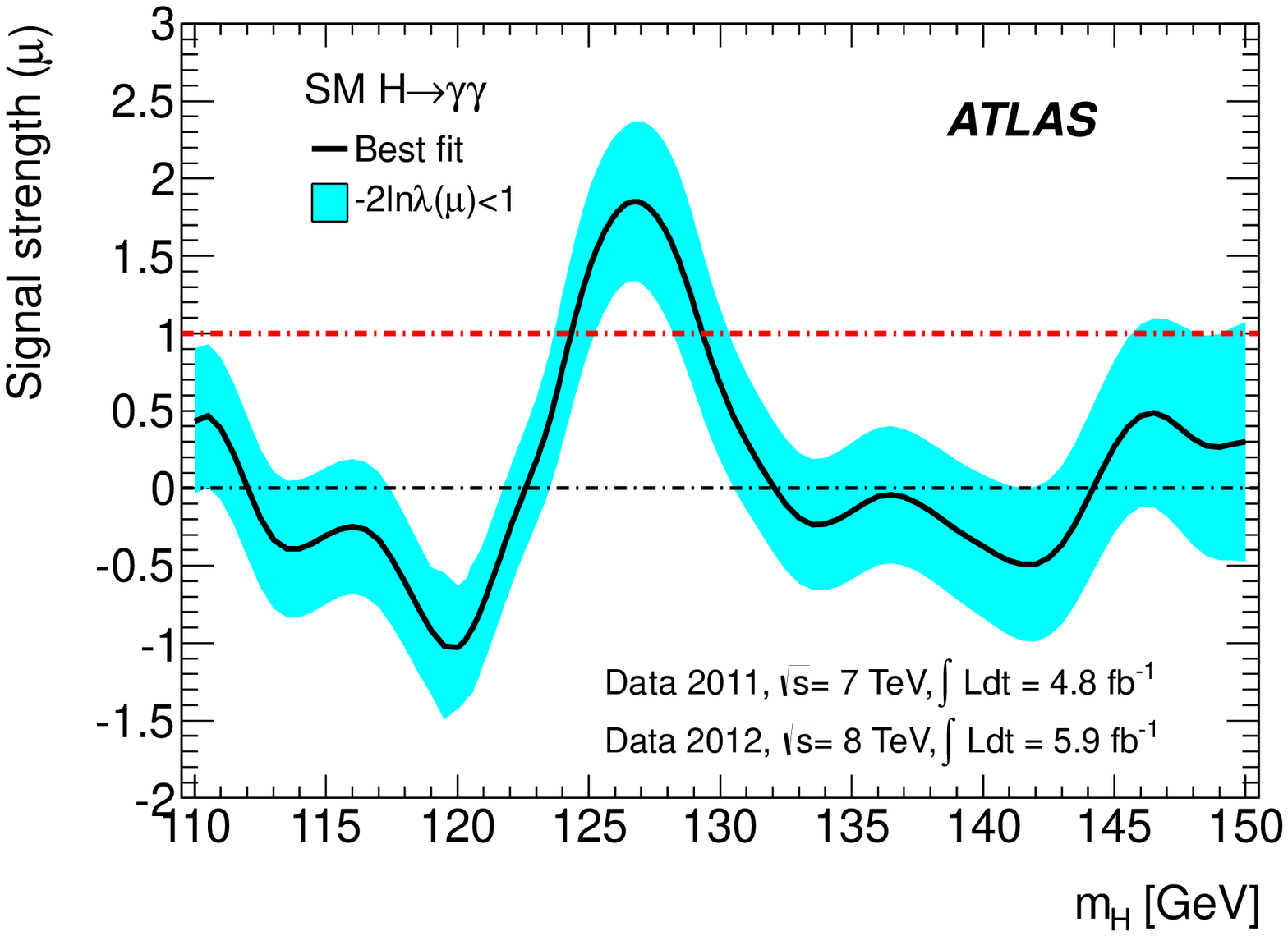}} 
\put(-95,160){\includegraphics[scale=0.38]{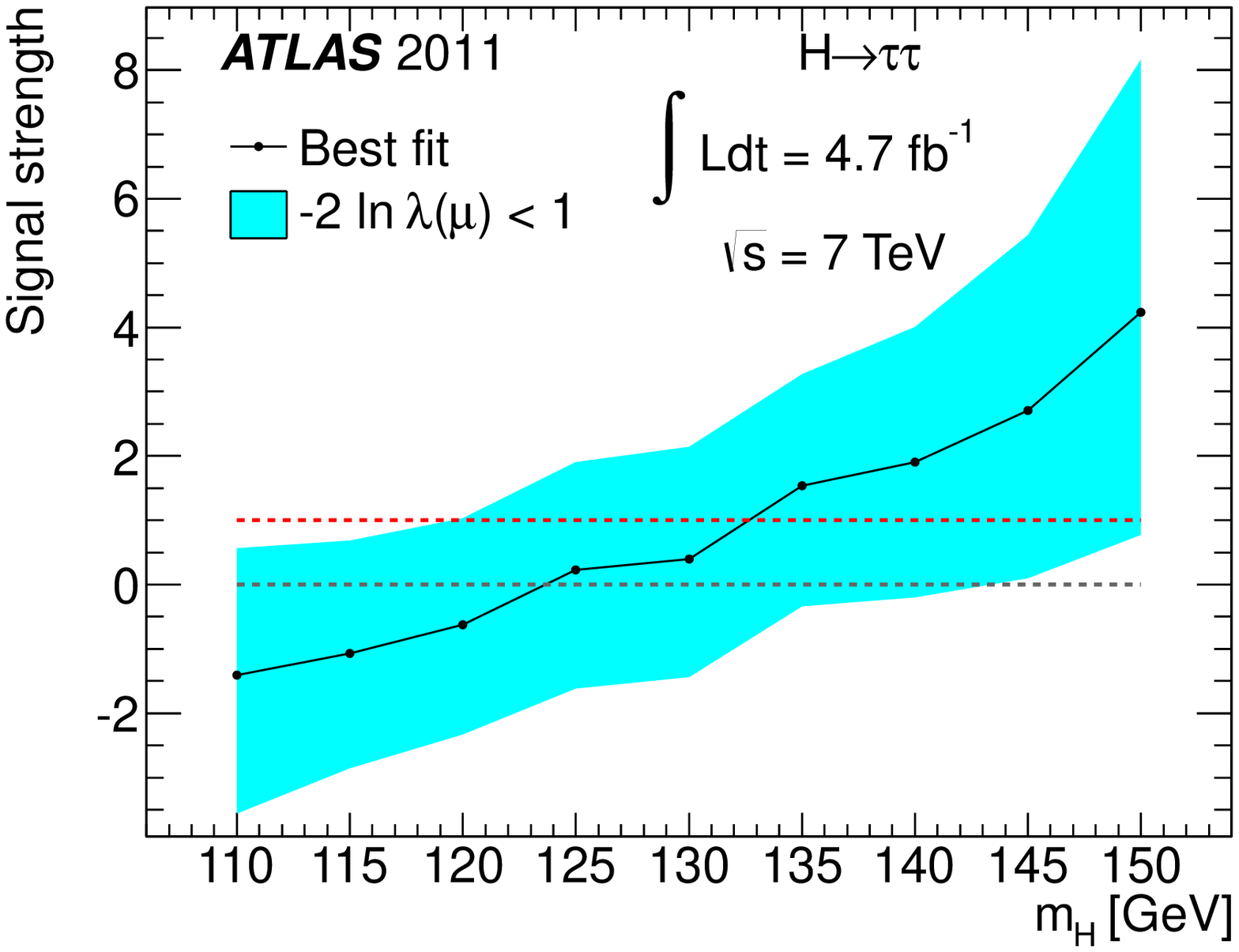}} 
\put(140.7,110){\includegraphics[scale=0.233]{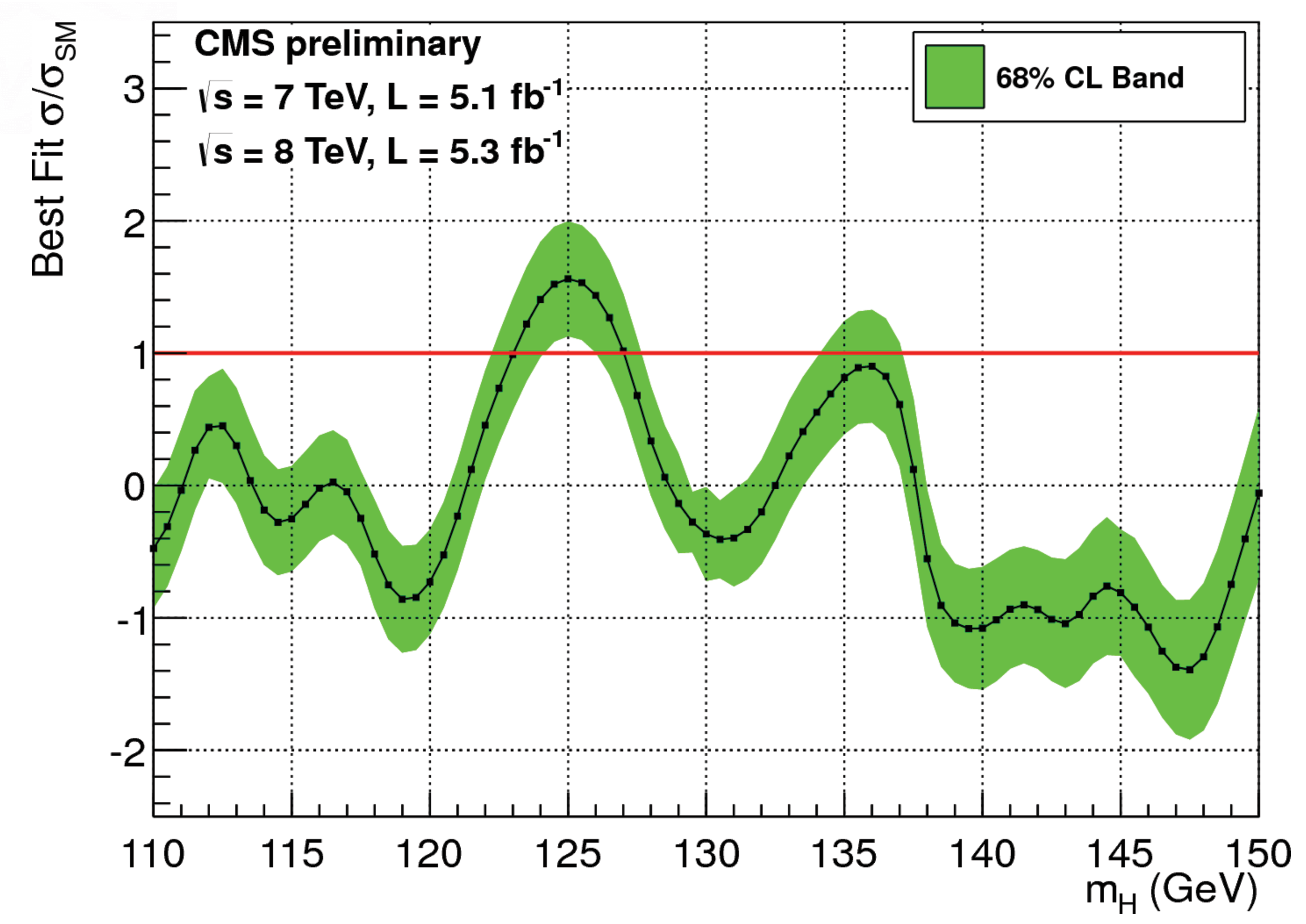}} 
\put(-95,0){\includegraphics[scale=0.38]{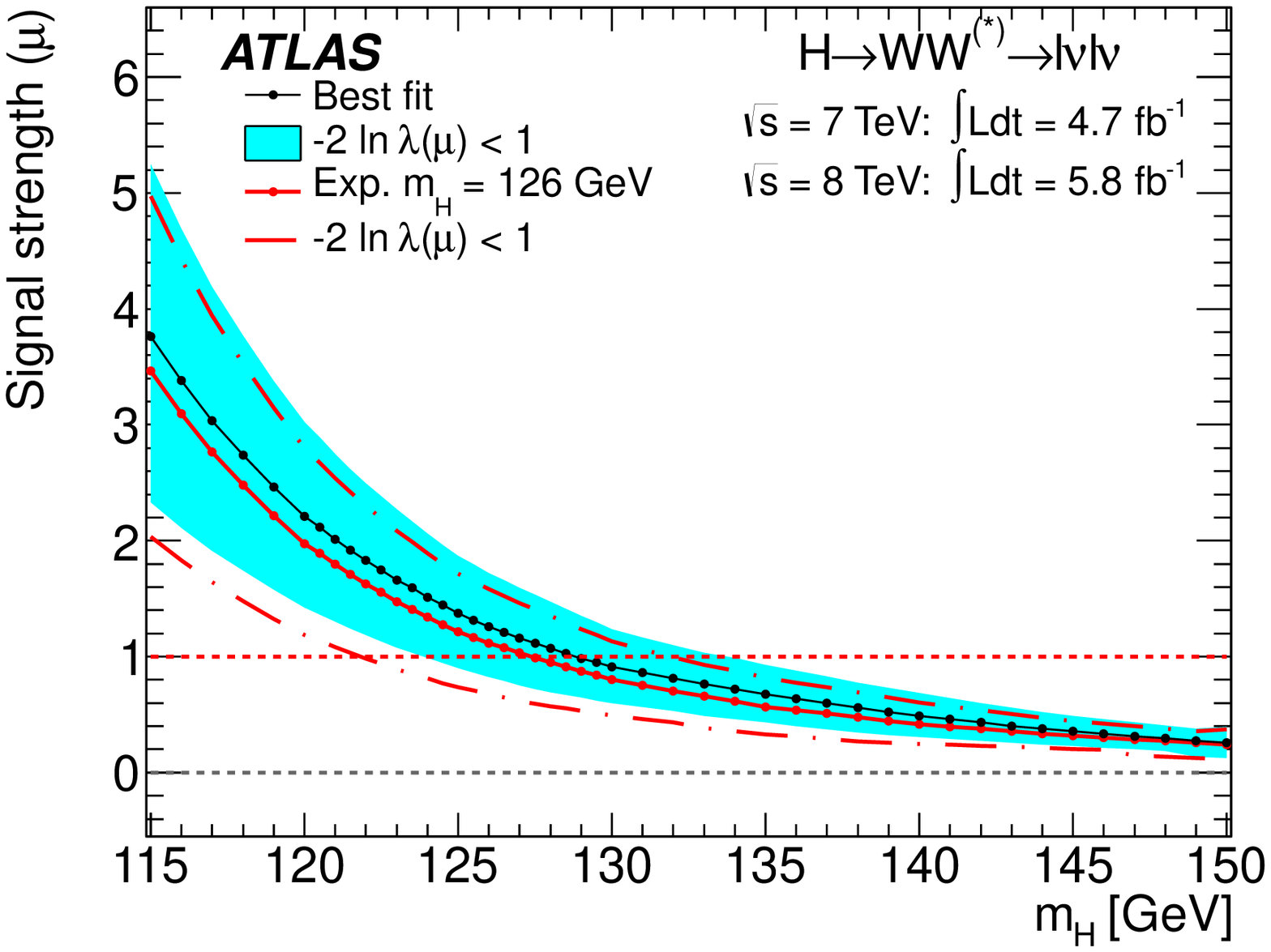}} 
\put(138,0){\includegraphics[scale=0.289]{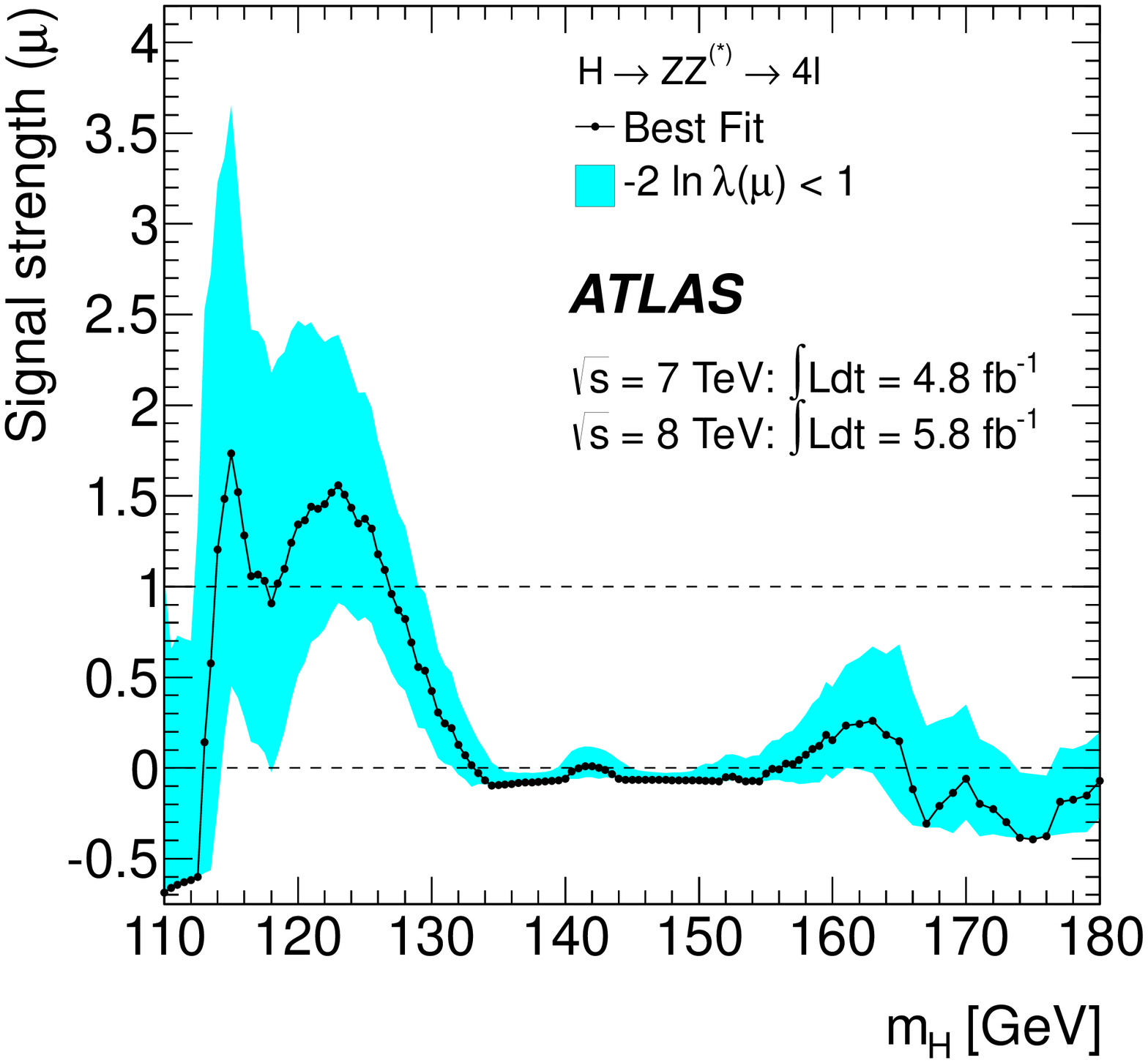}} 
\end{picture}
\caption[Higgs discovery signal strengths.]{Higgs signal strengths as functions of $m_H$ after the summer conferences in 2012. They stem from the CDF, D${\emptyset}$, ATLAS and CMS detectors \cite{Group:2012zca,Aad:2012tfa,Aad:2012an,CMS-PAS-HIG-12-015} and serve as inputs for the Higgs mass scan in Fig.\ \ref{fig:bluebandSM4}(b). (The Tevatron combination was normalized to the red SM expectation.)}
\label{fig:signalstrengths2012}
\end{figure}
\vspace*{10pt}
In Fig.\ \ref{fig:bluebandSM4}\ottooo{(b)}, I show the explicit Higgs mass dependence of the SM and SM4 fits to the signal strength that were publicly available after the Higgs discovery in July 2012. They are based on the digitalized plots shown in Fig.\ \ref{fig:signalstrengths2012}.

\newpage

\begin{figure}[htbp]
 \centering
 \begin{picture}(180,260)(0,0)
 \put(-50,0){\includegraphics[width=0.65\linewidth]{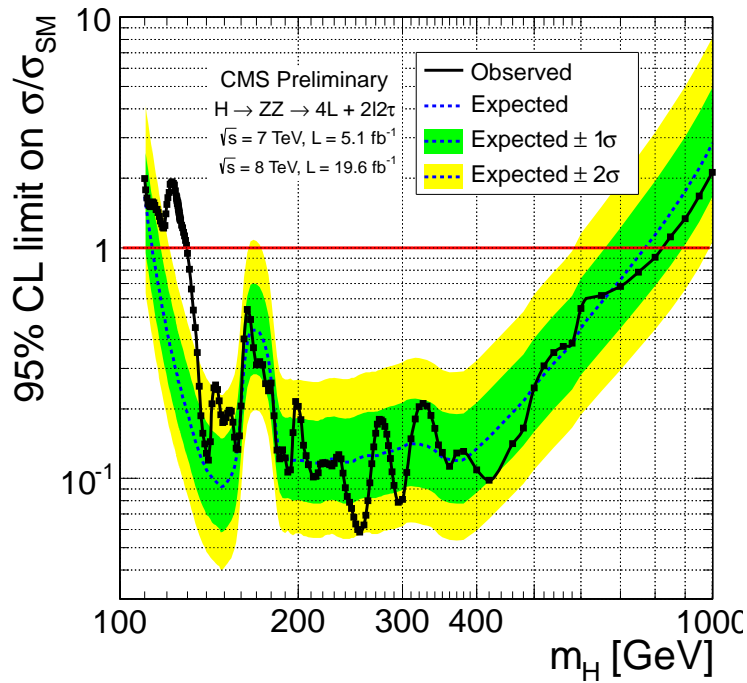}}
 \end{picture}
 \caption[Heavy Higgs searches at CMS.]{An $H^0$ lighter than about $800$\:GeV can be excluded at $95\%$ CL by CMS searches in $H\to ZZ$ decays \cite{CMS-PAS-HIG-13-002}.}
 \label{fig:heavyhiggssearches}
\end{figure}

In the 2HDM fits information on non-observation of heavy Higgs bosons were included. I digitalised the CMS signal strength exclusion plot for the $H\to ZZ$ decay (Fig.\ \ref{fig:heavyhiggssearches}) and extrapolated the data up to 10 TeV in order to ameliorate the convergence of the fit minimization. I did not take into account the update published in May 2013 \cite{CMS-PAS-HIG-13-014}. Instead of the actually observed exclusion limits, I use the expected \otto{limits} because above $150$\:GeV, both lines agree within $2\sigma$ and I want to prevent the fit from being sensitive to background fluctuations.

Furthermore, the SM Higgs decay branching ratios are of major importance. Since uncertainties on them are already included in the experimental signal strength errors, I only need to take the fixed central values from \cite{LHCXSWG:2013aa}. (Some of the values were already shown in Fig.\ \ref{fig:higgsproddecay}; here I list them ordered by magnitude.)

{\small
\begin{longtable}{ll|ll}
${\cal B}(H\to b\oline{b})$ & $0.561$ & ${\cal B}(H\to c\oline{c})$ & $0.0283$\\
${\cal B}(H\to WW^*)$ & $0.231$ & ${\cal B}(H\to \gamma \gamma )$ & $2.28\cdot 10^{-3}$\\
${\cal B}(H\to gg)$ & $0.0848$ & ${\cal B}(H\to Z\gamma )$ & $1.62\cdot 10^{-3}$\\
${\cal B}(H\to \tau \tau )$ & $0.0615$ & ${\cal B}(H\to \mu \mu )$ & $2.14\cdot 10^{-4}$\\
${\cal B}(H\to ZZ^*)$ & $0.0289$ & \\
\caption{Branching ratios of a Higgs boson with a mass of $126$\:GeV \cite{LHCXSWG:2013aa}.}
\label{tab:SMBranchingRatios}
\end{longtable}
}

\newpage
Finally, the inputs for the flavour observables ${\cal B}(\oline{B}\to X_s\gamma)$ and $\Delta m_{B_s}$, which I need for the 2HDM fit, are presented in Table \ref{tab:Flavourinputs}. The inputs for ${\cal B}(\oline{B}\to X_s\gamma)$ were fixed for the parametrization; their systematic errors are accounted for by an additional theoretical error on the observable. 

\begin{center}
\begin{longtable}{lll}
Quantity & Input value & Source\\
\hline
\hline
$\Delta m_{B_s}$ & $17.768\pm 0.023\pm 0.006 \hbar $\:ps$^{-1}$ & \cite{Aaij:2013mpa}\\
$G_F$ & $1.16638 \cdot 10^{-5}$\:GeV$^{-2}$ & \cite{Beringer:1900zz}\\
$f_{B_s}$ & $225\pm 0\pm 4$\:GeV & \cite{McNeile:2011ng,CKMfitter:2012mo}\\
$\eta _B$ & $0.5510\pm 0\pm 0.0022$ & \cite{Buras:1990fn,Deschamps:2009rh}\\
$\hat B_s$ & $1.322\pm 0.040\pm 0.035$ & \cite{Gamiz:2009ku,CKMfitter:2012mo}\\
$m_{B_s}$ & $5.367$\:GeV & \cite{Beringer:1900zz}\\
$\oline{m}_t$ & $166.6$\:GeV & SM fit\\
\hline
${\cal B}(\oline{B}\to X_s\gamma)$ & $(3.43 \pm 0.21 \pm 0.55)\cdot 10^{-4}$ & \cite{HFAG:2012bs,Misiak:2006ab,Hermann:2012fc}\\
${\cal B}(\oline{B}\to X_c e \oline{\nu})_{\text{\tiny exp}}$ & $0.1061$ & \cite{Misiak:2006ab}\\
$C$ & $0.580$ & \cite{Misiak:2006ab}\\
\caption{Flavour inputs}
\label{tab:Flavourinputs}
\end{longtable}
\end{center}

%% file: fitresults.tex
\chapter{Prediction fits} \label{fitresults}

The larger the deviation of an observable, the larger is its potential to exclude a specific model. In this appendix I want to illustrate the individual impact of all observables used in this thesis on the three discussed models. In the following plots, the one-dimensional scans over all important observables are shown with the corresponding naive p-value as defined in \eqref{eq:naivepvalue}: the blue curves show the SM prediction, the red ones the SM4 prediction, and the green ones the 2HDM prediction. Furthermore, the experimental values are shown at the $1\sigma$ level ($p\approx 0.31$) with systematic and statistical errors, where the former correspond to the inner error bars and the latter to the outer error bars.

\begin{figure}[bhtp]
 \centering
  \subfigure[]{\begin{picture}(210,140)(0,10)
	      \put(-2,0){\includegraphics[width=0.5\linewidth]{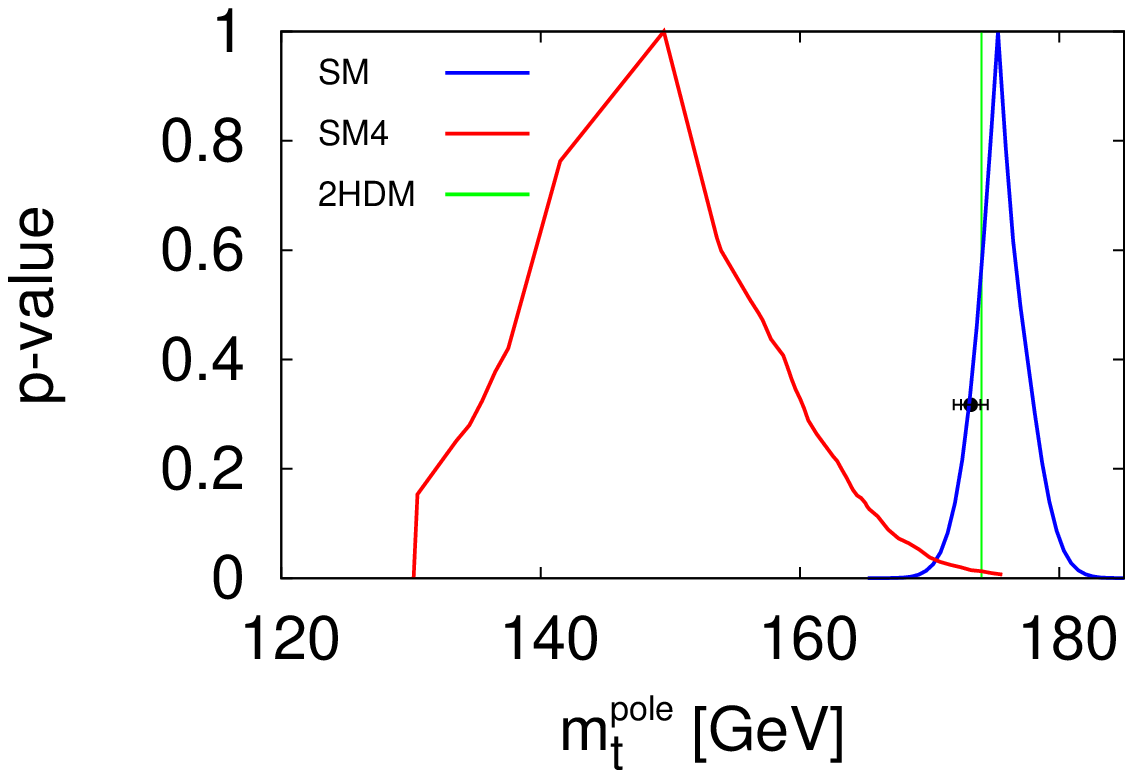}}
	      \put(126,116.1){\includegraphics[width=0.08\linewidth]{Images/CKMfitterPackage.eps}}
	      \end{picture}
	     }
 \qquad
 \subfigure[]{\begin{picture}(180,140)(0,10)
	      \put(-12,0){\includegraphics[width=0.5\linewidth]{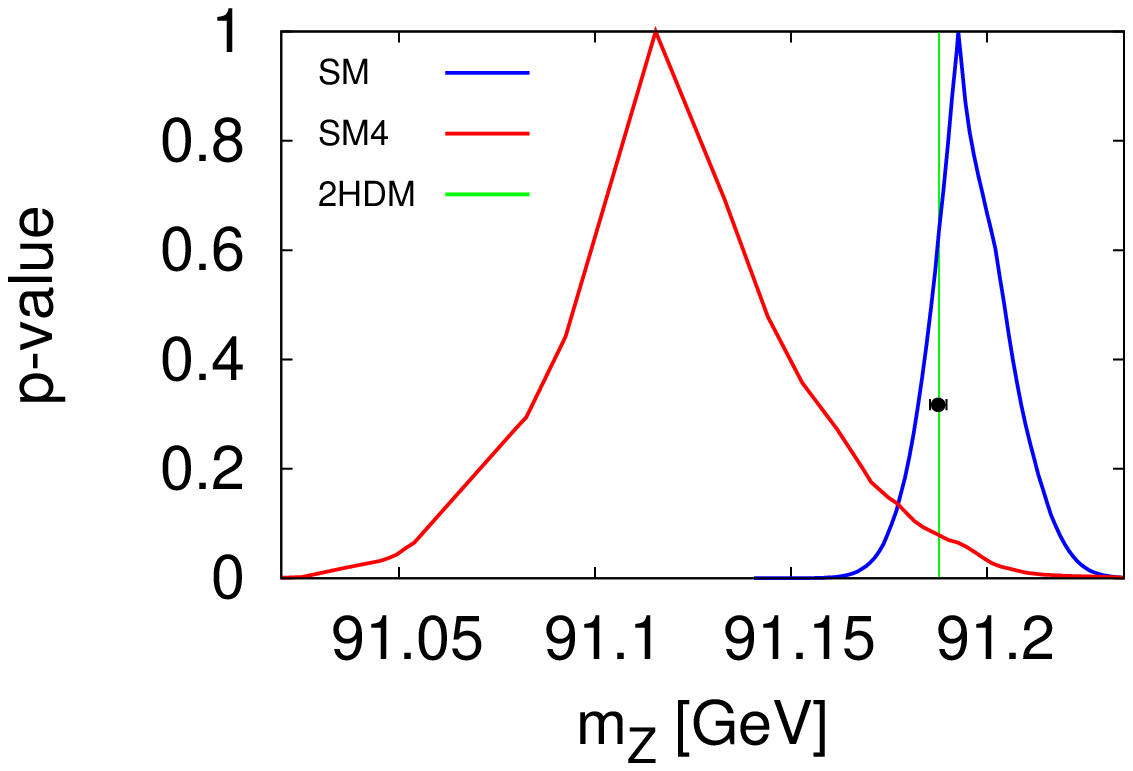}}
	      \put(114,116.1){\includegraphics[width=0.08\linewidth]{Images/CKMfitterPackage.eps}}
	      \end{picture}
	     }
  \subfigure[]{\begin{picture}(210,140)(0,10)
	      \put(-2,0){\includegraphics[width=0.5\linewidth]{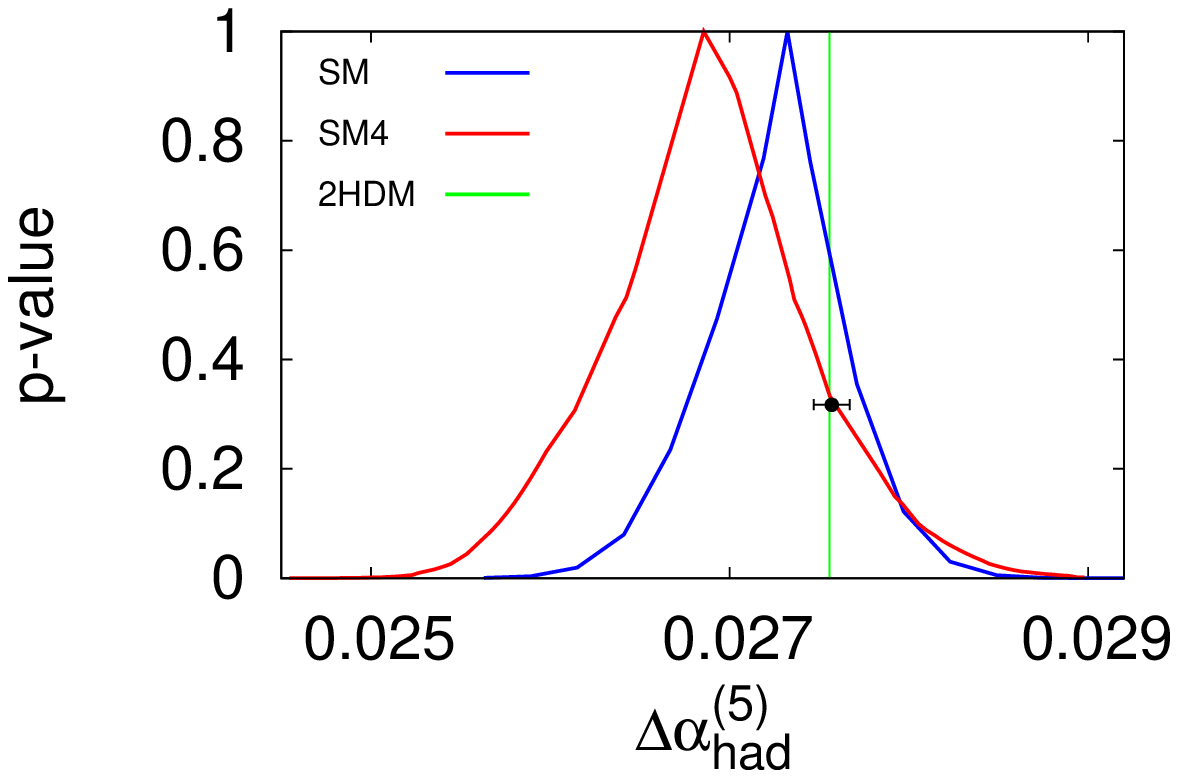}}
	      \put(157.9,116.1){\includegraphics[width=0.08\linewidth]{Images/CKMfitterPackage.eps}}
	      \end{picture}
	     }
 \qquad
 \subfigure[]{\begin{picture}(180,140)(0,10)
	      \put(-12,0){\includegraphics[width=0.5\linewidth]{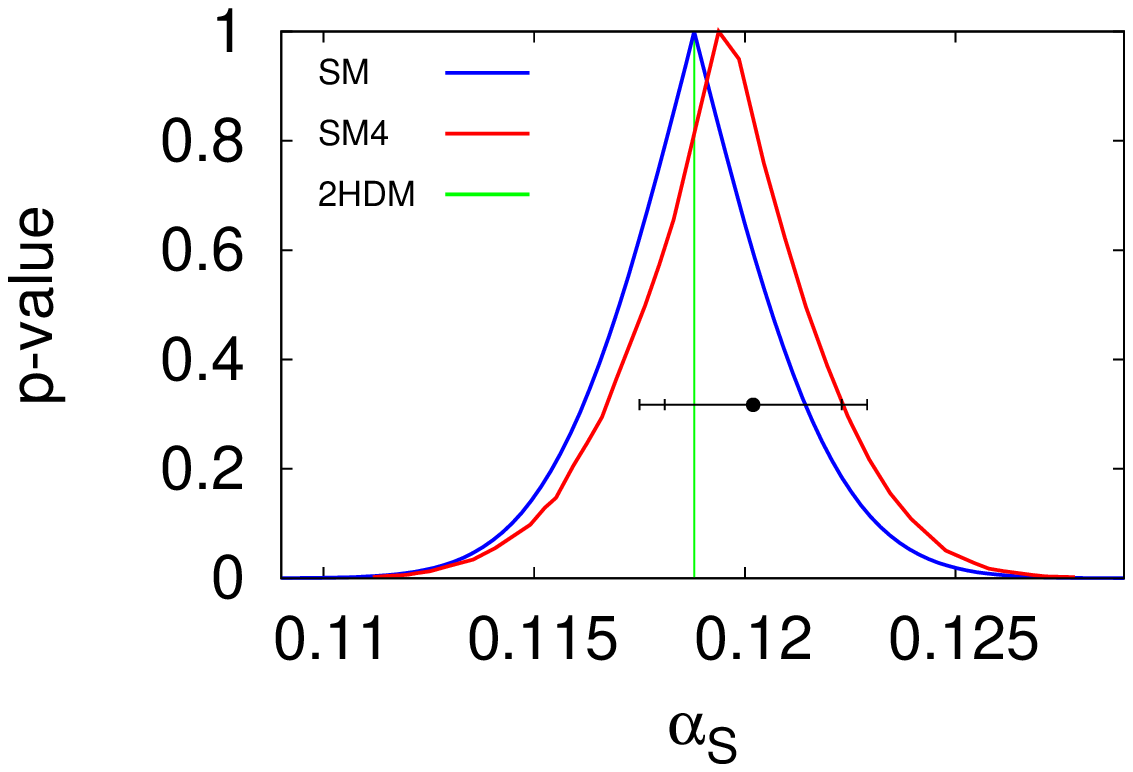}}
	      \put(147.9,116.1){\includegraphics[width=0.08\linewidth]{Images/CKMfitterPackage.eps}}
	      \end{picture}
	     }
 \caption[Prediction scans for $m_t^{\text{\tiny pole}}$, $m_Z$, $\Delta\alpha_\text{had}^{(5)} $ and $\alpha_s $.]{$p$-value scans predicting $m_t^{\text{\tiny pole}}$, $m_Z$, $\Delta\alpha_\text{had}^{(5)} $ and $\alpha_s $ in the SM \otto{and in the SM4}.}
 \label{fig:FitEWPOI}
\end{figure}

\begin{figure}[htbp]
 \centering
 \subfigure[]{\begin{picture}(210,140)(0,10)
	      \put(0,0){\includegraphics[width=0.5\linewidth]{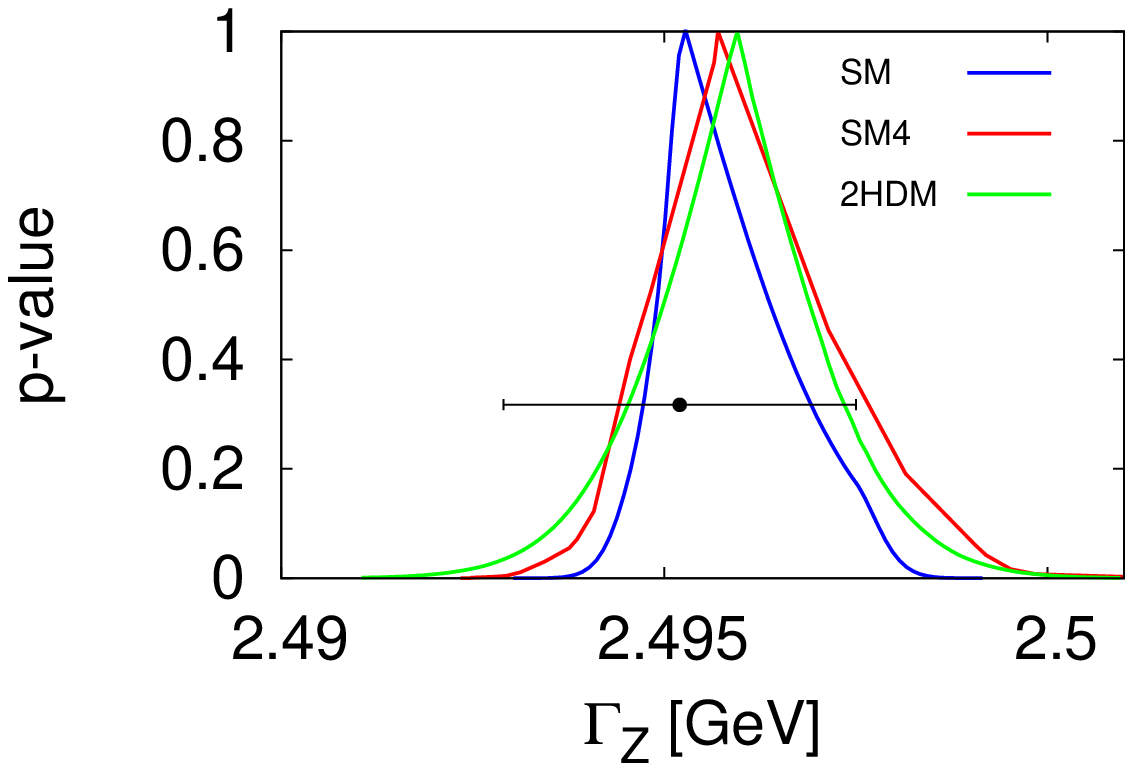}}
	      \put(51.3,116.0){\includegraphics[width=0.08\linewidth]{Images/CKMfitterPackage.eps}}
	      \end{picture}
	     }
 \qquad
 \subfigure[]{\begin{picture}(180,140)(0,10)
	      \put(0,0){\includegraphics[width=0.5\linewidth]{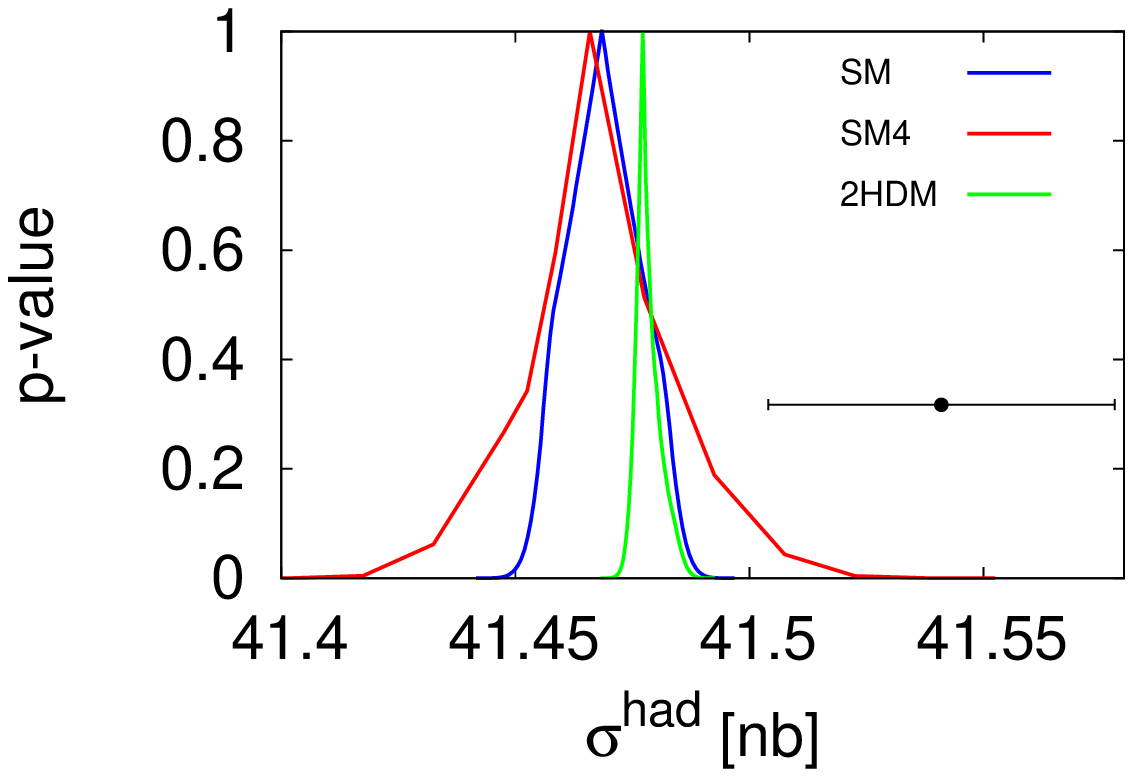}}
	      \put(51.3,116.0){\includegraphics[width=0.08\linewidth]{Images/CKMfitterPackage.eps}}
	      \end{picture}
	     }
 \subfigure[]{\begin{picture}(210,140)(0,10)
	      \put(0,0){\includegraphics[width=0.5\linewidth]{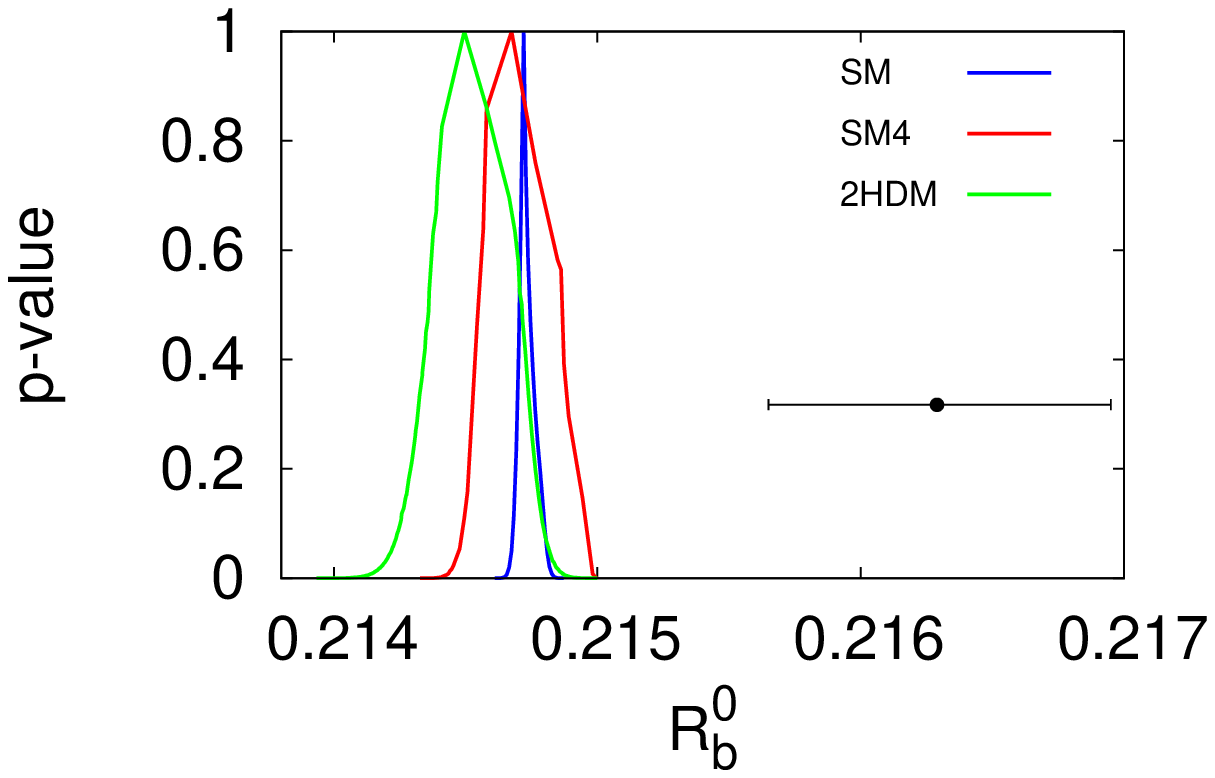}}
	      \put(105,116.0){\includegraphics[width=0.08\linewidth]{Images/CKMfitterPackage.eps}}
	      \end{picture}
	     }
 \qquad
 \subfigure[]{\begin{picture}(180,140)(0,10)
	      \put(0,0){\includegraphics[width=0.5\linewidth]{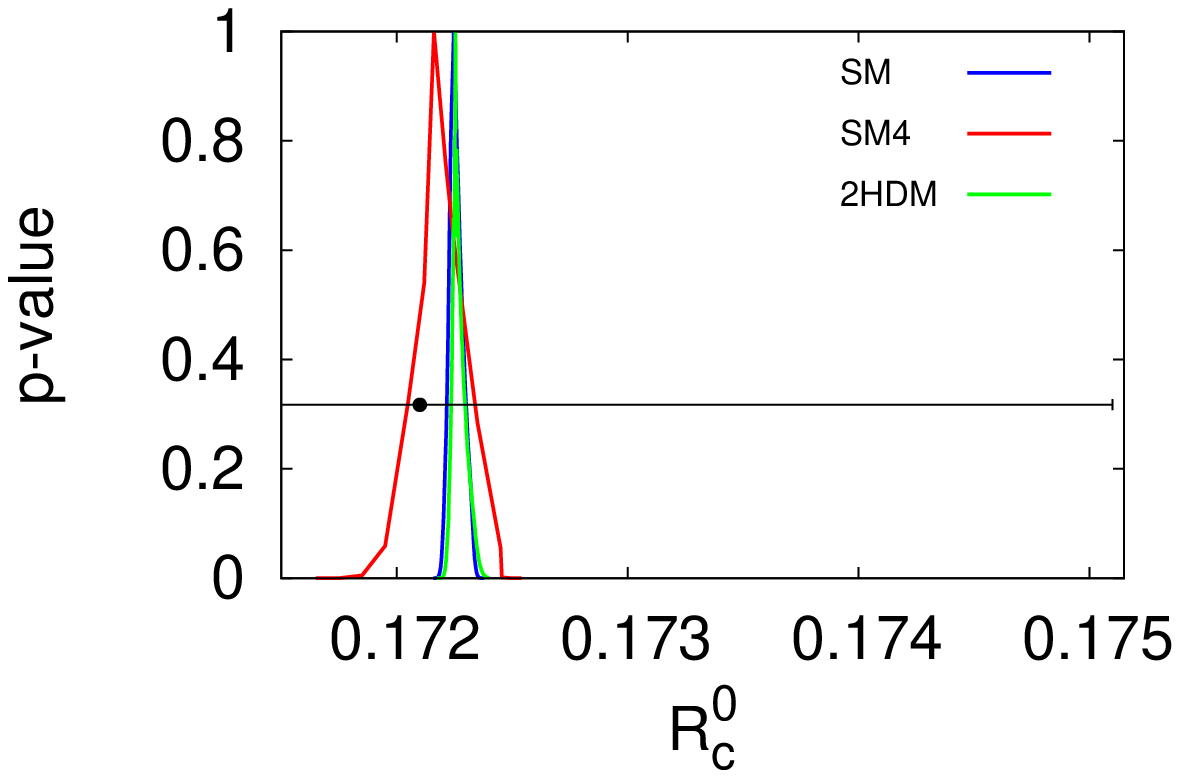}}
	      \put(100,116.0){\includegraphics[width=0.08\linewidth]{Images/CKMfitterPackage.eps}}
	      \end{picture}
	     }
 \subfigure[]{\begin{picture}(210,140)(0,10)
	      \put(0,0){\includegraphics[width=0.5\linewidth]{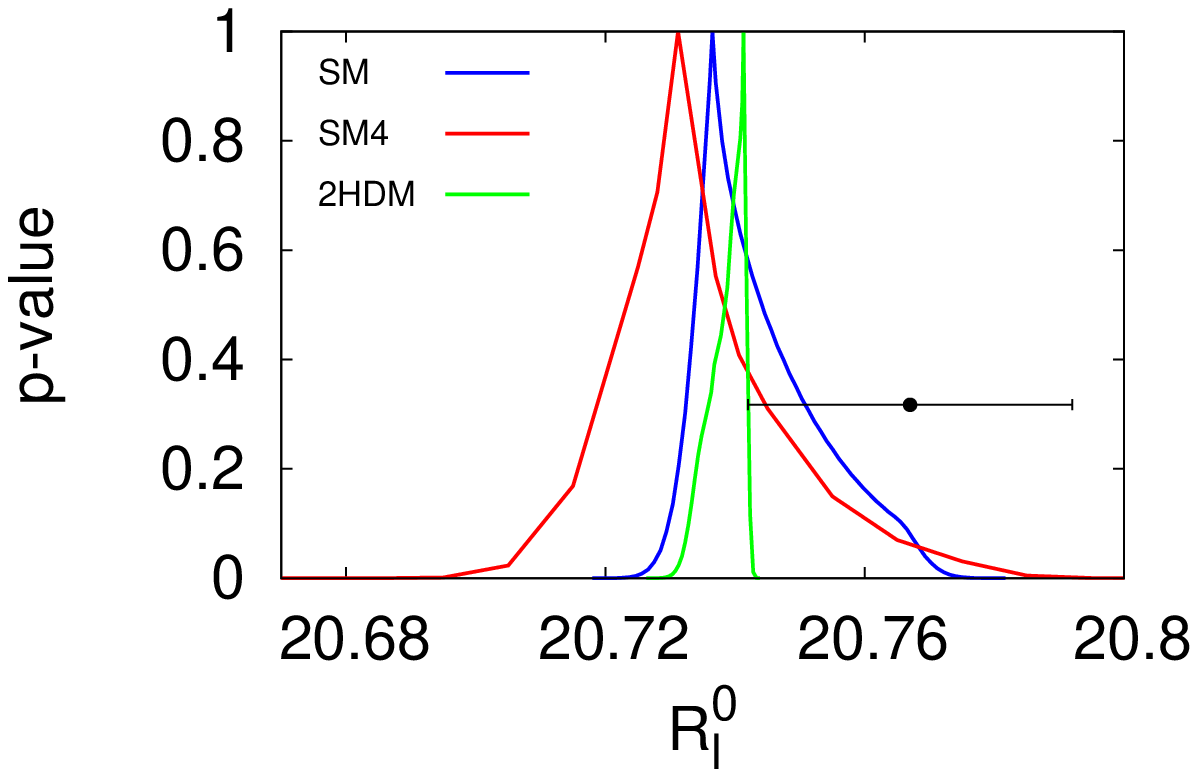}}
	      \put(160.1,115.9){\includegraphics[width=0.08\linewidth]{Images/CKMfitterPackage.eps}}
	      \end{picture}
	     }
 \qquad
 \subfigure[]{\begin{picture}(180,140)(0,10)
	      \put(0,0){\includegraphics[width=0.5\linewidth]{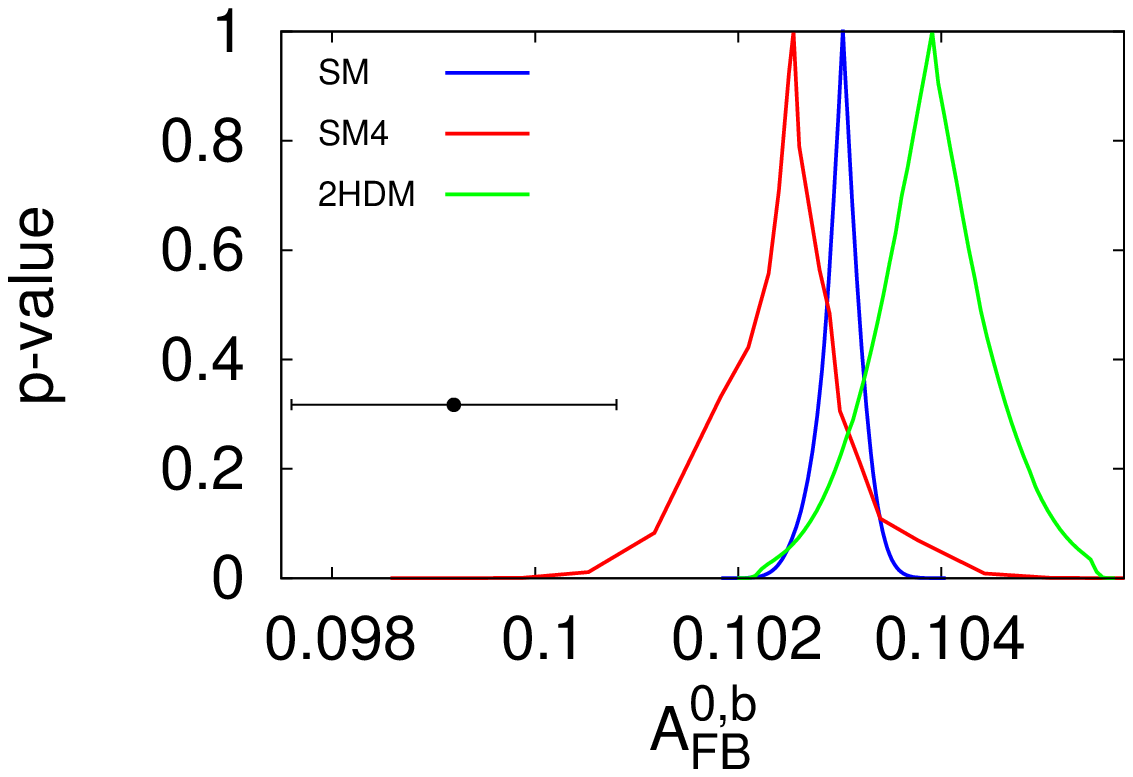}}
	      \put(97,115.9){\includegraphics[width=0.08\linewidth]{Images/CKMfitterPackage.eps}}
	      \end{picture}
	     }
 \subfigure[]{\begin{picture}(210,140)(0,10)
	      \put(0,0){\includegraphics[width=0.5\linewidth]{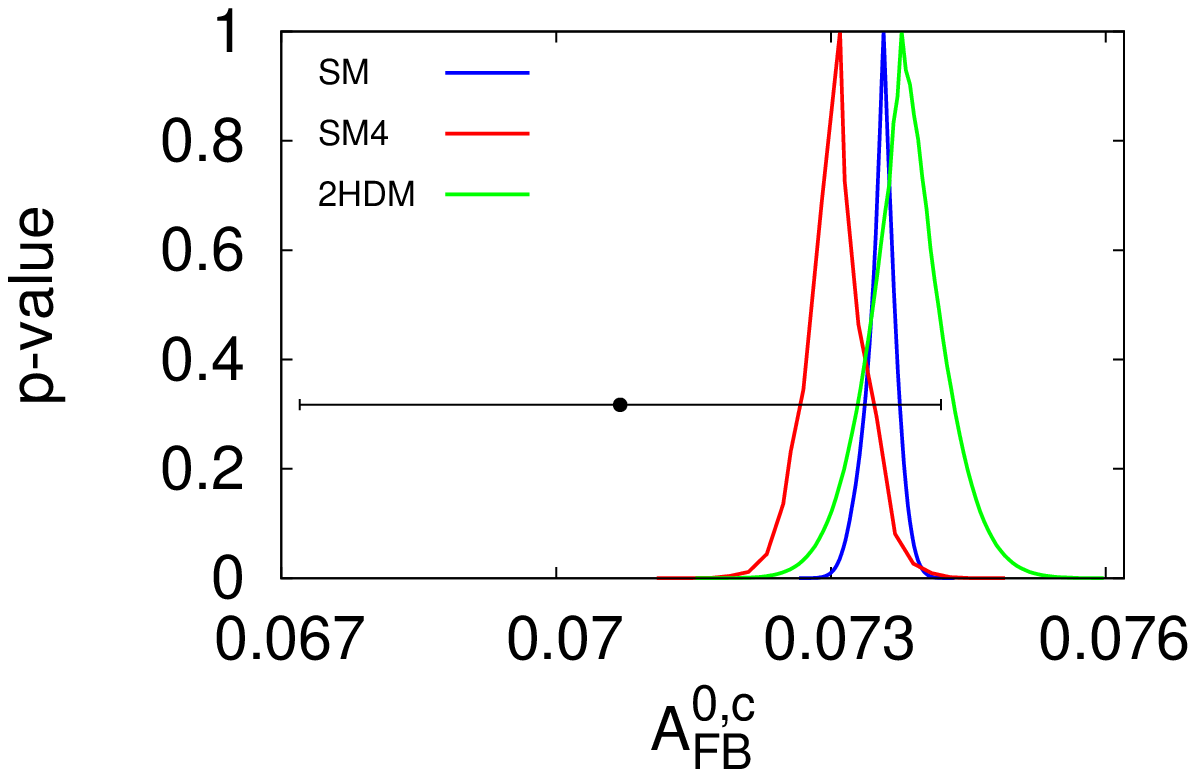}}
	      \put(102,115.9){\includegraphics[width=0.08\linewidth]{Images/CKMfitterPackage.eps}}
	      \end{picture}
	     }
 \qquad
 \subfigure[]{\begin{picture}(180,140)(0,10)
	      \put(0,0){\includegraphics[width=0.5\linewidth]{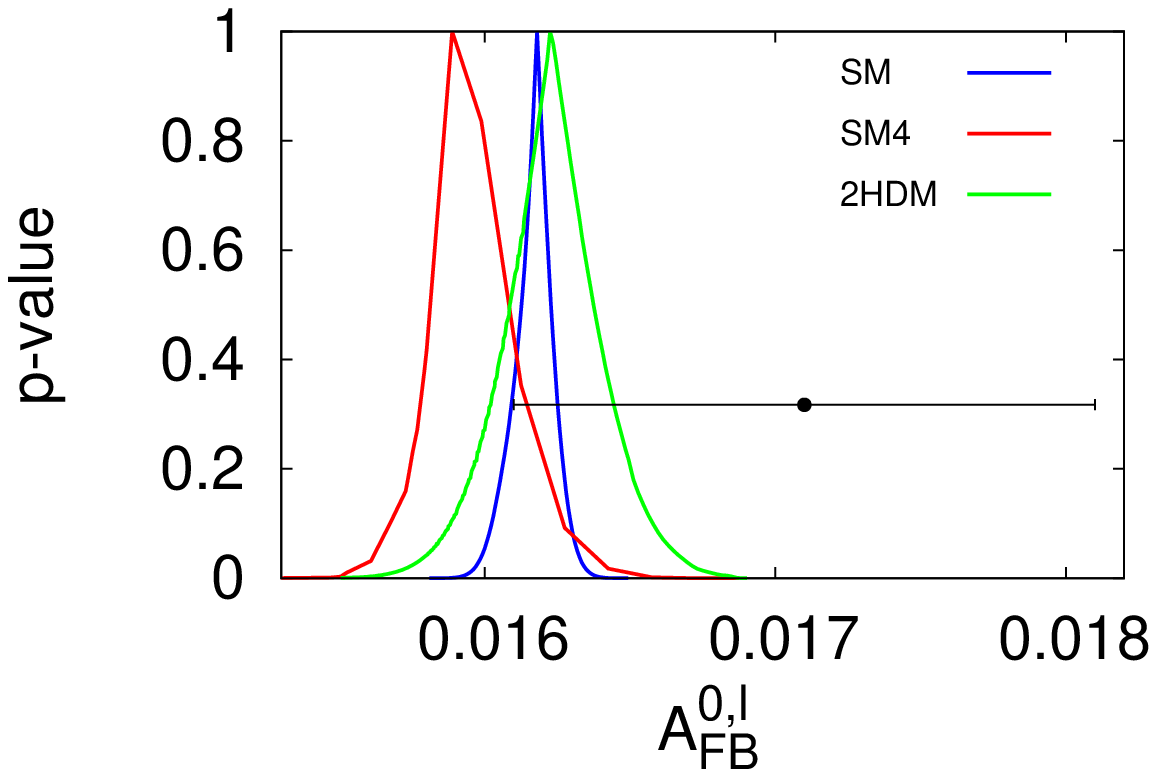}}
	      \put(105,115.9){\includegraphics[width=0.08\linewidth]{Images/CKMfitterPackage.eps}}
	      \end{picture}
	     }
 \caption[Prediction scans for $\Gamma_Z $, $\sigma^0_\text{had} $, $R_b^0 $, $R_c^0 $, $R_\ell^0 $, $A_\text{FB}^{0,b} $, $A_\text{FB}^{0,c} $ and $A_\text{FB}^{0,\ell} $.]{$p$-value scans predicting $\Gamma_Z $, $\sigma^0_\text{had} $, $R_b^0 $, $R_c^0 $, $R_\ell^0 $, $A_\text{FB}^{0,b} $, $A_\text{FB}^{0,c} $ and $A_\text{FB}^{0,\ell} $ in the SM, \otto{in} the SM4 and \otto{in} the 2HDM.}
 \label{fig:FitEWPOII}
\end{figure}

\begin{figure}[htbp]
 \centering
 \subfigure[]{\begin{picture}(210,140)(0,10)
	      \put(0,0){\includegraphics[width=0.5\linewidth]{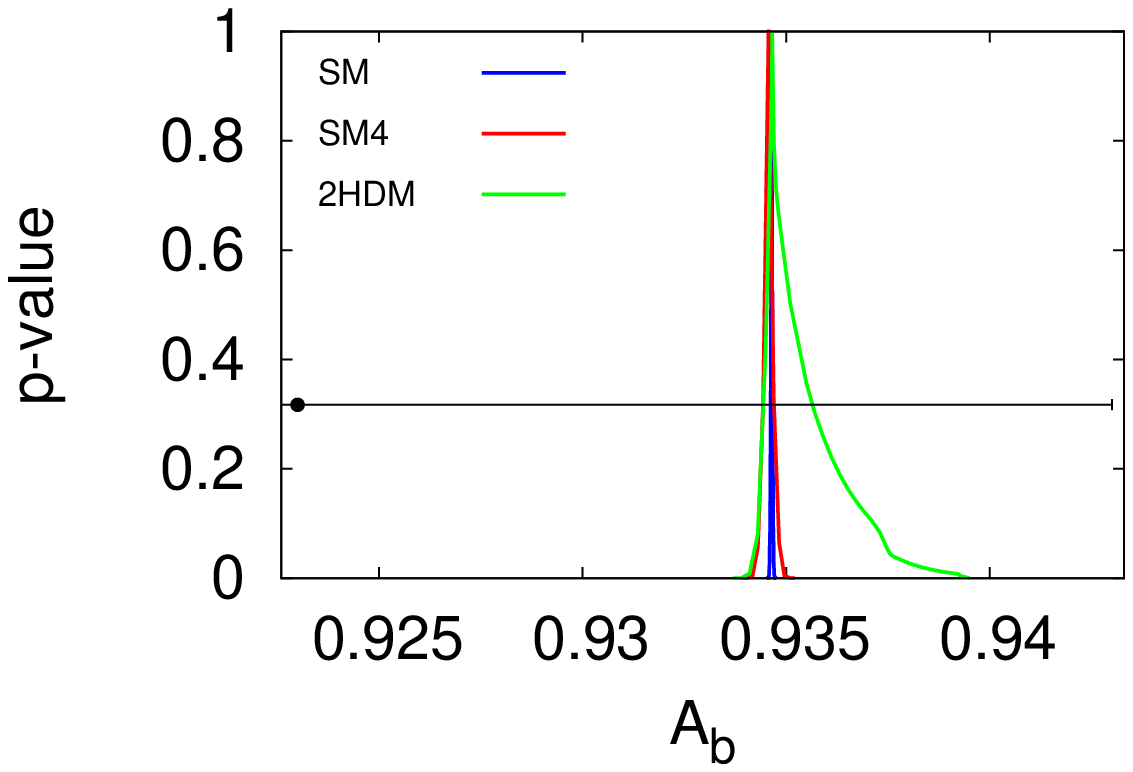}}
	      \put(160.0,116.0){\includegraphics[width=0.08\linewidth]{Images/CKMfitterPackage.eps}}
	      \end{picture}
	     }
 \qquad
 \subfigure[]{\begin{picture}(180,140)(0,10)
	      \put(0,0){\includegraphics[width=0.5\linewidth]{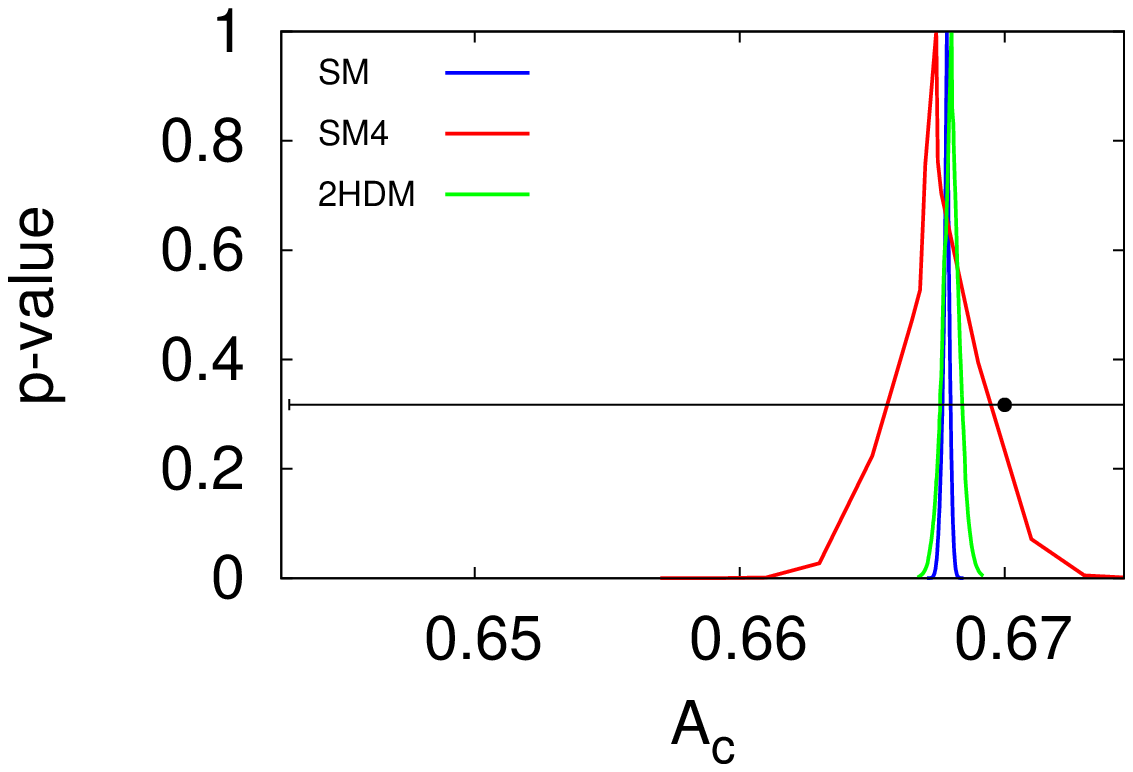}}
	      \put(102,116.0){\includegraphics[width=0.08\linewidth]{Images/CKMfitterPackage.eps}}
	      \end{picture}
	     }
 \subfigure[]{\begin{picture}(210,140)(0,10)
	      \put(0,0){\includegraphics[width=0.5\linewidth]{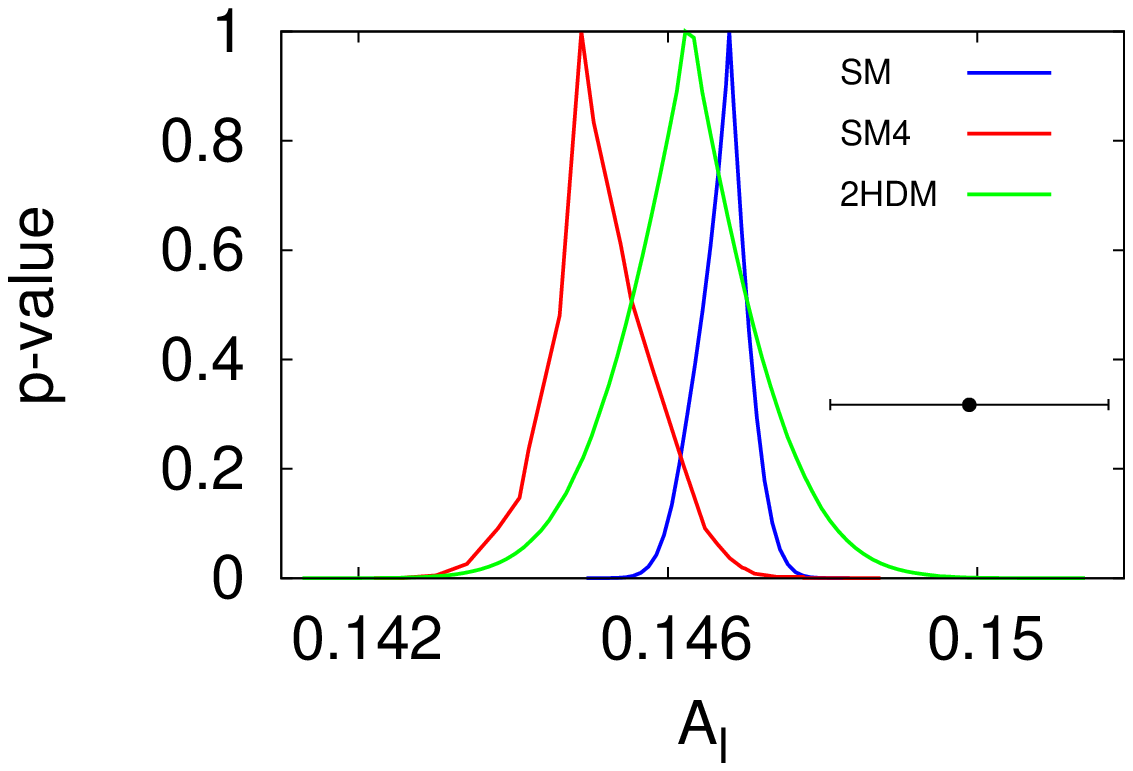}}
	      \put(51.3,116.0){\includegraphics[width=0.08\linewidth]{Images/CKMfitterPackage.eps}}
	      \end{picture}
	     }
 \qquad
 \subfigure[]{\begin{picture}(180,140)(0,10)
	      \put(0,0){\includegraphics[width=0.5\linewidth]{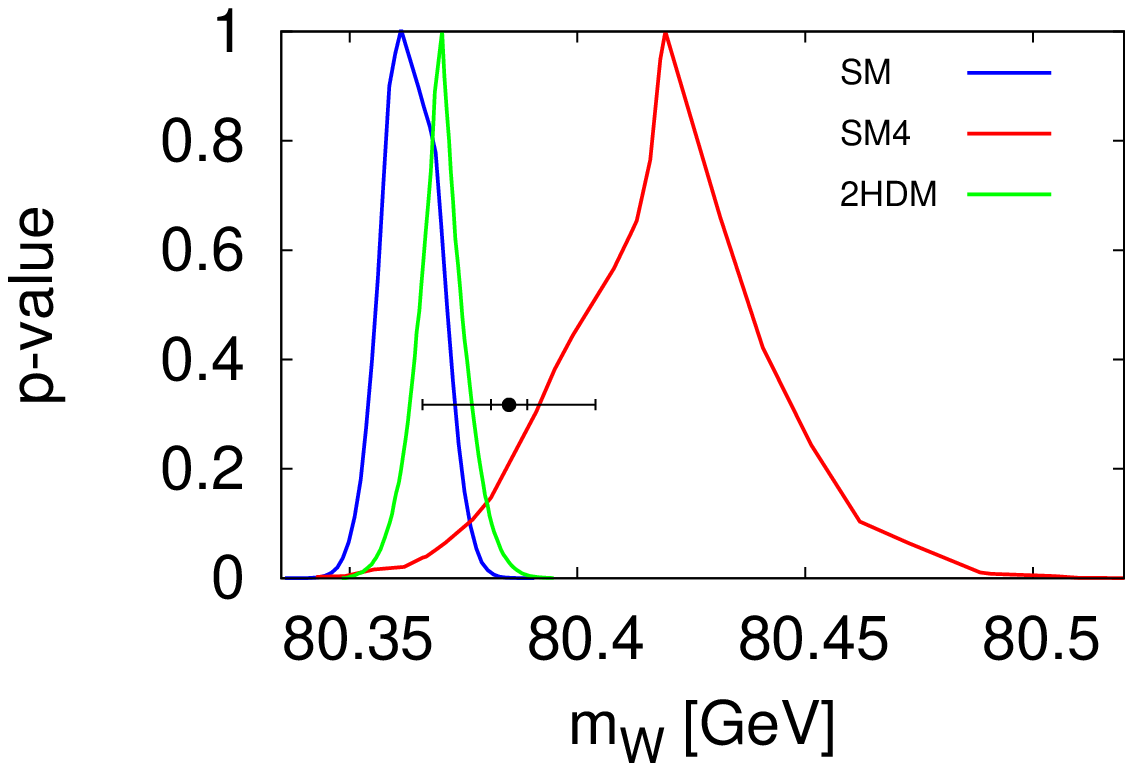}}
	      \put(159.8,82){\includegraphics[width=0.08\linewidth]{Images/CKMfitterPackage.eps}}
	      \end{picture}
	     }
 \subfigure[]{\begin{picture}(210,140)(0,10)
	      \put(0,0){\includegraphics[width=0.5\linewidth]{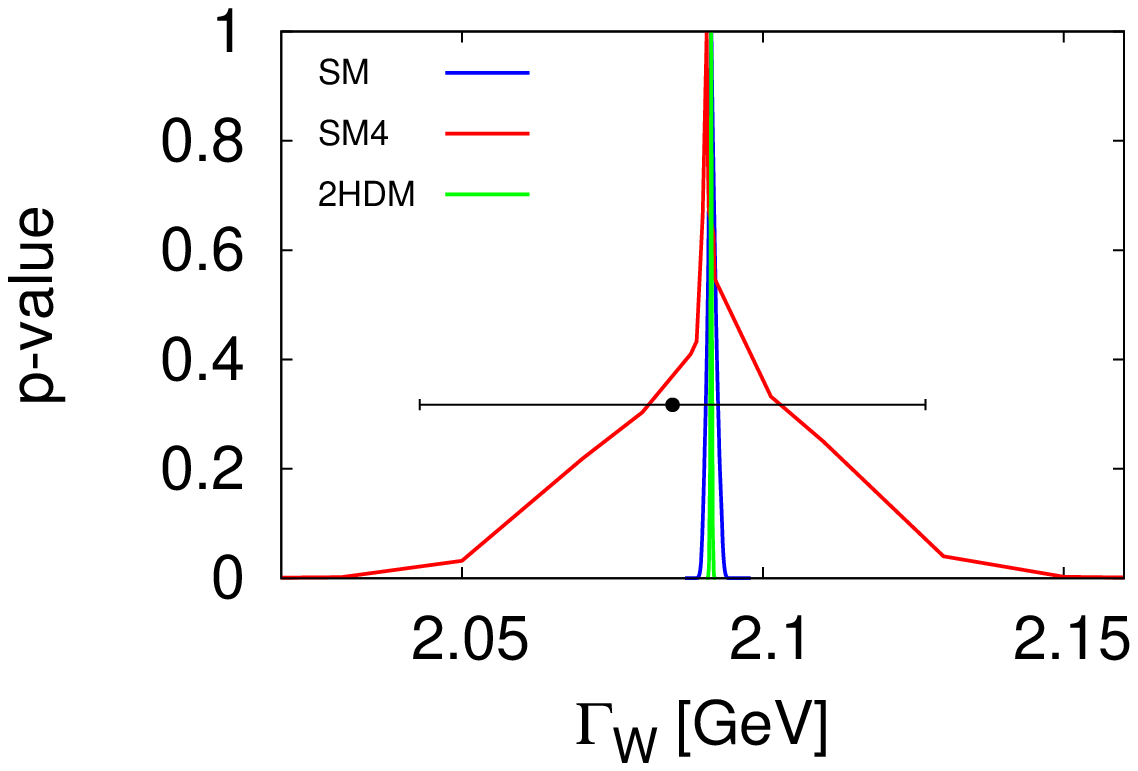}}
	      \put(160.0,115.9){\includegraphics[width=0.08\linewidth]{Images/CKMfitterPackage.eps}}
	      \end{picture}
	     }
 \qquad
 \subfigure[]{\begin{picture}(180,140)(0,10)
	      \put(0,0){\includegraphics[width=0.5\linewidth]{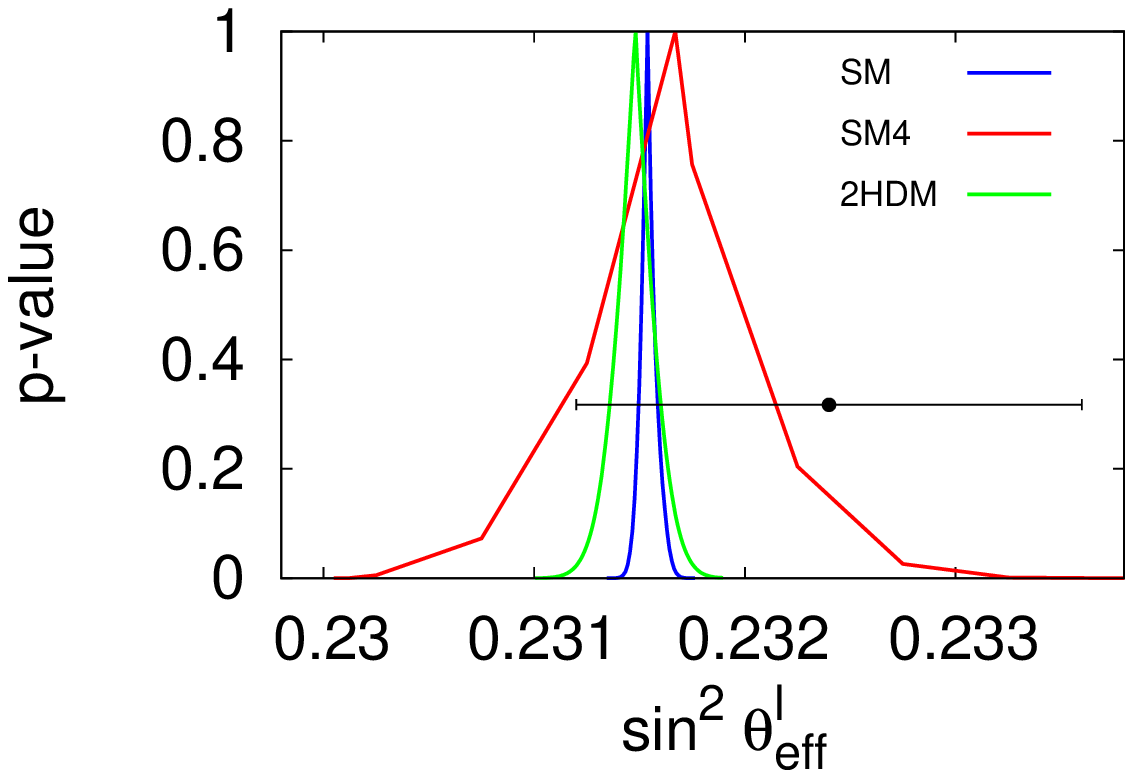}}
	      \put(51.2,115.9){\includegraphics[width=0.08\linewidth]{Images/CKMfitterPackage.eps}}
	      \end{picture}
	     }
 \subfigure[]{\begin{picture}(210,140)(0,10)
	      \put(0,0){\includegraphics[width=0.5\linewidth]{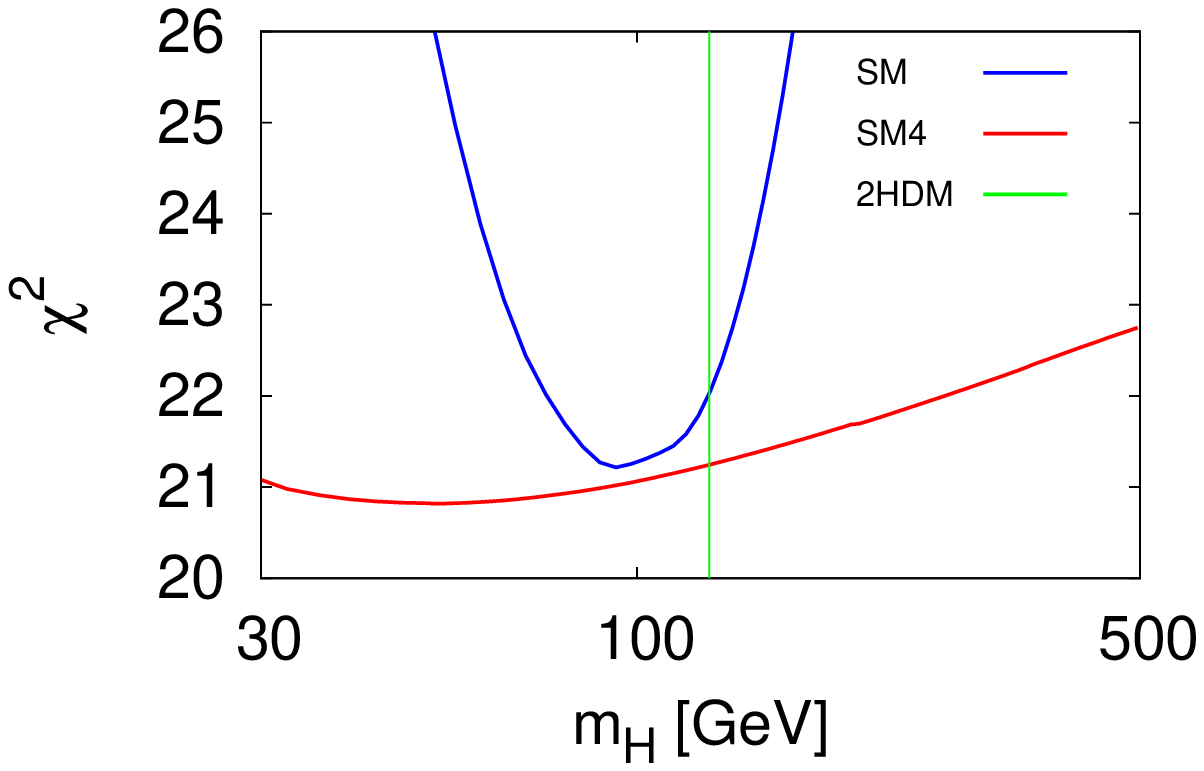}}
	      \put(160,44){\includegraphics[width=0.08\linewidth]{Images/CKMfitterPackage.eps}}
	      \end{picture}
	     }
 \caption[Prediction scans for ${\cal A}_b $, ${\cal A}_c $, ${\cal A}_\ell $, $m_W $, $\Gamma_W $ and $\sin^2\theta_\ell^\text{eff} $ and $m_H$.]{$p$-value scans predicting ${\cal A}_b $, ${\cal A}_c $, ${\cal A}_\ell $, $m_W $, $\Gamma_W $ and $\sin^2\theta_\ell^\text{eff} $ in the SM, \otto{in} the SM4 and \otto{in} the 2HDM\otto{,} and a $\chi^2$ scan of the Higgs mass prediction by the EWPO in the respective models.}
 \label{fig:FitEWPOIII}
\end{figure}

\begin{figure}[htbp]
 \centering
  \subfigure[]{\begin{picture}(210,140)(0,0)
	      \put(0,0){\includegraphics[width=0.5\linewidth]{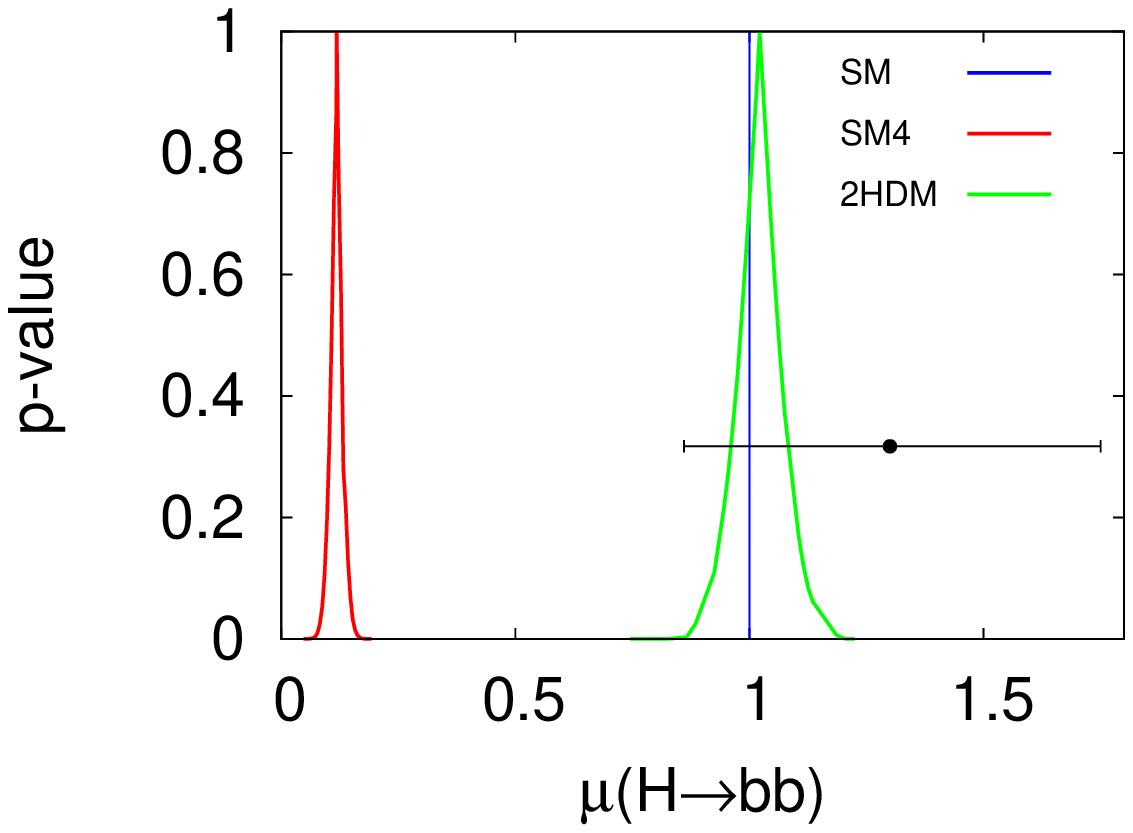}}
	      \put(80,115.9){\includegraphics[width=0.08\linewidth]{Images/CKMfitterPackage.eps}}
	      \end{picture}
	     }
 \qquad
 \subfigure[]{\begin{picture}(180,140)(0,0)
	      \put(0,0){\includegraphics[width=0.5\linewidth]{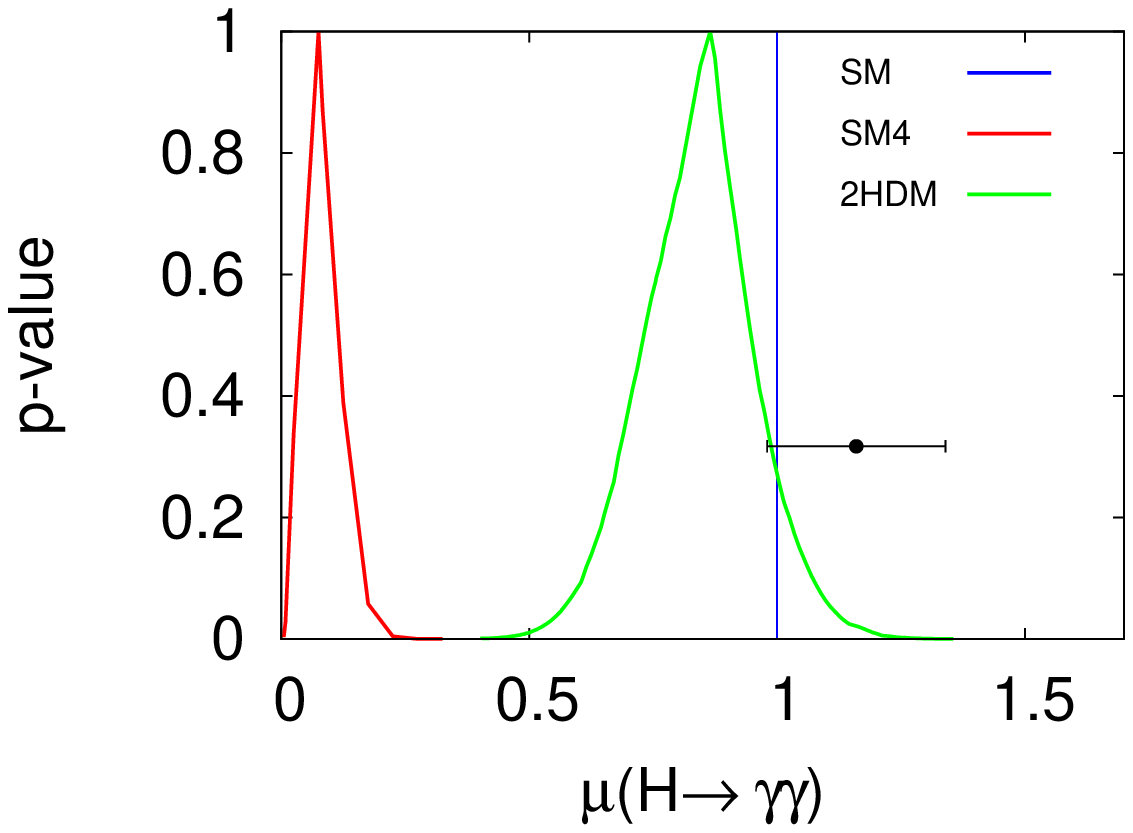}}
	      \put(70,115.9){\includegraphics[width=0.08\linewidth]{Images/CKMfitterPackage.eps}}
	      \end{picture}
	     }
 \subfigure[]{\begin{picture}(210,140)(0,0)
	      \put(0,0){\includegraphics[width=0.5\linewidth]{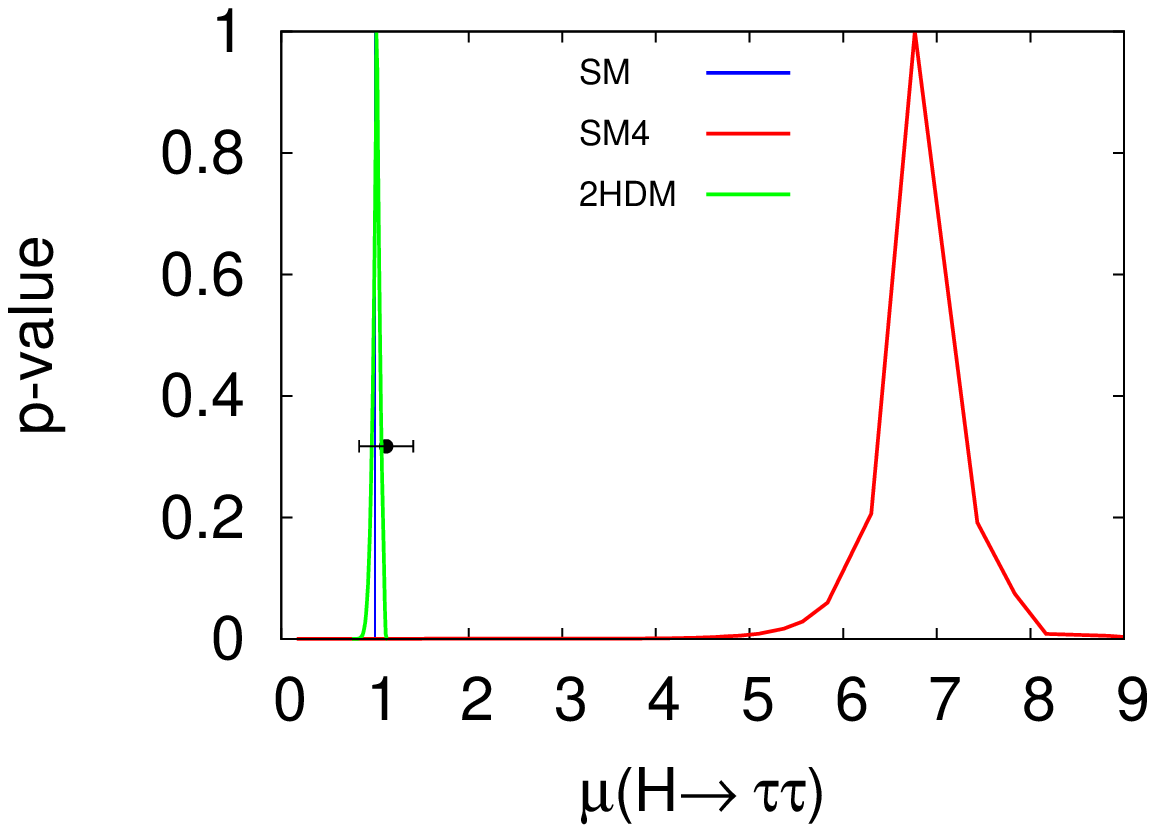}}
	      \put(101,82){\includegraphics[width=0.08\linewidth]{Images/CKMfitterPackage.eps}}
	      \end{picture}
	     }
 \qquad
 \subfigure[]{\begin{picture}(180,140)(0,0)
	      \put(0,0){\includegraphics[width=0.5\linewidth]{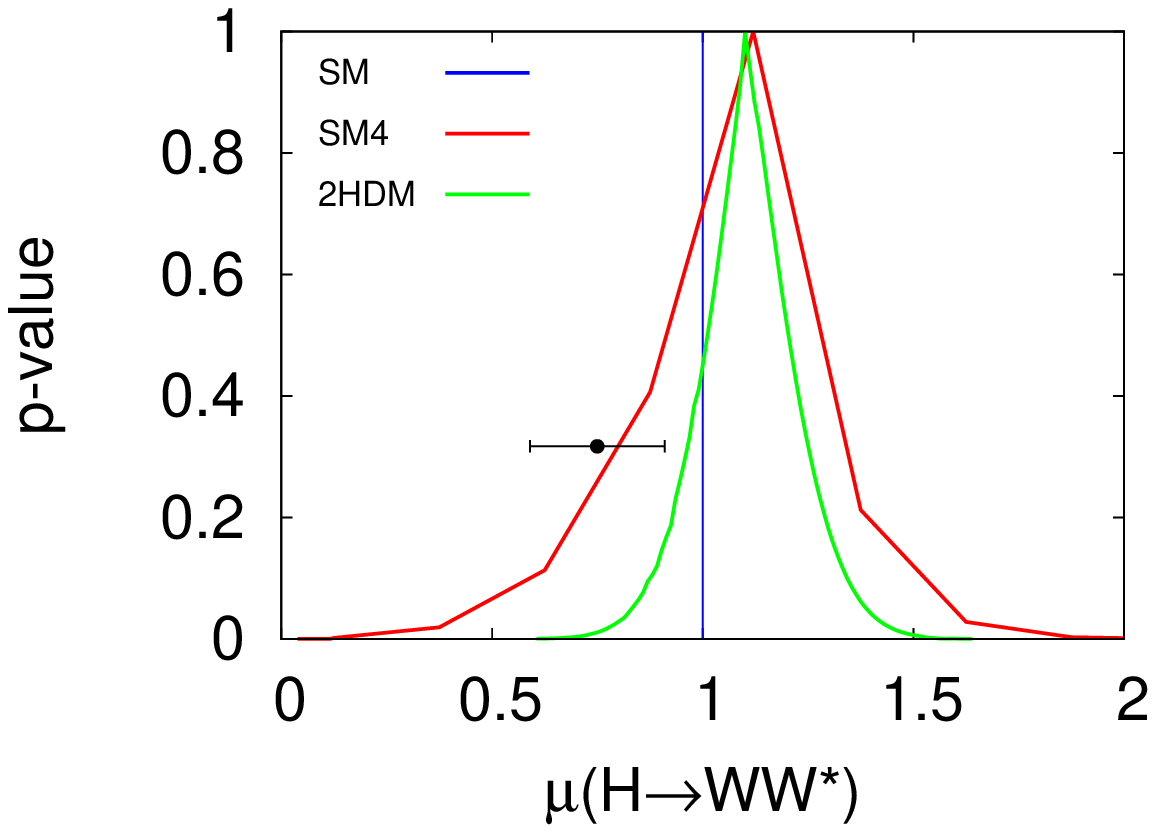}}
	      \put(160,115.9){\includegraphics[width=0.08\linewidth]{Images/CKMfitterPackage.eps}}
	      \end{picture}
	     }
 \subfigure[]{\begin{picture}(210,140)(0,0)
	      \put(0,0){\includegraphics[width=0.5\linewidth]{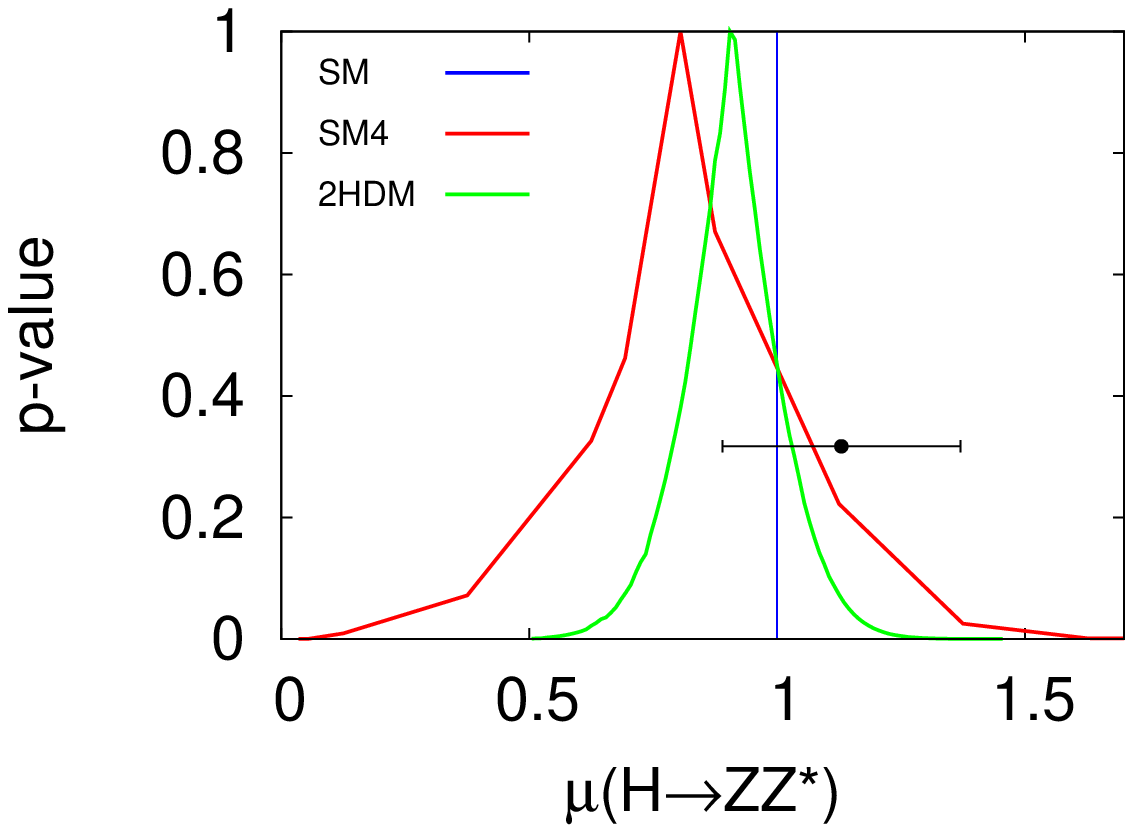}}
	      \put(159.8,115.9){\includegraphics[width=0.08\linewidth]{Images/CKMfitterPackage.eps}}
	      \end{picture}
	     }
 \caption[Prediction scans for the Higgs signal strengths.]{$p$-value scans predicting the \otto{combined} Higgs signal strengths in the SM4 and \otto{in} the 2HDM.}
 \label{fig:FitSignalStrengths}
\end{figure}

\ottoo{All SM predictions of the EWPO \ottooo{largely} agree with the predicted $1\sigma$ confidence intervals of the Bayesian analysis by Ciuchini et al. \cite{Ciuchini:2013pca}.} \ottoo{The top quark pole mass prediction in the SM4 features remarkably small allowed values. Due to technical reasons, I required $m_t^{\text{\tiny pole}} >130$\:GeV, so the $p$-value vanishes at that point; however, a top mass below $130$\:GeV \ottooo{would} be allowed at $95\%$ CL.}
For $\sigma^0_\text{had}$, $R_b^0$, $A_\text{FB}^{0,b}$ and ${\cal A}_\ell$\otto{, which were introduced in Sect.\ \ref{SMEWPO},} none of the prediction fits agrees at one standard deviation with the experiment.
The fits of $R_c^0$, ${\cal A}_b$ and ${\cal A}_c$ show that the measurements are less precise than the predictions, so only a limited range of the $1\sigma$ interval is shown. However, in some models the observables might be important constraints. A good example is $\Gamma _W$, for which the SM prediction is within a narrow interval, but in the SM4, the $W$ decay width is allowed to deviate sizeably from the SM value. In the 2HDM fits, the SM parameters from \eqref{eq:SMparams} were fixed to the SM best-fit values (not to the SM predictions!) as mentioned in Chapter \ref{2HDM}.
Without Higgs discovery data, the Higgs mass fit of the SM4 prefers lighter values for $m_H$ than the SM fit does. Around the measured value of the Higgs mass, the $\chi^2_{\rm min}$ of the SM4 is slightly lower than the one of the SM.
All signal strength predictions agree well with the SM and with the 2HDM. The fermionic and the diphoton Higgs decays interpreted in the SM4 are incompatible with the measured signal strength values. \ottoo{Here, the prediction scans reveal complementary information to the deviations Fig.\ \ref{fig:deviationsSM4}: $\mu (H\to \tau \tau )$ features a best-fit deviation smaller than one. However, its prediction is in conflict with the best-fit value of $1.34$. Without tauonic Higgs decay measurements, the best-fit parameters are completely different. The fourth generation neutrino mass is pushed to $m_H/2$, which changes all Higgs decay branching ratios sizeably; $m_{t'}$ and $m_{b'}$ are at the upper end of their allowed range.
In the complete SM4 fit, $\mu (H\to \tau \tau )$ is as small as possible, which in turn requires a high invisible Higgs decay probability. Shifting $m_{\nu _4}$ to $56.7$\:GeV suppresses the Higgs branching ratio to $\tau$ leptons by a factor of five.
The reduction of the $\mu (H\to \tau \tau )$ deviation is compensated by large deviations of $\mu (H\to \gamma \gamma )$ and $\mu (H\to b\oline{b})$, however, the tauonic Higgs decay observables exhibit the largest $\Delta \chi ^2$, even larger than the ones of the diphoton decay observables.}

%% file: 2hdmrelations.tex
\chapter{2HDM relations} \label{thdmrelations}

There are different parametrizations of the 2HDM of type II (see for instance \cite{Gunion:2002zf}). Starting from the quadratic and quartic couplings from the Lagrangian \eqref{eq:thdmlagrangian}, I want to list the relations connecting $m_{ij}$ and $\lambda_i$ to the parameters that I used in my fits, i.e.\ the four physical Higgs masses $m_{h^0}$, $m_{H^0}$, $m_{A^0}$, $m_{H^+}$, the vacuum expectation value $v$, the quartic coupling $\lambda_5$, and the two mixing angles $\alpha $ and $\beta $.

\begin{align*}
m_{11}^2 &= -\frac{1}{2}\left[ m_{H^0}^2 \cos ^2 \alpha + m_{h^0}^2 \sin ^2\!\alpha +\left( m_{H^0}^2 - m_{h^0}^2 \right) \cos \alpha \sin \alpha \tan \beta\right] +\frac{\tan ^2\!\beta \left( m_{A^0}^2+v^2\lambda_5 \right) }{1+\tan ^2\!\beta }\\
m_{22}^2 &= -\frac{1}{2}\left[ m_{H^0}^2 \sin ^2\!\alpha + m_{h^0}^2 \cos ^2\!\alpha +\left( m_{H^0}^2 - m_{h^0}^2 \right) \frac{\cos \alpha \sin \alpha}{\tan \beta}\right] +\frac{m_{A^0}^2+v^2\lambda_5 }{1+\tan ^2\!\beta }\\
m_{12}^2 &= \frac{\tan \beta }{1+\tan ^2\!\beta }\left( m_{A^0}^2+v^2\lambda_5 \right) \\
 \lambda_1 &= \frac{1}{v^2}\left[ \left( 1+\tan ^2\!\beta \right) \left( m_{H^0}^2 \cos ^2\!\alpha + m_{h^0}^2 \sin ^2\!\alpha \right) -\tan ^2\!\beta \left( m_{A^0}^2+v^2\lambda_5 \right) \right] \\
 \lambda_2 &= \frac{1}{v^2\tan ^2\!\beta}\left[ \left( 1+\tan ^2\!\beta \right) \left( m_{h^0}^2 \cos ^2\!\alpha + m_{H^0}^2 \sin ^2\!\alpha \right) -m_{A^0}^2-v^2\lambda_5 \right] \\
 \lambda_3 &= \frac{1+\tan ^2\!\beta }{v^2\tan ^2\!\beta} \cos \alpha \sin \alpha \left( m_{H^0}^2 - m_{h^0}^2 \right) - \frac{m_{A^0}^2-2m_{H^+}^2}{v^2} -\lambda_5 \\
 \lambda_4 &= \frac{2(m_{A^0}^2-m_{H^+}^2)}{v^2} +\lambda_5
\end{align*}

%% file: parallelizer.tex
\chapter{External programs and parallelization with CKMfitter} \label{parallelize}

In this appendix, I want to elaborate on the possibilities of adding external programs to the CKMfitter package and parallelizing two-dimensional CKMfitter scans, which I developed in the last two years.

\section{External programs}

When I wanted to use the \texttt{Zfitter} subroutine \texttt{DIZET} to calculate electroweak precision observables in the SM, I encountered a problem that was new to CKMfitter users: up to this point, everybody had added theoretical formulae as analytic expressions. Fortunately, one can switch off the analytic gradient for the global minimum searches in CKMfitter, such that the minimization routine can be used even for numerical input. (As mentioned in Sect.\ \ref{systematicerrors}, the minimization is performed by a Fortran part.) So I had to establish an interface that linked \texttt{DIZET} to the CKMfitter code. For this purpose, I had to assign on Mathematica level one Fortran function for each of the EWPO, from which in turn I called \texttt{DIZET} with a given set of parameters.
The \texttt{Zfitter} subroutine returns all EWPO values at once, so I interposed a query that tests whether the input parameters have changed since the last call of \texttt{DIZET}, and if not only reads the stored values from the last call. As we have $14$ EWPO, this yields approximately a factor of $1/14$ in run time because the \texttt{DIZET} call is the bottleneck of the fits.\\
Generalizing this interface, I provided a method to include any external programs to the CKMfitter package. Only a few months later, when I wanted to add the Higgs signal strengths to the SM4 fits, I decided to use \texttt{HDecay} to calculate the Higgs decay width and the branching ratios. Although it was not possible to link \texttt{HDecay} directly like I did with \texttt{DIZET}, I ran \texttt{HDecay} for many parameter values and collected its output in look-up tables. (Neglecting flavour mixing, the relevant SM4 parameters are the fourth generation fermion masses and the Higgs mass; thus, the corresponding tables have five parameter dimensions.) With the help of an interpolation algorithm provided by Martin Wiebusch, I was now able to connect these tables to predefined observables in the code, following the same pattern as in the EWPO case.

\section{Parallelization}

As discussed in Sect.\ \ref{systematicerrors}, I wrote a parallelization program for two-dimensional scans with CKMfitter. If the granularity for a scan is $N$ and supposed one is equipped with $N$ processors, this program called \texttt{parallelize} can speed up the scan time by a factor of $1/N$. (Note that the \textit{fit} time of the initial global minimization is not affected. So if a complete scan is dominated by the first fit, the parallelization will not ameliorate the run time.) The two-dimensional 2HDM scans in Chapter \ref{2HDM}, for instance, have a granularity of $200$. With the run time of a one-dimensional fit of up to 50 minutes, I saved up to one week for each plot, even though some of the \otto{1D fits} did not converge properly, and I had \otto{t}o re-run them.\\
\texttt{parallelize} is a simple bash script. One needs to hard-code the path leading to a generic job description for the cluster computer (``qsub script'') as well as the command that sends the qsub script to the computing cluster. CKMfitter stores input data, theory expressions and job instructions in different files; the latter are called \textit{analysis cards} and are the only files that \texttt{parallelize} needs to access: it reads the relevant job parameters such as the granularity, the scan quantities and their ranges from the specified analysis card and creates $N$ subdirectories in the working directory.
Each of these new directories is provided with its own analysis card, in which the 2D scan has been changed to a 1D scan by setting one of the scan quantities to a fixed value. This value is different for each subdirectory. Moreover, each subdirectory gets its own modified qsub script. If successful, \texttt{parallelize} then submits all $N$ jobs to the cluster. (Several security queries such as a maximal allowed granularity and 
syntax checks along this procedure have been implemented.)\\
When all parallelized jobs have finished, we need to unite all data from the subfolders. For that purpose, I wrote a second bash script, called \texttt{unify}. The crux of the unification is the correct calculation of the (naive) $p$-value.
If we combined all one-dimensional scan points by simply concatenating the data files, we would get at least $N$ points with an assigned $p$-value of $1$ because every 1D scan has been normalized to its own $\chi^2$ minimum. So the relevant information are the $\chi^2$ values. \texttt{unify} finds the global $\chi^2_{\rm min}$ in all data files and subsequently re-calculates the $p$-value for all scan points.
At the end, \texttt{ckmgnuplot} is called, which is another bash script written by Andreas Menzel and myself to visualize CKMfitter data using the open source plotting program \texttt{gnuplot}.
(\texttt{unify} can be called if at least one of the $N$ parallelized jobs has finished -- in that case a fake $p$-value of $-1$ is assigned to the missing data points, which is converted into white space by \texttt{ckmgnuplot}.)\\
The drawback of the parallelization is that for the constrained scan fits at the individual scan points, CKMfitter cannot resort to the information of all neighbouring scan points. Hence, some of the 1D scan fits may not converge to its global minimum; in the resulting plot one gets a few ``streaks'' from one-dimensional scans where the minimum has obviously not been found.
In order to cope with this, I additionally wrote the short bash script \texttt{redo} that saves the old data files and restarts on request single 1D scans. \texttt{unify} has been modified such that it automatically merges the old and the new data file and for every scan point takes the smaller $\chi^2$ value. Making use of the three parallelization tools, CKMfitter users can save a lot of time; the scripts are available on the CKMfitter software system.

%% file: acknowledgements.tex
\section*{Acknowledgements}

I would like to thank everybody who supported me and my work in the last years.\\[5pt]
First and most all, I would like to thank Ulrich Nierste for the supervision of this thesis. He gave me as much freedom in research as I wanted, but he always took the time for inspiring discussions about physical problems and encouraged me, when I required help. He taught me a lot about the exciting fields of flavour and high energy physics, but also about the academic life.\\
Next, I want to thank Martin Wiebusch, who patiently advised me in all technical, statistical and physical details that I encountered, and modified his illustrations to fit my notation.
I am grateful to Alexander Lenz for his continued support even after my Diploma thesis and his help with my applications. In this context I also want to express my gratitude to Michael Chanowitz.\\
Also the rest of our little collaboration needs to be mentioned: regular fruitful discussions with Heiko Lacker, Andreas Menzel and Geoffrey Herbert about physical details but also concerning technical topics steadily pushed forward the progress of this work. Moreover, I would like to thank the CKMfitter group, especially J{\'e}r{\^o}me Charles, for technical support with the CKMfitter package.\\[5pt]
I am grateful to Dieter Zeppenfeld, who was my second supervisor and advisor.\\
I had the privilege to spend the past years at the TTP surrounded by many nice colleagues (also from the ITP), who contributed to a pleasant and productive atmosphere. In particular, I want to thank Markus Bobrowski and Philipp Frings for sharing an office and lots of cake. In many seminars and lectures, I learned a lot about various aspects of high energy physics; I would like to thank Johann K{\"u}hn and Matthias Steinhauser, who organized most of them. Thomas Hermann deserves a special mention for providing me with his $b\to s\gamma$ code. Furthermore, he helped together with Peter Marquard and Jens Hoff to maintain our computer system, which constitutes an essential part of the working conditions. I also appreciate Martina Schorn's assistance when she helped me filling in forms.\\[5pt]
This thesis would contain many errors if it had not been proofread by Martin Wiebusch, Markus Bobrowski, Alexander Lenz, Andreas Scholz, Elio K{\"o}nig and my father. Many thanks to all of them!\\
Furthermore, I want to thank the Deutsche Forschungsgemeinschaft, which supported the work presented in this thesis under grant no. NI 1105/2-1.\\[5pt]
Finally, I want to express my deepest gratitude to my family and my friends, who made me the person I am, and especially to my school teachers Gerhard Forster and Trevor Plant, who sparked my enthusiasm for physics and thus made me the physicist I am.

%% file: Thesis.bbl
\begingroup\raggedright\begin{thebibliography}{159}
\expandafter\ifx\csname natexlab\endcsname\relax\def\natexlab#1{#1}\fi

\bibitem[Higgs(1964)]{Higgs:1964pj}
P.~W. Higgs, ``{BROKEN SYMMETRIES AND THE MASSES OF GAUGE BOSONS}'', {\em
  Phys.Rev.Lett.} {\bfseries 13} (1964)
508--509.

\bibitem[Englert and Brout(1964)]{Englert:1964et}
F.~Englert and R.~Brout, ``{BROKEN SYMMETRY AND THE MASS OF GAUGE VECTOR
  MESONS}'', {\em Phys.Rev.Lett.} {\bfseries 13} (1964)
321--323.

\bibitem[Guralnik et~al.(1964)Guralnik, Hagen, and Kibble]{Guralnik:1964eu}
G.~Guralnik, C.~Hagen, and T.~Kibble, ``{GLOBAL CONSERVATION LAWS AND MASSLESS
  PARTICLES}'', {\em Phys.Rev.Lett.} {\bfseries 13} (1964)
585--587.

\bibitem[Aad et~al.(2012)]{Aad:2012tfa}
{\bfseries ATLAS Collaboration}, G.~Aad {\em et~al.}, ``{Observation of a New
  Particle in the Search for the Standard Model Higgs Boson with the ATLAS
  Detector at the LHC}'', {\em Phys.Lett.} {\bfseries B716} (2012) 1--29,
  \href{http://xxx.lanl.gov/abs/1207.7214}{{\ttfamily arXiv:1207.7214}},
{additional auxiliary plots on the website}.

\bibitem[Chatrchyan et~al.(2012)]{Chatrchyan:2012ufa}
{\bfseries CMS Collaboration}, S.~Chatrchyan {\em et~al.}, ``{Observation of a
  new boson at a mass of 125 GeV with the CMS experiment at the LHC}'', {\em
  Phys.Lett.} {\bfseries B716} (2012) 30--61,
 \href{http://xxx.lanl.gov/abs/1207.7235}{{\ttfamily arXiv:1207.7235}}.

\bibitem[Hou(2009)]{Hou:2008xd}
W.-S. Hou, ``{Source of \textit{CP} Violation for the Baryon Asymmetry of the
  Universe}'', {\em Chin.J.Phys.} {\bfseries 47} (2009) 134,
 \href{http://xxx.lanl.gov/abs/0803.1234}{{\ttfamily arXiv:0803.1234}}.

\bibitem[Hou et~al.(2005)Hou, Nagashima, and Soddu]{Hou:2005yb}
W.-S. Hou, M.~Nagashima, and A.~Soddu, ``{Enhanced $K_L \to \pi^0 \nu
  \oline{\nu}$ from Direct \textit{CP} Violation in $B \to K \pi$ with Four
  Generations}'', {\em Phys.Rev.} {\bfseries D72} (2005) 115007,
 \href{http://xxx.lanl.gov/abs/hep-ph/0508237}{{\ttfamily
  arXiv:hep-ph/0508237}}.

\bibitem[Soni et~al.(2010)Soni, Alok, Giri, Mohanta, and Nandi]{Soni:2008bc}
A.~Soni, A.~K. Alok, A.~Giri, R.~Mohanta, and S.~Nandi, ``{The fourth family: a
  simple explanation for the observed pattern of anomalies in
  \textit{B}-\textit{CP} asymmetries}'', {\em Phys.Lett.} {\bfseries B683}
  (2010) 302--305,
 \href{http://xxx.lanl.gov/abs/0807.1971}{{\ttfamily arXiv:0807.1971}}.

\bibitem[Raby and West(1988)]{Raby:1987ww}
S.~Raby and G.~B. West, ``{A FOURTH GENERATION NEUTRINO WITH A STANDARD HIGGS
  SCALAR SOLVES BOTH THE SOLAR NEUTRINO AND DARK MATTER PROBLEMS}'', {\em
  Phys.Lett.} {\bfseries B202} (1988)
47.

\bibitem[Lee et~al.(2011)Lee, Liu, and Soni]{Lee:2011jk}
H.-S. Lee, Z.~Liu, and A.~Soni, ``{Neutrino dark matter candidate in fourth
  generation scenarios}'', {\em Phys.Lett.} {\bfseries B704} (2011) 30--35,
 \href{http://xxx.lanl.gov/abs/1105.3490}{{\ttfamily arXiv:1105.3490}}.

\bibitem[{INSPIRE HEP Search}(9/5/2013)]{inspirehep:2013}
{INSPIRE HEP Search}. \url{http://www.inspirehep.net}, 9/5/2013.

\bibitem[Eberhardt et~al.(2012)Eberhardt, Herbert, Lacker, Lenz, Menzel,
  Nierste, and Wiebusch]{Eberhardt:2012gv}
O.~Eberhardt, G.~Herbert, H.~Lacker, A.~Lenz, A.~Menzel, U.~Nierste, and
  M.~Wiebusch, ``{Impact of a Higgs boson at a mass of 126 GeV on the standard
  model with three and four fermion generations}'', {\em Phys.Rev.Lett.}
  {\bfseries 109} (2012) 241802,
 \href{http://xxx.lanl.gov/abs/1209.1101}{{\ttfamily arXiv:1209.1101}}.

\bibitem[Beringer et~al.(2012)]{Beringer:1900zz}
{\bfseries Particle Data Group}, J.~Beringer {\em et~al.}, ``{Review of
  Particle Physics}'', {\em Phys.Rev.} {\bfseries D86} (2012)
010001.

\bibitem[H{\"o}cker et~al.(2001)H{\"o}cker, Lacker, Laplace, and
  Le~Diberder]{Hocker:2001xe}
A.~H{\"o}cker, H.~Lacker, S.~Laplace, and F.~Le~Diberder, ``{A NEW APPROACH TO
  A GLOBAL FIT OF THE CKM MATRIX}'', {\em Eur.Phys.J.} {\bfseries C21} (2001)
  225--259,
 \href{http://xxx.lanl.gov/abs/hep-ph/0104062}{{\ttfamily
  arXiv:hep-ph/0104062}}.

\bibitem[Wiebusch(????)]{Wiebusch:2012en}
M.~Wiebusch, ``{Numerical Computation of $p$-values with \textit{my}Fitter}'',
  \href{http://xxx.lanl.gov/abs/1207.1446}{{\ttfamily arXiv:1207.1446}},
{\text{accepted by} \textit{Comput.Phys.Commun.}}

\bibitem[Lyons(2013)]{Lyons:2013aa}
L.~Lyons, ``{Bayes and Frequentism: a particle physicist's perspective}'', {\em
  {Contemporary Physics}} {\bfseries 54} (2013), no.~{1}, {1--16},
  \href{http://xxx.lanl.gov/abs/1301.1273}{{\ttfamily arXiv:1301.1273}}.

\bibitem[Wilks(1938)]{Wilks:1938aa}
S.~S. Wilks, ``{THE LARGE-SAMPLE DISTRIBUTION OF THE LIKELIHOOD RATIO FOR
  TESTING COMPOSITE HYPOTHESES}'', {\em Ann. Math. Statist.} {\bfseries 9-1}
  (1938) 060062.

\bibitem[B{\"o}hm et~al.(2001)B{\"o}hm, Denner, and Joos]{Bohm:2001yx}
M.~B{\"o}hm, A.~Denner, and H.~Joos, ``{Gauge theories of the strong and
  electroweak interaction}'', {Vieweg+Teubner Verlag}, {Stuttgart, Leipzig,
  Wiesbaden}, {3rd rev.}~ed.,
2001.
\newblock

\bibitem[Novaes(2000)]{Novaes:1999yn}
S.~Novaes, ``{Standard Model: An Introduction'', published in: ``Particles and
  fields. Proceedings, 10th Jorge Andr\'{e} Swieca Summer School, Sao Paulo,
  Brazil, February 6-12, 1999}'',
 \href{http://xxx.lanl.gov/abs/hep-ph/0001283}{{\ttfamily
  arXiv:hep-ph/0001283}}.

\bibitem[Cabibbo(1963)]{Cabibbo:1963yz}
N.~Cabibbo, ``{UNITARY SYMMETRY AND LEPTONIC DECAYS}'', {\em Phys.Rev.Lett.}
  {\bfseries 10} (1963)
531--533.

\bibitem[Kobayashi and Maskawa(1973)]{Kobayashi:1973fv}
M.~Kobayashi and T.~Maskawa, ``{\textit{CP}-Violation in the Renormalizable
  Theory of Weak Interaction}'', {\em Prog.Theor.Phys.} {\bfseries 49} (1973)
652--657.

\bibitem[Wu et~al.(1957)Wu, Ambler, Hayward, Hoppes, and Hudson]{Wu:1957my}
C.~Wu, E.~Ambler, R.~Hayward, D.~Hoppes, and R.~Hudson, ``{Experimental Test of
  Parity Conservation in Beta Decay}'', {\em Phys.Rev.} {\bfseries 105} (1957)
1413--1414.

\bibitem[Christenson et~al.(1964)Christenson, Cronin, Fitch, and
  Turlay]{Christenson:1964fg}
J.~Christenson, J.~Cronin, V.~Fitch, and R.~Turlay, ``{EVIDENCE FOR THE 2$\pi$
  DECAY OF THE $K_2^0$ MESON}'', {\em Phys.Rev.Lett.} {\bfseries 13} (1964)
138--140.

\bibitem[Schael et~al.(2006)]{ALEPH:2005ab}
{\bfseries ALEPH Collaboration, DELPHI Collaboration, L3 Collaboration, OPAL
  Collaboration, SLD Collaboration, LEP Electroweak Working Group, SLD
  Electroweak Group, SLD Heavy Flavour Group}, S.~Schael {\em et~al.},
  ``{Precision Electroweak Measurements on the $Z$ Resonance}'', {\em
  Phys.Rept.} {\bfseries 427} (2006) 257--454,
 \href{http://xxx.lanl.gov/abs/hep-ex/0509008}{{\ttfamily
  arXiv:hep-ex/0509008}}.

\bibitem[{Interactive Tevatron timeline}(22/6/2013)]{Tevatron:2013tl}
{Interactive Tevatron timeline}.
  \url{http://www.fnal.gov/pub/tevatron/milestones/interactive-timeline.html},
  22/6/2013.

\bibitem[Fl{\"a}cher et~al.(2009)Fl{\"a}cher, Goebel, Haller, H{\"o}cker,
  M{\"o}nig, and Stelzer]{Flacher:2008zq}
{\bfseries The Gfitter Group}, H.~Fl{\"a}cher, M.~Goebel, J.~Haller,
  A.~H{\"o}cker, K.~M{\"o}nig, and J.~Stelzer, ``{Revisiting the Global
  Electroweak Fit of the Standard Model and Beyond with Gfitter}'', {\em
  Eur.Phys.J.} {\bfseries C60} (2009) 543--583,
 \href{http://xxx.lanl.gov/abs/0811.0009}{{\ttfamily arXiv:0811.0009}}.

\bibitem[Davier et~al.(2011)Davier, H{\"o}cker, Malaescu, and
  Zhang]{Davier:2010nc}
M.~Davier, A.~H{\"o}cker, B.~Malaescu, and Z.~Zhang, ``{Reevaluation of the
  Hadronic Contributions to the Muon $g-2$ and to $\alpha(M_Z^2)$}'', {\em
  Eur.Phys.J.} {\bfseries C71} (2011) 1515,
 \href{http://xxx.lanl.gov/abs/1010.4180}{{\ttfamily arXiv:1010.4180}}.

\bibitem[Baikov et~al.(2008)Baikov, Chetyrkin, and Kuhn]{Baikov:2008jh}
P.~Baikov, K.~Chetyrkin, and J.~H. Kuhn, ``{Hadronic $Z$- and $\tau$-Decays in
  Order $\alpha_s^4$}'', {\em Phys.Rev.Lett.} {\bfseries 101} (2008) 012002,
 \href{http://xxx.lanl.gov/abs/0801.1821}{{\ttfamily arXiv:0801.1821}}.

\bibitem[Bardin et~al.(1990)Bardin, Bilenky, Riemann, Sachwitz, and
  Vogt]{Bardin:1989tq}
D.~Y. Bardin, M.~S. Bilenky, T.~Riemann, M.~Sachwitz, and H.~Vogt, ``{DIZET --
  ELECTROWEAK ONE-LOOP CORRECTIONS FOR $e^+ +e^-\to f^+ +f^-$ AROUND THE Z$^0$
  PEAK}'', {\em Comput.Phys.Commun.} {\bfseries 59} (1990)
303--312.

\bibitem[Bardin et~al.(2001)Bardin, Christova, Jack, Kalinovskaya, Olchevski,
  et~al.]{Bardin:1999yd}
D.~Y. Bardin, P.~Christova, M.~Jack, L.~Kalinovskaya, A.~Olchevski, {\em
  et~al.}, ``{\texttt{ZFITTER} v.6.21. A Semi-Analytical Program for Fermion
  Pair Production in $e^+e^-$ Annihilation}'', {\em Comput.Phys.Commun.}
  {\bfseries 133} (2001) 229--395,
 \href{http://xxx.lanl.gov/abs/hep-ph/9908433}{{\ttfamily
  arXiv:hep-ph/9908433}}.

\bibitem[Arbuzov et~al.(2006)Arbuzov, Awramik, Czakon, Freitas, Grunewald,
  et~al.]{Arbuzov:2005ma}
A.~Arbuzov, M.~Awramik, M.~Czakon, A.~Freitas, M.~Grunewald, {\em et~al.},
  ``{\texttt{ZFITTER}: a semi-analytical program for fermion pair production in
  $e^+e^-$ annihilation, from version 6.21 to version 6.42}'', {\em
  Comput.Phys.Commun.} {\bfseries 174} (2006) 728--758,
 \href{http://xxx.lanl.gov/abs/hep-ph/0507146}{{\ttfamily
  arXiv:hep-ph/0507146}}.

\bibitem[Freitas and Huang(2012)]{Freitas:2012sy}
A.~Freitas and Y.-C. Huang, ``{Electroweak two-loop corrections to
  $sin^2{\theta}_{\text{eff}}^{b\bar{b}}$ and $R_b$ using numerical
  Mellin-Barnes integrals}'', {\em JHEP} {\bfseries 1208} (2012) 050,
 \href{http://xxx.lanl.gov/abs/1205.0299}{{\ttfamily arXiv:1205.0299}}.

\bibitem[Peskin and Takeuchi(1990)]{Peskin:1990zt}
M.~E. Peskin and T.~Takeuchi, ``{New Constraint on a Strongly Interacting Higgs
  Sector}'', {\em Phys.Rev.Lett.} {\bfseries 65} (1990)
964--967.

\bibitem[Peskin and Takeuchi(1992)]{Peskin:1991sw}
M.~E. Peskin and T.~Takeuchi, ``{Estimation of oblique electroweak
  corrections}'', {\em Phys.Rev.} {\bfseries D46} (1992)
381--409.

\bibitem[Veltman(1977)]{Veltman:1976rt}
M.~Veltman, ``{Second Threshold in Weak Interactions}'', {\em Acta Phys.Polon.}
  {\bfseries B8} (1977)
475.

\bibitem[Barate et~al.(2003)]{Barate:2003sz}
{\bfseries LEP Working Group for Higgs boson searches, ALEPH Collaboration,
  DELPHI Collaboration, L3 Collaboration, OPAL Collaboration}, R.~Barate {\em
  et~al.}, ``{Search for the Standard Model Higgs Boson at LEP}'', {\em
  Phys.Lett.} {\bfseries B565} (2003) 61--75,
 \href{http://xxx.lanl.gov/abs/hep-ex/0306033}{{\ttfamily
  arXiv:hep-ex/0306033}}.

\bibitem[Aaltonen et~al.(2010)]{Aaltonen:2010yv}
{\bfseries CDF Collaboration, D0 Collaboration}, T.~Aaltonen {\em et~al.},
  ``{Combination of Tevatron searches for the standard model Higgs boson in the
  $W+W-$ decay mode}'', {\em Phys.Rev.Lett.} {\bfseries 104} (2010) 061802,
 \href{http://xxx.lanl.gov/abs/1001.4162}{{\ttfamily arXiv:1001.4162}}.

\bibitem[CDF(2013)]{CDF:2013kxa}
{\bfseries D0 Collaborations}, CDF, ``{Higgs Boson Studies at the Tevatron}'',
 \href{http://xxx.lanl.gov/abs/1303.6346}{{\ttfamily arXiv:1303.6346}}.

\bibitem[ATL(23/6/2013{\natexlab{a}})]{ATLAS-CONF-2013-034}
{\bfseries ATLAS Collaboration}, ``{ATLAS Note CONF-2013-034}''.
  \url{http://cds.cern.ch/record/1528170}, 23/6/2013{\natexlab{a}}.

\bibitem[ATL(23/6/2013{\natexlab{b}})]{ATLAS-CONF-2013-040}
{\bfseries ATLAS Collaboration}, ``{ATLAS Note CONF-2013-040}''.
  \url{http://cds.cern.ch/record/1542341}, 23/6/2013{\natexlab{b}}.

\bibitem[CMS(23/6/2013)]{CMS-PAS-HIG-13-005}
{\bfseries CMS Collaboration}, ``{CMS Physics Analysis Summary HIG-13-005}''.
  \url{http://cds.cern.ch/record/1542387}, 23/6/2013.

\bibitem[LHC(22/6/2013)]{LHCXSWG:2013aa}
{\bfseries {LHC Higgs Cross Section Working Group}}.
  \url{https://twiki.cern.ch/twiki/bin/view/LHCPhysics/CrossSections},
  22/6/2013.

\bibitem[Baglio and Djouadi(2010)]{Baglio:2010um}
J.~Baglio and A.~Djouadi, ``{Predictions for Higgs production at the Tevatron
  and the associated uncertainties}'', {\em JHEP} {\bfseries 1010} (2010) 064,
 \href{http://xxx.lanl.gov/abs/1003.4266}{{\ttfamily arXiv:1003.4266}}.

\bibitem[Dittmaier et~al.(2012)Dittmaier, Mariotti, Passarino, Tanaka,
  et~al.]{Dittmaier:2012vm}
{\bfseries {LHC Higgs Cross Section Working Group}}, S.~Dittmaier, C.~Mariotti,
  G.~Passarino, R.~Tanaka, {\em et~al.}, ``{Handbook of LHC Higgs cross
  sections: 2. Differential Distributions}'',
 \href{http://xxx.lanl.gov/abs/1201.3084}{{\ttfamily arXiv:1201.3084}}.

\bibitem[Azatov et~al.(2012)Azatov, Contino, Del~Re, Galloway, Grassi,
  et~al.]{Azatov:2012rd}
A.~Azatov, R.~Contino, D.~Del~Re, J.~Galloway, M.~Grassi, {\em et~al.},
  ``{Determining Higgs couplings with a model-independent analysis of $h \to
  \gamma \gamma$}'', {\em JHEP} {\bfseries 1206} (2012) 134,
 \href{http://xxx.lanl.gov/abs/1204.4817}{{\ttfamily arXiv:1204.4817}}.

\bibitem[Klute et~al.(2012)Klute, Lafaye, Plehn, Rauch, and
  Zerwas]{Klute:2012pu}
M.~Klute, R.~Lafaye, T.~Plehn, M.~Rauch, and D.~Zerwas, ``{Measuring Higgs
  Couplings from LHC Data}'', {\em Phys.Rev.Lett.} {\bfseries 109} (2012)
  101801,
 \href{http://xxx.lanl.gov/abs/1205.2699}{{\ttfamily arXiv:1205.2699}}.

\bibitem[Espinosa et~al.(2012)Espinosa, Grojean, M{\"u}hlleitner, and
  Trott]{Espinosa:2012im}
J.~Espinosa, C.~Grojean, M.~M{\"u}hlleitner, and M.~Trott, ``{First Glimpses at
  Higgs' face}'', {\em JHEP} {\bfseries 1212} (2012) 045,
 \href{http://xxx.lanl.gov/abs/1207.1717}{{\ttfamily arXiv:1207.1717}}.

\bibitem[Carmi et~al.(2012)Carmi, Falkowski, Kuflik, Volansky, and
  Zupan]{Carmi:2012in}
D.~Carmi, A.~Falkowski, E.~Kuflik, T.~Volansky, and J.~Zupan, ``{Higgs After
  the Discovery: A Status Report}'', {\em JHEP} {\bfseries 1210} (2012) 196,
 \href{http://xxx.lanl.gov/abs/1207.1718}{{\ttfamily arXiv:1207.1718}}.

\bibitem[Baak et~al.(2012)Baak, Goebel, Haller, H{\"o}cker, Kennedy,
  et~al.]{Baak:2012kk}
M.~Baak, M.~Goebel, J.~Haller, A.~H{\"o}cker, D.~Kennedy, {\em et~al.}, ``{The
  Electroweak Fit of the Standard Model after the Discovery of a New Boson at
  the LHC}'', {\em Eur.Phys.J.} {\bfseries C72} (2012) 2205,
 \href{http://xxx.lanl.gov/abs/1209.2716}{{\ttfamily arXiv:1209.2716}}.

\bibitem[Ciuchini et~al.(2013)Ciuchini, Franco, Mishima, and
  Silvestrini]{Ciuchini:2013pca}
M.~Ciuchini, E.~Franco, S.~Mishima, and L.~Silvestrini, ``{Electroweak
  Precision Observables, New Physics and the Nature of a 126 GeV Higgs
  Boson}'',
 \href{http://xxx.lanl.gov/abs/1306.4644}{{\ttfamily arXiv:1306.4644}}.

\bibitem[Chanowitz et~al.(1978)Chanowitz, Furman, and
  Hinchliffe]{Chanowitz:1978uj}
M.~S. Chanowitz, M.~Furman, and I.~Hinchliffe, ``{WEAK INTERACTIONS OF ULTRA
  HEAVY FERMIONS}'', {\em Phys.Lett.} {\bfseries B78} (1978)
285.

\bibitem[Chanowitz et~al.(1979)Chanowitz, Furman, and
  Hinchliffe]{Chanowitz:1978mv}
M.~S. Chanowitz, M.~Furman, and I.~Hinchliffe, ``{WEAK INTERACTIONS OF ULTRA
  HEAVY FERMIONS (II)}'', {\em Nucl.Phys.} {\bfseries B153} (1979)
402.

\bibitem[Denner et~al.(2012)Denner, Dittmaier, Muck, Passarino, Spira,
  et~al.]{Denner:2011vt}
A.~Denner, S.~Dittmaier, A.~Muck, G.~Passarino, M.~Spira, {\em et~al.},
  ``{Higgs production and decay with a fourth Standard-Model-like fermion
  generation}'', {\em Eur.Phys.J.} {\bfseries C72} (2012) 1992,
 \href{http://xxx.lanl.gov/abs/1111.6395}{{\ttfamily arXiv:1111.6395}}.

\bibitem[Sakharov(1967)]{Sakharov:1967dj}
A.~Sakharov, ``{Violation of \textit{CP} invariance, \textit{C} asymmetry, and
  baryon asymmetry of the universe}'', {\em Pisma Zh.Eksp.Teor.Fiz.} {\bfseries
  5} (1967)
32--35.

\bibitem[Buras et~al.(2010)Buras, Duling, Feldmann, Heidsieck, Promberger,
  et~al.]{Buras:2010pi}
A.~J. Buras, B.~Duling, T.~Feldmann, T.~Heidsieck, C.~Promberger, {\em et~al.},
  ``{Patterns of flavour violation in the presence of a fourth generation of
  quarks and leptons}'', {\em JHEP} {\bfseries 1009} (2010) 106,
 \href{http://xxx.lanl.gov/abs/1002.2126}{{\ttfamily arXiv:1002.2126}}.

\bibitem[Aad et~al.(2013)]{ATLAS:2012qe}
{\bfseries ATLAS Collaboration}, G.~Aad {\em et~al.}, ``{Search for pair
  production of heavy top-like quarks decaying to a high-$p_T$ $W$ boson and a
  $b$ quark in the lepton plus jets final state at $\sqrt{s}=7$ TeV with the
  ATLAS detector}'', {\em Phys.Lett.} {\bfseries B718} (2013) 1284--1302,
 \href{http://xxx.lanl.gov/abs/1210.5468}{{\ttfamily arXiv:1210.5468}}.

\bibitem[Chatrchyan et~al.(2013)]{Chatrchyan:2012af}
{\bfseries CMS Collaboration}, S.~Chatrchyan {\em et~al.}, ``{Search for heavy
  quarks decaying into a top quark and a W or Z boson using lepton + jets
  events in pp collisions at $\sqrt{s}=7$ TeV}'', {\em JHEP} {\bfseries 01}
  (2013) 154,
 \href{http://xxx.lanl.gov/abs/1210.7471}{{\ttfamily arXiv:1210.7471}}.

\bibitem[Chatrchyan et~al.(2012)]{Chatrchyan:2012fp}
{\bfseries CMS Collaboration}, S.~Chatrchyan {\em et~al.}, ``{Combined search
  for the quarks of a sequential fourth generation}'', {\em Phys.Rev.}
  {\bfseries D86} (2012) 112003,
 \href{http://xxx.lanl.gov/abs/1209.1062}{{\ttfamily arXiv:1209.1062}}.

\bibitem[Achard et~al.(2001)]{Achard:2001qw}
{\bfseries L3 Collaboration}, P.~Achard {\em et~al.}, ``{Search for Heavy
  Neutral and Charged Leptons in $e^{+} e^{-}$ Annihilation at LEP}'', {\em
  Phys.Lett.} {\bfseries B517} (2001) 75--85,
 \href{http://xxx.lanl.gov/abs/hep-ex/0107015}{{\ttfamily
  arXiv:hep-ex/0107015}}.

\bibitem[Abreu et~al.(1992)]{Abreu:1991pr}
{\bfseries DELPHI Collaboration}, P.~Abreu {\em et~al.}, ``{Searches for heavy
  neutrinos from $Z$ decays}'', {\em Phys.Lett.} {\bfseries B274} (1992)
230--238.

\bibitem[Bobrowski et~al.(2009)Bobrowski, Lenz, Riedl, and
  Rohrwild]{Bobrowski:2009ng}
M.~Bobrowski, A.~Lenz, J.~Riedl, and J.~Rohrwild, ``{How much space is left for
  a new family of fermions?}'', {\em Phys.Rev.} {\bfseries D79} (2009) 113006,
 \href{http://xxx.lanl.gov/abs/0902.4883}{{\ttfamily arXiv:0902.4883}}.

\bibitem[Pontecorvo(1957)]{Pontecorvo:1957cp}
B.~Pontecorvo, ``{Mesonium and anti-mesonium}'', {\em Sov.Phys.JETP} {\bfseries
  6} (1957)
429.

\bibitem[Maki et~al.(1962)Maki, Nakagawa, and Sakata]{Maki:1962mu}
Z.~Maki, M.~Nakagawa, and S.~Sakata, ``{Remarks on the Unified Model of
  Elementary Particles}'', {\em Prog.Theor.Phys.} {\bfseries 28} (1962)
870--880.

\bibitem[Pontecorvo(1968)]{Pontecorvo:1967fh}
B.~Pontecorvo, ``{Neutrino Experiments and the Problem of Conservation of
  Leptonic Charge}'', {\em Sov.Phys.JETP} {\bfseries 26} (1968)
984--988.

\bibitem[Gerhold et~al.(2011)Gerhold, Jansen, and Kallarackal]{Gerhold:2010wv}
P.~Gerhold, K.~Jansen, and J.~Kallarackal, ``{Higgs boson mass bounds in the
  presence of a very heavy fourth quark generation}'', {\em JHEP} {\bfseries
  1101} (2011) 143,
 \href{http://xxx.lanl.gov/abs/1011.1648}{{\ttfamily arXiv:1011.1648}}.

\bibitem[Bulava et~al.(2013)Bulava, Jansen, and Nagy]{Bulava:2013ep}
J.~Bulava, K.~Jansen, and A.~Nagy, ``{Constraining a fourth generation of
  quarks: non-perturbative Higgs boson mass bounds}'', {\em Phys.Lett.}
  {\bfseries B723} (2013) 95--99,
 \href{http://xxx.lanl.gov/abs/1301.3416}{{\ttfamily arXiv:1301.3416}}.

\bibitem[He et~al.(2001)He, Polonsky, and Su]{He:2001tp}
H.-J. He, N.~Polonsky, and S.~Su, ``{Extra Families, Higgs Spectrum and Oblique
  Corrections}'', {\em Phys.Rev.} {\bfseries D64} (2001) 053004,
 \href{http://xxx.lanl.gov/abs/hep-ph/0102144}{{\ttfamily
  arXiv:hep-ph/0102144}}.

\bibitem[Kribs et~al.(2007)Kribs, Plehn, Spannowsky, and Tait]{Kribs:2007nz}
G.~D. Kribs, T.~Plehn, M.~Spannowsky, and T.~M. Tait, ``{Four Generations and
  Higgs Physics}'', {\em Phys.Rev.} {\bfseries D76} (2007) 075016,
  \href{http://xxx.lanl.gov/abs/0706.3718}{{\ttfamily arXiv:0706.3718}}.

\bibitem[Eberhardt et~al.(2010)Eberhardt, Lenz, and Rohrwild]{Eberhardt:2010bm}
O.~Eberhardt, A.~Lenz, and J.~Rohrwild, ``{Less space for a new family of
  fermions}'', {\em Phys.Rev.} {\bfseries D82} (2010) 095006,
 \href{http://xxx.lanl.gov/abs/1005.3505}{{\ttfamily arXiv:1005.3505}}.

\bibitem[Amsler et~al.(2008)]{Amsler:2008zzb}
{\bfseries Particle Data Group}, C.~Amsler {\em et~al.}, ``{Review of Particle
  Physics}'', {\em Phys.Lett.} {\bfseries B667} (2008)
1--1340.

\bibitem[Gonz{\'a}lez et~al.(2012)Gonz{\'a}lez, Rohrwild, and
  Wiebusch]{Gonzalez:2011he}
P.~Gonz{\'a}lez, J.~Rohrwild, and M.~Wiebusch, ``{Electroweak Precision
  Observables in a Fourth Generation Model with General Flavour Structure}'',
  {\em Eur.Phys.J.} {\bfseries C72} (2012) 2007,
 \href{http://xxx.lanl.gov/abs/1105.3434}{{\ttfamily arXiv:1105.3434}}.

\bibitem[Hahn and P{\'e}rez-Victoria(1999)]{Hahn:1998yk}
T.~Hahn and M.~P{\'e}rez-Victoria, ``{Automatized One-Loop Calculations in four
  and \textit{D} dimensions}'', {\em Comput.Phys.Commun.} {\bfseries 118}
  (1999) 153--165,
 \href{http://xxx.lanl.gov/abs/hep-ph/9807565}{{\ttfamily
  arXiv:hep-ph/9807565}}.

\bibitem[Hahn(2001)]{Hahn:2000kx}
T.~Hahn, ``{Generating Feynman Diagrams and Amplitudes with \textit{FeynArts}
  3}'', {\em Comput.Phys.Commun.} {\bfseries 140} (2001) 418--431,
 \href{http://xxx.lanl.gov/abs/hep-ph/0012260}{{\ttfamily
  arXiv:hep-ph/0012260}}.

\bibitem[Hahn and Rauch(2006)]{Hahn:2006qw}
T.~Hahn and M.~Rauch, ``{News from FormCalc and LoopTools}'', {\em
  Nucl.Phys.Proc.Suppl.} {\bfseries 157} (2006) 236--240,
 \href{http://xxx.lanl.gov/abs/hep-ph/0601248}{{\ttfamily
  arXiv:hep-ph/0601248}}.

\bibitem[Khoze(2001)]{Khoze:2001ug}
V.~A. Khoze, ``{Comment on an invisible Higgs boson and 50 GeV neutrino}'',
 \href{http://xxx.lanl.gov/abs/hep-ph/0105069}{{\ttfamily
  arXiv:hep-ph/0105069}}.

\bibitem[Djouadi et~al.(1998)Djouadi, Kalinowski, and Spira]{Djouadi:1997yw}
A.~Djouadi, J.~Kalinowski, and M.~Spira, ``{HDECAY: a Program for Higgs Boson
  Decays in the Standard Model and its Supersymmetric Extension}'', {\em
  Comput.Phys.Commun.} {\bfseries 108} (1998) 56--74,
 \href{http://xxx.lanl.gov/abs/hep-ph/9704448}{{\ttfamily
  arXiv:hep-ph/9704448}}.

\bibitem[Djouadi and Gambino(1995)]{Djouadi:1994gf}
A.~Djouadi and P.~Gambino, ``{QCD corrections to Higgs boson self-energies and
  fermionic decay widths}'', {\em Phys.Rev.} {\bfseries D51} (1995) 218--228,
 \href{http://xxx.lanl.gov/abs/hep-ph/9406431}{{\ttfamily
  arXiv:hep-ph/9406431}}.

\bibitem[Djouadi and Gambino(1994)]{Djouadi:1994ge}
A.~Djouadi and P.~Gambino, ``{Leading electroweak correction to Higgs boson
  production at proton colliders.}'', {\em Phys.Rev.Lett.} {\bfseries 73}
  (1994) 2528--2531,
 \href{http://xxx.lanl.gov/abs/hep-ph/9406432}{{\ttfamily
  arXiv:hep-ph/9406432}}.

\bibitem[Passarino et~al.(2011)Passarino, Sturm, and
  Uccirati]{Passarino:2011kv}
G.~Passarino, C.~Sturm, and S.~Uccirati, ``{Complete Electroweak Corrections to
  Higgs production in a Standard Model with four generations at the LHC}'',
  {\em Phys.Lett.} {\bfseries B706} (2011) 195--199,
 \href{http://xxx.lanl.gov/abs/1108.2025}{{\ttfamily arXiv:1108.2025}}.

\bibitem[Gunion et~al.(1996)Gunion, McKay, and Pois]{Gunion:1995tp}
J.~Gunion, D.~W. McKay, and H.~Pois, ``{A MINIMAL FOUR-FAMILY SUPERGRAVITY
  MODEL}'', {\em Phys.Rev.} {\bfseries D53} (1996) 1616--1647,
 \href{http://xxx.lanl.gov/abs/hep-ph/9507323}{{\ttfamily
  arXiv:hep-ph/9507323}}.

\bibitem[Gunion and Geer(1993)]{Gunion:1993bj}
J.~Gunion and S.~Geer, ``{PROGRESS IN SSC HIGGS PHYSICS: REPORT OF THE HIGGS
  WORKING GROUP'', published in: ``Proceedings of the `Workshop on Physics at
  Current Accelerators and the Supercollider', eds.\ J.~Hewett, A.~White, and
  D.~Zeppenfeld, Argonne National Laboratory, 2-5 June (1993)}'',
 \href{http://xxx.lanl.gov/abs/hep-ph/9310333}{{\ttfamily
  arXiv:hep-ph/9310333}}.

\bibitem[Caso et~al.(1998)]{Caso:1998tx}
{\bfseries Particle Data Group}, C.~Caso {\em et~al.}, ``{Review of particle
  physics}'', {\em Eur.Phys.J.} {\bfseries C3} (1998)
1--794.

\bibitem[Frampton et~al.(2000)Frampton, Hung, and Sher]{Frampton:1999xi}
P.~H. Frampton, P.~Hung, and M.~Sher, ``{Quarks and Leptons Beyond the Third
  Generation.}'', {\em Phys.Rept.} {\bfseries 330} (2000) 263,
 \href{http://xxx.lanl.gov/abs/hep-ph/9903387}{{\ttfamily
  arXiv:hep-ph/9903387}}.

\bibitem[Novikov et~al.(2002{\natexlab{a}})Novikov, Okun, Rozanov, and
  Vysotsky]{Novikov:2001md}
V.~Novikov, L.~Okun, A.~N. Rozanov, and M.~Vysotsky, ``{Extra generations and
  discrepancies of electroweak precision data}'', {\em Phys.Lett.} {\bfseries
  B529} (2002){\natexlab{a}} 111--116,
 \href{http://xxx.lanl.gov/abs/hep-ph/0111028}{{\ttfamily
  arXiv:hep-ph/0111028}}.

\bibitem[Novikov et~al.(2002{\natexlab{b}})Novikov, Okun, Rozanov, and
  Vysotsky]{Novikov:2002tk}
V.~Novikov, L.~Okun, A.~N. Rozanov, and M.~Vysotsky, ``{Mass of the higgs
  versus fourth generation masses}'', {\em JETP Lett.} {\bfseries 76}
  (2002){\natexlab{b}} 127--130,
 \href{http://xxx.lanl.gov/abs/hep-ph/0203132}{{\ttfamily
  arXiv:hep-ph/0203132}}.

\bibitem[Chanowitz(2009)]{Chanowitz:2009mz}
M.~S. Chanowitz, ``{Bounding CKM Mixing with a Fourth Family}'', {\em
  Phys.Rev.} {\bfseries D79} (2009) 113008,
 \href{http://xxx.lanl.gov/abs/0904.3570}{{\ttfamily arXiv:0904.3570}}.

\bibitem[Erler and Langacker(2010)]{Erler:2010sk}
J.~Erler and P.~Langacker, ``{Precision Constraints on Extra Fermion
  Generations}'', {\em Phys.Rev.Lett.} {\bfseries 105} (2010) 031801,
 \href{http://xxx.lanl.gov/abs/1003.3211}{{\ttfamily arXiv:1003.3211}}.

\bibitem[Djouadi and Lenz(2012)]{Djouadi:2012ae}
A.~Djouadi and A.~Lenz, ``{Sealing the fate of a fourth generation of
  fermions}'', {\em Phys.Lett.} {\bfseries B715} (2012) 310--314,
 \href{http://xxx.lanl.gov/abs/1204.1252}{{\ttfamily arXiv:1204.1252}}.

\bibitem[Kuflik et~al.(2012)Kuflik, Nir, and Volansky]{Kuflik:2012ai}
E.~Kuflik, Y.~Nir, and T.~Volansky, ``{Implications of Higgs Searches on the
  Four Generation Standard Model}'',
 \href{http://xxx.lanl.gov/abs/1204.1975}{{\ttfamily arXiv:1204.1975}}.

\bibitem[Buchkremer et~al.(2012)Buchkremer, G\'{e}rard, and
  Maltoni]{Buchkremer:2012yy}
M.~Buchkremer, J.-M. G\'{e}rard, and F.~Maltoni, ``{Closing in on a
  perturbative fourth generation}'', {\em JHEP} {\bfseries 1206} (2012) 135,
 \href{http://xxx.lanl.gov/abs/1204.5403}{{\ttfamily arXiv:1204.5403}}.

\bibitem[Eberhardt et~al.(2012{\natexlab{a}})Eberhardt, Lenz, Menzel, Nierste,
  and Wiebusch]{Eberhardt:2012ck}
O.~Eberhardt, A.~Lenz, A.~Menzel, U.~Nierste, and M.~Wiebusch, ``{Status of the
  fourth fermion generation before ICHEP2012: Higgs data and electroweak
  precision observables}'', {\em Phys.Rev.} {\bfseries D86}
  (2012){\natexlab{a}} 074014,
 \href{http://xxx.lanl.gov/abs/1207.0438}{{\ttfamily arXiv:1207.0438}}.

\bibitem[Eberhardt et~al.(2012{\natexlab{b}})Eberhardt, Herbert, Lacker, Lenz,
  Menzel, Nierste, and Wiebusch]{Eberhardt:2012sb}
O.~Eberhardt, G.~Herbert, H.~Lacker, A.~Lenz, A.~Menzel, U.~Nierste, and
  M.~Wiebusch, ``{Joint analysis of Higgs decays and electroweak precision
  observables in the Standard Model with a sequential fourth generation}'',
  {\em Phys.Rev.} {\bfseries D86} (2012){\natexlab{b}} 013011,
 \href{http://xxx.lanl.gov/abs/1204.3872}{{\ttfamily arXiv:1204.3872}}.

\bibitem[Lee(1973)]{Lee:1973iz}
T.~Lee, ``{A Theory of Spontaneous \textit{T} Violation}'', {\em Phys.Rev.}
  {\bfseries D8} (1973)
1226--1239.

\bibitem[Gunion and Haber(2003)]{Gunion:2002zf}
J.~F. Gunion and H.~E. Haber, ``{The CP-conserving two-Higgs-doublet model: the
  approach to the decoupling limit}'', {\em Phys.Rev.} {\bfseries D67} (2003)
  075019,
 \href{http://xxx.lanl.gov/abs/hep-ph/0207010}{{\ttfamily
  arXiv:hep-ph/0207010}}.

\bibitem[Branco et~al.(2012)Branco, Ferreira, Lavoura, Rebelo, Sher,
  et~al.]{Branco:2011iw}
G.~Branco, P.~Ferreira, L.~Lavoura, M.~Rebelo, M.~Sher, {\em et~al.}, ``{Theory
  and phenomenology of two-Higgs-doublet models}'', {\em Phys.Rept.} {\bfseries
  516} (2012) 1--102,
 \href{http://xxx.lanl.gov/abs/1106.0034}{{\ttfamily arXiv:1106.0034}}.

\bibitem[Glashow and Weinberg(1977)]{Glashow:1976nt}
S.~L. Glashow and S.~Weinberg, ``{Natural conservation laws for neutral
  currents}'', {\em Phys.Rev.} {\bfseries D15} (1977)
1958.

\bibitem[Donoghue and Li(1979)]{Donoghue:1978cj}
J.~F. Donoghue and L.~F. Li, ``{Properties of charged Higgs bosons}'', {\em
  Phys.Rev.} {\bfseries D19} (1979)
945.

\bibitem[Djouadi(2008)]{Djouadi:2005gj}
A.~Djouadi, ``{The anatomy of electroweak symmetry breaking Tome II: The Higgs
  bosons in the Minimal Supersymmetric Model}'', {\em Phys.Rept.} {\bfseries
  459} (2008) 1--241,
 \href{http://xxx.lanl.gov/abs/hep-ph/0503173}{{\ttfamily
  arXiv:hep-ph/0503173}}.

\bibitem[Barroso et~al.(2013)Barroso, Ferreira, Ivanov, and
  Santos]{Barroso:2013awa}
A.~Barroso, P.~Ferreira, I.~Ivanov, and R.~Santos, ``{Metastability bounds on
  the two Higgs doublet model}'', {\em JHEP} {\bfseries 1306} (2013) 045,
 \href{http://xxx.lanl.gov/abs/1303.5098}{{\ttfamily arXiv:1303.5098}}.

\bibitem[Nierste and Riesselmann(1996)]{Nierste:1995zx}
U.~Nierste and K.~Riesselmann, ``{Higgs Sector Renormalization Group in the
  $\overline{\rm MS}$ and ${\rm OMS}$ Scheme: The Breakdown of Perturbation
  Theory for a Heavy Higgs}'', {\em Phys.Rev.} {\bfseries D53} (1996)
  6638--6652,
 \href{http://xxx.lanl.gov/abs/hep-ph/9511407}{{\ttfamily
  arXiv:hep-ph/9511407}}.

\bibitem[Heister et~al.(2002)]{Heister:2002ev}
{\bfseries ALEPH Collaboration}, A.~Heister {\em et~al.}, ``{Search for charged
  Higgs bosons in $e^{+} e^{-}$ collisions at energies up to $\sqrt{s}$ = 209
  GeV}'', {\em Phys.Lett.} {\bfseries B543} (2002) 1--13,
 \href{http://xxx.lanl.gov/abs/hep-ex/0207054}{{\ttfamily
  arXiv:hep-ex/0207054}}.

\bibitem[Hollik(1986)]{Hollik:1986gg}
W.~Hollik, ``{Non-Standard Higgs Bosons in \textit{SU}(2)$\times$\textit{U}(1)
  Radiative Corrections}'', {\em Z.Phys.} {\bfseries C32} (1986)
291.

\bibitem[Hollik(1988)]{Hollik:1987fg}
W.~Hollik, ``{Radiative corrections with two Higgs doublets at LEP/SLC and
  HERA}'', {\em Z.Phys.} {\bfseries C37} (1988)
569.

\bibitem[Haber and Logan(2000)]{Haber:1999zh}
H.~E. Haber and H.~E. Logan, ``{Radiative Corrections to the $Zb\oline{b}$
  Vertex and Constraints on Extended Higgs Sectors}'', {\em Phys.Rev.}
  {\bfseries D62} (2000) 015011,
 \href{http://xxx.lanl.gov/abs/hep-ph/9909335}{{\ttfamily
  arXiv:hep-ph/9909335}}.

\bibitem[Ferreira et~al.(2012)Ferreira, Santos, Sher, and
  Silva]{Ferreira:2012my}
P.~Ferreira, R.~Santos, M.~Sher, and J.~P. Silva, ``{Could the LHC two-photon
  signal correspond to the heavier scalar in two-Higgs-doublet models?}'', {\em
  Phys.Rev.} {\bfseries D85} (2012) 035020,
 \href{http://xxx.lanl.gov/abs/1201.0019}{{\ttfamily arXiv:1201.0019}}.

\bibitem[Ferreira et~al.(2013)Ferreira, Santos, Haber, and
  Silva]{Ferreira:2012nv}
P.~Ferreira, R.~Santos, H.~E. Haber, and J.~P. Silva, ``{Mass-degenerate Higgs
  bosons at 125 GeV in the Two-Higgs-Doublet Model}'', {\em Phys.Rev.}
  {\bfseries D87} (2013) 055009,
 \href{http://xxx.lanl.gov/abs/1211.3131}{{\ttfamily arXiv:1211.3131}}.

\bibitem[CMS(23/6/2013)]{CMS-PAS-HIG-13-002}
{\bfseries CMS Collaboration}, ``{CMS Physics Analysis Summary HIG-13-002}''.
  \url{http://cds.cern.ch/record/1523767}, 23/6/2013.

\bibitem[Eberhardt et~al.(2013)Eberhardt, Nierste, and
  Wiebusch]{Eberhardt:2013uba}
O.~Eberhardt, U.~Nierste, and M.~Wiebusch, ``{Status of the two-Higgs-doublet
  model of type II}'', {\em JHEP} {\bfseries 1307} (2013) 118,
 \href{http://xxx.lanl.gov/abs/1305.1649}{{\ttfamily arXiv:1305.1649}}.

\bibitem[Crivellin et~al.(2012)Crivellin, Greub, and Kokulu]{Crivellin:2012ye}
A.~Crivellin, C.~Greub, and A.~Kokulu, ``{Explaining $B\to D\tau\nu$ , $B\to
  D^*\tau\nu$ and $B\to \tau\nu$ in a two Higgs doublet model of type III}'',
  {\em Phys.Rev.} {\bfseries D86} (2012) 054014,
 \href{http://xxx.lanl.gov/abs/1206.2634}{{\ttfamily arXiv:1206.2634}}.

\bibitem[Abazov et~al.(2006)]{Abazov:2006dm}
{\bfseries D0 Collaboration}, V.~Abazov {\em et~al.}, ``{Direct Limits on the
  $B^0_{s}$ Oscillation Frequency}'', {\em Phys.Rev.Lett.} {\bfseries 97}
  (2006) 021802,
 \href{http://xxx.lanl.gov/abs/hep-ex/0603029}{{\ttfamily
  arXiv:hep-ex/0603029}}.

\bibitem[Aaij et~al.(2013)]{Aaij:2013mpa}
{\bfseries LHCb collaboration}, R.~Aaij {\em et~al.}, ``{Precision measurement
  of the $B^{0}_{s}$-$\oline{B}^{0}_{s}$ oscillation frequency with the decay
  $B^{0}_{s}\rightarrow D^{-}_{s}\pi^{+}$}'', {\em New J.Phys.} {\bfseries 15}
  (2013) 053021,
 \href{http://xxx.lanl.gov/abs/1304.4741}{{\ttfamily arXiv:1304.4741}}.

\bibitem[Abbott et~al.(1980)Abbott, Sikivie, and Wise]{Abbott:1979dt}
L.~Abbott, P.~Sikivie, and M.~B. Wise, ``{Constraints on charged-Higgs-boson
  couplings}'', {\em Phys.Rev.} {\bfseries D21} (1980)
1393.

\bibitem[Branco et~al.(1985)Branco, Buras, and G\'{e}rard]{Branco:1985pf}
G.~Branco, A.~Buras, and J.-M. G\'{e}rard, ``{\textit{CP} VIOLATION IN MODELS
  WITH TWO- AND THREE-SCALAR DOUBLETS}'', {\em Nucl.Phys.} {\bfseries B259}
  (1985)
306.

\bibitem[Geng and Ng(1988)]{Geng:1988bq}
C.~Geng and J.~N. Ng, ``{Charged-Higgs-boson effect in $B_d^0$-$\oline{B}_d^0$
  mixing, $K\to \pi \nu \oline{\nu}$ decay, and rare decays of $B$ mesons}'',
  {\em Phys.Rev.} {\bfseries D38} (1988)
2857.

\bibitem[Buras et~al.(1990)Buras, Krawczyk, Lautenbacher, and
  Salazar]{Buras:1989ui}
A.~J. Buras, P.~Krawczyk, M.~E. Lautenbacher, and C.~Salazar,
  ``{$B^0$--$\oline{B}^0$ MIXING, \textit{CP} VIOLATION, $K^+ \to \pi^+ \nu
  \oline{\nu}$ AND $B \to K \gamma X$ IN A TWO-HIGGS-DOUBLET MODEL}'', {\em
  Nucl.Phys.} {\bfseries B337} (1990)
284--312.

\bibitem[Deschamps et~al.(2010)Deschamps, Descotes-Genon, Monteil, Niess,
  T'Jampens, et~al.]{Deschamps:2009rh}
O.~Deschamps, S.~Descotes-Genon, S.~Monteil, V.~Niess, S.~T'Jampens, {\em
  et~al.}, ``{The Two Higgs Doublet Model of Type II facing flavour physics
  data}'', {\em Phys.Rev.} {\bfseries D82} (2010) 073012,
 \href{http://xxx.lanl.gov/abs/0907.5135}{{\ttfamily arXiv:0907.5135}}.

\bibitem[Hermann et~al.(2012)Hermann, Misiak, and Steinhauser]{Hermann:2012fc}
T.~Hermann, M.~Misiak, and M.~Steinhauser, ``{$\oline{B}\to X_s \gamma$ in the
  Two Higgs Doublet Model up to Next-to-Next-to-Leading Order in QCD}'', {\em
  JHEP} {\bfseries 1211} (2012) 036,
 \href{http://xxx.lanl.gov/abs/1208.2788}{{\ttfamily arXiv:1208.2788}}.

\bibitem[Misiak and Steinhauser(2007)]{Misiak:2006ab}
M.~Misiak and M.~Steinhauser, ``{NNLO QCD Corrections to the $\oline{B}\to X_s
  \gamma$ Matrix Elements Using Interpolation in $m_c$}'', {\em Nucl.Phys.}
  {\bfseries B764} (2007) 62--82,
 \href{http://xxx.lanl.gov/abs/hep-ph/0609241}{{\ttfamily
  arXiv:hep-ph/0609241}}.

\bibitem[Ferreira et~al.(2012)Ferreira, Santos, Sher, and
  Silva]{Ferreira:2011aa}
P.~Ferreira, R.~Santos, M.~Sher, and J.~P. Silva, ``{Implications of the LHC
  two-photon signal for two-Higgs-doublet models}'', {\em Phys.Rev.} {\bfseries
  D85} (2012) 077703,
 \href{http://xxx.lanl.gov/abs/1112.3277}{{\ttfamily arXiv:1112.3277}}.

\bibitem[Basso et~al.(2012)Basso, Lipniacka, Mahmoudi, Moretti, Osland,
  et~al.]{Basso:2012st}
L.~Basso, A.~Lipniacka, F.~Mahmoudi, S.~Moretti, P.~Osland, {\em et~al.},
  ``{Probing the charged Higgs boson at the LHC in the CP-violating type-II
  2HDM}'', {\em JHEP} {\bfseries 1211} (2012) 011,
 \href{http://xxx.lanl.gov/abs/1205.6569}{{\ttfamily arXiv:1205.6569}}.

\bibitem[Cheon and Kang(2012)]{Cheon:2012rh}
H.~Cheon and S.~K. Kang, ``{Constraining parameter space in type-II two-Higgs
  doublet model in light of a 125 GeV Higgs boson}'',
 \href{http://xxx.lanl.gov/abs/1207.1083}{{\ttfamily arXiv:1207.1083}}.

\bibitem[Drozd et~al.(2013)Drozd, Grzadkowski, Gunion, and Jiang]{Drozd:2012vf}
A.~Drozd, B.~Grzadkowski, J.~F. Gunion, and Y.~Jiang, ``{Two-Higgs-Doublet
  Models and Enhanced Rates for a 125 GeV Higgs}'', {\em JHEP} {\bfseries 1305}
  (2013) 072,
 \href{http://xxx.lanl.gov/abs/1211.3580}{{\ttfamily arXiv:1211.3580}}.

\bibitem[Chen and Dawson(2013)]{Chen:2013kt}
C.-Y. Chen and S.~Dawson, ``{Exploring Two Higgs Doublet Models Through Higgs
  Production}'', {\em Phys. Rev. D} {\bfseries 87} (2013) 055016,
 \href{http://xxx.lanl.gov/abs/1301.0309}{{\ttfamily arXiv:1301.0309}}.

\bibitem[Celis et~al.(2013)Celis, Ilisie, and Pich]{Celis:2013rcs}
A.~Celis, V.~Ilisie, and A.~Pich, ``{LHC constraints on two-Higgs doublet
  models}'', {\em JHEP} {\bfseries 1307} (2013) 053,
 \href{http://xxx.lanl.gov/abs/1302.4022}{{\ttfamily arXiv:1302.4022}}.

\bibitem[Giardino et~al.(2013)Giardino, Kannike, Masina, Raidal, and
  Strumia]{Giardino:2013bma}
P.~P. Giardino, K.~Kannike, I.~Masina, M.~Raidal, and A.~Strumia, ``{The
  universal Higgs fit}'',
 \href{http://xxx.lanl.gov/abs/1303.3570}{{\ttfamily arXiv:1303.3570}}.

\bibitem[Grinstein and Uttayarat(2013)]{Grinstein:2013npa}
B.~Grinstein and P.~Uttayarat, ``{Carving Out Parameter Space in Type-II Two
  Higgs Doublets Model}'', {\em JHEP} {\bfseries 1306} (2013) 094,
 \href{http://xxx.lanl.gov/abs/1304.0028}{{\ttfamily arXiv:1304.0028}}.

\bibitem[Barroso et~al.(2013)Barroso, Ferreira, Santos, Sher, and
  Silva]{Barroso:2013zxa}
A.~Barroso, P.~Ferreira, R.~Santos, M.~Sher, and J.~P. Silva, ``{2HDM at the
  LHC - the story so far}'',
 \href{http://xxx.lanl.gov/abs/1304.5225}{{\ttfamily arXiv:1304.5225}}.

\bibitem[Coleppa et~al.(2013)Coleppa, Kling, and Su]{Coleppa:2013dya}
B.~Coleppa, F.~Kling, and S.~Su, ``{Constraining Type II 2HDM in Light of LHC
  Higgs Searches}'',
 \href{http://xxx.lanl.gov/abs/1305.0002}{{\ttfamily arXiv:1305.0002}}.

\bibitem[Burdman et~al.(2012)Burdman, Haluch, and Matheus]{Burdman:2011ki}
G.~Burdman, C.~E. Haluch, and R.~D. Matheus, ``{Is the LHC observing the
  pseudoscalar state of a two-Higgs-doublet model?}'', {\em Phys.Rev.}
  {\bfseries D85} (2012) 095016,
 \href{http://xxx.lanl.gov/abs/1112.3961}{{\ttfamily arXiv:1112.3961}}.

\bibitem[Bar-Shalom et~al.(2011)Bar-Shalom, Nandi, and Soni]{BarShalom:2011zj}
S.~Bar-Shalom, S.~Nandi, and A.~Soni, ``{Two Higgs doublets with 4th generation
  fermions - models for TeV-scale compositeness}'', {\em Phys.Rev.} {\bfseries
  D84} (2011) 053009,
 \href{http://xxx.lanl.gov/abs/1105.6095}{{\ttfamily arXiv:1105.6095}}.

\bibitem[Chen and He(2012)]{Chen:2012wz}
N.~Chen and H.-J. He, ``{LHC Signatures of Two-Higgs-Doublets with Fourth
  Family}'', {\em JHEP} {\bfseries 1204} (2012) 062,
 \href{http://xxx.lanl.gov/abs/1202.3072}{{\ttfamily arXiv:1202.3072}}.

\bibitem[Hardy and Towner(2009)]{Hardy:2008gy}
J.~Hardy and I.~Towner, ``{Superallowed $0^+\to 0^+$ nuclear $\beta$ decays: A
  new survey with precision tests of the conserved vector current hypothesis
  and the standard model}'', {\em Phys.Rev.} {\bfseries C79} (2009) 055502,
 \href{http://xxx.lanl.gov/abs/0812.1202}{{\ttfamily arXiv:0812.1202}}.

\bibitem[{Private communication with the CKMfitter
  group}(????)]{CKMfitter:2012mo}
{Private communication with the CKMfitter group}.

\bibitem[Aaltonen et~al.(2012)]{Aaltonen:2012ra}
{\bfseries CDF Collaboration, D0 Collaboration}, T.~Aaltonen {\em et~al.},
  ``{Combination of the top-quark mass measurements from the Tevatron
  collider}'', {\em Phys.Rev.} {\bfseries D86} (2012) 092003,
 \href{http://xxx.lanl.gov/abs/1207.1069}{{\ttfamily arXiv:1207.1069}}.

\bibitem[Gro(2012)]{Group:2012gb}
{\bfseries {Tevatron Electroweak Working Group, CDF Collaboration, D0
  Collaboration}}, ``{2012 Update of the Combination of CDF and D0 Results for
  the Mass of the $W$ Boson}'',
 \href{http://xxx.lanl.gov/abs/1204.0042}{{\ttfamily arXiv:1204.0042}}.

\bibitem[Awramik et~al.(2004)Awramik, Czakon, Freitas, and
  Weiglein]{Awramik:2003rn}
M.~Awramik, M.~Czakon, A.~Freitas, and G.~Weiglein, ``{Precise prediction for
  the \textit{W}-boson mass in the standard model}'', {\em Phys.Rev.}
  {\bfseries D69} (2004) 053006,
 \href{http://xxx.lanl.gov/abs/hep-ph/0311148}{{\ttfamily
  arXiv:hep-ph/0311148}}.

\bibitem[848(2010)848923]{TEW:2010aj}
{\bfseries Tevatron Electroweak Working Group}, ``{Combination of CDF and D0
  Results on the Width of the $W$ boson}'',
 \href{http://xxx.lanl.gov/abs/1003.2826}{{\ttfamily arXiv:1003.2826}}.

\bibitem[CDF(2013)]{CDF:2013jga}
{\bfseries Tevatron Electroweak Working Group, D0 Collaborations}, CDF,
  ``{Combination of CDF and D\O{} results on the mass of the top quark using up
  to 8.7 fb$^{-1}$ at the Tevatron}'',
 \href{http://xxx.lanl.gov/abs/1305.3929}{{\ttfamily arXiv:1305.3929}}.

\bibitem[ATL(23/6/2013)]{ATLAS-CONF-2013-012}
{\bfseries ATLAS Collaboration}, ``{ATLAS Note CONF-2013-012}''.
  \url{http://cds.cern.ch/record/1523698}, 23/6/2013.

\bibitem[CMS(23/6/2013)]{CMS-PAS-HIG-13-001}
{\bfseries CMS Collaboration}, ``{CMS Physics Analysis Summary HIG-13-001}''.
  \url{http://cds.cern.ch/record/1530524}, 23/6/2013.

\bibitem[ATL(23/6/2013)]{ATLAS-CONF-2013-013}
{\bfseries ATLAS Collaboration}, ``{ATLAS Note CONF-2013-013}''.
  \url{http://cds.cern.ch/record/1523699}, 23/6/2013.

\bibitem[Aad et~al.(2012)]{Aad:2012an}
{\bfseries ATLAS Collaboration}, G.~Aad {\em et~al.}, ``{Combined search for
  the Standard Model Higgs boson in $pp$ collisions at $\sqrt{s} = 7$\:TeV with
  the ATLAS detector}'', {\em Phys.Rev.} {\bfseries D86} (2012) 032003,
 \href{http://xxx.lanl.gov/abs/1207.0319}{{\ttfamily arXiv:1207.0319}}.

\bibitem[ATL(23/6/2013{\natexlab{a}})]{ATLAS-CONF-2012-091}
{\bfseries ATLAS Collaboration}, ``{ATLAS Note CONF-2012-091, Fig. 14}''.
  \url{http://cds.cern.ch/record/1460410}, 23/6/2013{\natexlab{a}}.

\bibitem[ATL(23/6/2013{\natexlab{b}})]{ATLAS-CONF-2013-030}
{\bfseries ATLAS Collaboration}, ``{ATLAS Note CONF-2013-030}''.
  \url{http://cds.cern.ch/record/1527126}, 23/6/2013{\natexlab{b}}.

\bibitem[ATL(23/6/2013{\natexlab{c}})]{ATLAS-CONF-2012-161}
{\bfseries ATLAS Collaboration}, ``{ATLAS Note CONF-2012-161}''.
  \url{http://cds.cern.ch/record/1493625}, 23/6/2013{\natexlab{c}}.

\bibitem[ATL(23/6/2013{\natexlab{d}})]{ATLAS-CONF-2012-160}
{\bfseries ATLAS Collaboration}, ``{ATLAS Note CONF-2012-160}''.
  \url{http://cds.cern.ch/record/1493624}, 23/6/2013{\natexlab{d}}.

\bibitem[CMS(23/6/2013{\natexlab{a}})]{CMS-PAS-HIG-12-020}
{\bfseries CMS Collaboration}, ``{CMS Physics Analysis Summary HIG-12-020}''.
  \url{http://cds.cern.ch/record/1460438}, 23/6/2013{\natexlab{a}}.

\bibitem[CMS(23/6/2013{\natexlab{b}})]{CMS-PAS-HIG-13-003}
{\bfseries CMS Collaboration}, ``{CMS Physics Analysis Summary HIG-13-003}''.
  \url{http://cds.cern.ch/record/1523673}, 23/6/2013{\natexlab{b}}.

\bibitem[CMS(23/6/2013{\natexlab{c}})]{CMS-PAS-HIG-12-045}
{\bfseries CMS Collaboration}, ``{CMS Physics Analysis Summary HIG-12-045}''.
  \url{http://cds.cern.ch/record/1494149}, 23/6/2013{\natexlab{c}}.

\bibitem[CMS(23/6/2013{\natexlab{d}})]{CMS-PAS-HIG-13-004}
{\bfseries CMS Collaboration}, ``{CMS Physics Analysis Summary HIG-13-004}''.
  \url{http://cds.cern.ch/record/1528271}, 23/6/2013{\natexlab{d}}.

\bibitem[CMS(23/6/2013{\natexlab{e}})]{CMS-PAS-HIG-12-043}
{\bfseries CMS Collaboration}, ``{CMS Physics Analysis Summary HIG-12-043}''.
  \url{http://cds.cern.ch/record/1493615}, 23/6/2013{\natexlab{e}}.

\bibitem[ATL(23/6/2013)]{ATLAS-CONF-2012-091a}
{\bfseries ATLAS Collaboration}, ``{ATLAS Note CONF-2012-091, Table 6}''.
  \url{http://cds.cern.ch/record/1460410}, 23/6/2013.

\bibitem[Gro(2012)]{Group:2012zca}
{\bfseries Tevatron New Physics Higgs Working Group, CDF Collaboration, D0
  Collaboration}, ``{Updated Combination of CDF and D0 Searches for Standard
  Model Higgs Boson Production with up to 10.0 fb$^{-1}$ of Data}'',
 \href{http://xxx.lanl.gov/abs/1207.0449}{{\ttfamily arXiv:1207.0449}}.

\bibitem[CMS(23/6/2013{\natexlab{a}})]{CMS-PAS-HIG-12-015}
{\bfseries CMS Collaboration}, ``{CMS Physics Analysis Summary HIG-12-015}''.
  \url{http://cds.cern.ch/record/1460419}, 23/6/2013{\natexlab{a}}.

\bibitem[CMS(23/6/2013{\natexlab{b}})]{CMS-PAS-HIG-13-014}
{\bfseries CMS Collaboration}, ``{CMS Physics Analysis Summary HIG-13-014}''.
  \url{http://cds.cern.ch/record/1546776}, 23/6/2013{\natexlab{b}}.

\bibitem[McNeile et~al.(2012)McNeile, Davies, Follana, Hornbostel, and
  Lepage]{McNeile:2011ng}
C.~McNeile, C.~Davies, E.~Follana, K.~Hornbostel, and G.~Lepage,
  ``{High-Precision $f_{B_s}$ and HQET from Relativistic Lattice QCD}'', {\em
  Phys.Rev.} {\bfseries D85} (2012) 031503,
 \href{http://xxx.lanl.gov/abs/1110.4510}{{\ttfamily arXiv:1110.4510}}.

\bibitem[Buras et~al.(1990)Buras, Jamin, and Weisz]{Buras:1990fn}
A.~J. Buras, M.~Jamin, and P.~H. Weisz, ``{LEADING AND NEXT-TO-LEADING QCD
  CORRECTIONS TO $\varepsilon$-PARAMETER AND $B^0$--$\oline{B}^0$ MIXING IN THE
  PRESENCE OF A HEAVY TOP QUARK}'', {\em Nucl.Phys.} {\bfseries B347} (1990)
491--536.

\bibitem[G{\'a}miz et~al.(2009)G{\'a}miz, Davies, Lepage, Shigemitsu, and
  Wingate]{Gamiz:2009ku}
{\bfseries HPQCD Collaboration}, E.~G{\'a}miz, C.~T. Davies, G.~P. Lepage,
  J.~Shigemitsu, and M.~Wingate, ``{Neutral $B$ Meson Mixing in Unquenched
  Lattice QCD}'', {\em Phys.Rev.} {\bfseries D80} (2009) 014503,
 \href{http://xxx.lanl.gov/abs/0902.1815}{{\ttfamily arXiv:0902.1815}}.

\bibitem[{Heavy Flavor Averaging Group}(23/6/2013)]{HFAG:2012bs}
{Heavy Flavor Averaging Group}.
  \url{http://www.slac.stanford.edu/xorg/hfag/rare/}, 23/6/2013.

\end{thebibliography}\endgroup
